\documentstyle[preprint,aps,eqsecnum]{revtex}
\tightenlines
\begin{document}
\draft

\newcommand{\beq}{\begin{equation}}
\newcommand{\eeq}{\end{equation}}
\newcommand{\bea}{\begin{eqnarray}}
\newcommand{\eea}{\end{eqnarray}}
\newcommand{\cir}{{\buildrel \circ \over =}}

\title{Dirac's Observables for the Rest-Frame Instant Form of
Tetrad Gravity in a Completely Fixed 3-Orthogonal Gauge.}

\author{Roberto De Pietri}

\address{Dipartimento di Fisica, Campus Universitario\\
               Universita' di Parma\\
               Viale delle Scienze\\
               43100 Parma, Italy\\
               E-mail: DEPIETRI@PR.INFN.IT}

\author{and}

\author{Luca Lusanna}

\address{
Sezione INFN di Firenze\\
L.go E.Fermi 2 (Arcetri)\\
50125 Firenze, Italy\\
E-mail LUSANNA@FI.INFN.IT}

\author{and}

\author{Luca Martucci}

\address{Dipartimento di Fisica\\
Universita' di Milano I\\
 via G.Celoria  16\\
 20133 Milano\\
 E-mail MARTUCCI@MI.INFN.IT}

\author{and}

\author{Stefano Russo}

\address
{Condominio dei Pioppi 16\\
6916 Grancia (Lugano)\\
Svizzera}

\maketitle
\begin{abstract}

We define the {\it rest-frame instant form} of tetrad gravity
restricted to Christodoulou-Klainermann spacetimes. After a study
of the Hamiltonian group of gauge transformations generated by the
14 first class constraints of the theory, we define and solve the
multitemporal equations associated with the rotation and space
diffeomorphism constraints, finding how the cotriads and their
momenta depend on the corresponding gauge variables. This allows
to find a quasi-Shanmugadhasan canonical transformation to the
class of 3-orthogonal gauges and to find the Dirac observables for
superspace in these gauges. The construction of the explicit form
of the transformation and of the solution of the rotation and
supermomentum constraints is reduced to solve a system of elliptic
linear and quasi-linear partial differential equations. We then
show that the superhamiltonian constraint becomes the Lichnerowicz
equation for the conformal factor of the 3-metric and that the
last gauge variable is the momentum conjugated to the conformal
factor. The gauge transformations generated by the
superhamiltonian constraint perform the transitions among the
allowed foliations of spacetime, so that the theory is independent
from its 3+1 splittings. In the special 3-orthogonal gauge defined
by the vanishing of the conformal factor momentum we determine the
final Dirac observables for the gravitational field even if we are
not able to solve the Lichnerowicz equation. The final Hamiltonian
is the weak ADM energy restricted to this completely fixed gauge.

\vskip 1truecm
\noindent \today
\vskip 1truecm

\end{abstract}

\newpage

\vfill\eject

\section
{Introduction}

In a previous paper\cite{ru11}\footnote{See the
papers\cite{russo1,russo2,russo3}, quoted as I, II, III in what
follows, for a preliminary presentation of many results.} a new
parametrization of arbitrary cotetrads has been introduced to
achieve a simplification of the 14 first class constraints of
tetrad gravity starting from the ADM action for metric
gravity\cite{adm} with the 4-metric expressed in terms of
cotetrads.

As said in Ref.\cite{ru11}, this has been done in the special
class of non-compact spacetimes $M^4$, which:

i) are {\it globally hyperbolic} 4-manifolds \footnote{$\Sigma$ is an
abstract model of spacelike Cauchy surface. These spacetimes admit
regular foliations with orientable, complete, non-intersecting
spacelike 3-manifolds: the leaves of the foliation are the embeddings
$i_{\tau}:\Sigma \rightarrow \Sigma_{\tau} \subset M^4$, $\vec \sigma
\mapsto z^{\mu}(\tau ,\vec \sigma )$, where $\vec \sigma =\{ \sigma^r
\}$, r=1,2,3, are local coordinates in a chart of the
$C^{\infty}$-atlas of the abstract 3-manifold $\Sigma$ and $\tau :M^4
\rightarrow R$, $z^{\mu} \mapsto \tau (z^{\mu})$, is a global timelike
future-oriented function labelling the leaves (surfaces of
simultaneity). In this way, one obtains 3+1 splittings of $M^4$ and
the possibility of a Hamiltonian formulation.} $M^4 \approx R\times
\Sigma$, so that a Hamiltonian formulation is possible;

ii) are {\it asymptotically flat at spatial infinity}, so that
Poincar\'e charges may be defined\cite{restmg};

iii) admit a {\it spin structure}, i.e. are parallelizable and have a
trivial orthonormal frame bundle $F(M^4)=M^4\times SO(3,1)$ and
coframe bundle $L(M^4)=M^4\times SO(3,1)$, so that tetrads and
cotetrads are globally defined\footnote{The parallelizable spacelike
hypersurfaces $\Sigma_{\tau}$ of simultaneity have trivial orthonormal
frame bundle $F\Sigma_{\tau}=\Sigma_{\tau}\times SO(3)$ and coframe
bundle $L\Sigma_{\tau}=\Sigma_{\tau}\times SO(3)$, so that triads and
cotriads are globally defined.};

iv) have the Cauchy surfaces $\Sigma_{\tau}$ that are {\it
topologically trivial}, {\it geodesically complete} Riemannian
3-manifolds $(\Sigma_{\tau}, {}^3g)$ and {\it
diffeomorphic\footnote{Therefore the 3-manifolds $\Sigma_{\tau}$
admit global coordinate systems.} to $R^3$}, $\Sigma_{\tau}\approx
R^3$;

v) have the non-compact Riemannian 3-manifolds $(\Sigma_{\tau},
{}^3g)$ {\it not admitting isometries}\footnote{This requires that
triads, cotriads and 3-spin connections belong to suited weighted
Sobolev spaces to avoid Gribov ambiguities.}.

This new formulation of tetrad gravity has been introduced with
the aim to study the Hamiltonian group of gauge transformations
and to perform the canonical reduction of the theory to a
completely fixed gauge with the identification of the physical
degrees of freedom (Dirac observables\cite{dirac,ber,ber1}) of the
gravitational field. This would conclude the research program
aiming to express the four interactions only in terms of canonical
cases of Dirac's observables\footnote{See Ref.\cite{india} for
such a canonical reduction of the electromagnetic, weak and strong
interactions in Minkowski spacetime.}. The program is based on the
Shanmugadhasan canonical transformations \cite{sha}: if a system
has first class constraints at the Hamiltonian level \footnote{So
that its dynamics is restricted to a presymplectic submanifold of
phase space.}, then, at least locally, one can find a canonical
basis with as many new momenta as first class constraints
(Abelianization of first class constraints), with their conjugate
canonical variables as Abelianized gauge variables and with the
remaining pairs of canonical variables as pairs of canonically
conjugate Dirac's observables \footnote{Canonical basis of
physical variables adapted to the chosen Abelianization; they give
a trivialization of the BRST construction of observables.}.
Putting equal to zero the Abelianized gauge variables  defines a
local gauge of the model. If a system with constraints admits one
(or more) {\it global} Shanmugadhasan canonical transformations,
one obtains one (or more) privileged {\it global gauges} in which
the physical Dirac observables are globally defined and globally
separated from the gauge degrees of freedom \footnote{For systems
with a compact configuration space this is in general
impossible.}. These privileged gauges (when they exist) can be
called {\it generalized Coulomb or radiation gauges}. Second class
constraints\cite{chai}, when present, are also taken into account
by the Shanmugadhasan canonical transformation\cite{sha}.

In flat spacetime the problem of how to covariantize this kind  of
canonical reduction is solved by using Dirac reformulation (see
the book in Ref.\cite{dirac}) of classical field theory on
spacelike hypersurfaces foliating \footnote{The foliation is
defined by an embedding $R\times \Sigma \rightarrow M^4$, $(\tau
,\vec \sigma ) \mapsto z^{(\mu )}(\tau ,\vec \sigma )$ [$(\mu )$
are flat Minkowski indices], with $\Sigma$ an abstract 3-surface
diffeomorphic to $R^3$: this is the classical basis of
Tomonaga-Schwinger quantum field theory.} Minkowski spacetime
$M^4$. In this way one gets parametrized Minkowski field theory
with a covariant 3+1 splitting of flat spacetime and already in a
form suited to the transition to general relativity in its ADM
canonical formulation \footnote{See also Ref.\cite{kuchar} , where
a theoretical study of this problem is done in curved
spacetimes.}. The price is that one has to add as new
configuration variables  the embeddings $z^{(\mu )}(\tau ,\vec
\sigma )$ of the spacelike hypersurface $\Sigma_{\tau}$
\footnote{The only ones carrying Lorentz indices; the scalar
parameter $\tau$ labels the leaves of the foliation and $\vec
\sigma$ are curvilinear coordinates on $\Sigma_{\tau}$.} and then
define the fields on $\Sigma_{\tau}$ so that they know  the
hypersurface $\Sigma_{\tau}$ of $\tau$-simultaneity \footnote{For
a Klein-Gordon field $\phi (x)$, this new field is $\tilde \phi
(\tau ,\vec \sigma )=\phi (z(\tau ,\vec \sigma ))$: it contains
the non-local information about the embedding.}. Then one rewrites
the Lagrangian of the given isolated system in the form required
by the coupling to an external gravitational field, makes the
previous 3+1 splitting of Minkowski spacetime and interpretes all
the fields of the system as the new fields on $\Sigma_{\tau}$
(they are Lorentz scalars, having only surface indices). Instead
of considering the 4-metric as describing a gravitational field
\footnote{Therefore as an independent field as it is done in
metric gravity, where one adds the Hilbert action to the action
for the matter fields.}, here one replaces the 4-metric with the
the induced metric $g_{ AB}[z] =z^{(\mu )}_{A}\eta_{(\mu )(\nu
)}z^{(\nu )}_{B}$ on $\Sigma_{\tau}$ \footnote{A functional of
$z^{(\mu )}$; here we use the notation $\sigma^{A}=(\tau
,\sigma^{r})$; $(\mu )$ is a flat Minkowski index; $z^{(\mu
)}_{A}=\partial z^{(\mu )}/\partial \sigma^{A}$ are flat cotetrad
fields on Minkowski spacetime with the $z^{(\mu )}_r$'s tangent to
$\Sigma_{\tau}$.} and considers the embedding coordinates $z^{(\mu
)}(\tau ,\vec \sigma )$ as independent fields \footnote{This is
not possible in metric gravity, because in curved spacetimes,
given the embeddings $z^{\mu}(\tau ,\vec \sigma )$, $\Sigma
\rightarrow \Sigma_{\tau}$, the $z^{\mu}_{A}\not=
\partial z^{\mu}/\partial \sigma^{A}$'s are not cotetrad fields.
In tetrad gravity, given a resolution of the 4-metric in
non-holonomic cotetrads (the configurational degrees of freedom of
tetrad gravity), ${}^4g_{\mu\nu}={}^4E^{(\alpha )}_{\mu}\,
{}^4\eta_{(\alpha )(\beta )}\, {}^4E^{(\beta )}_{\nu}$, and the
$\Sigma_{\tau}$-adapted 4-metric ${}^4g_{AB}(\tau ,\vec \sigma
)=z^{\mu}_A(\tau ,\vec \sigma )\, {}^4g_{\mu\nu}(z(\tau ,\vec
\sigma ))\, z^{\nu}_B(\tau ,\vec \sigma )$, the
$\Sigma_{\tau}$-adapted cotetrads replacing the $z^{(\mu )}_A$'s
of the flat case are ${}^4F^{(\alpha )}_A(\tau ,\vec \sigma
)=z^{\mu}_A(\tau ,\vec \sigma )\, {}^4E^{(\alpha )}_{\mu}(z(\tau
,\vec \sigma ))$: they depend simultaneously on the embedding and
on the non-holonomic cotetrads and can be found only a
posteriori.}. From this Lagrangian, besides a Lorentz-scalar form
of the constraints of the given system, we get four extra primary
first class constraints ${\cal H}_{\mu}(\tau ,\vec \sigma )
\approx 0$ implying the independence of the description from the
choice of the foliation with spacelike hypersufaces. Therefore the
embedding variables $z^{(\mu )}(\tau ,\vec \sigma )$ are the {\it
gauge} variables associated with this kind of general covariance.
If we interpret the unit  normal $l^{\mu}(\tau ,\vec \sigma )$ at
$\Sigma_{\tau}$ as the unit 4-velocity of a timelike observer at
$(\tau ,\vec \sigma )$, the foliation with leaves $\Sigma_{\tau}$
identifies a congruence of timelike surface-forming accelerated
observers. In special relativity, it is convenient to restrict
ourselves to arbitrary spacelike hyperplanes $z^{(\mu )} (\tau
,\vec \sigma )=x^{(\mu )}_s(\tau )+b^{(\mu )}_{r}(\tau )
\sigma^{r}$, which are associated with a congruence of timelike
inertial observers. Since the hyperplanes are described by only 10
variables \footnote{An origin $x^{(\mu )}_s(\tau )$ and, on it,
three orthogonal spacelike unit vectors $b^{(\mu )}_r(\tau )$
generating the fixed constant timelike unit normal $l^{(\mu
)}=b_{\tau}^{(\mu )}=\epsilon ^{(\mu )}{}_{(\nu )(\rho )(\sigma
)}b^{(\nu )}_{\check 1}(\tau ) b^{(\rho )}_{\check 2}(\tau
)b^{(\sigma )}_{\check 3}(\tau )$ to the hyperplane.}, we remain
only with {\it ten} first class constraints determining the 10
variables conjugate to the hyperplane \footnote{They are a
4-momentum $p^{(\mu )}_s$ and the six independent degrees of
freedom hidden in a spin tensor $S^{(\mu )(\nu )}_s$.} in terms of
the variables of the system.

If we now consider only the set of configurations of the isolated system with
 timelike \footnote{$\epsilon p^2_s > 0$; $\epsilon =\pm 1$ according to the chosen
 convention for the Lorentz signature of the metric $\eta^{(\mu )(\nu )}
 =\epsilon (+---)$.} 4-momenta, we can
restrict the description to the so-called {\it Wigner hyperplanes}
orthogonal to $p^{(\mu )}_s$ itself. To get this result, we must
boost at rest all the variables with Lorentz indices by using the
standard Wigner boost $L^{(\mu )}{} _{(\nu )}(p_s,{\buildrel \circ
\over p}_s)$ for timelike Poincar\'e orbits, and then add the
gauge-fixings $b^{(\mu )}_{r}(\tau )-L^{(\mu )}{}_{r}(p_s,
{\buildrel \circ \over p}_s)\approx 0$. Since these gauge-fixings
depend on $p^{(\mu )}_s$, the final canonical variables, apart
$p^{(\mu )}_s$ itself, are of 3 types: i) there is a non-covariant
canonical {\it external} center-of-mass variable ${\tilde x}^{(\mu
)}_s (\tau )$ \footnote{It is only covariant under the little
group of timelike Poincar\'e orbits like the Newton-Wigner
position operator.}; ii) all the 3-vector variables become Wigner
spin 1 3-vectors\footnote{Boosts in $M^4$ induce Wigner rotations
on them.}; iii) all the other variables are Lorentz scalars. Only
{\it four} 1st class constraints are left: one of them identifies
the invariant mass of the isolated system, to be used as
Hamiltonian, while the other three are the rest-frame conditions
implying the vanishing of the {\it internal} (i.e. inside the
Wigner hyperplane) total 3-momentum.

We obtain in this way a new kind of instant form of the dynamics (see
Ref.\cite{dira2}), the  {\it Wigner-covariant 1-time rest-frame
instant form}\cite{lus1,india,crater} with a universal breaking of
Lorentz covariance independent from the isolated system under
investigation. It is the special relativistic generalization of the
non-relativistic separation of the center of mass from the relative
motions [$H={{ {\vec P}^2}\over {2M}}+H_{rel}$].

 As shown in Refs.\cite{lus1,lusa}, the rest-frame instant form
of dynamics automatically gives a physical ultraviolet cutoff in
the spirit of Dirac and Yukawa for all the rotating configurations
of an isolated system: it is the {\it M$\o$ller radius}\cite{mol}
$\rho =\sqrt{-\epsilon W^2}/\epsilon P^2=|\vec S|/\sqrt{\epsilon
P^2}$ \footnote{$W^2=-\epsilon P^2{\vec S}^2$ is the
Pauli-Lubanski Casimir when $\epsilon P^2 > 0$.}, namely the
classical intrinsic radius of the worldtube, around the covariant
non-canonical Fokker-Pryce {\it 4-center of inertia} $Y^{(\mu
)}_s$, inside which the non-covariance of the {\it canonical
4-center of mass} ${\tilde x}_s^{(\mu )}$ is concentrated. At the
quantum level $\rho$ becomes the Compton wavelength of the
isolated system multiplied its spin eigenvalue $\sqrt{s(s+1)}$ ,
$\rho \mapsto \hat \rho = \sqrt{s(s+1)} \hbar /M=\sqrt{s(s+1)}
\lambda_M$ with $M=\sqrt{\epsilon P^2}$ the invariant mass and
$\lambda_M=\hbar /M$ its Compton wavelength. Therefore, the {\it
criticism} to classical relativistic physics, based on quantum
{\it pair production}, concerns the testing of distances where,
due to the Lorentz signature of spacetime, one has intrinsic
classical covariance problems: it is impossible to localize the
canonical 4-center of mass ${\tilde x}^{(\mu )}_s$ of the system
in a frame independent way. Let us remember \cite{lus1} that
$\rho$ is also a remnant in flat Minkowski spacetime of the {\it
energy conditions} of general relativity: since the M$\o$ller
non-canonical, non-covariant {\it 4-center of energy} $R_s^{(\mu
)}$ has its non-covariance localized inside the same worldtube
with radius $\rho$ (it was discovered in this way) \cite{mol}, it
turns out that for an extended relativistic system with the
material radius smaller than its intrinsic radius $\rho$ one has:
i) its peripheral rotation velocity can exceed the velocity of
light; ii) its classical energy density cannot be positive
definite everywhere in every frame.

Now, the real relevant point is that this ultraviolet cutoff
determined by $\rho$ exists also in Einstein's general relativity
(which is not power counting renormalizable) in the case of
asymptotically flat spacetimes, taking into account the Poincar\'e
Casimirs of its asymptotic ADM Poincar\'e charges \footnote{When
supertranslations are eliminated with suitable boundary
conditions; let us remark that Einstein and Wheeler use closed
universes because they don't want to introduce (non Machian)
boundary conditions, but in this way they loose Poincar\'e charges
and the possibility to make contact  with particle physics and to
define spin.} at spatial infinity.

Therefore in Ref.\cite{restmg},  after a review of ADM metric
gravity, of spacetimes asymptotically flat at spatial infinity, of
supertranslations and of ADM strong and weak asymptotic Poincar\'e
charges, the definition of the rest-frame instant form of metric
gravity was given. This is possible only when the requirement of
absence of supertranslations, needed for the existence of a well
defined and unique asymptotic Poincar\'e group ${\cal P}_{(\infty
)}$, is imposed. In this way the allowed foliations of the
spacetime $M^4$ have the leaves $\Sigma_{\tau}$ approaching
Minkowski hyperplanes at spatial infinity in an angle-independent
way, the allowed atlas of coordinates of $M^4$ must be compatible
with asymptotic Minkowski Cartesian coordinates and the allowed
diffeomorphisms of $M^4$ in $Diff\, M^4$ are restricted to tend to
$Diff_I\, M^4 \times {\cal P}_{(\infty )}$ at spatial infinity in
an angle-independent way \footnote{$Diff_I\, M^4$ are the
diffeomorphisms which reduce to the identity at spatial
infinity.}. The class of Christodoulou-Klainermann spacetimes
\cite{ckl} is selected: in them the strong ADM 3-momentum vanishes
identically so that the vanishing of the weak ADM 3-momentum is
equivalent to three first class constraints defining the {\it rest
frame} of the universe. Therefore, the {\it rest-frame instant
form of metric gravity} may be consistently defined. The {\it
Wigner hyperplanes} of Minkowski spacetime are replaced by the
so-called {\it Wigner-Sen-Witten hypersurfaces}, which
asymptotically tend in a direction-independent way to Minkowski
hyperplanes orthogonal to the weak ADM 4-momentum.  In presence of
matter the Wigner-Sen-Witten hypersurfaces tend to the Wigner
hyperplanes for the same matter in Minkowski spacetime when the
Newton constant vanishes: in this way we get the {\it
deparametrization} of general relativity.

It will be shown in this paper that in a similar way we can define the
{\it rest-frame instant form of tetrad gravity} restricted to
Christodoulou-Klainermann spacetimes.

Then we can study the component connected with the identity of the
Hamiltonian group of gauge transformations, whose generators are the
14 first class constraints of tetrad gravity given in Ref.\cite{ru11},
following the scheme developed for Yang-Mills theory \cite{lusa}.
Since seven constraints are already Abelianized, this study is
concentrated on rotations, space diffeomorphisms and gauge
transformations generated by the superhamiltonian constraint. The main
problem in general relativity is the lack of control on the group
manifold of diffeomorphisms groups. Also the interpretation of the
gauge transformations generated by the superhamiltonian constraint is
given following Ref.\cite{restmg}: they change the foliation, so that
the theory is independent from the 3+1 splittings of spacetime like it
happens in parametrized Minkowski theories.

The next step is the study of the gauge transformations generated
by the rotation and space diffeomorphism constraints. By solving
the associated multitemporal equations\cite{india,lusa} we can
find how the cotriads and their momenta depend upon the gauge
angles and the gauge parameters of pseudo-diffeomorphisms ({\it
passive} diffeomorphisms), which are the Abelianized gauge
variables associated with these six constraints. This allows to
Abelianize these six constraints and to find the
quasi-Shanmugadhasan canonical transformation (it is a point
canonical transformation) corresponding to the class of
3-orthogonal gauges: only the superhamiltonian constraint has not
been Abelianized at this stage. We are able to find a canonical
basis of Dirac observables with respect to these six constraints
and to write a system of {\it elliptic linear and quasi-linear}
partial differential equations, whose solution would give the
expression of the cotriad momenta in terms of the gauge variables
and of the Dirac observables in these gauges. To solve these
equations is equivalent to the solution of the supermomentum
constraints of metric gravity, namely to find York's
gravitomagnetic potential and how the extrinsic curvature of the
3-surfaces depends on it in the 3-orthogonal gauges. We write the
equations for the determination of the shift functions in the
3-orthogonal gauges.

Then with a canonical transformation the previous canonical basis of
Dirac observables (it is a canonical basis for superspace) is put in a
form convenient for starting the search of the final gauge variable
conjugate to the superhamiltonian constraint. As already said in
Ref.\cite{restmg}, from the study of the Gauss law associated with the
superhamiltonian constraint (the ADM strong and weak energies are the
charges) it turns out that this gauge variable is the momentum
conjugate to the conformal factor of the 3-metric. Therefore, the
superhamiltonian constraint is an equation for the conformal factor of
the 3-metric, the Lichnerowicz equation.

A special 3-orthogonal gauge, replacing the maximal slicing
condition in our approach, is defined by putting equal to zero the
momentum conjugate to the conformal factor of the 3-metric as a
gauge fixing constraint. Even if we do not know the solution of
the Lichnerowicz equation, in this gauge we can identify a
canonical basis of the final Dirac observables, namely two pairs
of conjugate variables describing the gravitational field in this
special completely fixed 3-orthogonal gauge. This is the first
time that the canonical reduction of gravity can be pushed till
the end: the weak ADM energy restricted to this gauge is the
Hamiltonian for the Dirac observables. Finally we write the
equation for the lapse and shift functions associated with this
special gauge.

Then we study the Wigner-Sen-Witten hypersurfaces of the
rest-frame instant form of tetrad gravity and we write the
equations, whose solution would allow to find the embedding of
these hypersurfaces into the spacetime. We also show the existence
of {\it preferred dynamical asymptotic inertial observers} to be
identified with the {\it fixed stars}.

We refer to Ref\cite{restmg} for the discussion of the
interpretational problems with the observables in general relativity,
for the problem of time and for the quantization in a completely fixed
gauge, because the treatment of these topics is the same in metric and
tetrad gravity.

Finally some comments on inertial effets in Minkowski spacetime,
like the ones in non-inertial frames in Newtonian gravity, as
distinct from the gravitational field can be made by imposing the
vanishing of the Dirac observables of the gravitational field:
this defines the {\it void spacetimes} including Minkowski
spacetime in Cartesian coordinates.

In Section II, after a review of Hamitonian tetrad gravity and of the
rest-frame instant form of metric gravity, we define the rest-frame
instant form of tetrad gravity.

In Section III we study the Hamiltonian group of gauge transformations
of tetrad gravity, whose component connected to the identity is
generated by its 14 first class constraints.

In Section IV we define and solve the multitemporal equations
associated with rotations and space diffeomorphism constraints,
finding how the cotriads and their momenta depend on the corresponding
gauge variables.

In Section V we find the quasi-Shanmugadhasan canonical transformation
adapted to 3-orthogonal gauges.

In Section VI, after a further canonical transformation, we rewrite
the superhamiltonian constraint, restricted to 3-orthogonal gauges, as
the Lichnerowicz equation for the conformal factor of the 3-metric and
we define a special completely fixed 3-orthogonal gauge, whose final
Dirac observables for the gravitational field are then identified.

In Section VII we study the embedding of the Wigner-Sen-Witten
hypersurfaces in the given spacetime.

In Section VIII we define and study void spacetimes as those
spacetimes in which there is no gravitational field (meant as source
of tidal effects), but only inertial effects like in non-inertial
frames in Newton gravity.

In the Conclusions we make some comments about completely fixed gauges
and on the open problems and perspectives.

In Appendix A there is the Hamiltonian tetrad gravity expression of
relevant 3-tensor  in the special 3-orthogonal gauge.

In Appendix B there is the Hamiltonian tetrad gravity expression of
the ADM Poincar\'e charges in the special 3-orthogonal gauge.

\vfill\eject

\section{The Rest-Frame Instant Form of Tetrad Gravity.}

\subsection{Review of the New Parametrization of Tetrad Gravity and of its Constraints.}

Refs. \cite{naka,oneil,blee} are used for the background in
differential geometry. A spacetime is a time-oriented
pseudo-Riemannian (or Lorentzian) 4-manifold $(M^4,{}^4g)$ with
signature $\epsilon \, (+---)$ ($\epsilon =\pm 1$) and with a choice
of time orientation
\footnote{I.e. there exists a continuous, nowhere vanishing timelike
vector field which is used to separate the non-spacelike vectors at
each point of $M^4$ in either future- or past-directed vectors.}. In
Appendix A of Ref.\cite{ru11}we give a review of notions on
4-dimensional pseudo-Riemannian manifolds, tetrads on them and triads
on 3-manifolds, which unifies many results, scattered in the
literature, needed not only for a well posed formulation of tetrad
gravity but also for the further study of its canonical reduction.

As shown in Section II  and in Appendix A of Ref.\cite{ru11}, in the
family of $\Sigma_{\tau}$-adapted frames and coframes on $M^4$, we can
select special tetrads and cotetrads ${}^4_{(\Sigma )}{\check
E}_{(\alpha )}$ and ${}^4_{(\Sigma )}{\check
\theta}^{(\alpha )}$ also adapted to a given set of triads
${}^3e^r_{(a)}$ and cotriads ${}^3e^{(a)}_r={}^3e_{(a)r}$ on
$\Sigma_{\tau}$

\begin{eqnarray}
{}^4_{(\Sigma )}{\check E}^{\mu}_{(\alpha )}&=&\lbrace {}^4_{(\Sigma )}{\check
E}^{\mu}_{(o)}=l^{\mu}= {\hat b}^{\mu}_l={1\over N}(b^{\mu}
_{\tau}-N^rb^{\mu}_r);
\,\, {}^4_{(\Sigma )}{\check E}^{\mu}_{(a)}={}^3e^s_{(a)}
b^{\mu}_s \rbrace ,\nonumber \\
{}^4_{(\Sigma )}{\check E}_{\mu}^{(\alpha )}&=&\lbrace {}^4_{(\Sigma )}{\check
E}_{\mu}^{(o)}=\epsilon l_{\mu}= {\hat b}^l_{\mu}= N b^{\tau}
_{\mu};\,\, {}^4_{(\Sigma )}{\check E}_{\mu}^{(a)}={}^3e_s^{(a)}
{\hat b}^s_{\mu}\rbrace ,\nonumber \\
&&{}\nonumber \\
{}^4_{(\Sigma )}{\check E}^{\mu}_{(\alpha )}&& {}^4g_{\mu\nu}\,\,\,
{}^4_{(\Sigma )}{\check E}^{\nu}_{(\beta )} = {}^4\eta_{(\alpha )(\beta )},
\nonumber \\
 &&{}\nonumber \\
{}^4_{(\Sigma )}{\check {\tilde E}}^A_{(\alpha )}&=&{}^4_{(\Sigma )}{\check E}
^{\mu}_{(\alpha )}\, b^A_{\mu},\quad \Rightarrow {}^4_{(\Sigma )}{\check
{\tilde E}}^A_{(o)}=\epsilon l^A,\nonumber \\
 &&{}\nonumber \\
&&{}^4_{(\Sigma )}{\check {\tilde E}}^{\tau}_{(o)}={1\over N},\quad\quad
{}^4_{(\Sigma )}{\check {\tilde E}}^{\tau}_{(a)}=0,\nonumber \\
&&{}^4_{(\Sigma )}{\check {\tilde E}}^r_{(o)}=-{{N^r}\over N},\quad\quad
{}^4_{(\Sigma )}{\check {\tilde E}}^r_{(a)}={}^3e^r_{(a)};\nonumber \\
{}^4_{(\Sigma )}{\check {\tilde E}}_A^{(\alpha )}&=&{}^4_{(\Sigma )}{\check E}
_{\mu}^{(\alpha )}\, b^{\mu}_A,\quad \Rightarrow {}^4_{(\Sigma )}{\check
{\tilde E}}_A^{(o)} = l_A,\nonumber \\
 &&{}\nonumber \\
 &&{}^4_{(\Sigma )}{\check {\tilde E}}^{(o)}_{\tau}=N,\quad\quad
{}^4_{(\Sigma )}{\check {\tilde E}}^{(a)}_{\tau}=N^r\, {}^3e^{(a)}_r=N^{(a)},
\nonumber \\
 &&{}^4_{(\Sigma )}{\check {\tilde E}}^{(o)}_r=0,\quad\quad
{}^4_{(\Sigma )}{\check {\tilde E}}^{(a)}_r={}^3e^{(a)}_r,\nonumber \\
 &&{}\nonumber \\
 &&{}^4_{(\Sigma )}{\check E}^A_{(\alpha )}\,
{}^4g_{AB}\, {}^4_{(\Sigma )}{\check E}^B_{(\beta )}={}^4\eta_{(\alpha
)(\beta )}.
\label{II1}
\end{eqnarray}

\noindent Here $b^{\mu}_r$ and $b_{\mu}^r$ are defined in Eqs.(A3) of Ref\cite{ru11}
\footnote{Instead of local coordinates $x^{\mu}$ for $M^4$,
we use local coordinates $\sigma^A$ on
$R\times \Sigma \approx M^4$ [$x^{\mu}=z^{\mu}(\sigma )$ with inverse $\sigma^A=
\sigma^A(x)$], i.e. a {\it $\Sigma_{\tau}$-adapted holonomic coordinate basis} for
vector fields $\partial_A={{\partial}\over {\partial \sigma^A}}\in
T(R\times \Sigma ) \mapsto b^{\mu}_A(\sigma ) \partial_{\mu} ={{\partial z
^{\mu}(\sigma )}\over {\partial \sigma^A}} \partial_{\mu} \in TM^4$, and for
differential one-forms $dx^{\mu}\in T^{*}M^4 \mapsto d\sigma^A=b^A
_{\mu}(\sigma )dx^{\mu}={{\partial \sigma^A(z)}\over {\partial z^{\mu}}} dx
^{\mu} \in T^{*}(R\times \Sigma )$. The induced 4-metric and inverse 4-metric
become in the new basis
${}^4g_{\mu\nu}=b^A_{\mu}\, {}^4g_{AB} b^B_{\nu} =
\epsilon \, (N^2-{}^3g_{rs}N^rN^s)\partial_{\mu}\tau
\partial_{\nu}\tau -
\epsilon \, {}^3g_{rs}N^s(\partial_{\mu}\tau \partial_{\nu}\sigma^r+
\partial_{\nu}\tau \partial_{\mu}\sigma^r)-\epsilon \, {}^3g_{rs}
\partial_{\mu}\sigma^r \partial_{\nu}\sigma^s=
 \epsilon \, l_{\mu} l_{\nu} -\epsilon \, {}^3g_{rs} (\partial_{\mu}
\sigma^r +N^r\, \partial_{\mu}\tau ) (\partial_{\nu}\sigma^s+N^s\,
\partial_{\nu}\tau )$,
${}^4g_{AB}=\lbrace {}^4g_{\tau\tau}=
\epsilon (N^2-{}^3g_{rs}N^rN^s); {}^4g_{\tau r}=-
\epsilon \, {}^3g_{rs}N^s; {}^4g_{rs}=-\epsilon \, {}^3g_{rs}\rbrace =
\epsilon [l_Al_B-{}^3g_{rs}(\delta^r_A+N^r\delta^{\tau}_A)(\delta^s_B+
N^s\delta^{\tau}_B)]$, ${}^4g^{\mu\nu}= b^{\mu}_A {}^4g^{AB}
b^{\nu}_B={{\epsilon}\over {N^2}}
\partial_{\tau}z^{\mu}\partial_{\tau}z^{\nu}- {{\epsilon \, N^r}\over
{N^2}} (\partial_{\tau}z^{\mu}\partial_rz^{\nu}+
\partial_{\tau}z^{\nu}\partial_rz^{\mu}) -\epsilon ({}^3g^{rs}-{{N^rN^s}\over
{N^2}})\partial_rz^{\mu}\partial_sz^{\nu}= \epsilon [\, l^{\mu}
l^{\nu} - \, {}^3g^{rs} \partial_rz^{\mu}
\partial_sz^{\nu}]$,
${}^4g^{AB}=\lbrace {}^4g^{\tau\tau}= {{\epsilon}\over {N^2}};
{}^4g^{\tau r}=-{{\epsilon \, N^r}
\over {N^2}}; {}^4g^{rs}=-\epsilon ({}^3g^{rs} - {{N^rN^s}\over {N^2}})
\rbrace
=\epsilon [l^Al^B -{}^3g^{rs}\delta^A_r\delta^B_s]$. For the unit normals we have
$l^A=l^{\mu} b^A_{\mu} = N \, {}^4g^{A\tau}={{\epsilon}\over N} (1;
-N^r)$ and $l_A=l_{\mu} b_A^{\mu} = N \partial_A \tau =N \delta^{\tau}_A = (N;
\vec 0)$.
We introduced the 3-metric  of $\Sigma_{\tau}$: $\,
{}^3g_{rs}=-\epsilon \, {}^4g_{rs}$ with signature (+++). If
${}^4\gamma^{rs}$ is the inverse of the spatial part of the 4-metric
(${}^4\gamma^{ru}\, {}^4g_{us}=\delta^r_s$), the inverse of the
3-metric is ${}^3g^{rs}=-\epsilon \, {}^4\gamma^{rs}$ (${}^3g^{ru}\,
{}^3g_{us}=\delta^r_s$). ${}^3g_{rs}(\tau ,\vec \sigma )$ are the
components of the {\it first fundamental form} of the Riemann
3-manifold $(\Sigma_{\tau},{}^3g)$ and we have the following form for
the line element in $M^4$:
 $ds^2={}^4g_{\mu\nu} dx^{\mu} dx^{\nu}=
\epsilon (N^2-{}^3g_{rs}N^rN^s) (d\tau )^2-2\epsilon
\, {}^3g_{rs}N^s d\tau d\sigma^r -\epsilon \, {}^3g_{rs} d\sigma^rd\sigma^s=
 \epsilon \Big[ N^2(d\tau )^2 -{}^3g_{rs}(d\sigma^r+N^rd\tau )(d\sigma^s+
N^sd\tau )\Big]$.} and $l^{\mu}(\tau ,\vec \sigma )$ is the unit
normal to $\Sigma_{\tau}$ at $\vec \sigma$. $N$ and $N^r$ are the
standard lapse and shift functions.

We have also shown the components of these tetrads and cotetrads in
the holonomic basis (${}^4_{(\Sigma )}{\check {\tilde E}}^{(o)}_r=0$
is the {\it Schwinger time gauge condition}\cite{schw}).

We define our class of  {\it arbitrary cotretads} ${}^4E^{(\alpha
)}_{\mu}(z(\sigma ))$ on $M^4$ starting from the special
$\Sigma_{\tau}$- and cotriad-adapted cotetrads ${}^4_{(\Sigma
)}{\check E}^{(\alpha )}_{\mu} (z(\sigma ))$ by means of the formula

\bea
{}^4E^{(\alpha )}_{\mu}(z(\sigma ))&=&L^{(\alpha )}{}_{(\beta
)}(V(z(\sigma )); {\buildrel \circ \over V})\, {}^4_{(\Sigma )}{\check
E}^{(\beta )}_{\mu} (z(\sigma )),\nonumber \\
 &&{}\nonumber \\
\left( \begin{array}{l} {}^4E^{(o)}_{\mu}\\ {}^4E^{(a)}_{\mu} \end{array}
\right) (z(\sigma ))&=&\left( \begin{array}{cc}  \sqrt{1+\sum_{(c)}
\varphi^{(c) 2}} &-\epsilon \varphi_{(b)}\\  \varphi^{(a)} &
\delta^{(a)}_{(b)}-\epsilon {{\varphi^{(a)}\varphi_{(b)} }\over {1+
\sqrt{1+\sum_{(c)}\varphi^{(c) 2}} }} \end{array} \right) (z(\sigma ))
\left( \begin{array}{l} l_{\mu} \\ {}^3e^{(b)}_s\, b^s_{\mu} \end{array}
\right) (\sigma ),\nonumber \\
 &&{}\nonumber \\
 {}^4E^{(o)}_{\tau}(z(\sigma ))&=& \sqrt{1+\sum_{(c)}\varphi^{(c) 2}
(\sigma )}\, N(\sigma )+\sum_{(a)}\varphi^{(a)}(\sigma )
N^{(a)}(\sigma ).
\label{II2}
\eea

Here $L^{(\alpha )}{}_{(\beta )}(V(z(\sigma )); {\buildrel \circ
\over V})$ are the components of   the standard Wigner boost for
timelike Poincar\'e orbits (see Ref.\cite{longhi}), which in the
tangent space at each point of $M^4$ connects the timelike
4-vectors ${\buildrel \circ \over V}^{(\alpha )}=l^{\mu}(z(\sigma
))\, {}^4_{(\Sigma )}{\check E}^{(\alpha )}_{\mu}(z(\sigma ))= (1
;\vec 0)$ and $V^{(\alpha )}(z(\sigma )) = L^{(\alpha )}{}_{(\beta
)}(V(z(\sigma ));{\buildrel \circ \over V})\, {\buildrel \circ
\over V}^{(\beta )} {\buildrel {def} \over =}\, l^{\mu}(z(\sigma
))\, {}^4E^{(\alpha )}_{\mu} (z(\sigma ))$.

Let ${}^4E^{\mu}_{(\alpha )}(z)$ and ${}^4E^{(\alpha )}_{\mu}(z)$
be arbitrary tetrads and cotetrads on $M^4$. Let us define the
point-dependent Minkowski 4-vector\footnote{The point-dependent
flat 4-vector $V^{(\alpha )}(z(\sigma )) =( V^{(o)}(z(\sigma ))$
$=+\sqrt{1+\sum_rV^{(r) 2}(z(\sigma ))}; V^{(r)}(z(\sigma ))
{\buildrel {def}\over =}\, \varphi^{(r)}(\sigma ) )$ depends only
on the three functions $\varphi^{(r)}(\sigma )$. One has
$\varphi^{(r)}(\sigma )=-\epsilon \varphi_{(r)}(\sigma )$ since
${}^4\eta_{rs}=-\epsilon \, \delta_{rs}$; having the Euclidean
signature (+++) for both $\epsilon =\pm 1$, we shall define the
 Kronecker delta as $\delta^{(i)(j)}=\delta^{(i)}_{(j)}=\delta_{(i)(j)}$.}
$V^{(\alpha )}(z(\sigma ))=l^{\mu}(z(\sigma ))\, {}^4E^{(\alpha )}
_{\mu}(z(\sigma ))$ (assumed to be future-pointing), which
satisfies $V^{(\alpha )}(z(\sigma ))\, {}^4\eta_{(\alpha )(\beta
)}\, V^{(\beta )} (z(\sigma ))=\epsilon$.

The $\varphi^{(a)}(\sigma )=V^{(a)}(z(\sigma ))=l^{\mu}(z(\sigma ))\,
{}^4E^{(a)}_{\mu}(z(\sigma ))$ are the three parameters of the Wigner
boost.

If we go to $\Sigma_{\tau}$-adapted bases, ${}^4E^{(\alpha
)}_A(z(\sigma ))={}^4E^{(\alpha )} _{\mu}(z(\sigma ))\,
b^{\mu}_A(\sigma )$ and ${}^4_{(\Sigma )}{\check {\tilde
E}}^{(\alpha )}_A(z(\sigma ))={}^4_{(\Sigma )}{\check E}^{(\alpha
)}_{\mu} (z(\sigma ))\, b^{\mu}_A(\sigma )$, one has

\begin{eqnarray}
\left( \begin{array}{l} {}^4E^{(o)}_A \\ {}^4E^{(a)}_A \end{array} \right) &=&
\left( \begin{array}{cc}  \sqrt{1+\sum_{(c)}
\varphi^{(c) 2}} &-\epsilon \varphi_{(b)}\\  \varphi^{(a)} &
\delta^{(a)}_{(b)}-\epsilon {{\varphi^{(a)}\varphi_{(b)} }\over {1+
\sqrt{1+\sum_{(c)}\varphi^{(c) 2}} }} \end{array} \right) \times\nonumber \\
&&\left( \begin{array}{l} {}^4_{(\Sigma )}{\check {\tilde
E}}^{(o)}_A=(N;\vec 0)\\
 {}^4_{(\Sigma )}{\check {\tilde
E}}^{(b)}_A=(N^{(b)}={}^3e^{(b)}_rN^r; {}^3e^{(b)}_r) \end{array}
\right) ,\nonumber \\
 &&{}\nonumber \\
 &&{}\nonumber \\
&&{}^4E^{(o)}_r(z(\sigma ))= \sum_{(a)}\varphi^{(a)}(\sigma )\,
{}^3e^{(a)}_r(\sigma ),\nonumber \\
&&{}^4E^{(a)}_{\tau}(z(\sigma ))= \varphi^{(a)}(\sigma )N(\sigma )+
\sum_{(b)}[\delta^{(a)}_{(b)}-\epsilon {{\varphi^{(a)}(\sigma )\varphi_{(b)}
(\sigma )}\over {1+\sqrt{1+\sum_{(c)}\varphi^{(c) 2}(\sigma )} }}]N^{(b)}
(\sigma ),\nonumber \\
&&{}^4E^{(a)}_r(z(\sigma ))=\sum_{(b)}[\delta^{(a)}_{(b)}-\epsilon
{{\varphi^{(a)}(\sigma )\varphi_{(b)}
(\sigma )}\over {1+\sqrt{1+\sum_{(c)}\varphi^{(c) 2}(\sigma )} }}] {}^3e
^{(b)}_r(\sigma ),\nonumber \\
 &&{}\nonumber \\
 &&{}\nonumber \\
 \Rightarrow {}^4g_{AB}&=&{}^4E^{(\alpha )}_A\,
{}^4\eta_{(\alpha )(\beta )}\, {}^4E^{(\beta )}_B= {}^4_{(\Sigma
)}{\check E}^{(\alpha )}_A\, {}^4\eta_{(\alpha )(\beta )}\,
{}^4_{(\Sigma )}{\check E}^{(\beta )}_B=\nonumber \\
 &=&\epsilon \left( \begin{array}{cc} (N^2- {}^3g_{rs}N^rN^s) &
-{}^3g_{st}N^t\\ -{}^3g_{rt}N^t & -{}^3g
_{rs} \end{array} \right) ,\nonumber \\
 &&{}\nonumber \\
  &&{}\nonumber \\
\left( \begin{array}{l} {}^4E^{\mu}_{(o)}\\
{}^4E^{\mu}_{(a)}\end{array}
\right) &=&\left( \begin{array}{cc}  \sqrt{1+\sum_{(c)}
\varphi^{(c) 2}} &- \varphi^{(b)}\\ \epsilon \varphi_{(a)} &
\delta_{(a)}^{(b)}-\epsilon {{\varphi_{(a)}\varphi^{(b)} }\over {1+
\sqrt{1+\sum_{(c)}\varphi^{(c) 2}} }} \end{array} \right) \,
\left( \begin{array}{l} l^{\mu} \\ b^{\mu}_s\, {}^3e^s_{(b)} \end{array}
\right) , \nonumber \\
 &&{}\nonumber \\
 \left( \begin{array}{l} {}^4E^A_{(o)} \\ {}^4E^A_{(a)} \end{array} \right)
&=&\left( \begin{array}{cc}  \sqrt{1+\sum_{(c)}
\varphi^{(c) 2}} &- \varphi^{(b)}\\ \epsilon \varphi^{(a)} &
\delta_{(a)}^{(b)}-\epsilon {{\varphi_{(a)}\varphi^{(b)} }\over {1+
\sqrt{1+\sum_{(c)}\varphi^{(c) 2}} }} \end{array} \right) \,
\left( \begin{array}{l} {}^4_{(\Sigma )}{\check {\tilde E}}^A_{(o)}=(1/N;
-N^r/N) \\ {}^4_{(\Sigma )}{\check {\tilde E}}^A_{(b)}=(0; {}^3e^r_{(b)})
\end{array} \right) ,\nonumber \\
 &&{}\nonumber \\
 &&{}\nonumber \\
 &&{}^4E^{\tau}_{(o)}(z(\sigma ))= \sqrt{1+\sum_{(c)}\varphi^{(c) 2}
(\sigma )} {1\over {N(\sigma )}},\nonumber \\
&&{}^4E^r_{(o)}(z(\sigma )=- \sqrt{1+\sum_{(c)}\varphi^{(c) 2}
(\sigma )} {{N^r(\sigma )}\over {N(\sigma )}}-\varphi^{(b)}(\sigma )
\, {}^3e^r_{(b)}(\sigma ),\nonumber \\
&&{}^4E^{\tau}_{(a)}(z(\sigma ))=\epsilon {{\varphi_{(a)}(\sigma )}\over
{N(\sigma )}},\nonumber \\
&&{}^4E^r_{(a)}(z(\sigma ))=-\epsilon \varphi_{(a)}(\sigma ) {{N^r(\sigma )}
\over {N(\sigma )}}+\sum_{(b)}[\delta_{(a)}^{(b)}-\epsilon
{{\varphi_{(a)}(\sigma )\varphi^{(b)}(\sigma )}\over {1+\sqrt{1+\sum_{(c)}
\varphi^{(c) 2}(\sigma )}}}]\, {}^3e^r_{(b)}(\sigma ),\nonumber \\
 &&{}\nonumber \\
 \Rightarrow {}^4g^{AB}&=&{}^4E^A_{(\alpha )}\, {}^4\eta^{(\alpha
)(\beta )}\, {}^4E^B_{(\beta )}= {}^4_{(\Sigma )}{\check E}_{(\alpha
)}^A\, {}^4\eta_{(\alpha )(\beta )}\, {}^4_{(\Sigma )}{\check
E}_{(\beta )}^B=\nonumber \\
 &=&\epsilon \left( \begin{array}{cc} {1\over {N^2}} & -
{{N^s}\over {N^2}} \\ - {{N^r}\over {N^2}} &
- ({}^3g^{rs}-{{N^rN^s}\over {N^2}})\end{array} \right).
\label{II3}
\end{eqnarray}

We get that the cotetrad in the $\Sigma_{\tau}$-adapted basis can
be expressed in terms of N, $N^{(a)}={}^3e^{(a)}_sN^s=N_{(a)}$,
$\varphi^{(a)}$ and ${}^3e^{(a)}_r$ [${}^3g_{rs}=\sum_{(a)}
{}^3e_{(a)r}\, {}^3e_{(a)s}$]

From ${}^4_{(\Sigma )}{\check {\tilde E}}^{(\alpha )}_A(z(\sigma
))=(L^{-1}) ^{(\alpha )}{}_{(\beta )}(V(z(\sigma ));{\buildrel
\circ \over V})\, {}^4E ^{(\beta )}_A(z(\sigma ))$ and
${}^4_{(\Sigma )}{\check {\tilde E}}^A_{(\alpha )}(z(\sigma
))={}^4E^A_{(\beta )}\, (L^{-1})^{(\beta )}{}_{(\alpha )}
(V(z(\sigma ));{\buildrel \circ \over V})$ it turns
out\cite{longhi} that the flat indices $(a)$ of the adapted
tetrads ${}^4_{(\Sigma )}{\check E}^{\mu} _{(a)}$ and of the
triads ${}^3e^r_{(a)}$ and cotriads ${}^3e_r^{(a)}$ on
$\Sigma_{\tau}$ transform as Wigner spin 1 indices under
point-dependent SO(3) Wigner rotations $R^{(a)}{}_{(b)}(V(z(\sigma
));\Lambda (z(\sigma )))$ associated with Lorentz transformations
$\Lambda^{(\alpha )}{}_{(\beta )}(z)$ in the tangent plane to
$M^4$ in the same point \footnote{$R^{(\alpha )}{}_{(\beta )}
(\Lambda (z(\sigma ));V(z(\sigma )))=[L({\buildrel \circ \over
V};V(z(\sigma )) )\, \Lambda^{-1}(z(\sigma ))\, L(\Lambda
(z(\sigma ))V(z(\sigma ));{\buildrel \circ \over V})]^{(\alpha
)}{}_{(\beta )}=\left(
\begin{array}{cc} 1&0\\ 0& R^{(a)}{}_{(b)}(V(z(\sigma ));\Lambda
(z(\sigma )))\end{array}\right)$.}. Instead the index ${(o)}$ of
the adapted tetrads ${}^4_{(\Sigma )}{\check E} ^{\mu}_{(o)}$ is a
local Lorentz scalar in each point. Therefore, the adapted tetrads
in the $\Sigma_{\tau}$-adapted basis should be denoted as
${}^4_{(\Sigma )}{\check {\tilde E}}^A_{(\bar \alpha )}$, with
$(\bar o)$ and $A=(\tau ,r)$ Lorentz scalar indices and with
$(\bar a)$ Wigner spin 1 indices; we shall go on with the indices
$(o),(a)$ without the overbar for the sake of simplicity. In this
way the tangent planes to $\Sigma_{\tau}$ in $M^4$ are described
in a Wigner covariant way, reminiscent of the flat rest-frame
covariant instant form of dynamics introduced in Minkowski
spacetime in Ref.\cite{lus1}.

Therefore, an arbitrary tetrad field, namely a (in general
non-geodesic) congruence of observers' timelike worldlines with
4-velocity field $u^A(\tau ,\vec \sigma )={}^4E^A_{(o)}(\tau ,\vec
\sigma )$, can be obtained with a pointwise Wigner boost from the
special surface-forming timelike congruence whose 4-velocity field
is the normal to $\Sigma_{\tau}$, $l^A(\tau ,\vec \sigma )
=\epsilon \, {}^4_{(\Sigma )}{\check {\tilde E}}^A_{(o)}(\tau
,\vec \sigma )$ \footnote{It is associated with the 3+1 splitting
of $M^4$ with leaves $\Sigma_{\tau}$.}.

We can invert Eqs.(\ref{II3}) to get N, $N^r={}^3e^r_{(a)}N^{(a)}$,
$\varphi^{(a)}$ and ${}^3e^r_{(a)}$ in terms of the tetrads
${}^4E^A_{(\alpha )}$

\begin{eqnarray}
N&=&  \frac{1}{ \sqrt{[{}^4E^{\tau}_{(o)}]^2
          -\sum_{(c)} [{}^4E^{\tau}_{(c)}]^2}}.
 \nonumber \\
N^r&=& - \frac{{}^4E^{\tau}_{(o)}\, {}^4E^{r}_{(0)} -\sum_{(c)}{}^4E^{\tau}
_{(c)}\, {}^4E^r_{(c)}}
{[{}^4E^{\tau}_{(0)}]^2-\sum_{(c)}[{}^4E^{\tau}_{(c)}]^2 }\nonumber \\
\varphi_{(a)}&=&\frac{\epsilon~ {}^4E^{\tau}_{(a)}  }{
          \sqrt{[{}^4E^{\tau}_{(o)}]^2
          -\sum_{(c)} [{}^4E^{\tau}_{(c)}]^2}}\nonumber\\
{}^3e^r_{(a)}&=&\sum_{(b)} B_{(a)(b)}
    \big( {}^4E^r_{(b)} + N^r~ {}^4E^\tau_{(b)} \big)
\nonumber \\
&&{}\nonumber \\
&& B_{(a)(b)} = \delta_{(a)(b)}
    -\frac{ {}^4E^{\tau}_{(a)} {}^4E^{\tau}_{(b)}  }{
           {}^4E^{\tau}_{(0)} \big[  {}^4E^{\tau}_{(0)}
       + \sqrt{[{}^4E^{\tau}_{(0)}]^2
          -\sum_{(c)} [{}^4E^{\tau}_{(c)}]^2}] }.
\label{II4}
\end{eqnarray}

If ${}^3e^{-1}=det\, ({}^3e^r_{(a)})$, then from the orthonormality
condition we get ${}^3e_{(a)r}= {}^3e ({}^3e^s_{(b)}\,
{}^3e^t_{(c)}-{}^3e^t_{(b)}\, {}^3e^s_{(c)})$ \footnote{With
$(a),(b),(c)$ and $r,s,t$ cyclic.} and it allows to express the
cotriads in terms of the tetrads ${}^4E^A_{(\alpha )}$. Therefore,
given the tetrads ${}^4E^A_{(\alpha )}$ (or equivalently the cotetrads
${}^4E_A^{(\alpha )}$) on $M^4$, an equivalent set of variables with
the local Lorentz covariance replaced with local Wigner covariance are
the lapse N, the shifts $N^{(a)}=N_{(a)}={}^3e_{(a)r}N^r$, the
Wigner-boost parameters $\varphi^{(a)}=-\epsilon \varphi_{(a)}$ and
either the triads ${}^3e^r_{(a)}$ or the cotriads ${}^3e_{(a)r}$.

The independent variables in metric gravity have now the following
expression in terms of N, $N^{(a)}=N_{(a)}={}^3e^r_{(a)}N_r$,
$\varphi^{(a)}=-\epsilon \varphi_{(a)}$, ${}^3e^{(a)}_r={}^3e_{(a)r}$
\footnote{$\gamma =det\, ({}^3g_{rs})= ({}^3e)^2=(det\, (e_{(a)r}))^2$.}

\begin{eqnarray}
N,&&{}\quad\quad
N_r={}^3e_r^{(a)}N_{(a)}={}^3e_{(a)r}N_{(a)},\nonumber \\
{}^3g_{rs}&=&{}^3e^{(a)}_r\, \delta_{(a)(b)}\, {}^3e^{(b)}_s={}^3e_{(a)r}\,
{}^3e_{(a)s},
\label{II5}
\end{eqnarray}

\noindent so that the line element of $M^4$ becomes

\bea
ds^2 &=&\epsilon (N^2- N_{(a)}N_{(a)})(d\tau )^2-2\epsilon N_{(a)}\,
{}^3e_{(a)r} d\tau d\sigma^r-
\epsilon \, {}^3e_{(a)r}\, {}^3e_{(a)s} d\sigma^rd\sigma^s=\nonumber \\
 &=&\epsilon \Big[ N^2
(d\tau )^2-({}^3e_{(a)r}d\sigma^r +N_{(a)}d\tau
)({}^3e_{(a)s}d\sigma^s +N_{(a)}d\tau )\Big].
\label{II6}
\eea

The extrinsic curvature takes the form
\footnote{$N_{(a)|r}={}^3e^s_{(a)}N_{s|r}=
\partial_r N_{(a)}-\epsilon_{(a)(b)(c)}\, {}^3\omega_{r(b)}N_{(c)}$ with ${}^3\omega_{r(b)}$
being the 3-spin connection, see Eq.(A22) of Ref.\cite{ru11}.}

\begin{eqnarray}
{}^3K_{rs}&=& {\hat b}^{\mu}_r{\hat b}^{\nu}_s {}^3K_{\mu\nu}={1\over {2N}}
(N_{r | s}+N_{s | r}-\partial_{\tau} {}^3g_{rs})=\nonumber \\
&=&{1\over {2N}} ({}^3e_{(a)r} \delta^w_s+{}^3e_{(a)s}\delta^w_r)(N_{(a) | w}-
\partial_{\tau}\, {}^3e_{(a)w}),\nonumber \\
{}^3K_{r(a)}&=&{}^3K_{rs}\, {}^3e^s_{(a)}={1\over {2N}}(\delta_{(a)(b)}\delta
^w_r+{}^3e^w_{(a)}\, {}^3e_{(b)r})(N_{(b)|w}-\partial_{\tau}\, {}^3e_{(c)w}),
\nonumber \\
{}^3K&=&{1\over N}\, {}^3e^r_{(a)} (N_{(a)|r}-\partial_{\tau}\, {}^3e_{(a)r}),
\label{II7}
\end{eqnarray}

The ADM action in the new variables is obtained from the metric
gravity  ADM action and has the form

\begin{eqnarray}
{\hat S}_{ADMT}&=&\int d\tau {\hat L}_{ADMT}=\nonumber \\
&=&-{{\epsilon c^3}\over {16\pi G}} \int d\tau d^3\sigma \lbrace
N\, {}^3e\, \epsilon_{(a)(b)(c)}\, {}^3e^r_{(a)}\, {}^3e^s_{(b)}\,
{}^3\Omega_{rs(c)}+\nonumber \\ &+&{{{}^3e}\over {2N}}
({}^3G_o^{-1})_{(a)(b)(c)(d)} {}^3e^r_{(b)}(N_{(a) | r}-
\partial_{\tau}\, {}^3e_{(a)r})\, {}^3e^s_{(d)}(N_{(c) | s}-\partial_{\tau}
\, {}^3e_{(c) \ s})\rbrace,
\label{II8}
\end{eqnarray}

\noindent where we introduced the flat inverse Wheeler-DeWitt supermetric

\begin{eqnarray}
({}^3G_o^{-1})_{(a)(b)(c)(d)}&=&\delta_{(a)(c)}\delta_{(b)(d)}+\delta_{(a)(d)}
\delta_{(b)(c)}-2\delta_{(a)(b)}\delta_{(c)(d)},\nonumber \\
&&\Downarrow \nonumber \\
{}^3G_{o(a)(b)(c)(d)}&=&{}^3G_{o(b)(a)(c)(d)}={}^3G_{o(a)(b)(d)(c)}=
{}^3G_{o(c)(d)(a)(b)}=\nonumber \\
&=&\delta_{(a)(c)}\delta_{(b)(d)}+\delta_{(a)(d)}\delta_{(b)(c)}-\delta
_{(a)(b)}\delta_{(c)(d)},\nonumber \\
&&{}\nonumber \\
&&{1\over 2}\, {}^3G_{o(a)(b)(e)(f)}\, {1\over 2}\, {}^3G^{-1}_{o(e)(f)(c)(d)}
={1\over 2}[\delta_{(a)(c)}\delta_{(b)(d)}+\delta_{(a)(d)}\delta_{(b)(c)}].
\label{II9}
\end{eqnarray}

The new action does not depend on the 3 boost variables
$\varphi^{(a)}$, contains lapse N and modified shifts $N_{(a)}$ as
Lagrange multipliers, and is a functional independent from the second
time derivatives of the fields.

After the definition of the canonical momenta, whose  Poisson brackets
are

\begin{eqnarray}
&&\lbrace N(\tau ,\vec
\sigma ),{\tilde \pi}^N(\tau ,{\vec
\sigma}^{'} )
\rbrace = \delta^3(\vec \sigma ,{\vec \sigma}^{'}),\nonumber \\
&&\lbrace N_{(a)}(\tau ,\vec \sigma ),{\tilde \pi}^{\vec N}_{(b)}(\tau ,{\vec
\sigma}^{'} )\rbrace =\delta_{(a)(b)} \delta^3(\vec \sigma ,{\vec \sigma}^{'}),
\nonumber \\
&&\lbrace \varphi_{(a)}(\tau ,\vec \sigma ),{\tilde \pi}^{\vec \varphi}_{(b)}
(\tau ,{\vec \sigma}^{'} )\rbrace = \delta_{(a)(b)} \delta^3(\vec \sigma ,
{\vec \sigma}^{'}),\nonumber \\
&&\lbrace {}^3e_{(a)r}(\tau ,\vec \sigma ),{}^3{\tilde \pi}^s_{(b)}(\tau ,
{\vec \sigma}^{'} )\rbrace =\delta_{(a)(b)} \delta^s_r \delta^3(\vec \sigma ,
{\vec \sigma}^{'}),\nonumber \\
&&{}\nonumber \\
&&\lbrace {}^3e^r_{(a)}(\tau ,\vec \sigma),{}^3{\tilde \pi}^s_{(b)}(\tau ,
{\vec \sigma}^{'})\rbrace =-{}^3e^r_{(b)}(\tau ,\vec \sigma )\, {}^3e^s
_{(a)}(\tau ,\vec \sigma ) \delta^3(\vec \sigma ,{\vec \sigma}^{'}),
\nonumber \\
&&\lbrace {}^3e(\tau ,\vec \sigma ), {}^3{\tilde \pi}^r_{(a)}(\tau ,{\vec
\sigma}^{'})\rbrace ={}^3e(\tau ,\vec \sigma )\, {}^3e^r_{(a)}(\tau ,\vec
\sigma )\, \delta^3(\vec \sigma ,{\vec \sigma}^{'}),
\label{II10}
\end{eqnarray}

\noindent we find ten primary constraints  and four secondary ones

\bea
&&{\tilde \pi}^{\vec \varphi}_{(a)} (\tau ,\vec \sigma )\approx
0,\nonumber \\
 &&{\tilde \pi}^N(\tau ,\vec \sigma )\approx 0,\nonumber \\
 && {\tilde \pi}^{\vec N}_{(a)}(\tau ,\vec \sigma )\approx 0,\nonumber \\
{}^3{\tilde M}_{(a)}(\tau ,\vec \sigma )&=&\epsilon_{(a)(b)(c)}\, {}^3e_{(b)r}
(\tau ,\vec \sigma )\, {}^3{\tilde \pi}^r_{(c)}(\tau ,\vec \sigma )={1\over 2}
\epsilon_{(a)(b)(c)}\, {}^3{\tilde M}_{(b)(c)}(\tau ,\vec \sigma )\approx 0,
\nonumber \\
&\Rightarrow& {}^3{\tilde M}_{(a)(b)}(\tau ,\vec \sigma )=\epsilon_{(a)(b)(c)}
\, {}^3{\tilde M}_{(c)}(\tau ,\vec \sigma )=\nonumber \\
&=&{}^3e_{(a)r}(\tau ,\vec \sigma )\, {}^3{\tilde \pi}^r_{(b)}(\tau
,\vec \sigma )-{}^3e_{(b)r}(\tau ,\vec \sigma )\, {}^3{\tilde
\pi}^r_{(a)}(\tau ,\vec \sigma )\approx 0,\nonumber \\
 &&{}\nonumber \\
{\hat {\cal H}}(\tau ,\vec \sigma )&=& \epsilon \Big[
 {{c^3}\over {16\pi G}} \, {}^3e\, \epsilon_{(a)(b)(c)}\, {}^3e^r _{(a)}\,
{}^3e^s_{(b)}\, {}^3\Omega_{rs(c)}-\nonumber \\
 &-&{{2\pi G}\over {c^3\,\,
{}^3e}} {}^3G_{o(a)(b)(c)(d)}\, {}^3e_{(a)r}\, {}^3{\tilde
\pi}^r_{(b)}\, {}^3e_{(c)s}\, {}^3{\tilde \pi}^s_{(d)}\Big] (\tau
,\vec \sigma )=\nonumber \\
 &=&\epsilon \Big[  {{c^3}\over {16\pi G}}\, {}^3e\,
{}^3R-{{2\pi G}\over {c^3\,\, {}^3e}} {}^3G_{o(a)(b)(c)(d)}\,
{}^3e_{(a)r}\, {}^3{\tilde \pi}^r_{(b)}\, {}^3e_{(c)s}\,
{}^3{\tilde \pi}^s_{(d)}\Big] (\tau ,\vec \sigma ) \approx
0,\nonumber \\
 &&{}\nonumber \\
{\hat {\cal H}}_{(a)}(\tau ,\vec \sigma )&=&[\partial_r\, {}^3{\tilde
\pi}^r
_{(a)}-\epsilon_{(a)(b)(c)}\, {}^3\omega_{r(b)}\, {}^3{\tilde \pi}^r_{(c)}]
(\tau ,\vec \sigma )={}^3{\tilde \pi}^r_{(a)|r}(\tau ,\vec \sigma )
\approx 0,\nonumber \\
&&{}\nonumber \\ &\Rightarrow& {\hat H}_{(c)}= \int d^3\sigma [ N\,
{\hat {\cal H}}- N_{(a)}\, {\hat {\cal H}}_{(a)}](\tau ,\vec \sigma
)=\nonumber \\
 &=& -{}^3e^r_{(a)}(\tau ,\vec \sigma ) [{}^3{\tilde \Theta}_r +
{}^3\omega_{r(b)}\, {}^3{\tilde M}_{(b)}] (\tau ,\vec \sigma )\approx
0,\nonumber \\
 &&{}\nonumber \\
 &&\text{or}\nonumber \\
 &&{}\nonumber \\
{}^3{\tilde \Theta}_r(\tau ,\vec \sigma )&=&-[{}^3e_{(a)r}\, {\hat {\cal H}}
_{(a)}+{}^3\omega_{r(a)}\, {}^3{\tilde M}_{(a)}](\tau ,\vec \sigma )=
\nonumber \\
&=&[{}^3{\tilde \pi}^s_{(a)}\, \partial_r\, {}^3e_{(a)s}-\partial_s
({}^3e_{(a)r}\, {}^3{\tilde \pi}^s_{(a)})](\tau ,\vec \sigma )\approx
0.
\label{II11}
\eea

We see that the first seven constraints are already Abelianized: our
parametrization of the cotetrads is equivalent to a Shanmugadhasan
canonical transformation Abelianizing the Lorentz boosts.

It can be checked (see Ref.\cite{ru11}) that the superhamiltonian
constraint ${\hat {\cal H}}(\tau ,
\vec \sigma )\approx 0$ coincides with the ADM metric superhamiltonian one
${\tilde {\cal H}}(\tau ,\vec \sigma )\approx 0$, where also the ADM
metric supermomentum constraints is expressed in terms of the tetrad
gravity constraints.

It is convenient to replace the constraints ${\hat {\cal H}}_{(a)}(\tau ,\vec
\sigma )\approx 0$ \footnote{They are of the type of SO(3) Yang-Mills Gauss laws,
because they are the covariant divergence of a vector density.}  with
the 3 constraints ${}^3{\tilde \Theta}_r(\tau ,\vec \sigma )\approx 0$
generating space pseudo-diffeomorphisms on the cotriads and their
conjugate momenta.

We can get the following phase space expression of the extrinsic
curvature, of the ADM canonical  momentum ${}^3{\tilde \Pi}^{rs}$ and
the following decomposition of the cotriad momentum

\begin{eqnarray}
{}^3K_{rs}&=&{{\epsilon 4\pi G}\over {c^3\,\, {}^3e}}\,
{}^3G_{o(a)(b)(c)(d)}\, {}^3e _{(a)r}\, {}^3e_{(b)s}\,
{}^3e_{(c)u}\, {}^3{\tilde \pi}^u_{(d)},\nonumber \\
 {}^3K&=&-{{\epsilon 8\pi G}\over {c^3\, \sqrt{\gamma}}} {}^3{\tilde \Pi}=-
 {{\epsilon 4\pi G}\over {c^3\,\, {}^3e}} {}^3e_{(a)r}\, {}^3{\tilde
\pi}^r_{(a)},\nonumber \\
 &&{}\nonumber \\
 {}^3{\tilde \Pi}^{rs} &=& {1\over 4} \Big( {}^3e^r_{(a)}\,
 {}^3{\tilde \pi}^s_{(a)} + {}^3e^s_{(a)}\, {}^3{\tilde \pi}^r_{(a)}\Big),
 \nonumber \\
 &&{}\nonumber \\
{}^3{\tilde \pi}^r_{(a)}&=&{}^3e^r_{(b)}\, {}^3e_{(b)s}\,
{}^3{\tilde \pi}^s _{(a)}={1\over 2}{}^3e^r_{(b)}[{}^3e_{(a)s}\,
{}^3{\tilde \pi}^s_{(b)}+{}^3e _{(b)s}\, {}^3{\tilde
\pi}^s_{(a)}]-{1\over 2}{}^3{\tilde M}_{(a)(b)}\,
{}^3e^r_{(b)}\equiv \nonumber \\ &\equiv& {1\over
2}{}^3e^r_{(b)}[{}^3e_{(a)s}\, {}^3{\tilde \pi}^s_{(b)}+{}^3e
_{(b)s}\, {}^3{\tilde \pi}^s_{(a)}],\nonumber \\ &&{}\nonumber \\
{}^3{\tilde \pi}^r_{(a)} &\partial_{\tau}& {}^3e_{(a)r}\equiv
{1\over 2}[{}^3e_{(a)s}\, {}^3{\tilde \pi}^s_{(b)}+{}^3e _{(b)s}\,
{}^3{\tilde \pi}^s_{(a)}]{}^3e^r_{(b)}\, \partial_{\tau}\, {}^3e
_{(a)r}\equiv \nonumber \\ &\equiv& {}^3{\tilde \pi}^r_{(a)}\,
N_{(a) | r}-{{4\pi G\, N}\over {c^3\,\, {}^3e}}\,
{}^3G_{o(a)(b)(c)(d)}\, {}^3e_{(a)s}\, {}^3{\tilde \pi}^s_{(b)}\,
{}^3e_{(c)r}\, {}^3{\tilde \pi}^r _{(d)}.
 \label{II12}
\end{eqnarray}

After ignoring a surface term (see next Subsection) we get the
Dirac Hamiltonian (the $\lambda (\tau ,\vec \sigma )$'s are
arbitrary Dirac multipliers)

\bea
{\hat H}_{(D)}&=&{\hat H}_{(c)}+\int d^3\sigma [\lambda_N\, {\tilde
\pi}^N+\lambda^{\vec N}_{(a)}\, {\tilde \pi}^{\vec N}_{(a)}+\lambda^{\vec
\varphi}_{(a)}\, {\tilde \pi}^{\vec \varphi}_{(a)}+\mu_{(a)}\, {}^3{\tilde
M}_{(a)}](\tau ,\vec \sigma )=\nonumber \\
 &=&{\hat H}^{'}_{(c)}+\int
d^3\sigma [\lambda_N{\tilde \pi}^N+\lambda^{\vec N}_{(a)}{\tilde
\pi}^{\vec N}_{(a)}+\lambda^{\vec \varphi}_{(a)}{\tilde \pi}^{\vec
\varphi}_{(a)}+{\hat \mu}_{(a)}
\, {}^3{\tilde M}_{(a)}](\tau ,\vec \sigma ),\nonumber \\
 &&{}\nonumber \\
 &&{\hat H}_c= \int d^3\sigma [N {\hat {\cal H}}- N_{(a)}
 {\hat {\cal H}}_{(a)}](\tau ,\vec \sigma ),\nonumber \\
 && {\hat H}^{'}_{(c)}= \int d^3\sigma [ N {\hat {\cal H}} +N_{(a)}\,
{}^3e^r_{(a)}\, {}^3{\tilde \Theta}_r](\tau ,\vec \sigma ),
\label{II13}
\eea

\noindent where we replaced $[\mu_{(a)}- N_{(b)}\, {}^3e^r_{(b)}\,
{}^3\omega_{r(a)}](\tau ,\vec \sigma )$ with the new Dirac multipliers ${\hat
\mu}_{(a)}(\tau ,\vec \sigma )$.

All the constraints are first class because the only non-identically
vanishing Poisson brackets are

\begin{eqnarray}
\lbrace {}^3{\tilde M}_{(a)}(\tau ,\vec \sigma ),{}^3{\tilde M}_{(b)}
(\tau ,{\vec \sigma}^{'})\rbrace &=&\epsilon_{(a)(b)(c)}\, {}^3{\tilde M}_{(c)}
(\tau ,\vec \sigma ) \delta^3(\vec \sigma ,{\vec \sigma}^{'}),\nonumber \\
\lbrace {}^3{\tilde M}_{(a)}(\tau ,\vec \sigma ),{}^3{\tilde \Theta}_r
(\tau ,{\vec \sigma}^{'})\rbrace &=&{}^3{\tilde M}_{(a)}(\tau ,{\vec \sigma}
^{'})\, {{\partial \delta^3(\vec \sigma ,{\vec \sigma}^{'})}\over {\partial
\sigma^r}},\nonumber \\
\lbrace {}^3{\tilde \Theta}_r(\tau ,\vec \sigma ),{}^3{\tilde \Theta}_s
(\tau ,{\vec \sigma}^{'})\rbrace &=&[{}^3{\tilde \Theta}_r(\tau ,{\vec \sigma}
^{'}) {{\partial}\over {\partial \sigma^s}} +{}^3{\tilde \Theta}_s(\tau ,\vec
\sigma ) {{\partial}\over {\partial \sigma^r}}] \delta^3(\vec \sigma ,{\vec
\sigma}^{'}),\nonumber \\
\lbrace {\hat {\cal H}}(\tau ,\vec \sigma ),{}^3{\tilde \Theta}_r(\tau ,{\vec
\sigma}^{'})\rbrace &=& {\hat {\cal H}}(\tau ,{\vec \sigma}^{'}) {{\partial
\delta^3(\vec \sigma ,{\vec \sigma}^{'})}\over {\partial \sigma^r}},
\nonumber \\
\lbrace {\hat {\cal H}}(\tau ,\vec \sigma ),{\hat {\cal H}}(\tau ,{\vec
\sigma}^{'})\rbrace &=& [ {}^3e^r_{(a)}(\tau ,\vec \sigma )\, {\hat {\cal H}}
_{(a)}(\tau ,\vec \sigma ) +\nonumber \\
&+& {}^3e^r_{(a)}(\tau ,{\vec \sigma}^{'})\,
{\hat {\cal H}}_{(a)}(\tau ,{\vec \sigma}^{'}) ] {{\partial \delta^3(\vec \sigma
,{\vec \sigma}^{'})}\over {\partial \sigma^r}}=\nonumber \\
&=&\{ [{}^3e^r_{(a)}\, {}^3e^s_{(a)}\, [{}^3{\tilde \Theta}_s+{}^3\omega_{s(b)}
\, {}^3{\tilde M}_{(b)}]](\tau ,\vec \sigma ) + \nonumber \\
&+&[{}^3e^r_{(a)}\, {}^3e^s_{(a)}\, [{}^3{\tilde \Theta}_s+{}^3\omega_{s(b)}\,
{}^3{\tilde M}_{(b)}]](\tau ,{\vec \sigma}^{'}) \} \, {{\partial \delta^3(\vec
\sigma ,{\vec \sigma}^{'})}\over {\partial \sigma^r}}.
\label{II14}
\end{eqnarray}

The Hamiltonian gauge group has the 14 first class constraints as
generators of infinitesimal Hamiltonian gauge transformations
connected with the identity. In particular ${\tilde \pi}^{\vec
\varphi}_{(a)}(\tau ,\vec \sigma )\approx 0$ and ${}^3{\tilde
M}_{(a)}(\tau ,\vec \sigma )\approx 0$ are the generators of the
$R^3\times SO(3)$ subgroup replacing the Lorentz subgroup with our
parametrization, while ${}^3{\tilde \Theta}_r(\tau ,\vec \sigma
)\approx 0$ are the generators of the infinitesimal
pseudo-diffeomorphisms in $Diff\, \Sigma_{\tau}$.

\subsection{Review of the Rest-Frame Instant Form of Metric Gravity.}

In Ref.\cite{p22} and in the book in Ref.\cite{dirac} (see also
Ref.\cite{reg}), Dirac introduced asymptotic Minkowski Cartesian coordinates

\begin{equation}
z^{(\mu )}
_{(\infty )}(\tau ,\vec \sigma )=x^{(\mu )}_{(\infty )}(\tau )+b^{(\mu )}
_{(\infty )\, \check r}(\tau ) \sigma^{\check r}
\label{II15}
\end{equation}

\noindent in $M^4$ at spatial infinity $S_{\infty}=\cup_{\tau} S^2_{\tau
,\infty}$ \footnote{Here $\{ \sigma^{\check r} \}$ are  {\it global}
coordinate charts of the atlas ${\cal C}_{\tau}$ of $\Sigma_{\tau}$,
not matching the spatial coordinates $z^{(i)}_{(\infty )}(\tau ,\vec
\sigma )$.}. For each value of $\tau$, the coordinates $x^{(\mu )}
_{(\infty )}(\tau )$ labels an arbitrary point, near spatial infinity
chosen as {\it origin}. On it there is a flat tetrad $b^{(\mu
)}_{(\infty )\, A} (\tau )= (\, l^{(\mu )}_{(\infty )}=b^{(\mu
)}_{(\infty )\, \tau}=\epsilon^{(\mu )}{}_{(\alpha )(\beta )(\gamma )}
b^{(\alpha )}_{(\infty )\, \check 1} (\tau )b^{(\beta )}_{(\infty )\,
\check 2}(\tau )b^{(\gamma )}_{(\infty )\,
\check 3}(\tau );\, b^{(\mu )}_{(\infty )\, \check r}(\tau )\, )$, with
$l^{(\mu )}_{(\infty )}$ $\tau$-independent, satisfying $b^{(\mu
)}_{(\infty )\, A}\, {}^4\eta_{(\mu )(\nu )}\, b^{(\nu )}_{(\infty )\,
B}={}^4\eta_{AB}$ for every $\tau$. There will be transformation
coefficients $b^{\mu}_A(\tau ,\vec
\sigma )$ from the  adapted coordinates $\sigma^A=(\tau ,\sigma
^{\check r})$ to the coordinates $x^{\mu}=z^{\mu}(\sigma^A)$ in an atlas of $M^4$,
such that in a chart at spatial infinity one has $z^{\mu}(\tau ,\vec \sigma )
\rightarrow \delta^{\mu}_{(\mu )} z^{(\mu )}_{(\infty )}(\tau ,\vec \sigma )$ and $b^{\mu}
_A(\tau ,\vec \sigma ) \rightarrow
\delta^{\mu}_{(\mu )} b^{(\mu )}_{(\infty )A}(\tau )$
\footnote{For $r\, \rightarrow \, \infty$ one has ${}^4g_{\mu\nu}\,
\rightarrow \, \delta^{(\mu )}_{\mu}\delta^{(\nu )}_{\nu}{}^4\eta
_{(\mu )(\nu )}$ and ${}^4g_{AB}=b^{\mu}_A\, {}^4g_{\mu\nu} b^{\nu}_B\,
\rightarrow \, b^{(\mu )}_{(\infty )A}\, {}^4\eta_{(\mu )(\nu )}
b^{(\nu )}_{(\infty )B}= {}^4\eta_{AB}$.}. The atlas ${\cal C}$ of the
allowed coordinate systems of $M^4$ is assumed to have this property.

Dirac\cite{p22} and, then, Regge and Teitelboim\cite{reg} proposed
that the asymptotic Minkowski Cartesian coordinates $z^{(\mu
)}_{(\infty )}(\tau ,\vec \sigma )=x^{(\mu )}_{(\infty )} (\tau
)+b^{(\mu )}_{(\infty ) \check r}(\tau )\sigma^{\check r}$ should
define 10 new independent degrees of freedom at the spatial
boundary $S_{\infty}$ (with ten associated conjugate momenta), as
it happens for Minkowski parametrized
theories\cite{lus1,india,crater,iten,mate} (see Appendix A of
Ref.\cite{restmg} for a review), when the extra configurational
variables $z^{(\mu )}(\tau ,\vec \sigma )$ are reduced to 10
degrees of freedom by the restriction to spacelike hyperplanes,
defined by $z^{(\mu )}(\tau ,\vec \sigma )\approx x^{(\mu
)}_s(\tau )+b^{(\mu )}_{\check r}(\tau )\sigma^{\check r}$.

In Dirac's approach to metric gravity  the 20 extra variables of the
Dirac proposal can be chosen as the set: $x^{(\mu )}_{(\infty )}(\tau
)$, $p^{(\mu )}_{(\infty )}$, $b^{(\mu )}_{(\infty ) A}(\tau )$
\footnote{With $b^{(\mu )}_{(\infty ) \tau }=l^{(\mu )}_{(\infty )}$
$\tau$-independent and coinciding with the asymptotic normal to
$\Sigma_{\tau}$, tangent to $S_{\infty}$.}, $S^{(\mu )(\nu )}_{(\infty
)}$, with the non-vanishing  Dirac brackets $\{ b^{(\rho )}_A, S^{(\mu
)(\nu )}_{(\infty )} \}={}^4\eta^{(\rho )(\mu )} b^{(\nu )}_A
-{}^4\eta^{(\rho )(\nu )} b^{(\mu )}_A$,
 $\{ S^{(\mu )(\nu )}_{(\infty )},S^{(\alpha )(\beta )}_{(\infty )} \}
 = C^{(\mu )(\nu )(\alpha )(\beta )}_{(\gamma )(\delta )}
 S^{(\gamma )(\delta )}_{(\infty )}$, of
Ref.\cite{hanson,lus1}, implying the orthonormality constraints
$b^{(\mu )}_{(\infty ) A}\, {}^4\eta_{(\mu )(\nu )} b^{(\nu
)}_{(\infty ) B}={}^4\eta_{AB}$. Moreover, $p^{(\mu )}_{(\infty )}$
and $J^{(\mu )(\nu )}_{(\infty )}=x^{(\mu )}_{(\infty )}p^{(\nu
)}_{(\infty )}- x^{(\nu )}_{(\infty )}p^{(\mu )}_{(\infty )}+S^{(\mu
)(\nu )}_{(\infty )}$ satisfy a Poincar\'e algebra. In analogy with
Minkowski parametrized theories restricted to spacelike hyperplanes,
one introduces 10 extra first class constraints of the type

\beq
p^{(\mu )}_{(\infty )}-P^{(\mu )}_{ADM}\approx 0,\quad\quad S^{(\mu
)(\nu )}_{(\infty )}-S^{(\mu )(\nu )}_{ADM}\approx 0,
\label{II16}
\eeq

\noindent
with $P^{(\mu )}_{ADM}$, $S^{(\mu )(\nu )}_{ADM}$ related to the ADM
Poincar\'e charges\cite{adm} $P^A_{ADM}$, $J^{AB}_{ADM}$ and 10 extra
Dirac multipliers ${\tilde \lambda}_{(\mu )}(\tau )$, ${\tilde
\lambda}_{(\mu )(\nu )}(\tau )$, in front of them in the Dirac
Hamiltonian. The origin $x^{(\mu )}_{(\infty )}(\tau )$ is going to
play the role of an {\it external} decoupled observer with his
parametrized clock.

If we replace $p^{(\mu )}_{(\infty )}$ and $S^{(\mu )(\nu )}_{(\infty
)}$, whose Poisson algebra is the direct sum of an Abelian algebra of
translations and of a Lorentz algebra, with the new variables (with
indices adapted to $\Sigma_{\tau}$)

\beq
p^A_{(\infty )}=b^A_{(\infty )(\mu )}p^{(\mu )}
_{(\infty )}, \quad\quad
J^{AB}_{(\infty )}\, {\buildrel {def} \over =}\, b^A_{(\infty )(\mu
)}b^B_{(\infty )(\nu )} S^{(\mu )(\nu )}_{(\infty )} [\not=
b^A_{(\infty )(\mu )}b^B_{(\infty )(\nu )} J^{(\mu )(\nu )}_{(\infty
)}],
\label{II17}
\eeq

\noindent
the Poisson brackets for $p^{(\mu )}
_{(\infty )}$, $b^{(\mu )}_{(\infty ) A}$, $S^{(\mu )(\nu )}_{(\infty )}$, imply

\begin{eqnarray}
&&\lbrace p^A_{(\infty )},p^B_{(\infty )}\rbrace =0,\nonumber \\
&&\lbrace p^A_{(\infty )},J^{BC}_{(\infty )}\rbrace ={}^4g^{AC}_{(\infty )}p^B
_{(\infty )}-{}^4g^{AB}_{(\infty )} p^C_{(\infty )}, \nonumber \\
&&\lbrace J^{AB}
_{(\infty )},J^{CD}_{(\infty )}\rbrace =-(\delta^B_E\delta^C_F\, {}^4g^{AD}
_{(\infty )}+\delta^A_E\delta^D_F\, {}^4g^{BC}_{(\infty )}-\delta^B_E\delta^D
_F\, {}^4g^{AC}_{(\infty )}-\delta^A_E\delta^C_F\, {}^4g^{BD}_{(\infty )})J
^{EF}_{(\infty )}=\nonumber \\
&&=-C^{ABCD}_{EF} J^{EF}_{(\infty )},
\label{II18}
\end{eqnarray}

\noindent Therefore, we get the algebra of a realization of the Poincar\'e group
(this explains the notation $J^{AB}_{(\infty )}$) with all the
structure constants inverted in the sign (transition from a left to a
right action).

This  implies that, after the transition to the asymptotic Dirac
Cartesian coordinates the Poincar\'e generators ${P}^A_{ADM}$,
${J}^{AB}_{ADM}$ in $\Sigma_{\tau}$-adapted coordinates should
become a momentum ${P}^{(\mu )}_{ADM}=b_{(\infty )A}^{(\mu )}
P^A_{ADM}$ and only an ADM spin tensor ${S}^{(\mu )(\nu )}_{ADM}$
\footnote{To define an angular momentum tensor $J^{(\mu )(\nu
)}_{ADM}$ one should find an {\it external center of mass of the
gravitational field} $X^{(\mu )}_{ADM} [{}^3g, {}^3{\tilde \Pi}]$
(see Ref.\cite{lon,mate} for the Klein-Gordon case) conjugate to
$P^{(\mu )}_{ADM}$, so that $J^{(\mu )(\nu )}_{ADM}=X^{(\mu
)}_{ADM}P^{(\nu )} _{ADM}-X^{(\nu )}_{ADM}P^{(\mu )}_{ADM}+S^{(\mu
)(\nu )}_{ADM}$.}.

As shown in Ref.\cite{restmg}, the first problem with the ADM metric
gravity Hamiltonian, whose canonical part contains the
superhamiltonian and supermomentum secondary first class constraints,
is that it must be finite and differentiable\cite{reg1}. Since in
front of the secondary constraints there are the lapse and shift
functions, they are the parameters of the gauge transformations
generated by these constraints. To separate the {\it improper} gauge
transformations from the {\it proper} ones (like in Yang-Mills
theory\cite{lusa}), we shall assume the existence of a global
coordinate system $\{\sigma^{\check r} \}$ on $\Sigma_{\tau}$ , in
which we introduce the following decomposition of the lapse and shift
functions isolating their {\it asymptotic} part from the {\it bulk}
one

\begin{eqnarray}
N(\tau ,\vec \sigma )\, &=& N_{(as)}(\tau
,\vec \sigma )+m(\tau ,\vec \sigma ),\nonumber \\
N_{\check r}(\tau ,\vec \sigma )\, &=&
N_{(as) \check r}(\tau ,\vec \sigma )+m_{\check r}(\tau ,\vec \sigma ),
\nonumber \\
&&{}\nonumber \\
N_{(as)}(\tau ,\vec \sigma )&=&-{\tilde \lambda}_{(\mu )}(\tau )l^{(\mu )}
_{(\infty )}-l^{(\mu )}_{(\infty )}{\tilde \lambda}_{(\mu )(\nu )}(\tau )
b^{(\nu )}_{(\infty) \check s}(\tau ) \sigma^{\check s}=\nonumber \\
&=&-{\tilde \lambda}_{\tau}(\tau )-{1\over 2}{\tilde \lambda}_{\tau \check
s}(\tau )\sigma^{\check s},\nonumber \\
N_{(as) \check r}(\tau ,\vec \sigma )&=&-b^{(\mu )}_{(\infty ) \check r}(\tau )
{\tilde \lambda}_{(\mu )}(\tau )-b^{(\mu )}_{(\infty ) \check r}(\tau ){\tilde
\lambda}_{(\mu )(\nu )}(\tau ) b^{(\nu )}_{(\infty ) \check s}(\tau ) \sigma
^{\check s}=\nonumber \\
&=&-{\tilde \lambda}_{\check r}(\tau )-{1\over 2}{\tilde
\lambda}_{\check r\check s}(\tau ) \sigma^{\check s},
\label{II19}
\end{eqnarray}

\noindent with $m(\tau ,\vec \sigma )$, $m_{\check r}(\tau ,\vec \sigma )$,
given by Eqs.(3.8) of Ref.\cite{restmg}\footnote{They were obtained in
Ref.\cite{reg1} to ensure the differentiability of the Dirac
Hamiltonian of metric gravity. They satisfy the {\it parity
conditions} but in general they still contain odd supertranslations
({\it direction-dependent translations}, which are an obstacle to the
definition of angular momentum in general relativity as shown in the
review part of Ref.\cite{restmg}). Let us remark that the restriction
(\ref{II19}) on the lapse and shift functions of metric gravity is
analogous to the one introduced for the gauge parameters of the gauge
transformations generated by the Yang-Mills Gauss laws in gauge
theories \cite{lusa} (see Eq.(3.1) of Ref.\cite{restmg}).} and with
the asymptotic parts $N_{(as)}$, $N_{(as)\check r}$ equal to the lapse
and shift functions associated with Minkowski hyperplanes\footnote{

In general, there is the problem that in the gauges where ${\tilde
\lambda}_{(\mu )(\nu )}(\tau )$ are
different from zero the foliations with leaves $\Sigma_{\tau}$
associated to arbitrary 3+1 splittings of Minkowski spacetime are
geometrically {\it ill-defined} at spatial infinity so that the
variational principle describing the isolated system could make sense
only for those 3+1 splittings having these part of the Dirac's
multipliers vanishing. The problem is that, since on hyperplanes
${\dot l}^{(\mu )}=0$ and $l^{(\mu )}\, b_{\check r(\mu )}(\tau )=0$
imply $l^{(\mu )} {\dot b}_{\check r(\mu )}(\tau )=0$, then the
analogue of Eqs.(\ref{II19}) implies ${\tilde \lambda}_{\tau \check
r}(\tau )=0$ (i.e. only three ${\tilde \lambda}_{(\mu )(\nu )}(\tau )$
independent) on spacelike hyperplane, because otherwise Lorentz boosts
can create crossing of the leaves of the foliation. This points toward
the necessity of making  the reduction from arbitrary spacelike
hypersurfaces either directly to the Wigner hyperplanes or to
spacelike hypersurfaces approaching asymptotically Wigner
hyperplanes.}.

This very strong assumption implies that one is selecting
asymptotically at spatial infinity only coordinate systems in
which the lapse and shift functions have behaviours similar to
those of Minkowski spacelike hyperplanes, so that the {\it allowed
foliations} of the 3+1 splittings of the spacetime $M^4$ are
restricted to have the leaves $\Sigma_{\tau}$ approaching these
Minkowski hyperplanes at spatial infinity. But this is coherent
with Dirac's choice of asymptotic Cartesian coordinates (modulo
3-pseudo-diffeomorphisms not changing the nature of the
coordinates, namely tending to the identity at spatial infinity
like in Ref.\cite{reg2}). The request that  the decomposition
(\ref{II19}) holds in all the allowed coordinate systems of $M^4$
is also needed to eliminate coordinate transformations not
becoming the identity at spatial infinity, which are not
associated with the gravitational fields of isolated systems
\cite{ll}.

By replacing the ADM configuration variables $N(\tau ,\vec \sigma )$ and
$N_{\check r}(\tau ,\vec \sigma )$ with the new ones
${\tilde \lambda}_A(\tau )=\{ {\tilde \lambda}_{\tau}(\tau );
{\tilde \lambda}_{\check r}(\tau ) \}$, ${\tilde \lambda}_{AB}(\tau )=
-{\tilde \lambda}_{BA}(\tau )$, $n(\tau ,\vec \sigma )$, $n_{\check r}(\tau
,\vec \sigma )$ \footnote{We are using the notation $n$, $n_{\check
r}$ for $m$, $m_{\check r}$, because after the elimination of
supertranslation this the notation used in the rest of the paper.}
inside the ADM Lagrangian, one only gets the replacement of the
primary first class constraints of ADM metric gravity

\beq
{\tilde \pi}^N(\tau ,\vec \sigma )\approx 0,\quad\quad {\tilde \pi}
_{\vec N}^{\check r}(\tau ,\vec \sigma )\approx 0,
\label{II20}
\eeq

\noindent
with the new first class constraints

\beq
{\tilde \pi}^n(\tau ,\vec \sigma )\approx 0,\quad {\tilde \pi}
_{\vec n}^{\check r}(\tau ,\vec \sigma )\approx 0,\quad {\tilde \pi}^A(\tau )
\approx 0,\quad {\tilde \pi}^{AB}(\tau )=-{\tilde \pi}^{BA}(\tau )\approx 0,
\label{II21}
\eeq

\noindent
corresponding to the vanishing of the canonical momenta ${\tilde
\pi}^A$, ${\tilde \pi}^{AB}$ conjugate to the new configuration variables
\footnote{We assume the Poisson
brackets $\{ {\tilde \lambda}_A(\tau ),{\tilde \pi}^B(\tau )\}
=\delta^B_A$, $\{ {\tilde \lambda}_{AB}(\tau ), {\tilde \pi}^{CD}(\tau
) \} =\delta^C_A \delta^D_B-\delta^D_A \delta^C_B$.}. The only change in the
Dirac Hamiltonian of metric gravity $H_{(D)ADM}=H_{(c)ADM}+\int
d^3\sigma [\lambda_N{\tilde \pi}^N+\lambda^{\vec N}_{\check r}{\tilde
\pi}^{\check R}_{\vec N}](\tau ,\vec \sigma )$, $H_{(c)ADM}=\int d^3\sigma
[N{\tilde {\cal H}}+N_{\check r}{\tilde {\cal H}}^{\check r}](\tau
,\vec \sigma )$   is

\beq
\int d^3\sigma [\lambda_N {\tilde \pi}^N+\lambda
^{\vec N}_{\check r} {\tilde \pi}_{\vec N}^{\check r}](\tau ,\vec \sigma )\,
\mapsto \,
\zeta_A(\tau ) {\tilde \pi}^A(\tau )+\zeta_{AB}(\tau ) {\tilde \pi}^{AB}(\tau )
+\int d^3\sigma [\lambda_n {\tilde \pi}^n+\lambda^{\vec n}_{\check r} {\tilde
\pi}_{\vec n}^{\check r}](\tau ,\vec \sigma ),
\label{II22}
\eeq

\noindent
with $\zeta_A(\tau )$, $\zeta_{AB}(\tau )$ Dirac's multipliers.

The  presence of the terms $N_{(as)}$, $N_{(as)\check r}$ in
Eq.(\ref{II19}) makes $H_D$ not differentiable. In
Refs,\cite{reg,reg1}, following Refs.\cite{witt,dew}, it is shown
that the differentiability of the ADM canonical Hamiltonian
requires the introduction of a surface term
$H_{\infty}$\footnote{As shown in Ref.\cite{hh} $H_{\infty }$
contains the surface integrals neglected going from the Hilbert
action to the ADM action and then from the ADM action to the ADM
canonical Hamiltonian with the Legendre transformation.}
[$H_{(c)ADM}\, \rightarrow {\hat H}_{(c)ADM}+H_{\infty}$]. By
putting $N=N_{(as)}$, $N_{\check r}=N_{(as)\check r}$ into these
surface integrals,  the added term $H_{\infty}$ becomes the linear
combination ${\tilde \lambda}_A(\tau ) P^A_{ADM} + {1\over 2}
{\tilde \lambda}_{AB}(\tau )J^{AB}_{ADM}$ of the {\it strong ADM
Poincar\'e charges} $P^A_{ADM}$, $J^{AB}_{ADM}$ \cite{reg,reg1}
first identified in the linearized theory \cite{adm} (they are the
analogue of the strong Yang-Mills non-Abelian charges
\cite{lusa}):

\begin{eqnarray}
P^{\tau}_{ADM}&=&{{\epsilon c^3}\over {16\pi G}} \int_{S^2_{\tau
,\infty}}d^2\Sigma_{\check u} [\sqrt{\gamma}\,\, {}^3g^{\check
u\check v}\, {}^3g^{\check r\check s} (\partial_{\check r}\,
{}^3g_{\check v\check s}-\partial_{\check v}\, {}^3g _{\check
r\check s})](\tau ,\vec \sigma ),\nonumber \\
 P^{\check r}_{ADM}&=&-2 \int_{S^2_{\tau ,\infty}}d^2\Sigma _{\check u} \,
 {}^3{\tilde \Pi}^{\check r\check u}(\tau ,\vec \sigma ), \nonumber \\
  J_{ADM}^{\tau \check r}&=&{{\epsilon c^3}\over {16\pi G}} \int_{S^2_{\tau
,\infty}}d^2\Sigma_{\check u} \sqrt{\gamma}\,\, {}^3g^{\check
u\check v}\, {}^3g^{\check n\check s}\cdot \nonumber \\
 &\cdot&[\sigma^{\check r} (\partial_{\check n}\, {}^3g_{\check v\check
s}-\partial_{\check v}\, {}^3g_{\check n\check s})+\delta^{\check
r}_{\check v} ({}^3g_{\check n\check s}-\delta_{\check n\check
s})-\delta^{\check r} _{\check n}({}^3g_{\check s\check
v}-\delta_{\check s\check v})] (\tau ,\vec \sigma ),\nonumber \\
J_{ADM}^{\check r\check s}&=&\int_{S^2_{\tau
,\infty}}d^2\Sigma_{\check u} [\sigma^{\check r}\, {}^3{\tilde
\Pi}^{\check s\check u}- \sigma^{\check s}\, {}^3{\tilde
\Pi}^{\check r\check u}] (\tau ,\vec \sigma ),\nonumber \\
 &&{}\nonumber \\
 &&{}\nonumber \\
 P^{(\mu )}_{ADM}&=&
l^{(\mu )} P^{\tau}_{ADM}+b^{(\mu )}_{(\infty ) \check r}(\tau ) P^{\check
r}_{ADM}=b^{(\mu )}_{(\infty )A}(\tau ) P^A_{ADM},\nonumber \\
S^{(\mu )(\nu )}_{ADM}&=&
[l^{(\mu )}_{(\infty )}b^{(\nu )}_{(\infty
) \check r}(\tau )-l^{(\nu )}_{(\infty )}b^{(\mu )}_{(\infty ) \check r}(\tau
)] J^{\tau \check r}_{ADM}+\nonumber \\
&+&[b^{(\mu )}_{(\infty ) \check r}(\tau )b^{(\nu )}_{(\infty ) \check s}(\tau
)-b^{(\nu )}_{(\infty ) \check r}(\tau )b^{(\mu )}_{(\infty ) \check s}(\tau )]
J^{\check r\check s}_{ADM}=\nonumber \\
&=&[b^{(\mu )}_{(\infty )A}(\tau )b^{(\nu )}_{(\infty )B}(\tau )-b^{(\nu )}
_{(\infty )A}(\tau )b^{(\mu )}_{(\infty )B}(\tau )] J^{AB}_{ADM}.
\label{II23}
\end{eqnarray}

\noindent Here $J^{\tau \check r}_{ADM}=-J^{\check r\tau}_{ADM}$ by definition
and the inverse asymptotic tetrads are defined by $b^A_{(\infty )(\mu )}
b^{(\nu )}_{(\infty )B}=\delta^A_B$, $b^A_{(\infty )(\mu )}b^{(\nu )}
_{(\infty )A}=\delta^{(\nu )}_{(\mu )}$.
As shown in Ref.\cite{reg,reg1}, the {\it parity conditions}  of
Ref.\cite{reg1} are necessary to have a well defined and finite
3-angular-momentum $J_{ADM}^{\check r\check s}$. Moreover in
Ref.\cite{reg1} it is noted that, with the boundary conditions of
Refs. \cite{reg}, a regularization of the boosts is needed.

In Ref.\cite{restmg} it is shown that the surface term $H_{\infty}$
arises from a suitable splitting of the superhamiltonian and
supermomentum constraints of metric gravity.

By using Eqs.(\ref{II19}) the modified canonical Hamiltonian becomes

\begin{eqnarray}
{\hat H}_{(c)ADM}&=&\int d^3\sigma [N {\tilde {\cal H}}+N_{\check r}\,
{}^3{\tilde {\cal H}}^{\check r}](\tau ,\vec \sigma ) =\nonumber \\
&=&\int d^3\sigma [(N_{(as)}+m) {\tilde {\cal H}}+(N_{(as)\check r}+
m_{\check r})\, {}^3{\tilde {\cal H}}^{\check r}](\tau ,\vec \sigma )
\mapsto\nonumber \\
\mapsto {\hat H}^{'}_{(c)ADM}&&={\hat H}^{'}_{(c)ADM}[N,N^{\check r}]=
{\hat H}_{(c)ADM}+H_{\infty}=\nonumber \\
 &=&\int d^3\sigma
[(N_{(as)}+m) {\tilde {\cal H}}+(N_{(as)\check r}+ m_{\check r})\,
{}^3{\tilde {\cal H}}^{\check r}](\tau ,\vec \sigma )+
\nonumber \\
&+&{\tilde \lambda}_{(\mu )}(\tau ) P^{(\mu )}_{ADM}+
{\tilde \lambda}_{(\mu )(\nu )}(\tau ) S^{(\mu )(\nu )}_{ADM}=\nonumber \\
&=&\int d^3\sigma [(N_{(as)}+m) {\tilde {\cal H}}+(N_{(as)\check r}+
m_{\check r})\, {}^3{\tilde {\cal H}}^{\check r}](\tau ,\vec \sigma )+
\nonumber \\
&+&{\tilde \lambda}_A(\tau )P^A_{ADM}+{1\over 2}{\tilde
\lambda}_{AB}(\tau ) J^{AB}_{ADM} = \nonumber \\
&=& \int d^3\sigma [ m {\tilde {\cal H}}+m_{\check r} {}^3{\tilde
{\cal H}}^{\check r}](\tau ,\vec \sigma )+ {\tilde \lambda}_{(\mu
)}(\tau ) {\hat P}^{(\mu )}_{ADM}+{\tilde \lambda}
_{(\mu )(\nu )}(\tau ) {\hat S}^{(\mu )(\nu )}_{ADM} =\nonumber \\
&=&\int d^3\sigma [ m {\tilde {\cal H}}+m_{\check r}\,
{}^3{\tilde {\cal H}}^{\check r}](\tau ,\vec \sigma )+
{\tilde \lambda}_A(\tau ) {\hat P}^A_{ADM}+{1\over 2}{\tilde \lambda}
_{AB}(\tau ){\hat J}^{AB}_{ADM}, \nonumber \\
{\hat H}^{'}_{(D)ADM}&=&{\hat H}^{'}_{(c)ADM}[m,m^{\check
r}]+\nonumber \\ &+&\int d^3\sigma [\lambda_n {\tilde
\pi}^n+\lambda^{\vec n}_r {\tilde \pi}^r_{\vec n}](\tau ,\vec \sigma )+
\zeta_A(\tau ) {\tilde \pi}^A(\tau )+\zeta_{AB}(\tau ) {\tilde \pi}
^{AB}(\tau ).
\label{II24}
\end{eqnarray}

In  the last expression we introduced  the {\it weak conserved
improper Poincar\'e charges} ${\hat P}^A_{ADM}$, ${\hat J}_{ADM}^{AB}$
\footnote{These volume expressions (the analogue of the weak
Yang-Mills non Abelian charges) for the ADM 4-momentum are used in
Ref.\cite{positive} in the study of the positiviteness of the energy;
the weak charges are Noether charges.}

\begin{eqnarray}
{\hat P}^{\tau}_{ADM}&=& \int d^3\sigma \epsilon [ {{c^3}\over
{16\pi G}}\sqrt{\gamma}\,\, {}^3g^{\check r\check
s}({}^3\Gamma^{\check u} _{\check r\check v}\, {}^3\Gamma^{\check
v}_{\check s\check u}-{}^3\Gamma^{\check u}_{\check r\check s}\,
{}^3\Gamma^{\check v} _{\check v\check u})-\nonumber \\
 &-&{{8\pi G}\over {c^3\, \sqrt{\gamma} }} {}^3G_{\check r\check s\check u\check v}\,
{}^3{\tilde \Pi}^{\check r\check s}\, {}^3{\tilde \Pi}^{\check
u\check v}](\tau ,\vec \sigma ),\nonumber \\
 {\hat P}^{\check r}_{ADM}&=&- 2\int d^3\sigma \, {}^3\Gamma ^{\check r}_{\check
s\check u}(\tau ,\vec \sigma )\, {}^3{\tilde \Pi}^{\check s\check
u}(\tau ,\vec \sigma ),\nonumber \\
 {\hat J}^{\tau \check r}_{ADM}&=&-{\hat J}^{\check r\tau}_{ADM}= \int d^3\sigma
\epsilon \{ \sigma^{\check r}\nonumber \\
 &&[ {{c^3}\over {16\pi G}} \sqrt{\gamma}\,\,
{}^3g^{\check n\check s}({}^3\Gamma^{\check u}_{\check n\check
v}\, {}^3\Gamma^{\check v}_{\check s\check u}-{}^3\Gamma^{\check
u} _{\check n\check s}\, {}^3\Gamma^{\check v}_{\check v\check
u})-{{8\pi G}\over {c^3\, \sqrt{\gamma}}} {}^3G_{\check n\check
s\check u\check v}\, {}^3{\tilde \Pi}^{\check n\check s}\,
{}^3{\tilde \Pi}^{\check u\check v}]+\nonumber \\
 &+&  {{c^3}\over {16\pi G}}
\delta^{\check r}_{\check u}({}^3g_{\check v\check s}-\delta
_{\check v\check s})
\partial_{\check n}[\sqrt{\gamma}({}^3g^{\check n\check s} \,
{}^3g^{\check u\check v}-{}^3g^{\check n\check u}\, {}^3g^{\check
s\check v})] \} (\tau ,\vec \sigma ),\nonumber \\ {\hat J}^{\check
r\check s}_{ADM}&=& \int d^3\sigma  [(\sigma^{\check r}\,
{}^3\Gamma^{\check s} _{\check u\check v}-\sigma^{\check s}\,
{}^3\Gamma^{\check r}_{\check u\check v})\, {}^3{\tilde
\Pi}^{\check u\check v}](\tau ,\vec \sigma ),\nonumber \\
&&{}\nonumber \\ {\hat P}^{(\mu )}_{ADM}&=& l^{(\mu )}_{(\infty )}
{\hat P}^{\tau}_{ADM} + b^{(\mu )}_{(\infty ) \check r}(\tau )
{\hat P}^{\check r}_{ADM}=b^{(\mu )}_{(\infty )A}(\tau ) {\hat
P}^A _{ADM},\nonumber \\ {\hat S}^{(\mu )(\nu )}_{ADM}&=& [l^{(\mu
)}_{(\infty )}b^{(\nu )}_{(\infty ) \check r}(\tau )-l^{(\nu )}
_{(\infty )}b^{(\mu )}_{(\infty ) \check r}(\tau )] {\hat J}^{\tau
\check r} _{ADM} +\nonumber \\ &+&[b^{(\mu )}_{(\infty ) \check
r}(\tau )b^{(\nu )}_{(\infty ) \check s}(\tau )-b^{(\nu
)}_{(\infty ) \check r}(\tau )b^{(\mu )}_{(\infty ) \check s}(\tau
)] {\hat J}_{ADM}^{\check r\check s}=\nonumber \\ &=&[b^{(\mu
)}_{(\infty )A}b^{(\nu )}_{(\infty )B}-b^{(\nu )}_{(\infty )A}b
^{(\mu )}_{(\infty )B}](\tau ) {\hat J}^{AB}_{ADM},\nonumber \\
 &&{}\nonumber \\
P^{\tau}_{ADM}&=&{\hat P}^{\tau}_{ADM}+\int d^3\sigma {\tilde {\cal
H}}(\tau ,\vec \sigma )\approx {\hat P}^{\tau}_{ADM},\nonumber \\
P^{\check r}_{ADM}&=&{\hat P}^{\check r}_{ADM}+\int d^3\sigma \,
{}^3{\tilde {\cal H}}^{\check r}(\tau ,\vec \sigma )\approx {\hat
 P}^{\check r}_{ADM},\nonumber \\
 J^{\tau \check r}_{ADM}&=&{\hat J}^{\tau \check r}_{ADM}+{1\over 2}
\int d^3\sigma \sigma^{\check r}\, {\tilde {\cal H}}(\tau ,\vec \sigma )
\approx {\hat J}^{\tau \check r}_{ADM},\nonumber \\
J^{\check r\check s}_{ADM}&=&{\hat J}^{\check r\check s}_{ADM}+\int d^3\sigma
[\sigma^{\check s}\, {}^3{\tilde {\cal H}}^{\check r}(\tau ,\vec \sigma )-
\sigma^{\check r}\, {}^3{\tilde {\cal H}}^{\check s}(\tau ,\vec \sigma )]
\approx {\hat J}^{\check r\check s}_{ADM}.
\label{II25}
\end{eqnarray}

In both Refs.\cite{reg,reg1} it is shown that the canonical
Hamiltonian ${\hat H}^{'}_{(c)ADM}[N,N^{\check r}]$ of Eq.(\ref{II24})
with general $N$, $N^{\check r}={}^3g^{\check r\check s} N_{\check s}$
like the ones of Eqs.(\ref{II19}) (their asymptotic parts are the
parameters of {\it improper} gauge transformations), has the same
Poisson brackets as in the case of {\it proper} gauge transformations
with $N=m$, $N^{\check r}=m^{\check r}$ (see the third and fourth line
of Eqs.(\ref{II14}) for the universal Dirac algebra of the
superhamiltonian and supermomentum constraints).

This implies\cite{reg1,restmg}:
\hfill\break
i) the Poisson brackets of two {\it proper} gauge transformations [${\tilde
\lambda}_{iA}={\tilde \lambda}_{iAB}=0$, i=1,2] is a {\it proper} gauge
transformation [${\tilde \lambda}_{3A}={\tilde
\lambda}_{3AB}=0$];\hfill\break
 ii) if $N_2=m_2$, $N_{2\check
r}=m_{2\check r}$ [${\tilde \lambda}_{2A}= {\tilde \lambda}_{2AB}=0$]
correspond to a {\it proper} gauge transformation and $N_1,N_{1\check
r}$ [$m_1=m_{1\check r}=0$] to an {\it improper} one, then we get a
{\it proper} gauge transformation [${\tilde \lambda}_{3A}={\tilde
\lambda}_{3AB}=0$, $m_3\not= 0$, $m_{3\check r}\not= 0$]
and this may be interpreted as saying that the 10 Poincar\'e charges
are {\it gauge invariant} and {\it Noether constants of
motion}.\hfill\break
 iii) the Poisson bracket of two {\it improper}
gauge transformations [$m_i=m_{i\check r}=0$, i=1,2] is an {\it
improper} gauge transformation [${\tilde \lambda}_{3A}\not= 0$,
${\tilde \lambda}_{3AB}\not= 0$, $m_3\not= 0$, $m_{3\check r}\not=
0$]. This implies that the 10 strong Poincar\'e charges (and,
therefore, also the weak ones) satisfy the Poincar\'e algebra modulo
the first class constraints, namely modulo the Hamiltonian group of
gauge transformations

\begin{eqnarray}
\lbrace {\hat P}^A_{ADM},{\hat P}^B_{ADM} \rbrace &=& 0,
\nonumber \\
\lbrace {\hat P}^A_{ADM},{\hat J}^{BC}_{ADM} \rbrace &\approx& {}^4\eta^{AC}
{\hat P}^B_{ADM}-{}^4\eta^{AB} {\hat P}^C_{ADM},
\nonumber \\
\lbrace {\hat J}^{AB}_{ADM},{\hat J}^{CD}_{ADM}
\rbrace &\approx& - C^{ABCD}_{EF} {\hat J}^{EF}_{ADM},\nonumber \\
&&\Downarrow \nonumber \\
 &&{}\nonumber \\
  \lbrace P^A_{ADM},P^B_{ADM} \rbrace &\approx& 0,\nonumber \\
   \lbrace P^A_{ADM}, J^{BC}_{ADM} \rbrace &\approx& {}^4\eta^{AC} P^B_{ADM}-
{}^4\eta^{AB} P^C_{ADM},\nonumber \\
 \lbrace J^{AB}_{ADM},J^{CD}_{ADM} \rbrace &\approx& - C^{ABCD}_{EF} J^{EF}
_{ADM},
\label{II26}
\end{eqnarray}

\noindent in accord with Eqs. (\ref{II18}).

As shown in Ref.\cite{restmg}, the requirement of absence of
supertranslations, implying the existence of a well defined asymptotic
Poincar\'e group, may be satisfied by restricting all the fields to
have a {\it direction-independent} limit at spatial infinity. Let us
call $n(\tau ,\vec \sigma )$, $n_{\check r}(\tau ,\vec \sigma )$ the
lapse and shift functions $m(\tau ,\vec \sigma )$, $m_{\check r}(\tau
,\vec \sigma )$ with such a behaviour. In a suitable class ${\cal C}$
of coordinate systems for $M^4$
\footnote{Then transformed to  coordinates adapted to the 3+1
splitting of $M^4$ with an allowed  foliation with spacelike leaves
$\Sigma_{\tau}$, whose allowed coordinates systems are in the
previously defined atlas ${\cal C}_{\tau}$.} asymptotic to Minkowski
coordinates and with the general coordinate transformations suitably
restricted at spatial infinity so that it is not possible to go out
this class (they are {\it proper} gauge transformations which do not
introduce asymptotic angle-dependence), we must assume the following
direction-independent boundary conditions for the ADM variables for
$r\, \rightarrow \, \infty$ [$\epsilon > 0$]

\begin{eqnarray}
{}^3g_{\check r\check s}(\tau ,\vec \sigma )&=&(1+{M\over r})
\delta_{\check r\check s}+{}^3h_{\check r\check s}(\tau ,\vec \sigma ),
\quad\quad {}^3h_{\check r\check s}(\tau ,\vec \sigma )
 \, {\rightarrow}_{r \rightarrow \infty}\, O(r^{-(1+\epsilon )}),
\nonumber \\
{}^3{\tilde \Pi}^{\check r\check s}(\tau ,\vec \sigma
 )& \, {\rightarrow}_{r \rightarrow \infty}\,& O(r^{-(2+\epsilon )}),\nonumber \\
 &&{}\nonumber \\
  N(\tau ,\vec \sigma )&=& N_{(as)}(\tau ,\vec
\sigma ) +n(\tau ,\vec \sigma ),\quad\quad n(\tau ,\vec
\sigma )\,  \, {\rightarrow}_{r \rightarrow \infty}\, O(r^{-(2+\epsilon )}),\nonumber \\
N_{\check r}(\tau ,\vec \sigma )&=&N_{(as)\check r}(\tau ,\vec \sigma
)+ n_{\check r}(\tau ,\vec \sigma ),\quad\quad n_{\check r}(\tau ,\vec
\sigma )\,   \, {\rightarrow}_{r \rightarrow \infty}\,O(r^{-\epsilon}),\nonumber \\
 &&{}\nonumber \\
 N_{(as)}(\tau ,\vec \sigma )&=&
-{\tilde \lambda}_{\tau}(\tau )-{1\over 2}{\tilde \lambda}_{\tau \check
s}(\tau )\sigma^{\check s},\nonumber \\
N_{(as)\check r}(\tau ,\vec \sigma )&=&
-{\tilde \lambda}_{\check r}(\tau )-{1\over 2}{\tilde \lambda}_{\check r\check
s}(\tau ) \sigma^{\check s},\nonumber \\
\Rightarrow&& N_{(as)A}(\tau ,\vec \sigma )\, {\buildrel {def} \over =}\,
(N_{(as)}\, ;\, N_{(as) \check r}\, )(\tau ,\vec \sigma )
=-{\tilde \lambda}_A(\tau )-{1\over 2}{\tilde \lambda}_{A\check s}(\tau )
\sigma^{\check s},
\label{II27}
\end{eqnarray}

\noindent in accord with Regge-Teitelboim\cite{reg} and Beig-O'Murchadha
\cite{reg1}.

This implies the {\it vanishing} of the ADM momentum, $P^{\check
r}_{ADM}=0$, so that the {\it elimination of supertranslations} is
connected with a definition of {\it rest frame} in the asymptotic
Dirac coordinates $z^{(\mu )}_{(\infty )}(\tau ,\vec \sigma )$.
Therefore, the previous boundary conditions on ${}^3g$, ${}^3{\tilde
\Pi}$, are compatible and can be replaced with the
Christodoulou-Klainermann ones\cite{ckl}, but in general with {\it non
vanishing shift functions}.

The vanishing of the strong ADM 3-momentum $P^{\check r}_{ADM}=0$ and
Eq.(\ref{II25}) imply

\begin{eqnarray}
{\hat P}^{\check r}_{ADM}&\approx& 0,\nonumber \\
 &&{}\nonumber \\
P^{(\mu )}_{ADM}&=&b^{(\mu )}_{(\infty )\tau} P^{\tau}_{ADM}=l^{(\mu )}
_{(\infty )} P^{\tau}_{ADM},\nonumber \\
{\hat P}^{(\mu )}_{ADM}&\approx& l^{(\mu )}_{(\infty )} {\hat P}^{\tau}_{ADM}.
\label{II28}
\end{eqnarray}

Therefore, the boundary conditions (\ref{II27}) require three first
class constraints implying the vanishing of the weak ADM 3-momentum as
a {\it rest frame} condition.

Therefore, to have a formulation of metric gravity in which all the
fields and the gauge transformations have an angle-independent limit
at spatial infinity we have to add 6 gauge fixings on the $b^{(\mu
)}_{(\infty )A}(\tau )$ [see later on Eqs.(\ref{II39})] like we do in
parametrized Minkowski theory for going from arbitrary spacelike
hyperplanes to the Wigner ones (orthogonal to $p^{(\mu )}_s\approx
P^{(\mu )}_{sys}$, where $P^{(\mu )}_{sys}$ is the 4-momentum of the
isolated system under study): only on them we get the constraints
${\vec P}_{sys}\approx 0$ giving the rest-frame conditions.

Let us call Wigner-Sen-Witten (WSW) the so selected allowed spacelike
hypersurfaces $\Sigma^{(WSW)}_{\tau}$ (see Section  XII of
Ref.\cite{restmg} and Section VII for the justification of the name)
asymptotically orthogonal to the weak ADM 4-momentum. Since $b^{(\mu
)}_{(\infty )A}={{\partial z^{(\mu )}_{(\infty )}(\sigma )}\over
{\partial
\sigma^A}}$, this is a strong restriction on the coordinate systems
$x^{\mu}=z^{\mu}(\tau ,\vec
\sigma )\, \rightarrow \, \delta^{\mu}_{(\mu )} z^{(\mu )}
_{(\infty )}(\tau ,\vec \sigma )$ of $M^4$, which can be reached
from the $\Sigma^{(WSW)}_{\tau}$-adapted coordinates $\sigma^A=(\tau
,\vec
\sigma )$ without introducing asymptotic angle dependence (namely
supertranslations).

With these assumptions one has the following form of the line element
(it becomes Minkowskian in cartesian coordinates at spatial infinity)

\begin{eqnarray}
ds^2&=& \epsilon \Big( [N_{(as)}+n]^2 - [N_{(as)\check r}+n_{\check r}]
{}^3g^{\check r\check s}[N_{(as)\check s}+n_{\check s}] \Big) (d\tau )^2-
\nonumber \\
&-&2\epsilon [N_{(as)\check r}+n_{\check r}] d\tau d\sigma^{\check r} -
\epsilon \, {}^3g_{\check r\check s} d\sigma^{\check r}
d\sigma^{\check s}\Big) =
\nonumber \\
&=&\epsilon \Big( [N_{(as)}+n]^2 (d\tau )^2- \nonumber \\
&-&{}^3g^{\check r\check s} [{}^3g_{\check r\check u}d\sigma^{\check
u}+(N_{(as)\check r}+n_{\check r}) d\tau ] [{}^3g_{\check s\check
v}d\sigma^{\check v}+(N_{(as)\check s}+n_{\check s}) d\tau ]\Big) .
\label{II29}
\end{eqnarray}

Since we have ${\dot x}^{(\mu )}_{(\infty )}(\tau )\, {\buildrel \circ
\over =}\, b^{(\mu )}_{(\infty )A}(\tau ) {\tilde \lambda}^A(\tau )$, it
follows that for ${\tilde \lambda}_{\tau}(\tau )=\epsilon$, ${\tilde
\lambda}_r(\tau )=0$, the point ${\tilde x}^{(\mu )}_{(\infty )}(\tau )$
moves with 4-velocity $(\epsilon ;\vec 0)$ and has attached an
accelerated rotating coordinate system\cite{stephani}, which becomes
inertial when ${\tilde \lambda}_{AB}(\tau )=0$,namely when the
foliations become geometrically well defined at spatial infinity.

As a consequence of what has been said and of Eqs.(\ref{II27}), in
the allowed coordinate atlases ${\cal C}$ of $M^4$ and ${\cal
C}_{\tau}$ of $\Sigma_{\tau}$ the function space ${\cal W}$ (an
appropriate weighted Sobolev space as for Yang-Mills
theory\cite{lusa}) is needed for the field variables
${}^3g_{\check r\check s}(\tau ,\vec \sigma )$, ${}^3{\tilde
\Pi}^{\check r\check s}(\tau ,\vec \sigma )$, $n(\tau ,\vec \sigma
)$, $n_{\check r}(\tau ,\vec \sigma )$ and for the parameters
$\alpha (\tau ,\vec \sigma )$, $\alpha_{\check r}(\tau ,\vec
\sigma )$ (of which $n(\tau ,\vec \sigma )$, $n_{\check r}(\tau
,\vec \sigma )$ are special cases) of {\it allowed proper} gauge
transformations connected to the identity, should be defined by
angle-independent boundary conditions for $r\, \rightarrow \infty$
of the following form:

\begin{eqnarray}
&&{}^3g_{\check r\check s}(\tau ,\vec \sigma )\, {\rightarrow}_{r\,
\rightarrow \infty}\, (1+{M\over r})
\delta_{\check r\check s}+{}^3h_{\check r\check s}(\tau
,\vec \sigma )=(1+{M\over r})
\delta_{\check r\check s}+O(r^{-3/2}),\nonumber \\
&&{}^3{\tilde \Pi}^{\check r\check s}(\tau ,\vec \sigma )\, {\rightarrow}
_{r\, \rightarrow \infty}\, {}^3k^{\check r\check s}(\tau ,\vec \sigma )=
O(r^{-5/2}),\nonumber \\ &&n(\tau ,\vec \sigma )\, {\rightarrow}_{r\,
\rightarrow \infty}\, O(r^{-(2+\epsilon )}),\quad \epsilon > 0,\nonumber \\
&&n_{\check r}(\tau ,\vec \sigma )\,
{\rightarrow}_{r\, \rightarrow \infty}\, O(r^{-\epsilon}),\quad
\epsilon > 0,\nonumber \\
&&{\tilde \pi}_n(\tau ,\vec \sigma )\, {\rightarrow}_{r\,
\rightarrow \infty}\, O(r^{-3}),\nonumber \\
&&{\tilde \pi}^{\check r}_{\vec n}(\tau
,\vec \sigma )\, {\rightarrow}_{r\, \rightarrow \infty}\, O(r^{-3}),
\nonumber \\
&&\lambda_n(\tau ,\vec \sigma )\, {\rightarrow}_{r\, \rightarrow \infty}\,
O(r^{-(3+\epsilon )}),\nonumber \\
&&\lambda^{\vec n}_{\check r}(\tau ,\vec \sigma )\, {\rightarrow}_{r\,
\rightarrow \infty}\, O(r^{-\epsilon}),\nonumber \\
&&\alpha (\tau ,\vec \sigma )\, {\rightarrow}_{r\, \rightarrow
\infty}\, O(r^{-(2+\epsilon )}),\nonumber \\ &&\alpha_{\check r}(\tau
,\vec \sigma )\, {\rightarrow}_{r\, \rightarrow
\infty}\, O(r^{-\epsilon}),\nonumber \\
&&\Downarrow \nonumber \\
&&{\tilde {\cal H}}(\tau ,\vec \sigma )\, {\rightarrow}_{r\, \rightarrow
\infty}\, O(r^{-3}),\nonumber \\
&&{}^3{\tilde {\cal H}}^{\check r}(\tau ,\vec \sigma )\, {\rightarrow}_{r\,
\rightarrow \infty}\, O(r^{-3}).
\label{II30}
\end{eqnarray}

With these boundary conditions we have $\partial_{\check u}\,
{}^3g_{\check r\check s}=O(r^{-2})$ and not $O(r^{-(1+\epsilon )})$;
this is compatible with the definition of gravitational radiation
given by Christodoulou and Klainermann\cite{ckl}, but not with the one
of Ref.\cite{trautm}.

In this function space ${\cal W}$ supertranslations are not allowed by
definition and proper gauge transformations generated by the secondary
constraints map ${\cal W}$ into itself. A coordinate-independent
characterization of ${\cal W}$ (see Ref.\cite{reg4} for an attempt)
should be given through an intrinsic definition of a minimal atlas of
coordinate charts ${\cal C}_{\tau}$ of $\Sigma_{\tau}$ such that the
lifts to 3-tensors on $\Sigma_{\tau}$ in ${\cal W}$ of the
3-diffeomorphisms in $Diff\, \Sigma_{\tau}$  maps them into them.

Therefore, a unique asymptotic Poincar\'e group, modulo gauge
transformations, is selected. Moreover, in accord with Anderson\cite{reg2}
also $Diff\, M^4$ is restricted to $Diff_I\, M^4
\times {\cal P}_{(\infty )}$. Now in $Diff_I\, M^4\times {\cal P}_{(\infty )}$
the allowed  proper diffeomorphisms $Diff_I\, M^4$ are a normal
subgroup (they go to the identity in an angle-independent way at
spatial infinity), while the Poincar\'e group ${\cal P}_{(\infty )}$
describes the rigid improper gauge transformations (the non-rigid
improper ones are assumed to be absent) as in Bergmann''s
proposal\cite{be}. Finally, following Marolf\cite{p18}, the Poincar\'e
group ${\cal P}_{(\infty )}$ is not interpreted as a group of improper
gauge transformations but only as a source of superselection rules
(like it happens for the vanishing of the color charges for the
confinement of quarks), which however  are consistent only in the rest
frame $P^{\check r}_{ADM}=0$.

Since in Ref.\cite{restmg} it was shown that the gauge
transformations generated by the superhamiltonian constraint
produce a change in the extrinsic curvature of the spacelike
hypersurface $\Sigma_{\tau}$ transforming it in a different
spacelike hypersurface, one has the indication that, in absence of
supertranslations, the functions $N$, $\alpha$, $\lambda_N$,
should go like $O(r^{-(2+\epsilon )})$ and not like
$O(r^{-\epsilon })$ (in the case of proper gauge transformations).

The previous discussion points toward assuming the following Dirac
Hamiltonian

\begin{eqnarray}
{\hat H}^{"}_{(c)ADM} &=&\int d^3\sigma \Big[ (N_{(as)}+ n) {\tilde
{\cal H}}+(N_{(as)\check r}+ n_{\check r})\, {}^3{\tilde {\cal
H}}^{\check r}\Big] (\tau ,\vec \sigma )+
\nonumber \\
&+&{\tilde \lambda}_A(\tau ) P^A_{ADM}+{1\over 2}{\tilde
\lambda}_{AB}(\tau ) J^{AB}_{ADM}= \nonumber \\
 &=&\int d^3\sigma \Big[ n
{\tilde {\cal H}}+n_{\check r}\, {}^3{\tilde {\cal H}}^{\check r}\Big]
(\tau ,\vec \sigma )+ {\tilde \lambda}_A(\tau ) {\hat
P}^A_{ADM}+{1\over 2}{\tilde \lambda}
_{AB}(\tau ){\hat J}^{AB}_{ADM}, \nonumber \\
{\hat H}^{"}_{(D)ADM}&=&{\hat H}^{"}_{(c)ADM}+\int d^3\sigma [\lambda_n {\tilde
\pi}^n+\lambda^{\vec n}_r {\tilde \pi}^r_{\vec n}](\tau ,\vec \sigma )+
\zeta_A(\tau ) {\tilde \pi}^A(\tau )+\zeta_{AB}(\tau ) {\tilde \pi}
^{AB}(\tau ).
\label{II31}
\end{eqnarray}

\noindent However, the criticism of footnote 39 suggests that this
Hamiltonian is well defined only in the gauges where ${\tilde
\lambda}_{AB}(\tau )=0$.

After this modification of metric gravity at the canonical level two
possible Hamiltonian scenarios can be imagined:

a) Consider as configurational variables

\beq
n_A(\tau ,\vec \sigma )=(n\, ;\, n_{\check r}\, )(\tau ,\vec \sigma
),\quad {\tilde \lambda}_A(\tau ),\quad {\tilde
\lambda}_{AB}(\tau ),\quad {}^3g_{\check r\check s}(\tau ,\vec \sigma ),
\label{II32}
\eeq

\noindent with conjugate momenta\footnote{The vanishing momenta
are assumed to be the primary constraints.}

\beq
{\tilde \pi}^A_n(\tau ,\vec \sigma )=({\tilde \pi}^n\, ;\, {\tilde
\pi}^{\check r}_{\vec n}\, )(\tau ,\vec \sigma )\approx 0,\quad {\tilde
\pi}^A(\tau )\approx 0,\quad {\tilde \pi}^{AB} (\tau )\approx 0,\quad
{}^3{\tilde \Pi}^{\check r\check s} (\tau ,\vec \sigma ),
\label{II33}
\eeq

\noindent
 and
take the following Dirac Hamiltonian (it is finite and differentiable)
as the defining Hamiltonian:

\begin{eqnarray}
{\hat H}^{(1)}_{(D)ADM}&=&\int d^3\sigma [n_A\, {\tilde {\cal H}}^A+\lambda
_{n\, A} {\tilde \pi}^A_n](\tau ,\vec \sigma )+{\tilde \lambda}_A(\tau )
{\hat P}^A_{ADM}+
{1\over 2}{\tilde \lambda}_{AB}(\tau ) {\hat J}^{AB}_{ADM}+\nonumber \\
&+&\zeta_A(\tau ) {\tilde \pi}^A(\tau )+
\zeta_{AB}(\tau ) {\tilde \pi}^{AB}(\tau ),
\label{II34}
\end{eqnarray}

\noindent where $n_A=(n; n_{\check r})$,
${\tilde {\cal H}}^A=({\tilde {\cal H}}\, ;\, {}^3{\tilde
{\cal H}}^{\check r}\, )$ and where $\lambda_{n\, A}(\tau ,\vec \sigma )=
(\lambda_n\, ;\, \lambda^{\vec n}_{\check r}\, )(\tau ,\vec \sigma )$,
$\zeta_A(\tau )$, $\zeta_{AB}(\tau )$, are
Dirac multipliers associated with the primary constraints.\hfill\break
\hfill\break
For ${\tilde \lambda}
_{AB}(\tau ) =0$, ${\tilde \lambda}_A(\tau )=\epsilon \delta_{A\tau}$, one has
\cite{witt}:

\beq
{\hat H}^{(1)}_{(D)ADM}\approx -\epsilon {\hat P}^{\tau}_{ADM}.
\label{II35}
\eeq

The time constancy of the primary constraints implies the following
secondary ones

\bea
&&{\tilde {\cal H}}^A(\tau ,\vec \sigma )\approx 0\,  \nonumber \\
 &&{\hat P}^A_{ADM}\approx 0,\quad\quad {\hat J}^{AB}_{ADM}\approx 0,
\label{II36}
\eea

\noindent all of which are constants of the motion. While the
${\tilde {\cal H}}^A$'s are generators of proper gauge
transformations, the other 10 constraints are either generators of
improper gauge transformations (in this case 10 conjugate degrees
of freedom in the 3-metric are extra gauge variables) or,
following Marolf's proposal \cite{p18}, defining a superselection
sector.  All the constraints are first class, so that:
\hfill\break i) ${\tilde \lambda}_A(\tau )$, ${\tilde
\lambda}_{AB}(\tau )$ are arbitrary gauge variables conjugate to
${\tilde \pi}^A(\tau )\approx 0$, ${\tilde \pi}^{AB}(\tau )\approx
0$ \footnote{6 gauge fixings to the constraints ${\hat
J}^{AB}_{ADM} \approx 0$ are needed to get the induced result
${\tilde \lambda}_{AB}(\tau )=0$ which ensures foliations well
defined at spatial infinity.}; \hfill\break ii) the physical
reduced phase space of canonical metric gravity is restricted to
have {\it zero asymptotic Poincar\'e charges} so that there is no
natural Hamiltonian for the evolution in $\tau$. This corresponds
to the exceptional orbit ${\hat P}^A_{ADM}=0$ of the asymptotic
Poincar\'e group.

This is the natural interpretation of ADM metric gravity which leads
to the Wheeler-De Witt equation after quantization and, in a sense, it
is a Machian formulation of an asymptotically flat noncompact (with
boundary) spacetime $M^4$ in the same spirit of Barbour's
approach\cite{p20} and of the closed (without boundary)
Einstein-Wheeler universes. However, in this case there is no solution
to the problem of deparametrization of metric gravity and no
connection with parametrized Minkowski theories restricted to
spacelike hyperplanes.\hfill\break
\hfill\break

b) According to the suggestion of Dirac, modify ADM metric gravity by
adding the 10 new canonical pairs $x^{(\mu )}_{(\infty )}(\tau )$,
$p^{(\mu )}_{(\infty )}$, $b^{(\mu )}_{(\infty ) A}(\tau )$, $S^{(\mu
)(\nu )}_{\infty}$ to the metric gravity phase space with canonical
basis $n_A(\tau ,\vec
\sigma )
=(n\, ;\, n_{\check r}\, )(\tau ,\vec \sigma )$, ${\tilde \pi}^A_n(\tau ,\vec
\sigma )=({\tilde \pi}^n; {\tilde \pi}^{\check r}_{\vec n})
\approx 0$ (the primary constraints), ${}^3g_{\check r\check s}(\tau
,\vec \sigma )$, ${}^3{\tilde \Pi}^{\check r\check s}(\tau ,\vec
\sigma )$, and then: \hfill\break i) add the 10 new primary
constraints

\bea
\chi^A &=& p^A_{(\infty )}-{\hat P}^A_{ADM}=
b^A_{(\infty )(\mu )}(\tau ) [p^{(\mu )}_{(\infty )}-b^{(\mu )}
_{(\infty )B}(\tau ) {\hat P}^B_{ADM}] \approx 0,\nonumber \\
 \chi^{AB} &=& J^{AB}_{(\infty )}-{\hat J}^{AB}_{ADM}=b^A_{(\infty )
(\mu )}(\tau ) b^B_{(\infty )(\nu )}(\tau ) [S^{(\mu )(\nu )}_{(\infty )}-
b^{(\mu )}_{(\infty )C}(\tau ) b^{(\nu )}_{(\infty )D}(\tau ) {\hat J}^{CD}
_{ADM}] \approx 0,\nonumber \\
 &&{}\nonumber \\
&&\{ \chi^A(\tau ), \chi^{BC}(\tau ) \} \approx {}^4\eta^{AC}
\chi^B(\tau )- {}^4\eta^{AB} \chi^C(\tau ) \approx 0,\quad\quad
\{ \chi^A(\tau ),\chi^B(\tau ) \} \approx 0,\nonumber \\
&&\{ \chi^{AB}(\tau ), \chi^{CD}(\tau ) \} \approx - C^{ABCD}_{EF}
\chi^{EF}(\tau )\approx 0,\nonumber \\
 &&\{ \chi^A(\tau ), {\tilde \pi}^D_n(\tau ,\vec \sigma )\}=\{ \chi^{AB}(\tau ),
{\tilde \pi}^D_n(\tau ,\vec \sigma )\}=0,\nonumber \\
 &&\{ \chi^A(\tau ),
{\tilde {\cal H}}^D(\tau ,\vec \sigma )\} \approx 0,\quad
\quad \{ \chi^{AB}(\tau ), {\tilde {\cal H}}^D(\tau ,\vec \sigma )\} \approx
0,
\label{II37}
\eea

\noindent with
$p^A_{(\infty )}$, $J^{AB}_{(\infty )}$ of
Eqs.(\ref{II17});\hfill\break ii) consider ${\tilde
\lambda}_A(\tau )$, ${\tilde \lambda}_{AB}(\tau )$, as Dirac
multipliers [like $\lambda_{n A}(\tau ,\vec \sigma )$] for these 10
new primary constraints, and not as configurational (arbitrary gauge)
variables coming from the lapse and shift functions
\footnote{So that there are no conjugate momenta ${\tilde \pi}^A(\tau )$,
${\tilde \pi}^{AB}(\tau )$ and no associated Dirac multipliers
$\zeta_A(\tau )$, $\zeta_{AB}(\tau )$.}, in the assumed Dirac
Hamiltonian (it is finite and differentiable)

\begin{eqnarray}
H_{(D)ADM}&=& \int d^3\sigma [ n_A {\tilde {\cal H}}^A+\lambda_{n A} {\tilde
\pi}^A_n](\tau ,\vec \sigma )-\nonumber \\
&-&{\tilde \lambda}_A(\tau ) [p^A_{(\infty )}-{\hat P}^A
_{ADM}]-{1\over 2}{\tilde \lambda}_{AB}(\tau )[J^{AB}_{(\infty )}-
{\hat J}^{AB}_{ADM}]\approx 0.
\label{II38}
\end{eqnarray}

The reduced phase space is the ADM one and there is consistency with
Marolf's proposal regarding superselection sectors: on the ADM
variables there are only the secondary first class constraints
${\tilde {\cal H}}^A(\tau ,\vec \sigma )
\approx 0$ (generators of proper gauge transformations), because the other
first class constraints $p^A_{(\infty )}-{\hat P}^A_{ADM}\approx 0$, $J^{AB}
_{(\infty )}-{\hat J}^{AB}_{ADM}\approx 0$ do not generate improper gauge
transformations but eliminate 10 of the extra 20 variables.  One has
an asymptotically flat at spatial infinity noncompact (with boundary
$S_{\infty}$) spacetime $M^4$ with non-vanishing asymptotic Poincar\'e
charges  and the possibility to deparametrize metric gravity so to
obtain the connection with parametrized Minkowski theories restricted
to Wigner hyperplanes.

Scenario b) contains the rest-frame instant form of ADM metric
gravity.

To go to the WSW hypersurfaces \footnote{The analogue of the Minkowski
Wigner hyperplanes with the asymptotic normal $l^{(\mu )}_{(\infty
)}=l^{(\mu )}_{(\infty )\Sigma}$ parallel to ${\hat P}
^{(\mu )}_{ADM}$ (i.e. $l^{(\mu )}_{(\infty )}={\hat b}^{(\mu )}_{(\infty ) l}=
{\hat P}^{(\mu )}_{ADM}/\sqrt{\epsilon {\hat P}^2_{ADM}}$).} one
follows the procedure defined for Minkowski spacetime:
\hfill\break

i) one restricts oneself to spacetimes with $\epsilon p^2_{(\infty )}
={}^4\eta_{(\mu )(\nu )} p^{(\mu )}_{(\infty )}p^{(\nu )}
_{(\infty )} > 0$ \footnote{This is possible, because the positivity theorems for the
ADM energy imply that one has only timelike or light-like orbits of
the asymptotic Poincar\'e group.}; \hfill\break

ii) one boosts at rest $b^{(\mu )}_{(\infty )A}(\tau )$ and $S^{(\mu
)(\nu )}_{(\infty )}$ with the Wigner boost $L^{(\mu )}{}_{(\nu )}
(p_{(\infty )}, {\buildrel \circ \over p}_{(\infty )})$; \hfill\break

iii) one adds the gauge-fixings ( $u^{(\mu )}(p_{(\infty )})=p^{(\mu )}_{(\infty )}/
\pm \sqrt{\epsilon p^2_{(\infty )} }$)

\bea
b^{(\mu )}_{(\infty )A}(\tau ) &\approx& L^{(\mu )}{}_{(\nu )=A}
(p_{(\infty )}, {\buildrel \circ \over p}_{(\infty )})=\epsilon ^{(\mu
)}_A (u(p_{(\infty )})), \nonumber \\
 &&{}\nonumber \\
  &&\text{implying}\quad\quad {\tilde \lambda}_{AB}(\tau ) =0,
\label{II39}
\eea

\noindent  to the constraints $\chi^{AB}(\tau )\approx 0$ and goes to Dirac brackets.\hfill\break
 In this way one gets

\bea
S^{(\mu )(\nu )}_{(\infty )} &\equiv& \epsilon^{(\mu )}_C(u(p_{(\infty
)}))\epsilon_D^{(\nu )}(u(p_{(\infty )})) {\hat J}^{CD}_{ADM}=S^{(\mu
)(\nu )}_{ADM},\nonumber \\
 z^{(\mu )}_{(\infty )}(\tau ,\vec \sigma )&=&x^{(\mu )}_{(\infty )}(\tau
)+\epsilon^{(\mu )}_r(u(p_{(\infty )})) \sigma^r,
\label{II40}
\eea

\noindent so that $z^{(\mu )}_{(\infty )}(\tau ,\vec \sigma )$ becomes
equal to the embedding identifying a Wigner hyperplane in Minkowski
spacetime.

The origin $x^{(\mu )}_{(\infty )}$ is now replaced by the not
covariant {\it external} center-of-mass canonical variable

\beq
{\tilde x}^{(\mu )}_{(\infty )}=x^{(\mu )}_{(\infty )}+{1\over 2}
\epsilon^A_{(\nu )}(u(p_{(\infty )})) \eta_{AB} {{\partial \epsilon^B_{(\rho )}(u(p
_{(\infty )}))}\over {\partial p_{(\infty )(\mu )}}} S^{(\nu )(\rho )}
_{(\infty )},
 \label{II41}
 \eeq

\noindent and one has

\beq
J^{(\mu )(\nu )}_{(\infty )}= {\tilde x}^{(\mu )}_{(\infty )}p^{(\nu
)}_{(\infty )}-{\tilde x}^{(\nu )}
_{(\infty )}p^{(\mu )}_{(\infty )}+{\tilde S}^{(\mu )(\nu )}_{(\infty )},
\label{II42}
\eeq

\noindent
with ${\tilde S}^{(\mu )(\nu )}_{(\infty )}=S^{(\mu )(\nu )}_{(\infty )}-
{1\over 2} \epsilon^A_{(\rho )}(u(p_{(\infty )})) \eta_{AB} ({{\partial
\epsilon^B_{(\sigma )}(u(p_{(\infty )}))}\over {\partial p_{(\infty )(\mu )}}}
p^{(\nu )}_{(\infty )}-{{\partial \epsilon^B_{(\sigma )}(u(p_{(\infty )}))}
\over {\partial p_{(\infty )(\nu )}}} p^{(\mu )}_{(\infty )} ) S^{(\rho
)(\sigma )}_{(\infty )}$. \hfill\break
\hfill\break
As in the Minkowski case one defines

\beq
{\bar S}^{AB}_{(\infty )}=\epsilon^A_{(\mu )} (u(p_{(\infty
)}))\epsilon^B_{(\nu )}(u(p_{(\infty )})) {\tilde S}^{(\mu ) (\nu
)}_{(\infty )},
 \label{II43}
 \eeq

\noindent and one obtains at the level of Dirac brackets [$\epsilon_{(\infty )}
=-\epsilon \sqrt{\epsilon p^2_{(\infty )}}$]

\begin{eqnarray}
{\bar S}^{\check r\check s}_{(\infty )}&\equiv& {\hat J}^{\check r\check s}
_{ADM},\qquad {\tilde \lambda}_{AB}(\tau )=0,\nonumber \\
 &&{}\nonumber \\
-{\tilde \lambda}_A(\tau ) \chi^A &=&-{\tilde
\lambda}_A(\tau )\epsilon^A_{(\mu )}(u(p_{(\infty )})) [p^{(\mu )}_{(\infty )}
-\epsilon^{(\mu )}_B(u(p_{(\infty )})) {\hat P}^B_{AM}]=\nonumber \\
&=&-{\tilde \lambda}_A(\tau )\epsilon^A_{(\mu )}(u(p_{(\infty )})) [u^{(\mu )}
(p_{(\infty )}) (\epsilon_{(\infty )}-{\hat P}^{\tau}_{ADM})-\epsilon^{(\mu )}
_{\check r}(p_{(\infty )}){\hat P}^{\check r}_{ADM}]=\nonumber \\
&=&-{\tilde \lambda}_{\tau}(\tau ) [\epsilon_{(\infty )}-{\hat P}^{\tau}_{ADM}]
+{\tilde \lambda}_{\check r}(\tau ) {\hat P}^{\check r}_{ADM},\nonumber \\
&&{}\nonumber \\
\Rightarrow&& \epsilon_{(\infty )}-{\hat P}^{\tau}_{ADM} \approx 0,\quad\quad
{\hat P}^{\check r}_{ADM}\approx 0,\nonumber \\
 &&{}\nonumber \\
 H_{(D)ADM}&=& \int d^3\sigma \Big[ n_A {\cal H}^A +\lambda_{n A}
 {\tilde \pi}^A_n\Big] (\tau ,\vec \sigma ) -{\tilde \lambda}_{\tau}(\tau )
 [\epsilon_{(\infty )}-{\hat P}^{\tau}_{ADM}] +
 {\tilde \lambda}_{\check r}(\tau ) {\hat P}^{\check r}_{ADM},
\label{II44}
\end{eqnarray}

\noindent  in accord with Eq.(\ref{II28}).

Therefore, on the WSW hypersurfaces (whose 3-coordinates are denoted
$\{
\sigma^r \}$), which define the {\it intrinsic asymptotic rest frame
of the gravitational field}, the remaining four extra constraints are:

\bea
&&{\hat P}^{\check r}_{ADM}\approx 0,\nonumber \\
 &&\epsilon_{(\infty )}=-\epsilon \sqrt{\epsilon p^2
_{(\infty )}}\approx {\hat P}^{\tau }_{ADM} \approx -\epsilon M_{ADM}=
-\epsilon \sqrt{\epsilon {\hat P}^2_{AM}}.
\label{II45}
\eea

 Now the spatial indices have become spin-1 Wigner indices (they
transform with Wigner rotations under asymptotic Lorentz
transformations). As for parametrized theories in Minkowski spacetime,
in this special gauge 3 degrees of freedom of the gravitational field
become gauge variables, while ${\tilde x}^{(\mu )}_{(\infty )}$
becomes a decoupled observer with his clock near spatial infinity.
These 3 degrees of freedom represent an {\it internal} center-of-mass
3-variable ${\vec \sigma}_{ADM}[{}^3g,{}^3{\tilde
\Pi}]$ inside the WSW hypersurface; $\sigma^{
r}=\sigma^{r}_{ADM}$ is a variable representing the {\it 3-center of
mass} of the 3-metric of the slice $\Sigma_{\tau}$ of the
asymptotically flat spacetime $M^4$ and is obtainable from the weak
Poincar\'e charges with the group-theoretical methods of
Ref.\cite{pauri} as it is done in Ref.\cite{mate} for the Klein-Gordon
field on the Wigner hyperplane. Due to ${\hat P}^r_{ADM}\approx 0$ we
have

\bea
\sigma^r_{ADM} &=& -{{ {\hat J}^{\tau r} }\over { \sqrt{({\hat P}^{\tau}_{ADM})^2-
({\hat {\vec P}}_{ADM})^2} }}+ \nonumber \\
 &+&{{ ({\hat {\vec J}}_{ADM}
\times {\hat {\vec P}}_{ADM})^r }\over { \sqrt{({\hat
P}^{\tau}_{ADM})^2- ({\hat {\vec P}}_{ADM})^2} ({\hat P}^{\tau}_{ADM}
+\sqrt{({\hat P}^{\tau}_{ADM})^2- ({\hat {\vec P}}_{ADM})^2})}}+
\nonumber \\
 &+& {{({\hat J}^{\tau s}_{ADM} {\hat P}^s_{ADM}) {\hat P}^r_{ADM}}\over
 {{\hat P}^{\tau}_{ADM}\sqrt{({\hat P}^{\tau}_{ADM})^2-
({\hat {\vec P}}_{ADM})^2} ({\hat P}^{\tau}_{ADM} +\sqrt{({\hat
P}^{\tau}_{ADM})^2- ({\hat {\vec P}}_{ADM})^2})}}\approx \nonumber \\
&\approx& -{\hat J}^{\tau r}_{ADM}/ {\hat P}^{\tau}_{ADM},\nonumber \\
&&{}\nonumber \\
 &&\{ \sigma^r_{ADM},\sigma^s_{ADM} \} =0,\quad
\{ \sigma^r_{ADM},{\hat P}^s_{ADM} \} = \delta^{rs},
\label{II46}
\eea

\noindent
so that ${\vec \sigma}_{ADM}\approx 0$ is equivalent to the
requirement that the weak ADM boosts vanish: this is the way out from
the {\it boost problem} in the framework of the rest-frame instant
form.

When $\epsilon {\hat P}^2_{ADM} > 0$, with the asymptotic Poincar\'e
Casimirs ${\hat P}^2_{ADM}$, ${\hat W}^2_{ADM}$ one can build the
 M\o ller radius $\rho_{AMD}=\sqrt{-\epsilon {\hat W}^2_{ADM}}/\epsilon {\hat
P}^2_{ADM}c$, which is an intrinsic classical unit of length like in
parametrized Minkowski theories, to be used as an ultraviolet cutoff
in a future attempt of quantization.

By going from ${\tilde x}^{(\mu )}
_{(\infty )}$ and $p^{(\mu )}_{(\infty )}$ to the canonical basis \cite{lus1}

\bea
&&T_{(\infty )}=p_{(\infty )(\mu )}{\tilde x}^{(\mu )}_{(\infty
)}/\epsilon_{(\infty )}=p_{(\infty )(\mu )}x^{(\mu )}_{(\infty
)}/\epsilon_{(\infty )} {}{}{},\nonumber \\
 && \epsilon_{(\infty )},\nonumber \\
 &&z^{(i)}_{(\infty )}=\epsilon_{(\infty )} ({\tilde x}^{(i)}_{(\infty
)}-p^{(i)}_{(\infty )}{\tilde x}^{(o)}_{(\infty )}
/p^{(o)}_{(\infty )}), \nonumber \\
 &&k^{(i)}_{(\infty )}=p^{(i)}_{(\infty )}/\epsilon
_{(\infty )}=u^{(i)}(p^{(\rho )}_{(\infty )}),
\label{II47}
\eea

\noindent
one finds that the final reduction requires the gauge-fixings

\beq
T_{(\infty )}-\tau \approx 0,\quad\quad \sigma^{\check r}_{ADM}\approx
0\quad (or\, {\hat J}^{\tau r}_{ADM}\approx 0).
\label{II48}
\eeq

Since $\{ T_{(\infty )},\epsilon_{(\infty )} \}=-\epsilon$, with the
gauge fixing $T_{(\infty )}-\tau \approx 0$ one gets ${\tilde
\lambda}_{\tau} (\tau )\approx \epsilon$, $\epsilon_{(\infty )}\equiv {\hat P}^{\tau}_{ADM}$
and $H_{(D)ADM}={\tilde \lambda}_{\check r}(\tau ) {\hat P}^{\check
r}_{ADM}$. This is the frozen picture of the reduced phase space, like
it happens in the standard Hamilton-Jacobi theory: there is no time
evolution. To reintroduce an evolution in $T_{(\infty )}\equiv \tau$
we must use   the energy $M_{ADM}=-\epsilon {\hat P}^{\tau}_{ADM}$
(the ADM {\it mass of the universe}) as the natural physical
Hamiltonian. Therefore the final Dirac Hamiltonian is

\begin{eqnarray}
H_D&=&M_{ADM}+{\tilde \lambda}_{\check r}(\tau ) {\hat P}^{\check
r}_{ADM}+ \int d^3\sigma [ n_A {\tilde {\cal H}}^A+\lambda_{n A}
{\tilde \pi}^A_n](\tau ,\vec \sigma ) \approx \nonumber \\
&\approx& M_{ADM}= -\epsilon {\hat P}^{\tau}_{ADM}. \label{II49}
\end{eqnarray}

\noindent That $M_{ADM}$ is the correct Hamiltonian for getting a
$\tau$-evolution equivalent to Einstein's equations in spacetimes
asymptotically flat at spatial infinity is also shown in
Ref.\cite{fermi}. In the rest-frame the mathematical time is
identified with the parameter $\tau$ labelling the leaves
$\Sigma_{\tau}$ of the foliation of $M^4$.

The final gauge fixings $\sigma^{\check r}_{ADM}\approx 0$   [or
${\hat J}^{\tau r}_{ADM}\approx 0$] imply ${\tilde \lambda}_{\check
r}(\tau )\approx 0$, $H_D=M_{ADM}$ and a reduced theory with the {\it
external} 3-center-of-mass variables $z^{(i)}_{(\infty )}$,
$k^{(i)}_{(\infty )}$ decoupled (therefore the choice of the origin
$x^{(\mu )}_{(\infty )}$ becomes irrelevant) and playing the role of a
{\it point particle clock} for the time $T_{(\infty )}
\equiv \tau$. There would be a weak form of
Mach's principle, because only relative degrees of freedom would be present.

The condition ${\tilde \lambda}_{AB}(\tau )=0$ with ${\tilde
\lambda}_{\tau} (\tau )=\epsilon$, ${\tilde \lambda}_r(\tau )=0$ means
that at spatial infinity there are no local (direction dependent)
accelerations and/or rotations. The asymptotic line element for ${\vec
{\tilde \lambda}}(\tau )=0$ reduces to the line element of an inertial
system near spatial infinity: it defines the {\it preferred asymptotic
inertial observers}, for instance the {\it fixed stars} \cite{soffel}.

While the asymptotic {\it internal} realization of the Poincar\'e
algebra has the weak Poincar\'e charges ${\hat P}^{\tau}_{ADM}\approx
-\epsilon M_{ADM}$, ${\hat P}^r_{ADM}\approx 0$, ${\hat J}^{rs}_{ADM}$, ${\hat
K}_{ADM}^r={\hat J}^{\tau  r}_{ADM}\approx 0$ as generators, the
rest-frame instant form asymptotic {\it external} realization of the
Poincar\'e generators becomes

\begin{eqnarray}
&&\epsilon_{(\infty )}=M_{ADM},\nonumber \\
&&p^{(i)}_{(\infty )},\nonumber \\
&&J^{(i)(j)}_{(\infty )}={\tilde x}^{(i)}_{(\infty )}p^{(j)}_{(\infty )}-
{\tilde x}^{(j)}_{(\infty )} p^{(i)}_{(\infty )} +\delta^{(i)\check r}\delta
^{(j)\check s}{\hat J}^{\check r\check s}_{ADM},\nonumber \\
&&J^{(o)(i)}_{(\infty )}=p^{(i)}_{(\infty )} {\tilde x}^{(o)}_{(\infty )}-
\sqrt{M^2_{ADM}+{\vec p}^2_{(\infty )}} {\tilde x}^{(i)}_{(\infty )}-
{ {\delta^{(i)\check r}{\hat J}^{\check r\check s}_{ADM} \delta^{(\check s(j)}
p^{(j)}_{(\infty )} }\over
{M_{ADM}+\sqrt{M^2_{ADM}+{\vec p}^2_{(\infty )}} } } .
\label{II50}
\end{eqnarray}

\subsection{Tetrad Gravity and its Rest-Frame Instant Form.}

Since we have used the  ADM action of metric gravity in our
formulation of tetrad gravity, all the discussion about the
differentiability of the Hamiltonian, the definition of Poisson
brackets, the definition of proper and improper gauge
transformations can be directly reformulated in tetrad gravity.
The only difference inside tetrad gravity in the Hamiltonian
treatment of quantities depending upon ${}^3g_{rs}(\tau ,\vec
\sigma )={}^3e_{(a)r}\, {}^3e_{(a)s}$, ${}^3{\tilde \Pi}^{rs}(\tau
,\vec \sigma )={1\over 4} [ {}^3e^r_{(a)}\, {}^3{\tilde
\pi}^s_{(a)}+ {}^3e^s_{(a)}\, {}^3{\tilde \pi}^r_{(a)}]$ (they are
now derived quantities), is that now we have $\{ {}^3{\tilde
\Pi}^{rs}(\tau ,\vec \sigma ), {}^3{\tilde \Pi}^{uv}(\tau ,{\vec
\sigma}^{'}) \} = \delta^3(\vec \sigma ,{\vec \sigma}^{'})
F^{rsuv}_{(a)(b)}(\tau ,\vec \sigma ) \, {}^3{\tilde
M}_{(a)(b)}(\tau ,\vec \sigma ) \approx 0$ (see Eqs.(4.14) of
Ref.\cite{ru11}) and not $=0$. Therefore, constants of motion
(functional $F[{}^3g_{rs},{}^3{\tilde \Pi}^{rs}]$) of metric
gravity remain such in tetrad gravity, since they have weakly zero
Poisson brackets with ${\tilde {\cal H}}(\tau ,\vec \sigma )$,
${}^3{\tilde {\cal H}}^r(\tau ,\vec \sigma )$ [and, therefore,
with ${}^3{\tilde \Theta}_r(\tau ,\vec \sigma )$ and ${\hat
{\tilde {\cal H}}}_{(a)}(\tau ,\vec \sigma )$] and also with the
other first class constraints ${\tilde \pi}^{\vec
\varphi}_{(a)}(\tau ,\vec \sigma )\approx 0$, ${}^3{\tilde
M}_{(a)}(\tau ,\vec \sigma )\approx 0$.

As a consequence the weak and strong Poincar\'e charges are still
constants of motion in tetrad gravity and their weak Poincar\'e
algebra under Poisson brackets may only be modified by extra terms
containing ${}^3{\tilde M}_{(a)}(\tau ,\vec \sigma )\approx 0$. In
particular, after having added the gauge fixings to these constraints
and after having gone to Dirac brackets, the weak and strong
Poincar\'e algebras coincide with those of metric gravity. A more
complete study of these properties would require the study of the
quasi-invariances of the Lagrangian  (\ref{II8}) of tetrad gravity
under the gauge transformations generated by the 14 first class
constraints of the theory \footnote{Using the second Noether theorem
as it was done in Appendix A of III for metric gravity.}.

The only lacking ingredients are the definition of {\it proper} gauge
transformations generated by the primary (without associated
secondary) first class constraints ${\tilde \pi}^{\vec
\varphi}_{(a)}(\tau ,\vec \sigma )\approx 0$, ${}^3{\tilde
M}_{(a)}(\tau ,\vec \sigma )\approx 0$, and the boundary conditions
for cotriads ${}^3e_{(a)r}(\tau ,\vec \sigma )$, because the lapse and
shift functions $N(\tau ,\vec \sigma )$, $N_{(a)} (\tau ,\vec \sigma
)={}^3e_{(a)}^r(\tau ,\vec \sigma ) N_r(\tau ,\vec \sigma )$ are
treated in the same way as in metric gravity, namely we assume the
validity of Eqs. (\ref{II19}) in the form

\bea
N(\tau ,\vec \sigma ) &=& N_{(as)}(\tau ,\vec \sigma ) + m(\tau ,\vec
\sigma ),\nonumber \\
 N_{(a)}(\tau ,\vec \sigma ) &=& {}^3e^{\check r}_{(a)}(\tau ,\vec \sigma )
 [N_{(as)\check r}(\tau ,\vec \sigma ) + m_{\check r}(\tau ,\vec \sigma )]=\nonumber \\
  &=& {}^3e^{\check r}_{(a)}(\tau ,\vec \sigma ) N_{(as)\check r}(\tau ,\vec \sigma ) +
  m_{(a)}(\tau ,\vec \sigma ),\nonumber \\
  &&{}\nonumber \\
  N_{(as)}(\tau ,\vec \sigma ) &=& -{\tilde \lambda}_{\tau}(\tau )-{1\over 2}
  {\tilde \lambda}_{\tau \check s}(\tau ) \sigma^{\check s},\nonumber \\
  N_{(as)\check r}(\tau ,\vec \sigma ) &=& -{\tilde \lambda}_{\check r}(\tau ) -{1\over 2}
  {\tilde \lambda}_{\check r\check s}(\tau ) \sigma^{\check s}.
  \label{II51}
  \eea

Therefore, we shall assume that there exist the same coordinate
systems of $M^4$ and $\Sigma_{\tau}$ as in metric gravity and that the
$\Sigma_{\tau}$-adapted tetrads of Eqs.(\ref{II1}), whose expression
is ${}^4_{(\Sigma )}{\check E}^{\mu}_{(\mu )}$
with\footnote{$b^{\mu}_A$ are the transformation coefficients to
$\Sigma_{\tau}$-adapted coordinates.}

\beq
{}^4_{(\Sigma )}{\check E}^{\mu}_{(o)}=l^{\mu}={\hat
b}^{\mu}_l={1\over N}[b^{\mu}_{\tau}-N^{\check r} b^{\mu}_{\check
r}],\quad\quad {}^4_{(\Sigma )}{\check E}^{\mu}_{(a )}={}^3e^{\check
s}_{(a)} b^{\mu}_{\check s},
\label{II52}
\eeq

\noindent
 have a well defined angle-independent limit ${}^4
_{(\Sigma )}{\check E}^{\mu}_{(\infty )(\mu )}$ at spatial infinity,
such that

\bea
&&{}^4_{(\Sigma )}{\check E}^{\mu}_{(\infty )(o )}=\delta^{\mu}_{(\mu
)} l^{(\mu )}_{(\infty )}=
\delta^{\mu}_{(\mu )} {\hat b}^{(\mu )}_{(\infty )l}=\delta^{\mu}_{(\mu )}
{1\over {N_{(as)}}}( b^{(\mu )}_{(\infty )\tau}-N_{(as)}^{\check
r}b^{(\mu )}_{(\infty )\check r}),\nonumber \\
 && {}^4_{(\Sigma )}{\check
 E}^{\mu}_{(\infty )(a)}=\delta^{\check s}_{(a)}\delta^{\mu}_{(\mu )} b^{(\mu
)}_{(\infty )\check s}(\tau ),
\label{II53}
\eea

\noindent
with the same asymptotic $b^{(\mu )}_{(\infty )A}(\tau )$'s of
Eq.(\ref{II15}).

Let us remark that the $\Sigma_{\tau}$-adapted tetrads in adapted
coordinates of Eqs.(\ref{II2}), are ${}^4_{(\Sigma )}{\check {\tilde
E}}^A_{(\mu )}$ with

\bea
&&{}^4_{(\Sigma )}{\check {\tilde E}}^A_{(o )}={1\over N} \Big( 1;
-{}^3e^{\check r}_{(a)}\, N_{(a)} \Big) ,\nonumber \\
 &&{}^4_{(\Sigma )}{\check {\tilde E}}^A_{(a)}= \Big( 0;{}^3e^{\check r}_{(a)}\Big).
\label{II54}
\eea

Due to the presence of the lapse function in the denominator which
is linearly increasing in $\vec \sigma$ (to have the possibility
of defining $J^{AB}_{ADM}$), these adapted tetrads exist without
singularities at spatial infinity only if ${\tilde
\lambda}_{AB}(\tau )=0$, i.e. on WSW hypersurfaces\footnote{This
is connected with the criticism in footnote 39.}. The same happens
for the adapted cotetrads ${}^4_{(\Sigma )}{\check {\tilde
E}}^{(\mu )}_A$ with ${}^4_{(\Sigma )}{\check {\tilde E}}^{(o
)}_A=(N; 0)$, ${}^4_{(\Sigma )}{\check {\tilde E}}^{(a
)}_A=(N^{(a)}=N_{(a)}; {}^3e^{(a)}_r ={}^3e_{(a)r})$. Also the
concept of proper time of Eulerian observers
 connected with the lapse function is divergent at spatial infinity.
Therefore, tetrad gravity without supertranslations [${\tilde
\lambda}_{AB}(\tau )=0$, $m=n$, $m_{(a)}=n_{(a)}={}^3e^r_{(a)}n_r$]
and with Poincar\'e charges, admits well defined adapted tetrads and
cotetrads (with components in adapted holonomic coordinates) only
after having been restricted to WSW hypersurfaces ({\it rest frame}),
whose asymptotic normals $l^{(\mu )}_{(\infty )}=l^{(\mu )}_{(\infty
)\Sigma}$ , tangent to $S_{\infty}$, are parallel to ${\hat P}^{(\mu
)}_{ADM}=b^{(\mu )}_{(\infty )A} {\hat P}^A_{ADM}$ with ${\hat
P}^r_{ADM}\approx 0$ \footnote{Namely when one is inside the
Christodoulou-Klainermann class of solutions\cite{ckl}, but in general
with non vanishing shift functions.}. This again implies the existence
of an inertial system at spatial infinity when ${\tilde
\lambda}_A(\tau )=(\epsilon ;\vec 0)$ and ${\tilde \lambda}_{AB}(\tau
)=0$, namely the absence of accelerations and rotations there
\footnote{When ${\tilde
\lambda}_A(\tau )\not= 0$ there is a direction independent global
acceleration of the origin $x^{(\mu )}_{(\infty )}(\tau )$, since
${\dot x}^{(\mu )}_{(\infty )}(\tau )= b^{(\mu )}_{(\infty )A}{\tilde
\lambda}_A(\tau )$.}.

In tetrad gravity we shall assume the following boundary conditions
consistent with Eqs.(\ref{II27}) and (\ref{II30}) of metric gravity

\begin{eqnarray}
{}^3e_{(a)\check r}(\tau ,\vec \sigma )\,&& {\rightarrow}_{r\,
\rightarrow
\infty }\, (1+{M\over {2r}})
\delta_{(a)\check r}+{}^3w_{(a)\check r}(\tau ,\vec \sigma ),\quad\quad
{}^3w_{(a)\check r}(\tau ,\vec \sigma )=O(r^{-3/2}),\nonumber \\
{}^3e^{\check r}_{(a)}(\tau ,\vec \sigma )\,&& {\rightarrow}_{r\,
\rightarrow
\infty}\, (1-{M\over {2r}})
\delta_{(a)}^{\check r}+{}^3w^{\check r}_{(a)}(\tau ,\vec \sigma ),\quad\quad
{}^3w^{\check r}_{(a)}(\tau ,\vec \sigma )=O(r^{-3/2}),\nonumber \\
{}^3g_{\check r\check s}(\tau ,\vec \sigma )&&=[{}^3e_{(a)\check r}\,
{}^3e_{(a)\check s}] (\tau ,\vec \sigma )\, {\rightarrow}_{r\,
\rightarrow \infty}\, (1+{M\over r})
\delta_{\check r\check s}+{}^3h_{\check r\check s}(\tau ,\vec \sigma ),\nonumber \\
{}^3h_{\check r\check s}(\tau ,\vec \sigma )&&={1\over r}
[\delta_{(a)\check r}\, {}^3w
_{(as)(a)\check s}(\tau ,\vec \sigma )+{}^3w_{(as)(a)\check r}(\tau ,\vec \sigma )\delta
_{(a)\check s}]+O(r^{-2})=O(r^{-3/2}),\nonumber \\
&&{}\nonumber \\ {}^3{\tilde \pi}^{\check r}_{(a)}(\tau ,\vec \sigma
)\,&& {\rightarrow}_{r\,
\rightarrow \infty}\, O(r^{-5/2}),\nonumber \\
{}^3{\tilde \Pi}^{\check r\check s}(\tau ,\vec \sigma )&&={1\over
4}[{}^3e^{\check r}_{(a)}\, {}^3{\tilde
\pi}^{\check s}_{(a)}+{}^3e^{\check s}_{(a)}\, {}^3{\tilde \pi}
^{\check r}_{(a)}](\tau ,\vec \sigma )\, {\rightarrow}_{r\, \rightarrow
\infty}\, {}^3{\tilde k}^{\check r\check s}(\tau ,\vec \sigma )=O(r^{-5/2}),\nonumber \\
&&{}\nonumber \\
N(\tau ,\vec \sigma )&&=N_{(as)}(\tau ,\vec \sigma )+n(\tau ,\vec \sigma ),
\nonumber \\
n(\tau ,\vec \sigma )\,&& {\rightarrow}_{r\, \rightarrow \infty}\,
O(r^{-(2+\epsilon )}),\nonumber \\
 N_{(as)}(\tau ,\vec \sigma
)&&=-{\tilde \lambda}_{\tau}(\tau )-{1\over 2} {\tilde \lambda}_{\tau
\check s}(\tau ) \sigma^{\check s},\nonumber \\
 &&{}\nonumber \\
  N_{\check r}(\tau ,\vec
\sigma )&&=N_{(as)\check r}(\tau ,\vec \sigma )+ n_{\check r}(\tau ,\vec \sigma
),\nonumber \\
 n_{\check r}(\tau ,\vec \sigma )\,&& {\rightarrow}_{r\,
\rightarrow \infty}\, O(r^{-\epsilon}),\nonumber \\
 N_{(as)\check r}(\tau
,\vec \sigma )&&=-{\tilde \lambda}_{r}(\tau )- {1\over 2}{\tilde
\lambda}_{\check r\check s}(\tau ) \sigma^{\check s},
\nonumber \\
N_{(a)}(\tau ,\vec \sigma )&&={}^3e^{\check r}_{(a)}(\tau ,\vec \sigma
) N_{\check r}(\tau ,\vec \sigma )
 =N_{(as)(a)}(\tau ,\vec \sigma ) +n_{(a)}(\tau ,\vec \sigma ),\nonumber \\
 n_{(a)}(\tau ,\vec \sigma )&&=[{}^3e^{\check r}_{(a)}n_{\check r}](\tau ,\vec
\sigma )\, {\rightarrow}_{r\, \rightarrow \infty}\, O(r^{-\epsilon}),
\nonumber \\
&&{}\nonumber \\
{\tilde \pi}^n(\tau ,\vec \sigma )\,&& {\rightarrow}_{r\, \rightarrow \infty}\,
O(r^{-3}),\nonumber \\
{\tilde \pi}_{\vec n,(a)}(\tau ,\vec \sigma )\,&& {\rightarrow}_{r\,
\rightarrow \infty}\, O(r^{-3}),\nonumber \\
\lambda_n(\tau ,\vec \sigma )\,&& {\rightarrow}_{r\, \rightarrow \infty}\,
O(r^{-(3+\epsilon )}),\nonumber \\
\lambda_{\vec n,(a)}(\tau ,\vec \sigma )\,&& {\rightarrow}_{r\, \rightarrow
\infty}\, O(r^{-\epsilon}),\nonumber \\
\beta (\tau ,\vec \sigma )\,&& {\rightarrow}_{r\, \rightarrow \infty}\,
O(r^{-(3+\epsilon )}),\nonumber \\
\beta^{\check r}(\tau ,\vec \sigma )\,&& {\rightarrow}_{r\, \rightarrow \infty}\,
O(r^{-\epsilon}),\nonumber \\
&&{}\nonumber \\
{\hat {\cal H}}(\tau ,\vec \sigma )\,&& {\rightarrow}_{r\, \rightarrow \infty}\,
O(r^{-3}),\nonumber \\
{}^3{\tilde \Theta}_{\check r}(\tau ,\vec \sigma )\,&& {\rightarrow}_{r\,
\rightarrow \infty}\, O(r^{-3}),\nonumber \\
&&{}\nonumber \\
{}^3{\tilde M}_{(a)}(\tau ,\vec \sigma )\,&& {\rightarrow}_{r\, \rightarrow
\infty}\, O(r^{-6}),\nonumber \\
\alpha_{(a)}(\tau ,\vec \sigma )\,&& {\rightarrow}_{r\, \rightarrow \infty}\,
O(r^{-(1+\epsilon )}),\nonumber \\
{\hat \mu}_{(a)}(\tau ,\vec \sigma )\,&& {\rightarrow}_{r\, \rightarrow
\infty}\, O(r^{-(1+\epsilon )}),\nonumber \\
&&{}\nonumber \\
\varphi_{(a)}(\tau ,\vec \sigma )\,&& {\rightarrow}_{r\, \rightarrow \infty}\,
O(r^{-(1+\epsilon )}),\nonumber \\
{\tilde \pi}^{\vec \varphi}_{(a)}(\tau ,\vec \sigma )\,&& {\rightarrow}_{r\,
\rightarrow \infty}\, O(r^{-2}),\nonumber \\
\lambda^{\vec \varphi}_{(a)}(\tau ,\vec \sigma )\,&& {\rightarrow}_{r\,
\rightarrow \infty}\, O(r^{-(1+\epsilon )}),
\label{II55}
\end{eqnarray}

\noindent   with the asymptotic line element

\begin{eqnarray}
ds^2&=& \epsilon \Big( [N_{(as)}+n]^2 - [N_{(as)\check r}+n_{\check
r}] {}^3e^{\check r}_{(a)}\, {}^3e^{\check s}
_{(a)} [N_{(as)\check s}+n_{\check s}] \Big) (d\tau )^2-\nonumber \\
&-&2\epsilon [N_{(as)\check r}+n_{\check r}] d\tau d\sigma^{\check r}
-\epsilon
\, {}^3e_{(a)\check r}\, {}^3e_{(a)\check s} d\sigma^{\check r} d\sigma^{\check s}=\nonumber \\
&=&\epsilon \Big( [N_{(as)}+n]^2(d\tau )^2-[{}^3e_{(a)\check
r}d\sigma^{\check r}+(N_{(as)(a)}+ n_{(a)})d\tau
 ][{}^3e_{(a)\check s}d\sigma^{\check s}+(N_{(as)(a)}+n_{(a)})d\tau ]\Big) .\nonumber \\
  &&
\label{II56}
\end{eqnarray}

With these boundary conditions all proper gauge transformations
\footnote{Generated by ${\tilde {\cal H}}(\tau ,\vec \sigma )$ with parameter
$\beta (\tau ,\vec \sigma )\rightarrow O(r^{-(3+\epsilon )})$,
${\tilde \Theta}_r(\tau ,\vec \sigma )$  with $\beta^{\check r}(\tau
,\vec \sigma ) \rightarrow O(r^{-\epsilon}) $, ${}^3{\tilde
M}_{(a)}(\tau ,\vec \sigma )$ with $\alpha_{(a)}(\tau ,\vec \sigma
)\rightarrow O(r^{-(1+\epsilon )})$, ${\tilde \pi}_{(a)}
^{\vec \varphi}(\tau ,\vec \sigma )$ with $\varphi_{(a)}(\tau ,\vec \sigma )
\rightarrow O(r^{-(1+\epsilon )})$ for $r\, \rightarrow \infty$.} go
asymptotically to the identity.

Near spatial infinity there is a {\it dynamical preferred observer}
\footnote{Either the canonical non-covariant Newton-Wigner-like position
${\tilde x}^{(\mu )}_{(\infty )}(\tau )$ or the covariant
non-canonical origin of asymptotic Cartesian coordinates $x^{(\mu
)}_{(\infty )}(\tau )$.} with an associated asymptotic inertial
Lorentz reference frame given by the asymptotic limit of the
$\Sigma_{\tau}$-adapted tetrads of Eqs.(\ref{II2}): however, as said,
these asymptotic tetrads are well defined only in absence of
supertranslations on the rest-frame WSW hypersurfaces, where (modulo a
rigid 3-rotation) we get

\bea
&&{}^4_{(\infty \Sigma )}{\check {\tilde E}}^A_{(o)}=\Big( {1\over
{N_{(as)}(\tau )}}=-\epsilon; -{{\delta^r_{(a)} N_{(as)(a)}(\tau )}
\over {N_{(as)}(\tau )}} = \vec 0 \Big),\quad\quad
{}^4_{(\infty \Sigma )}{\check {\tilde E}}_A^{(a)}=\Big( 0;
\delta_{(a)}^r \Big),\nonumber \\
 &&{}\nonumber \\
 &&{}^4_{(\infty \Sigma )}{\check {\tilde E}}_A^{(o)}=\Big( N_{(as)}=-\epsilon ;
 \vec 0 \Big) ,\quad\quad {}^4_{(\infty \Sigma )}{\check {\tilde E}}_A^{(a)}=\Big(
 N_{(as)(a)}=0; \delta^{(a)}_r \Big) .
 \label{II57}
\eea

Then, following the scenario b), the differentiable and finite Dirac
Hamiltonian [replacing the one of Eqs.(\ref{II13})] is assumed to be

\begin{eqnarray}
{\hat H}_{(D)ADM}&=&\int d^3\sigma [n{\hat {\cal H}}+n_{(a)}\,
{}^3e^{\check r}_{(a)}\, {}^3{\tilde \Theta}_{\check r} +\lambda^{\vec
\varphi}_{(a)}{\tilde \pi}^{\vec \varphi}
_{(a)}+{\hat \mu}_{(a)}\, {}^3{\tilde M}_{(a)}+
\nonumber \\
&+&\lambda_n {\tilde \pi}^n +\lambda^{\vec n}_{(a)} {\tilde \pi}^{\vec n}
_{(a)}](\tau ,\vec \sigma )-\nonumber \\
&-&{\tilde \lambda}_A(\tau )[p^A_{(\infty )}-{\hat P}^A_{ADM}]-{1\over
2}{\tilde \lambda}_{AB}(\tau ) [J^{AB}_{(\infty )}-{\hat
J}^{AB}_{ADM}],
\label{II58}
\end{eqnarray}

\noindent with the same weak (and strong) Poincar\'e charges of metric gravity, Eqs.
(\ref{II23}) [(\ref{II25})], expressed in terms of cotriads
${}^3e_{(a)r}$ and their conjugate momenta ${}^3{\tilde \pi}^r_{(a)}$,
by using ${}^3g_{rs}={}^3e_{(a)r}\, {}^3e_{(a)s}$, ${}^3{\tilde
\Pi}^{rs}= {1\over 4} [{}^3e^r_{(a)}\, {}^3{\tilde
\pi}^s_{(a)}+{}^3e^s_{(a)}\, {}^3{\tilde \pi}^r_{(a)}]$. Let us remark that we have
 $n_{\check r}\, {}^3{\tilde {\cal H}}^{\check r}\approx
-n_{(a)}\, {}^3e^{\check s}_{(a)}\, {}^3{\tilde \Theta}_{\check s}\approx -n
^{\check s}\, {}^3{\tilde \Theta}_{\check s}=-n_{(a)}\, {}^3e^s_{(a)}
\,$ $ {}^3{\tilde \Theta}_s \approx -n_{(a)}\, {}^3e^s_{(a)} {{\partial \xi^r}
\over {\partial \sigma^s}} {\tilde \pi}^{\vec \xi}_r =-n_u\, {}^3g^{us}
{{\partial \xi^r}\over {\partial \sigma^s}} {\tilde \pi}^{\vec
\xi}_r=-{\tilde n}^r {\tilde \pi}^{\vec \xi}_r$.

However, as already said, we must restrict ourselves to gauges with
${\tilde \lambda}_{AB}(\tau )=0$, namely to WSW foliations, to avoid
inconsistencies at spatial infinity.

The rest-frame instant form of tetrad gravity on WSW hypersurfaces
is defined by Eqs. (\ref{II39})-(\ref{II49}). In this gauge the
final Hamiltonian for the evolution in $\tau \equiv T_s$ is weakly
$M_{ADM}=-\epsilon {\hat P}^{\tau}_{ADM}$ and  we also have
${\tilde \lambda}_A(\tau )=(\epsilon ;\vec 0)$, $N=-\epsilon +n$,
$N_r=n_r$ [$n_{(a)}={}^3e^r_{(a)}n_r$].

With the Dirac Hamiltonian (\ref{II58}) the Hamilton equations on WSW
hypersurfaces are

\begin{eqnarray}
\partial_{\tau} n(\tau ,\vec \sigma )\, &{\buildrel \circ \over =}\,&
\{ n(\tau ,\vec \sigma ), {\hat H}_{(D)ADM} \} =\lambda_n(\tau ,\vec \sigma ),\nonumber \\
\partial_{\tau} n_{(a)}(\tau ,\vec \sigma )\, &{\buildrel \circ \over =}\,&
\{ n_{(a)}(\tau ,\vec \sigma ), {\hat H}_{(D)ADM} \} =\lambda^{\vec n}_{(a)}(\tau ,\vec \sigma ),
\nonumber \\
\partial_{\tau} \varphi_{(a)}(\tau ,\vec \sigma )\, &{\buildrel \circ \over =}\,&
\{ \varphi_{(a)}(\tau ,\vec \sigma ), {\hat H}_{(D)ADM} \} =\lambda^{\vec \varphi}
_{(a)}(\tau ,\vec \sigma ),\nonumber \\
\partial_{\tau}\, {}^3e_{(a)r}(\tau ,\vec \sigma )\, &{\buildrel \circ \over =}\,&
 - {{\epsilon 4\pi G}\over {c^3}}\Big[ {n\over{{}^3e}}\,
{}^3G_{o(a)(b)(c)(d)}
\, {}^3e_{(b)r}\, {}^3e_{(c)s}\, {}^3{\tilde \pi}^s_{(d)}\Big] (\tau ,\vec
\sigma )+\nonumber \\
&+&\Big[ n_{(b)}\, {}^3e^s_{(b)}{{\partial \, {}^3e_{(a)r}}\over
{\partial
\sigma^s}}+{}^3e_{(a)s} {{\partial}\over {\partial \sigma^r}} (n_{(b)}\,
{}^3e^s_{(b)})\Big] (\tau ,\vec \sigma )+\nonumber \\
&+&\epsilon_{(a)(b)(c)}\, {\hat \mu}_{(b)}(\tau ,\vec \sigma )\,
{}^3e_{(c)r} (\tau ,\vec \sigma )+\nonumber \\
 &+&{\tilde \lambda}_A(\tau ) \{ {}^3e_{(a)r}(\tau ,\vec \sigma ),{\hat P}^A_{ADM}
\} ,\nonumber \\
 \partial_{\tau}\, {}^3{\tilde \pi}^r_{(a)}(\tau ,\vec \sigma )\, &{\buildrel \circ \over =}\,&
{{\epsilon c^3}\over {8\pi G}} \Big[ {}^3e\,  n\,
({}^3R^{rs}-{1\over 2} {}^3g^{rs}\, {}^3R){}^3e _{(a)s}+{}^3e
(n^{|r|s}-{}^3g^{rs}\, n^{|u}{}_{|u}){}^3e_{(a)s}\Big] (\tau ,\vec
\sigma )-\nonumber \\
 &-&\epsilon {{2\pi G\, n(\tau ,\vec \sigma )}\over
{c^3}} \Big[ {1\over {{}^3e}}\, {}^3G_{o(a)(b)(c)(d)}\,
{}^3{\tilde \pi} ^r_{(b)}\, {}^3e_{(c)s}\, {}^3{\tilde
\pi}^s_{(d)}-\nonumber
\\ &-&{2\over {{}^3e}}\, {}^3e^r_{(a)}\, {}^3G_{o(b)(c)(d)(e)}\,
{}^3e_{(b)u}\, {}^3{\tilde \pi}^u_{(c)}\, {}^3e_{(d)v}\,
{}^3{\tilde \pi}^v_{(e)}\Big] (\tau ,\vec \sigma )+\nonumber \\
&+&{{\partial}\over {\partial \sigma^s}} \Big[ n_{(b)}\,
{}^3e^s_{(b)}\, {}^3{\tilde \pi}^r_{(a)}\Big] (\tau ,\vec \sigma
)-{}^3{\tilde \pi}^u_{(a)} (\tau ,\vec \sigma ){{\partial}\over
{\partial \sigma^u}}\Big[ n_{(b)}\, {}^3e^r _{(b)}\Big] (\tau
,\vec \sigma )+\nonumber \\
 &+&\epsilon_{(a)(b)(c)}\, {\hat
\mu}_{(b)}(\tau ,\vec \sigma )\, {}^3{\tilde \pi}^r_{(c)}(\tau
,\vec \sigma )+\nonumber \\
 &+&{\tilde \lambda}_A(\tau ) \{
{}^3{\tilde \pi}^r_{(a)}(\tau ,\vec \sigma ), {\hat P}^A_{ADM} \}
,\nonumber \\
 &&{}\nonumber \\
 &&\text{with} \nonumber \\
 &&{}\nonumber \\
{\hat P}^{\tau}_{ADM} &=& \epsilon \int d^3\sigma \Big[
{{c^3}\over {16\pi G}}\, {}^3e\, {}^3e^r_{(a)}\, {}^3e^s_{(a)}
({}^3\Gamma^u_{rv}\, {}^3\Gamma^v_{su}- {}^3\Gamma^u_{rs}\,
{}^3\Gamma^v_{vu})-\nonumber \\
 &-& {{2\pi G}\over {c^3\,\, {}^3e}}\, {}^3G_{o(a)(b)(c)(d)}\, {}^3e_{(a)r}\,
 {}^3{\tilde \pi}^r_{(b)}\, {}^3e_{(c)s}\, {}^3{\tilde \pi}^s_{(d)} \Big]
 (\tau ,\vec \sigma ),\nonumber \\
 {\hat P}^r_{ADM} &=& -\int d^3\sigma \Big[ {}^3\Gamma^r_{uv}\, {}^3e^u_{(a)}\,
  {}^3{\tilde \pi}^v_{(a)} \Big] (\tau ,\vec \sigma ).
\label{II59}
\end{eqnarray}

In ${\hat P}^A_{ADM}$ the 3-Christoffel coefficients must be expressed
in terms of the cotriads, see Eq.(\ref{a1}) of Appendix A.

Let us remark that, since we
are using the ADM expression for the energy ${\hat P}^{\tau}_{ADM}$,
we have not to show that it is
definite positive, because the ADM canonical approach to metric gravity is
contained in the one to tetrad gravity.

\vfill\eject

\section{The Hamiltonian Group of Gauge Transformations.}

In this  Section we shall study the Hamiltonian group of gauge
transformations in the framework of scenario b) of the previous
Section following the scheme outlined in Ref.\cite{restmg} for metric
gravity. The generators of the infinitesimal Hamiltonian gauge
transformations connected with the identity are the first class
constraints. As shown in Appendix A of III for metric gravity, the ADM
action is quasi-invariant under the pull-back at the Lagrangian level
of the gauge transformations generated by the Hamiltonian group of
gauge transformations. As shown in Subsection B of Section IX of
Ref.\cite{restmg} it is only on the solutions of Einstein equations
that the Hamiltonian group of gauge transformations agrees with the
spacetime diffeomorphisms in $Diff\, M^4$, under which the Hilbert
action is invariant. Instead, {\it outside the solutions} the
Hamiltonian gauge group {\it connects different 4-geometries}
(4-metric modulo $Diff\, M^4$).

In the 32-dimensional functional phase space $T^{*}{\cal C}$ spanned
by the 16 field variables $n(\tau ,\vec
\sigma )$, $n_{(a)}(\tau ,\vec
\sigma )$, $\varphi_{(a)} (\tau ,\vec \sigma )$, ${}^3e_{(a)r}(\tau
,\vec \sigma )$ of the Lagrangian configuration space ${\cal C}$ and
by their 16 conjugate momenta, we have 14 first class constraints
${\tilde \pi}^n(\tau ,\vec \sigma )\approx 0$, ${\tilde
\pi}^{\vec n}_{(a)}(\tau ,\vec \sigma )\approx 0$, ${\tilde \pi}^{\vec
\varphi}_{(a)}(\tau ,\vec \sigma )\approx 0$, ${}^3{\tilde M}_{(a)}
(\tau ,\vec \sigma )\approx 0$, ${\hat {\cal H}}(\tau ,\vec
\sigma )\approx 0$ and either ${}^3{\tilde \Theta}_r(\tau ,\vec \sigma )\approx
0$ or ${\hat {\cal H}}_{(a)}(\tau ,\vec \sigma )\approx 0$. Seven pairs of
conjugate canonical variables, $\lbrace n(\tau ,\vec \sigma ), {\tilde \pi}^n
(\tau ,\vec \sigma ); n_{(a)}(\tau ,\vec \sigma ), {\tilde \pi}^{\vec n}_{(a)}
(\tau ,\vec \sigma ); \varphi_{(a)}(\tau ,\vec \sigma ), {\tilde \pi}^{\vec
\varphi}_{(a)}(\tau ,\vec \sigma )\rbrace$, are already decoupled from the
18-dimensional subspace spanned by $\lbrace {}^3e_{(a)r}(\tau ,\vec
\sigma ); {}^3{\tilde \pi}^r_{(a)}(\tau ,\vec \sigma ) \rbrace$. The
variables in ${\cal C}_g=\lbrace n(\tau ,\vec \sigma ), n_{(a)}(\tau
,\vec \sigma ),$ $\varphi_{(a)}(\tau ,\vec \sigma )
\rbrace$ are gauge variables, but due to the decoupling there is no
need to introduce gauge-fixing constraints to eliminate them
explicitly, at least at this stage. Therefore, let us concentrate on
the reduced 9-dimensional configuration function space ${\cal
C}_e=\lbrace {}^3e_{(a)r}(\tau ,\vec \sigma )\rbrace$ [${\cal C}={\cal
C}_g+{\cal C}_e$, $T^{*}{\cal C}=T^{*}{\cal C}_g+T^{*}{\cal C}_e$] and
on the 18-dimensional function phase space $T^{*}{\cal C}_e=
\lbrace {}^3e_{(a)r}(\tau ,\vec \sigma ), {}^3{\tilde \pi}^r_{(a)}(\tau ,\vec
\sigma ) \rbrace$, on which we have the seven first class constraints ${}^3{\tilde M}
_{(a)}(\tau ,\vec \sigma )\approx 0$, ${}^3{\tilde \Theta}_r(\tau ,\vec \sigma )
\approx 0$, ${\hat {\cal H}}(\tau ,\vec \sigma )\approx 0$, whose Poisson
brackets, defining an algebra $\bar g$, are given in Eqs.(\ref{II14}).

\subsection{Hamiltonian Gauge Transformations.}

Let us call ${\bar {\cal G}}$ the (component connected to the identity
of the) gauge group obtained from successions of gauge transformations
generated by the previous seven first class constraints. Since
${}^3{\tilde M}_{(a)}(\tau ,\vec \sigma )$ \footnote{The generators of
the inner gauge SO(3)-rotations.} and  ${}^3{\tilde \Theta}_r(\tau
,\vec \sigma )$ \footnote{The generators of space
pseudo-diffeomorphisms (passive diffeomorphisms) in $Diff\,
\Sigma_{\tau}$ extended to cotriads.} form a Lie subalgebra ${\bar
g}_R$ of $\bar g$ (the algebra of ${\bar {\cal G}}$), let ${\bar {\cal
G}}_R$ be the gauge group without the superhamiltonian constraint and
${\bar {\cal G}}_{ROT}$ its invariant subgroup containing only SO(3)
rotations. The addition to ${\bar g}_R$ of the superhamiltonian ${\hat
{\cal H}}(\tau ,\vec \sigma )$ introduces structure functions [the
last of Eqs.(\ref{II14})] as in the ADM Hamiltonian formulation of
metric gravity, so that $\bar g$ is not a Lie algebra.

The gauge group ${\bar {\cal G}}_R$ may be identified with the
automorphism group $Aut\, L\Sigma_{\tau}$ of the trivial principal
SO(3)-bundle $L\Sigma_{\tau} \approx \Sigma_{\tau}\times SO(3)$ of
orthonormal coframes, whose properties are studied in Ref.\cite{abba}.
The automorphism group $Aut\, L\Sigma_{\tau}$ contains the structure
group SO(3) of $L\Sigma_{\tau}$ as a subgroup, and, moreover, $Aut\,
L\Sigma_{\tau}$ is itself a principal bundle with base $Diff\,
\Sigma_{\tau}$ (which acts on the base $\Sigma_{\tau}$ of
$L\Sigma_{\tau}$) and structure group the group of gauge
transformations \footnote{$Gau\, L\Sigma_{\tau}$; see Ref.\cite{lusa}
for a review of the notations.} of the principal bundle
$L\Sigma_{\tau}$: therefore, locally $Aut\, L\Sigma
_{\tau}$ has the trivialization $[U\subset Diff\, \Sigma_{\tau}]\times SO(3)$
and we have

\begin{eqnarray}
\begin{array}{lcl}
Aut\, L\Sigma_{\tau} &\rightarrow & L\Sigma_{\tau}\approx \Sigma_{\tau}\times
SO(3)\\
\downarrow & {}& \downarrow \\
Diff\, \Sigma_{\tau} & \rightarrow & \Sigma_{\tau}
\end{array}
\label{III1}
\end{eqnarray}

Let us concentrate on the study of the non-Abelian Lie algebra
${\bar g}_R$ and of the associated group of gauge transformations
${\bar {\cal G}}_R$. Since ${\bar {\cal G}}_R$ contains the group
of space pseudo-diffeomorphisms $Diff\, \Sigma_{\tau}$ (or better
its action on the cotriads), it is not a Hilbert-Lie group, at
least in standard sense\cite{diff,abba} \footnote{Its differential
structure is defined in an inductive way.}; therefore, the
standard technology from the theory of Lie groups used for
Yang-Mills theory \footnote{See Ref.\cite{lusa} and the Appendix
of Ref.\cite{bao}.} is not directly available. However this
technology can be used for the invariant subgroup of gauge
SO(3)-rotations. The main problem is that it is not clear how to
parametrize the group manifold of $Diff\, \Sigma_{\tau}$: one only
knows that its algebra (the infinitesimal space diffeomorphisms)
is isomorphic to the tangent bundle $T\Sigma_{\tau}$\cite{diff}.

Moreover, while in a Lie (and also in a Hilbert-Lie) group the basic
tool is the group-theoretical exponential map, associated with the
one-parameter subgroups, which coincides with the geodesic exponential
map when the group manifold of a compact semisimple Lie group is
regarded as a symmetric Riemann manifold\cite{helga}, in $Diff\,
\Sigma_{\tau}$ this map does not produce a diffeomorphism between a
neighbourhood of zero in the algebra and a neighbourhood of the
identity in $Diff\, \Sigma_{\tau}$\cite{diff,abba}. Therefore, to
study the Riemannian 3-manifold $\Sigma_{\tau}$ we have to use the
geodesic exponential map as the main tool\cite{oneil,kobay}, even if
it is not clear its relationship with the differential structure of
$Diff\, \Sigma_{\tau}$. The {\it geodesic exponential map} at $p\in
M^4$ sends each vector ${}^4V_p={}^4V_p^{\mu}\partial_{\mu} \in
T_pM^4$ at p to the point of unit parameter distance along the unique
geodesic through p with tangent vector ${}^4V_p$ at p; in a small
neighbourhood U of p the exponential map has an inverse: $q\in
U\subset M^4 \Rightarrow q=Exp\, {}^4V_p$ for some ${}^4V_p\in
T_pM^4$. Then, ${}^4V^{\mu}_p$ are the {\it normal coordinates}
$x_2^{\mu}$ of $q$ and U is a {\it normal neighbourhood}. Let us
remark that in this way one defines an {\it inertial observer in free
fall} at $q$ in general relativity.

In Yang-Mills theory with trivial principal bundles $P(M,G)=M\times G$
\cite{lusa}, the abstract object behind the configuration space is the
{\it connection 1-form} $\omega$ on $P(M,G)=M\times G$ \footnote{G is
a compact, semisimple, connected, simply connected Lie group with
compact, semisimple real Lie algebra $g$.}; instead Yang-Mills
configuration space contains the {\it gauge potentials over the base}
M, ${}^{\sigma}A^{(\omega )}=\sigma^{*}\omega$, i.e. the pull-backs to
M of the connection 1-form through global cross sections $\sigma
:M\rightarrow P$. The group ${\cal G}$ of gauge transformations (its
component connected to the identity) acting on the gauge potentials on
M is interpreted in a {\it passive} sense as a change of global cross
section at fixed connection $\omega$, ${}^{\sigma_U}A^{(\omega
)}=U^{-1}\, {}^{\sigma}A^{(\omega )}\, U+U^{-1}dU$ \footnote{
$\sigma_U=\sigma \cdot U$ with $U:M\rightarrow G$.}: this formula
describes the {\it gauge orbit} associated with the given $\omega$. In
this case, the group manifold of ${\cal G}$  \footnote{It is the space
of the cross sections of the principal bundle P(M,G).} may be
considered the principal bundle $P(M,G)= M\times G$ itself
parametrized with a special connection-dependent family of global
cross sections, after having chosen {\it canonical coordinates of
first kind} on a reference fiber (a copy of the group manifold of G)
and having parallel (with respect to the given connection) transported
them to the other fibers (see the next Section). In this way we avoid
the overparametrization of ${\cal G}$ by means of the
infinite-dimensional space of all possible local and global cross
sections from M to P (this would be the standard description of ${\cal
G}$). The infinitesimal gauge transformations \footnote{The Lie
algebra $g_{\cal G}$ of ${\cal G}$: it is a vector bundle whose
standard fiber is the Lie algebra $g$.} in phase space are generated
by the first class constraints giving the Gauss laws $\Gamma_a\approx
0$. By Legendre pullback to configuration space, we find

\bea
{}^{\sigma +\delta \sigma}A
^{(\omega )}&=&{}^{\sigma}A^{(\omega )}+\delta_o\, {}^{\sigma}A^{(\omega )}=
{}^{\sigma}A^{(\omega )}+U^{-1}(dU+[{}^{\sigma}A^{(\omega )},U])={}^{\sigma}A
^{(\omega )}+{\hat D}^{(A)}\alpha =\nonumber \\
&=&{}^{\sigma}A^{(\omega )}+\lbrace {}^{\sigma}A
^{(\omega )},\int \alpha_a\Gamma_a \rbrace,\quad\quad if\,\, U=I+\alpha.
\label{III2}
\eea

In our formulation of tetrad gravity the relevant configuration
variables are {\it globally defined cotriads} ${}^3e_{(a)r}(\tau ,\vec
\sigma )$ on the hypersurface $\Sigma_{\tau} \approx R^3$, which is a
parallelizable Riemannian 3-manifold $(\Sigma_{\tau},
{}^3g_{rs}={}^3e_{(a)r}\, {}^3e_{(a)s})$ assumed asymptotically flat
(therefore noncompact) at spatial infinity and geodesically complete;
with these hypotheses we have $T\Sigma_{\tau}\approx
\Sigma_{\tau}\times R^3$ and the coframe orthogonal principal affine
SO(3)-bundle is also trivial $L\Sigma_{\tau}\approx
\Sigma_{\tau}\times SO(3)$ \footnote{Its points are the abstract
coframes ${}^3\theta_{(a)}\, (={}^3e_{(a)r}d\sigma^r\,$ in global
$\Sigma_{\tau}$-adapted coordinates).}. In the phase space of tetrad
gravity the rotations of the  gauge group SO(3) are generated by the
first class constraints ${}^3{\tilde M}_{(a)}(\tau ,\vec \sigma )
\approx 0$. Therefore, in this case the abstract object behind the configuration
space is the {\it so(3)-valued  soldering 1-form} ${}^3\theta ={\hat
R}^{(a)}\, {}^3\theta_{(a)}$ \footnote{${\hat R}^{(a)}$ are the
generators of the Lie algebra so(3).}. This shows that to identify the
global cotriads ${}^3e_{(a)r} (\tau ,\vec \sigma )$ we have to choose
an atlas of coordinate charts on $\Sigma_{\tau}$, so that in each
chart ${}^3\theta \mapsto {\hat R}^{(a)}\, {}^3e_{(a)r}(\tau ,\vec
\sigma ) d\sigma^r$. Since $\Sigma_{\tau}$ is assumed diffeomorphic to
$R^3$, global coordinate systems exist.

The general coordinate transformations or space
pseudo-diffeomorphisms of $Diff\, \Sigma_{\tau}$ are denoted as
$\vec \sigma \mapsto {\vec \sigma}^{'}(\vec \sigma )=\vec \xi
(\vec \sigma )=\vec \sigma +{\hat {\vec \xi}}(\vec \sigma )$; for
infinitesimal pseudo-diffeomorphisms, ${\hat {\vec \xi}}(\vec
\sigma )=\delta \vec \sigma (\vec \sigma )$ is an infinitesimal
quantity and the inverse infinitesimal pseudo-diffeomorphism is
$\vec \sigma ({\vec \sigma}^{'})={\vec \sigma} ^{'}-\delta \vec
\sigma ({\vec \sigma}^{'})={\vec \sigma}^{'}-{\hat {\vec
\xi}}({\vec \sigma}^{'})$. The cotriads ${}^3e_{(a)r}(\tau ,\vec
\sigma )$ and the 3-metric ${}^3g_{rs}(\tau ,\vec \sigma
)={}^3e_{(a)r}(\tau ,\vec \sigma )\, {}^3e_{(a)s}(\tau ,\vec
\sigma )$ transform as \footnote{${\hat V}(\vec \xi (\vec \sigma
))$ is the operator whose action on functions is ${\hat V}(\vec
\xi (\vec \sigma ))f(\vec \sigma )=f(\vec \xi (\vec \sigma ))$.}
[${\cal L}_X$ is the Lie derivative along the vector field $X$]

\begin{eqnarray}
{}^3e_{(a)r}(\tau ,\vec \sigma ) &\mapsto& {}^3e^{'}_{(a)r}(\tau ,{\vec \sigma}
^{'}(\vec \sigma ))={{\partial \sigma^s}\over {\partial \sigma^{{'}r}}}\,
{}^3e_{(a)s}(\tau ,\vec \sigma ),\nonumber \\
{} &\Rightarrow& {}^3e_{(a)r}(\tau ,\vec \sigma )={{\partial \xi^s(\vec \sigma )
}\over {\partial \sigma^r}}\, {}^3e^{'}_{(a)s}(\tau ,\vec \xi (\vec \sigma ))=
{{\partial \xi^s(\vec \sigma )}\over {\partial \sigma^r}}\, {\hat V}(\vec \xi
(\vec \sigma ))\, {}^3e^{'}_{(a)s}(\tau ,\vec \sigma ),\nonumber \\
&&{}\nonumber \\
{}^3g_{rs}(\tau ,\vec \sigma ) &\mapsto& {}^3g^{'}_{rs}(\tau ,{\vec \sigma}^{'}
(\vec \sigma ))={{\partial \sigma^u}\over {\partial \sigma^{{'}r}}}
{{\partial \sigma^v}\over {\partial \sigma^{{'}s}}}\, {}^3g_{uv}(\tau ,\vec
\sigma ),\nonumber \\
&&{}\nonumber \\
\delta \, {}^3e_{(a)r}(\tau ,\vec \sigma )&=&{}^3e^{'}_{(a)r}(\tau ,{\vec
\sigma}^{'}(\vec \sigma ))-{}^3e_{(a)r}(\tau ,\vec \sigma )=\delta_o\, {}^3e
_{(a)r}(\tau ,\vec \sigma )+{\hat \xi}^s(\vec \sigma )\partial_s\, {}^3e_{(a)r}
(\tau ,\vec \sigma )=\nonumber \\
&=&{{\partial \sigma^s}\over {\partial \sigma^{{'}r}}}\, {}^3e_{(a)s}(\tau ,
\vec \sigma )-{}^3e_{(a)r}(\tau ,\vec \sigma )=-\partial_r{\hat \xi}^s(\vec
\sigma )\, {}^3e_{(a)s}(\tau ,\vec \sigma ),\nonumber \\
\delta_o\, {}^3e_{(a)r}(\tau ,\vec \sigma )&=&{}^3e^{'}_{(a)r}(\vec \sigma )-
{}^3e_{(a)r}(\vec \sigma )=-[\partial_r{\hat \xi}^s(\vec \sigma )+\delta^s_r
{\hat \xi}^u(\vec \sigma )\partial_u]{}^3e_{(a)s}(\tau ,\vec \sigma )=
\nonumber \\
&=&[{\cal L}_{-{\hat \xi}^s\partial_s}\, {}^3e_{(a)u}(\tau ,\vec \sigma )
d\sigma^u]_r=-\lbrace {}^3e_{(a)r}(\tau ,\vec \sigma ), \int d^3\sigma_1 {\hat
\xi}^s({\vec \sigma}_1)\, {}^3{\tilde \Theta}_s(\tau ,{\vec \sigma}_1)
\rbrace ,\nonumber \\
&&{}\nonumber \\
\delta \, {}^3g_{rs}(\tau ,\vec \sigma )&=&{}^3g^{'}_{rs}(\tau ,{\vec \sigma}
^{'}(\vec \sigma ))-{}^3g_{rs}(\tau ,\vec \sigma )=\delta_o\, {}^3g_{rs}(\tau ,
\vec \sigma )+{\hat \xi}^u(\vec \sigma )\partial_u\, {}^3g_{rs}(\tau ,
\vec \sigma )=\nonumber \\
&=&{{\partial \sigma^u}\over {\partial \sigma^{{'}r}}}{{\partial \sigma^v}\over
{\partial \sigma^{{'}s}}}\, {}^3g_{uv}(\tau ,\vec \sigma )-{}^3g_{rs}(\tau ,
\vec \sigma )=-[\delta^u_r\partial_s{\hat \xi}^v(\vec \sigma )+\delta^v_s
\partial_r{\hat \xi}^u(\vec \sigma )]{}^3g_{uv}(\tau ,\vec \sigma ),\nonumber \\
\delta_o\, {}^3g_{rs}(\tau ,\vec \sigma )&=&{}^3g^{'}_{rs}(\tau ,\vec \sigma )-
{}^3g_{rs}(\tau ,\vec \sigma )=\nonumber \\
&=&-\Big[ \delta^u_r\partial_s{\hat \xi}^v(\vec \sigma )
+\delta^v_s\partial_r{\hat \xi}^u(\vec \sigma )+\delta^u_r\delta^v_s{\hat \xi}
^w(\vec \sigma )\partial_w\Big] {}^3g_{uv}(\tau ,\vec \sigma )=\nonumber \\
&=&[{\cal L}_{-{\hat \xi}^w\partial_w}\, {}^3g_{uv}(\tau ,\vec \sigma )d\sigma^u
\otimes d\sigma^v]_{rs}=-\lbrace {}^3g_{rs}(\tau ,\vec \sigma ),\int
d^3\sigma_1 {\hat \xi}^s({\vec \sigma}_1)\, {}^3{\tilde \Theta}_s(\tau ,{\vec
\sigma}_1)\rbrace .\nonumber \\
&&
\label{III3}
\end{eqnarray}

Instead the action of finite and infinitesimal gauge rotations of angles
$\alpha_{(c)}(\vec \sigma )$ and $\delta \alpha_{(c)}(\vec \sigma )$ is
respectively

\begin{eqnarray}
{}^3e_{(a)r}(\tau ,\vec \sigma )&\mapsto&
{}^3R_{(a)(b)}(\alpha_{(c)}(\vec \sigma ))\, {}^3e_{(b)r}(\tau
,\vec \sigma ),\nonumber \\
 \delta_o\, {}^3e_{(a)r}(\tau ,\vec
\sigma )&=&\lbrace {}^3e_{(a)r}(\tau , \vec \sigma ), \int
d^3\sigma_1 \delta \alpha_{(c)}({\vec \sigma}_1)\, {}^3{\tilde
M}_{(c)}(\tau ,{\vec \sigma}_1)\rbrace =\nonumber \\
&=&\epsilon_{(a)(b)(c)}\delta \alpha_{(b)}(\vec \sigma )\,
{}^3e_{(c)r}(\tau , \vec \sigma ). \label{III4}
\end{eqnarray}

To identify the algebra ${\bar g}_R$ of
${\bar {\cal G}}_R$, let us study its symplectic action on
$T^{*}{\cal C}_e$, i.e. the infinitesimal canonical transformations generated by
the first class constraints ${}^3{\tilde M}_{(a)}(\tau ,\vec \sigma )$,
${}^3{\tilde \Theta}_r(\tau ,\vec \sigma )$. Let us define the vector fields

\begin{eqnarray}
&&X_{(a)}(\tau ,\vec \sigma )=-\lbrace .,{}^3{\tilde M}_{(a)}(\tau ,\vec
\sigma )\rbrace ,\nonumber \\
&&Y_r(\tau ,\vec \sigma )=-\lbrace .,{}^3{\tilde \Theta}_r(\tau ,\vec \sigma )
\rbrace .
\label{III5}
\end{eqnarray}

\noindent Due to Eqs.(\ref{II14})  they close the algebra

\begin{eqnarray}
&&[X_{(a)}(\tau ,\vec \sigma ),X_{(b)}(\tau ,{\vec \sigma}^{'})]=\delta^3(\vec
\sigma ,{\vec \sigma}^{'}) \epsilon_{(a)(b)(c)} X_{(c)}(\tau ,\vec \sigma ),
\nonumber \\
&&[X_{(a)}(\tau ,\vec \sigma ),Y_r(\tau ,{\vec \sigma}^{'})]=-{{\partial
\delta^3(\vec \sigma ,{\vec \sigma}^{'})}\over {\partial \sigma^{{'}r}}}
X_{(a)}(\tau ,{\vec \sigma}^{'}),\nonumber \\
&&[Y_r(\tau ,\vec \sigma ),Y_s(\tau ,{\vec \sigma}^{'})]=-{{\partial \delta^3
(\vec \sigma ,{\vec \sigma}^{'})}\over {\partial \sigma^{{'}s}}}Y_r(\tau ,{\vec
\sigma}^{'})-{{\partial \delta^3(\vec \sigma ,{\vec \sigma}^{'})}\over
{\partial \sigma^{{'}r}}}Y_s(\tau ,\vec \sigma ).
\label{III6}
\end{eqnarray}

These six vector fields describe the symplectic action of rotation and
space pseudo-diffeomorphism gauge transformations on the subspace of
phase space containing cotriads ${}^3e_{(a)r}(\tau ,\vec \sigma )$ and
their conjugate momenta ${}^3{\tilde \pi}^r_{(a)}(\tau ,\vec \sigma
)$. The non commutativity of rotations and space
pseudo-diffeomorphisms means that the action of a space
pseudo-diffeomorphism on a rotated cotriad produces a cotriad which
differ by a rotation with modified angles from the action of the space
pseudo-diffeomorphism on the original cotriad: if $\vec \sigma
\rightarrow {\vec \sigma}^{'}(\vec
\sigma )$ is a space pseudo-diffeomorphism and ${}^3R_{(a)(b)}(\alpha_{(c)}(\vec
\sigma ))$ is a rotation matrix parametrized with angles $\alpha_{(c)}(\vec
\sigma )$, then

\begin{eqnarray}
{}^3e_{(a)r}(\tau ,\vec \sigma )&\mapsto& {}^3e^{'}_{(a)r}(\tau ,{\vec \sigma}
^{'}(\vec \sigma ))={{\partial \sigma^s}\over {\partial \sigma^{{'}r}}}\,
{}^3e_{(a)s}(\tau ,\vec \sigma ),\nonumber \\
{}^3R_{(a)(b)}(\alpha_{(c)}(\vec \sigma )) &\,& {}^3e_{(b)r}(\tau ,\vec \sigma )
\mapsto {}^3R_{(a)(b)}(\alpha^{'}_{(c)}({\vec \sigma}^{'}(\vec \sigma )))\,
{}^3e^{'}_{(b)r}(\tau ,{\vec \sigma}^{'}(\vec \sigma ))=\nonumber \\
&=&{{\partial \sigma^s}\over {\partial \sigma^{{'}r}}}\Big[{}^3R_{(a)(b)}(\alpha
_{(c)}(\vec \sigma ))\, {}^3e_{(b)s}(\tau ,\vec \sigma )\Big] =\nonumber \\
&=&{}^3R_{(a)(b)}(\alpha_{(c)}(\vec \sigma ))\, {}^3e^{'}_{(b)r}({\vec \sigma}
^{'}(\vec \sigma )),\nonumber \\
\Rightarrow &{}& \alpha^{'}_{(c)}({\vec \sigma}^{'}(\vec \sigma ))=\alpha_{(c)}
(\vec \sigma ),
\label{III7}
\end{eqnarray}

\noindent i.e. the rotation matrices, namely the angles $\alpha_{(c)}(\vec
\sigma )$, behave as {\it scalar fields} under space pseudo-diffeomorphisms. Under
infinitesimal rotations ${}^3R_{(a)(b)}(\delta \alpha_{(c)}(\vec \sigma ))=
\delta_{(a)(b)}+\delta \alpha_{(c)}(\vec \sigma )({\hat R}^{(c)})_{(a)(b)}=
\delta_{(a)(b)}+\epsilon_{(a)(b)(c)} \delta \alpha_{(c)}(\vec \sigma )$ and
space pseudo-diffeomorphisms ${\vec \sigma}^{'}(\vec \sigma )=\vec
\sigma +\delta
\vec \sigma (\vec \sigma )$ \footnote{${\hat R}^{(c)}$ are the SO(3) generators in the
adjoint representation; $\delta \alpha_{(c)}(\vec \sigma )$, $\delta \vec
\sigma (\vec \sigma )$ are infinitesimal variations.}, we have

\begin{eqnarray}
\int d^3\sigma_1d^3\sigma_2 &{}& \delta \sigma^s({\vec \sigma}_2)\delta \alpha
_{(c)}({\vec \sigma}_1) [Y_s(\tau ,{\vec \sigma}_2),X_{(c)}(\tau ,{\vec
\sigma}_1)] {}^3e_{(a)r}(\tau ,\vec \sigma )=\nonumber \\
&=&\int d^3\sigma_2 \delta \beta_{(c)}({\vec \sigma}_2) X_{(c)}(\tau ,{\vec
\sigma}_2) {}^3e_{(a)r}(\tau ,\vec \sigma ),\nonumber \\
\delta \beta_{(c)}(\vec \sigma )&=&\delta \sigma^s(\vec \sigma ) {{\partial
\alpha_{(c)}(\vec \sigma )}\over {\partial \sigma^s}},\nonumber \\
\Rightarrow &{}& \alpha^{'}_{(c)}(\vec \sigma )=\alpha_{(c)}(\vec \sigma -
\delta \vec \sigma (\vec \sigma ))=\alpha_{(c)}(\vec \sigma )-\delta \beta_{(c)}
(\vec \sigma )\, \Rightarrow \delta_o\alpha_{(c)}(\vec \sigma )=-\delta
\beta_{(c)}(\vec \sigma ).\nonumber \\
&&{}
\label{III8}
\end{eqnarray}

\subsection{What is Known on the Group Manifold of Gauge Transformations.}

The group manifold of the group ${\bar {\cal G}}_R$ of gauge
transformations (isomorphic to $Aut\, L\Sigma_{\tau}$) is locally
parametrized by three parameters $\vec \xi (\vec \sigma )$ and by
three angles $\alpha_{(c)}(\vec \sigma )$ (which are also functions of
$\tau$), which are scalar fields under pseudo-diffeomorphisms, and
contains an invariant subgroup ${\bar {\cal G}}
_{ROT}$ \footnote{The group of gauge transformations of the coframe bundle
$L\Sigma_{\tau}$; it is a splitting normal Lie subgroup of $Aut\,
L\Sigma_{\tau}$ \cite{abba}$\,\,$.}, whose group manifold (in the
passive interpretation) is the space of the cross sections of the
trivial principal bundle $\Sigma_{\tau} \times SO(3)\approx
L\Sigma_{\tau}$ over $\Sigma_{\tau}$, like in SO(3) Yang-Mills
theory\cite{lusa}, if $\Sigma_{\tau}$ is {\it topologically trivial}
(its homotopy groups $\pi_k(\Sigma_{\tau})$ all vanish); therefore, it
may be parametrized as said above. As affine function space of
connections on this principal SO(3)-bundle we shall take the space of
{\it spin connection 1-forms} ${}^3\omega_{(a)}$, whose pullback to
$\Sigma_{\tau}$ by means of cross sections $\sigma :\Sigma_{\tau}
\rightarrow \Sigma_{\tau}\times SO(3)$ are the
(Levi-Civita) {\it spin connections} (or {\it Ricci rotation
coefficients}) ${}^3\omega
_{r(a)}(\tau ,\vec \sigma )d\sigma^r=\sigma^{*}\, {}^3\omega_{(a)}$ built
with {\it cotriads} ${}^3e_{(a)r}(\tau ,\vec \sigma )$ \footnote{They
and not the spin connections are the independent variables of tetrad
gravity.} such that ${}^3g_{rs}={}^3e_{(a)r}\, {}^3e_{(a)s}$.

Due to our hypotheses on $\Sigma_{\tau}$ (parallelizable,
asymptotically flat, topologically trivial, geodesically complete),
the Hopf-Rinow theorem \cite{oneil} implies the existence of (at
least) one point $p\in \Sigma_{\tau}$ which can be chosen as reference
point and can be connected to every other point $q\in \Sigma_{\tau}$
with a minimizing geodesic segment $\gamma_{pq}$; moreover, the
theorem says that  there exists a point $p\in \Sigma_{\tau}$ from
which $\Sigma_{\tau}$ is geodesically complete and that the geodesic
exponential map $Exp_p$ is defined on all $T_p\Sigma_{\tau}$. If
$\Sigma_{\tau}$ is further restricted to have sectional curvature
${}^3K_p(\Pi ) \leq 0$ for each $p\in \Sigma_{\tau}$ and each tangent
plane $\Pi \subset T_p\Sigma_{\tau}$, the Hadamard theorem
\cite{oneil} says that for each $p\in \Sigma_{\tau}$ the geodesic
exponential map $Exp_p:T_p\Sigma_{\tau}\rightarrow \Sigma_{\tau}$ is a
diffeomorphism: therefore, there is a unique geodesic joining any pair
of points $p,q\in \Sigma_{\tau}$ and $\Sigma_{\tau}$ is diffeomorphic
to $R^3$ as we have assumed.

In absence of rotations, the group ${\bar {\cal G}}_R$ is reduced to
the group $Diff\,\Sigma_{\tau}$ of space pseudo-diffeomorphisms. In
the {\it active} point of view, diffeomorphisms are smooth mappings
(with smooth inverse) $\Sigma_{\tau}
\rightarrow \Sigma_{\tau}$: under $Diff\, \Sigma_{\tau}$ a point $p\in \Sigma
_{\tau}$ is sent (in many ways) in every point of $\Sigma_{\tau}$. In the
{\it passive} point of view, the action of the elements of $Diff\,
\Sigma_{\tau}$, called pseudo-diffeomorphisms, on a neighbourhood of a
point $p\in \Sigma_{\tau}$ is equivalent to all the possible
coordinatizations of the subsets of the neighbourhood of p
\footnote{I.e. to all possible changes of coordinate charts containing
p.}.

A {\it coordinate system} (or {\it chart}) $(U,\sigma )$ in
$\Sigma_{\tau}$ is a homeomorphism (which is also a diffeomorphism)
$\sigma$ of an open set $U\subset \Sigma_{\tau}$ onto an open set
$\sigma (U)$ of $R^3$: if $\sigma :U\rightarrow \sigma (U)$ and $p\in
U$, then $\sigma (p)=(\sigma^r(p))$, where the functions $\sigma^r$
are called the {\it coordinate functions} of $\sigma$. An {\it atlas}
on $\Sigma_{\tau}$ is a collection of charts in $\Sigma
_{\tau}$ such that: i) each point $p\in \Sigma_{\tau}$ is contained in the
domain of some chart; ii) any two charts overlap smoothly. Let ${\cal A}=
\lbrace (U_{\alpha},\sigma_{\alpha})\rbrace$ be the unique {\it complete} atlas on
$\Sigma_{\tau}$, i.e. an atlas by definition containing each coordinate system
$(U_{\alpha},\sigma_{\alpha})$ in $\Sigma_{\tau}$ that overlaps smoothly with
every coordinate system in ${\cal A}$.

Given an active diffeomorphism $\phi :\Sigma_{\tau} \rightarrow
\Sigma_{\tau}$ (i.e. a smooth mapping with smooth inverse) and any
chart $(U,\sigma )$ in ${\cal A}$, then $(\phi (U),
\sigma_{\phi}=\sigma \circ \phi)$ is another chart in ${\cal A}$
(the {\it dragged-along} chart) with $\sigma_{\phi}(p)\,
{\buildrel {def} \over =}\, \sigma(\phi (p))$ and, if $p\in U$.
Therefore, to each active diffeomorphism $\phi
:\Sigma_{\tau}\rightarrow \Sigma_{\tau}$ we can associate a
mapping $\phi_{\cal A}:{\cal A}\rightarrow {\cal A}$, i.e. a
pseudo-diffeomorphism. If we consider a point $p\in \Sigma_{\tau}$
and the set ${\cal A}_p=\lbrace (U^p
_{\beta},\sigma^p_{\beta})\rbrace$ of all charts in ${\cal A}$
containing p, then for each diffeomorphism $\phi
:\Sigma_{\tau}\rightarrow \Sigma_{\tau}$ we will have the
pseudo-diffeomorphism $\phi_{\cal A}:{\cal A}_p\rightarrow {\cal
A}_p$. This suggests that a {\it local parametrization} of $Diff\,
\Sigma_{\tau}$ around a point $p\in \Sigma_{\tau}$ \footnote{I.e.
local pseudo-diffeomorphisms defined on the open sets containing
p.}  may be done by choosing an arbitrary chart $(U^p_o,
\sigma^p_o)$ as the local identity of pseudo-diffeomorphisms
[$\vec \xi (\vec \sigma ) =\vec \sigma$] and associating with
every nontrivial diffeomorphism $\phi : \Sigma_{\tau}\rightarrow
\Sigma_{\tau}$, $\vec \sigma \mapsto {\vec \sigma}^{'} (\vec
\sigma )=\vec \xi (\vec \sigma )$, the chart $(U^p_{\beta}=\phi
(U^p_o),\sigma^p_{\beta}=\sigma^p_o\circ \phi)$. Since
$\Sigma_{\tau} \approx R^3$ admits global charts $\Xi$,  then the
group manifold of $Diff\, \Sigma_{\tau}$ may be tentatively
parametrized (in a nonredundant way)  with the space of smooth
global cross sections (global coordinate systems) in a {\it
fibration} $\Sigma_{\tau}\times \Sigma_{\tau}\rightarrow
\Sigma_{\tau}$ \footnote{Each global cross section of this
fibration is a copy $\Sigma^{(\Xi )}_{\tau}$ of $\Sigma_{\tau}$
with the given coordinate system $\Xi$.}: this is analogous to the
parametrization of the gauge group of Yang-Mills theory with a
family of global cross sections of the trivial principal bundle
$P(M,G)=M\times G$. The infinitesimal pseudo-diffeomorphisms
\footnote{The algebra $T\Sigma_{\tau}$ of $Diff\,
\Sigma_{\tau}$\cite{diff}; its generators in its symplectic action
on $T^{*} {\cal C}_e$ are the vector fields $Y_r(\tau ,\vec \sigma
)$.} would correctly correspond to the cross sections of the
fibration $\Sigma_{\tau}\times T\Sigma_{\tau}\rightarrow
\Sigma_{\tau}$. With more general $\Sigma_{\tau}$ the previous
description would hold only locally.

By remembering Eq.(\ref{III1}), the following picture
emerges:\hfill\break i) Choose a global coordinate system $\Xi$ on
$\Sigma_{\tau} \approx R^3$ (for instance 3-orthogonal
coordinates).\hfill\break ii) In the description of $Diff\,
\Sigma_{\tau}$ as $\Sigma_{\tau} \times
\Sigma_{\tau} \rightarrow \Sigma_{\tau}$ this corresponds to the choice of a
global cross section $\sigma_{\Xi}$ in $\Sigma_{\tau}\times
\Sigma_{\tau}$, chosen as conventional origin of the
pseudo-diffeomorphisms parametrized as $\vec \sigma \mapsto \vec \xi
(\vec \sigma )$.\hfill\break iii) This procedure identifies a cross
section ${\tilde \sigma}_{\Xi}$ of the principal bundle $Aut\,
L\Sigma_{\tau} \rightarrow Diff\, \Sigma_{\tau}$, whose action on
$L\Sigma_{\tau}$ will be the SO(3) gauge rotations in the chosen
coordinate system $\Xi$ on $\Sigma_{\tau}$.\hfill\break iv) This will
induce a $\Xi$-dependent trivialization of $L\Sigma_{\tau}$ to
$\Sigma_{\tau}^{(\Xi )}\times SO(3)$, in which $\Sigma_{\tau}$ has
$\Xi$ as coordinate system and the identity cross section
$\sigma^{(\Xi )}_I$ of $\Sigma_{\tau}^{(\Xi )}\times SO(3)$
corresponds to the origin of rotations in the coordinate system $\Xi$
\footnote{Remember that the angles are scalar fields under pseudo-diffeomorphisms
in $Diff\, \Sigma_{\tau}$.}.\hfill\break v) As we will see in the next
Section, it is possible to define new vector fields ${\tilde Y}_r(\tau
,\vec \sigma )$ which commute with the rotations ($[X_{(a)} (\tau
,\vec \sigma ),{\tilde Y}_r(\tau ,{\vec
\sigma}^{'})]=0$) and still satisfy the last line of Eqs.(\ref{III6}).
In this way the algebra ${\bar g}_R$ of the group ${\bar {\cal G}}_R$
is replaced (at least locally) by a new algebra ${\bar g}_R^{'}$,
which  defines a  group ${\bar {\cal G}}_R^{'}$, which is a (local)
trivialization of $Aut\, L\Sigma_{\tau}$. It is at this level that the
rotations in ${\bar {\cal G}}_{ROT}$ may be parametrized with a
special family of cross sections of the trivial orthogonal coframe
bundle $\Sigma_{\tau}^{(\Xi )}\times SO(3) \approx L\Sigma_{\tau}$, as
for SO(3) Yang-Mills theory, as said in iv).\hfill\break We do not
know whether these steps can be implemented rigorously in a global way
for $\Sigma_{\tau} \approx R^3$; if this is possible, then the
quasi-Shanmugadhasan canonical transformation of Section V can be
defined globally for global coordinate systems on $\Sigma_{\tau}$.

Both to study the singularity structure of De Witt superspace
\cite{dew,fis,ing} for the Riemannian 3-manifolds $\Sigma_{\tau}$
(the space of 3-metrics ${}^3g$ modulo $Diff\, \Sigma_{\tau}$),
for instance the cone over cone singularities of Ref.\cite{arms},
and the analogous phenomenon (called in this case Gribov
ambiguity) for the group ${\bar {\cal G}}_{ROT}$ of SO(3) gauge
transformations, we have to analyze the stability subgroups of the
group ${\bar {\cal G}}_R$ of gauge transformations for special
cotriads ${}^3e_{(a)r}(\tau ,\vec \sigma )$, the basic variables
in tetrad gravity. In metric gravity, where the metric is the
basic variable and pseudo-diffeomorphisms are the only gauge
transformations (we are ignoring the superhamiltonian constraint
at this stage), it is known that if the 3-metric ${}^3g$ over a
noncompact 3-manifold like $\Sigma_{\tau}$ satisfies boundary
conditions compatible with being a function in a Sobolev space
$W^{2,s}$ with $s > 3/2$, then there exist special metrics
admitting {\it isometries}. The {\it group
$Iso(\Sigma_{\tau},{}^3g)$ of isometries} of a 3-metric of a
Riemann 3-manifold $(\Sigma_{\tau},{}^3g)$ is the subgroup of
$Diff\, \Sigma_{\tau}$ which leaves the functional form of the
3-metric ${}^3g_{rs}(\tau ,\vec \sigma )$ invariant (its Lie
algebra is spanned by the Killing vector fields): the
pseudo-diffeomorphism $\vec \sigma \mapsto {\vec \sigma}^{'}(\vec
\sigma )=\vec \xi (\vec \sigma )$ in $Diff\, \Sigma_{\tau}$ is an
isometry in $Iso(\Sigma_{\tau},{}^3g)$ if

\begin{equation}
{}^3g_{rs}(\tau ,{\vec \sigma}^{'}(\vec \sigma ))={}^3g^{'}_{rs}(\tau ,
{\vec \sigma}^{'}(\vec \sigma ))={{\partial \sigma^u}\over {\partial
\sigma^{{'}r}}}{{\partial \sigma^v}\over {\partial \sigma^{{'}s}}}\,
{}^3g_{uv}(\tau ,\vec \sigma ).
\label{III9}
\end{equation}

In such a case the function space of 3-metrics turns out to be a {\it
stratified manifold  with singularities} \cite{fis}. Each stratum
contains all metrics ${}^3g$ with the same subgroup
$Iso(\Sigma_{\tau},{}^3g)\subset Diff\,
\Sigma_{\tau}$ \footnote{Isomorphic but not equivalent subgroups of $Diff\,
\Sigma_{\tau}$ produce different strata of 3-metrics.}; each point in a
stratum with n Killing vectors is the vertex of a cone, which is a
stratum with n-1 Killing vectors (the {\it cone over cone structure of
singularities}\cite{arms}).

{}From\footnote{At the level of cotriads a
pseudo-diffeomorphism-dependent rotation is allowed.}

\bea
{}^3g_{rs}(\tau ,{\vec \sigma}^{'}(\vec \sigma ))&=&{}^3g_{rs}^{'}
(\tau ,{\vec \sigma}^{'}(\vec \sigma ))={}^3e^{'}_{(a)r}(\tau ,{\vec
\sigma}
^{'}(\vec \sigma ))\, {}^3e^{'}_{(a)s}(\tau ,{\vec \sigma}^{'}(\vec \sigma ))=
  \nonumber \\
 &=&{{\partial \sigma^u}\over {\partial \sigma^{{'}r}}}{{\partial
\sigma^v}\over {\partial \sigma^{{'}s}}}\, {}^3g_{uv}(\tau ,\vec
\sigma )={{\partial
\sigma^u}\over {\partial \sigma^{{'}r}}}{{\partial \sigma^v}\over {\partial
\sigma^{{'}s}}}\, {}^3e_{(a)r}(\tau ,\vec \sigma )\, {}^3e_{(a)s}(\tau ,
\vec \sigma ),\nonumber \\
{}^3e^{'}_{(a)r}(\tau ,{\vec \sigma}^{'}(\vec \sigma ))&=&
R_{(a)(b)}(\gamma (\tau ,{\vec \sigma}^{'}(\vec \sigma ))) {{\partial
\sigma^u}\over {\partial \sigma^{{'}r}}}\, {}^3e_{(b)u}(\tau ,
\vec \sigma ),
\label{III10}
\eea

\noindent it follows that also the functional form of the associated cotriads is
invariant under $Iso(\Sigma_{\tau},{}^3g)$

\begin{equation}
{}^3e_{(a)r}(\tau ,{\vec \sigma}^{'}(\vec \sigma ))={}^3e^{'}_{(a)r} (\tau ,
{\vec \sigma}^{'}(\vec \sigma ))=R_{(a)(b)}(\gamma (\tau ,{\vec \sigma}
^{'}(\vec \sigma )))
{{\partial \sigma^s}\over {\partial \sigma^{{'}r}}}\,
{}^3e_{(b)s}(\tau ,\vec \sigma ).
\label{III11}
\end{equation}

Moreover, ${}^3g_{rs}(\tau ,{\vec \sigma}^{'}(\vec \sigma ))={}^3g^{'}_{rs}
(\tau ,{\vec \sigma}^{'}(\vec \sigma ))$ implies ${}^3\Gamma^{{'}u}_{rs}(\tau ,
{\vec \sigma}^{'}(\vec \sigma ))={}^3\Gamma^u_{rs}(\tau ,{\vec \sigma}^{'}
(\vec \sigma ))$ and
${}^3R^{{'}u}{}_{rst}(\tau ,{\vec \sigma}^{'}(\vec \sigma ))
={}^3R^u{}_{rst}(\tau ,{\vec \sigma}^{'}(\vec \sigma ))$, so that $Iso(\Sigma
_{\tau},{}^3g)$ is also the stability group for the associated Christoffel
symbols and Riemann tensor

\begin{eqnarray}
{}^3\Gamma^u_{rs}(\tau ,{\vec \sigma}^{'}(\vec \sigma ))&=&{}^3\Gamma^{{'}u}
_{rs}(\tau ,{\vec \sigma}^{'}(\vec \sigma ))=\nonumber \\
&=&{{\partial \sigma
^{{'}u}}\over {\partial \sigma^v}}{{\partial \sigma^m}\over {\partial
\sigma^{{'}r}}}{{\partial \sigma^n}\over {\partial \sigma^{{'}s}}}\,
{}^3\Gamma^v_{mn}(\tau ,\vec \sigma )+{{\partial^2\sigma^v}\over {\partial
\sigma^{{'}r}\partial \sigma^{{'}s}}}{{\partial \sigma^{{'}u}}\over
{\partial \sigma^v}},\nonumber \\
&&{}\nonumber \\
{}^3R^u{}_{rst}(\tau ,{\vec \sigma}^{'}(\vec \sigma ))&=&{}^3R^{{'}u}{}_{rst}
(\tau ,{\vec \sigma}^{'}(\vec \sigma ))=\nonumber \\
&=&{{\partial \sigma
^{{'}u}}\over {\partial \sigma^v}}{{\partial \sigma^l}\over {\partial
\sigma^{{'}r}}} {{\partial \sigma^m}\over {\partial \sigma^{{'}s}}}
{{\partial \sigma^n}\over {\partial \sigma^{{'}t}}}\, {}^3R^v{}_{lmn}(\tau ,
\vec \sigma ).
\label{III12}
\end{eqnarray}

\noindent Let us remark that in the Yang-Mills case (see Ref.\cite{lusa} and
the end of this Section) the field strengths have generically a larger
stability group (the {\it gauge copies} problem) than the gauge
potentials (the {\it gauge symmetry} problem). Here, one expects that
Riemann tensors (the field strengths) should have a stability group
${\cal S}_R(\Sigma_{\tau},{}^3g)$ generically larger of the one of the
Christoffel symbols (the connection) ${\cal S}_{\Gamma}(\Sigma
_{\tau},{}^3g)$, which in turn should be larger of the isometry group of the
metric: ${\cal S}_R(\Sigma_{\tau},{}^3g) \supseteq {\cal S}_{\Gamma}(\Sigma
_{\tau},{}^3g) \supseteq Iso(\Sigma_{\tau},{}^3g)$. However, these stability
groups do not seem to have been explored in the literature.

Since the most general transformation in ${\bar {\cal G}}_R$ for
cotriads ${}^3e_{(a)r}(\tau ,\vec \sigma )$, spin connections ${\hat
R}^{(a)}\, {}^3\omega_{r(a)}(\tau ,\vec \sigma )$ and field strengths
${\hat R}^{(a)}\, {}^3\Omega_{rs(a)}(\tau ,\vec \sigma )$ is
\footnote{We conform with the notations of Ref.\cite{lusa}.}

\begin{eqnarray}
{}^3e^{{'}R}_{(a)r}(\tau ,{\vec \sigma}^{'}(\vec \sigma ))&=&
{}^3R_{(a)(b)}(\alpha_{(c)}(\tau ,\vec \sigma )){{\partial \sigma^s}\over
{\partial \sigma^{{'}r}}}\,
{}^3e_{(b)s}(\tau ,\vec \sigma ),\nonumber \\
&&{}\nonumber \\
{\hat R}^{(a)}\, {}^3\omega^{{'}R}_{r(a)}(\tau ,{\vec \sigma}^{'}(\vec \sigma )
&=&{{\partial \sigma^u}\over {\partial \sigma^{{'}r}}}
\Big[ {}^3R^{-1}(\alpha_{(e)}
(\tau ,\vec \sigma ))\, {\hat R}^{(a)}\, {}^3\omega_{u(a)}(\tau ,\vec \sigma )\,
{}^3R(\alpha_{(e)}(\tau ,\vec \sigma ))+\nonumber \\
&+&{}^3R^{-1}(\alpha_{(e)}(\tau ,\vec \sigma ))
\partial_u\, {}^3R(\alpha_{(e)}(\tau ,\vec \sigma ))\Big] =\nonumber \\
&=&{{\partial \sigma^u}\over {\partial \sigma^{{'}r}}}\Big[ {\hat R}^{(a)}\,
{}^3\omega_{u(a)}(\tau ,\vec \sigma )+{}^3R^{-1}(\alpha_{(e)}(\tau ,\vec
\sigma ))\, {\hat D}^{(\omega )}_u\, {}^3R(\alpha_{(e)}(\tau ,\vec \sigma ))
\Big] =\nonumber \\
&=&{\hat R}^{(a)}\, {}^3\omega^{'}_{r(a)}(\tau ,{\vec \sigma}^{'}(\vec \sigma ))
+{}^3R^{-1}(\alpha_{(e)}(\tau ,\vec \sigma ))\, {\hat D}^{(\omega^{'})}_r\,
{}^3R(\alpha^{'}_{(e)}({\tau ,\vec \sigma}^{'}(\vec \sigma ))),\nonumber \\
&&{}\nonumber \\
{\hat R}^{(a)}\, {}^3\Omega^{{'}R}_{rs(a)}(\tau ,{\vec \sigma}^{'}(\vec
\sigma ))&=&{{\partial \sigma^u}\over {\partial \sigma^{{'}r}}}{{\partial
\sigma^v}\over {\partial \sigma^{{'}s}}}\, {}^3R^{-1}(\alpha_{(e)}(\tau ,\vec
\sigma ))\, {\hat R}^{(a)}\, {}^3\Omega_{uv(a)}(\tau ,\vec \sigma )\,
{}^3R(\alpha_{(e)}(\tau ,\vec \sigma ))=\nonumber \\
&=&{{\partial \sigma^u}\over {\partial \sigma^{{'}r}}}{{\partial
\sigma^v}\over {\partial \sigma^{{'}s}}}\, \Big( {\hat R}^{(a)}\, {}^3\Omega
_{uv(a)}(\tau ,\vec \sigma )+\nonumber \\
&+&{}^3R^{-1}(\alpha_{(e)}(\tau ,\vec
\sigma ))\, \Big[ {\hat R}^{(a)}\, {}^3\Omega_{uv(a)}(\tau ,\vec \sigma ),{}^3R
(\alpha_{(e)}(\tau ,\vec \sigma ))\Big]\, \Big) =\nonumber \\
&=&{\hat R}^{(a)}\, {}^3\Omega^{'}_{rs(a)}(\tau ,{\vec \sigma}^{'}(\vec \sigma
)) +\nonumber \\
&+&{}^3R^{-1}(\alpha_{(e)}^{'}(\tau ,{\vec \sigma}^{'}(\vec \sigma )))\,\Big[
{\hat R}^{(a)}\, {}^3\Omega^{'}_{rs(a)}(\tau ,{\vec \sigma}^{'}(\vec \sigma )),
{}^3R(\alpha_{(e)}^{'}(\tau ,{\vec \sigma}^{'}(\vec \sigma )))\Big] .
\label{III13}
\end{eqnarray}

\noindent where $({\hat D}^{(\omega )}_r)_{(a)(b)}={\hat D}^{(\omega )}
_{(a)(b)r}(\tau ,\vec \sigma )=\delta_{(a)(b)}\partial_r+\epsilon_{(a)(c)(b)}\,
{}^3\omega_{r(c)}(\tau ,\vec \sigma )$ and ${}^3R(\alpha_{(e)})$ are $3\times
3$ rotation matrices, the behaviour of spin connections and field strengths
under isometries can be studied.

\subsection{The Gribov Ambiguity.}

Let us now briefly review the Gribov ambiguity for the spin
connections and the field strengths following Ref.\cite{lusa}. All
spin connections are invariant under gauge transformations belonging
to the {\it center} $Z_3$ of SO(3): ${}^3R\in Z_3$ $\Rightarrow \,\,
{}^3\omega^R_{r(a)}= {}^3\omega_{r(a)}$.

As shown in Ref.\cite{lusa}, there can be special spin connections ${}^3\omega
_{r(a)}(\tau ,\vec \sigma )$, which admit a stability subgroup ${\bar {\cal
G}}^{\omega}_{ROT}$ ({\it gauge symmetries}) of ${\bar {\cal
G}}_{ROT}$, leaving them fixed

\begin{equation}
{}^3R(\alpha_{(e)}(\tau ,\vec \sigma ))\in {\bar {\cal G}}^{\omega}_R
\Rightarrow {\hat D}^{(\omega )}_r\, {}^3R(\alpha_{(e)}(\tau ,\vec \sigma ))
=0\, \Rightarrow \, {}^3\omega^R_{r(a)}(\tau ,\vec \sigma )={}^3\omega_{r(a)}
(\tau ,\vec \sigma ).
\label{III14}
\end{equation}

\noindent {}From Eq.(\ref{III11}), it follows that under an isometry in $Iso\,
(\Sigma_{\tau},{}^3g)$ we have ${}^3\omega^{'}_{r(a)}(\tau ,{\vec \sigma}
^{'}(\vec \sigma ))= {}^3\omega_{r(a)}(\tau ,{\vec \sigma}^{'}(\vec \sigma ))$,
namely the rotations ${}^3R(\gamma (\tau ,{\vec \sigma}^{'}(\vec \sigma )))$
are gauge symmetries.

When there are gauge symmetries, the spin connection is {\it
reducible}: its holonomy group $\Phi^{\omega}$ is a subgroup of SO(3)
[$\Phi^{\omega}\subset SO(3)$] and ${\bar {\cal G}}^{\omega}_{ROT}$
\footnote{It is always equal to the {\it centralizer of the holonomy group} in
SO(3), $Z_{SO(3)}(\Phi^{\omega})$>} satisfies ${\bar {\cal
G}}_{ROT}^{\omega}=Z_{SO(3)}(\Phi^{\omega})\supset Z_3$.

Moreover, there can be special field strengths ${}^3\Omega_{rs(a)}$
which admit a stability subgroup ${\bar {\cal G}}_{ROT}^{\Omega}$ of ${\bar
{\cal G}}_{ROT}$ leaving them fixed

\begin{eqnarray}
{}^3R(\alpha_{(e)}(\tau ,\vec \sigma )&&\in {\bar {\cal G}}_R^{\Omega}\,
\Rightarrow \,  [{\hat R}^{(a)}\, {}^3\Omega_{rs(a)}(\tau ,\vec \sigma ),{}^3R(
\alpha_{(e)}(\tau ,\vec \sigma )]=0\, \nonumber \\
&&\Rightarrow {}^3\Omega^R_{rs(a)}(\tau ,
\vec \sigma )={}^3\Omega_{rs(a)}(\tau ,\vec \sigma ).
\label{III15}
\end{eqnarray}

\noindent We have ${\bar {\cal G}}_{ROT}^{\Omega}\supseteq {\bar {\cal G}}_{ROT}
^{\omega}=Z_{SO(3)}(\Phi^{\omega})\supset Z_3$ and there is the problem of
{\it gauge copies}: there exist different spin connections
${}^3\omega_{r(a)}(\tau ,\vec \sigma )$ giving rise to the same field
strength ${}^3\Omega_{rs(a)}(\tau ,\vec \sigma )$.

A spin connection is {\it irreducible}, when its holonomy group
$\Phi^{\omega}$ is a {\it not closed} irreducible matrix subgroup of
SO(3). In this case we have ${\bar {\cal G}}_{ROT}^{\Omega}\supset
{\bar {\cal G}}_{ROT}^{\omega}=Z_{SO(3)}(\Phi^{\omega})=Z_3$ and there
are gauge copies, but not gauge symmetries.

Finally, a spin connection ${}^3\omega_{r(a)}(\tau ,\vec \sigma )$
is {\it fully irreducible} if $\Phi^{\omega}=SO(3)$: in this case
there are neither gauge symmetries nor gauge copies (${\bar {\cal
G}}_{ROT}^{\Omega}={\bar {\cal G}}_{ROT}^{\omega}=Z_3$) and the
holonomy bundle $P^{\omega}(p)$ of every point $p\in
\Sigma_{\tau}\times SO(3)$ coincides with $\Sigma_{\tau}\times
SO(3)$ itself, so that every two points in $\Sigma_{\tau}\times
SO(3)$ can be joined by a $\omega$-horizontal curve. Only in this
case  {\it the covariant divergence is an elliptic operator
without zero modes} (this requires the use of special {\it
weighted Sobolev spaces} for the spin connections to exclude the
irreducible and reducible ones) and its Green function can be
globally defined (absence of Gribov ambiguities).

In conclusion, the following diagram

\begin{eqnarray}
\begin{array}{ccccccc}
{}& {}& \rightarrow& {}& {}^3\omega_{r(a)}& \rightarrow& {}^3\Omega_{rs(a)}\\
{}^3e_{(a)r}& {}& {}& {}& {}& {}& \Updownarrow \\
{}& \rightarrow& {}^3g_{rs}& \rightarrow& {}^3\Gamma^u_{rs}& \rightarrow&
{}^3R^u{}_{vrs}, \end{array}
\label{III16}
\end{eqnarray}

\noindent together with Eqs.(\ref{III11}), (\ref{III13}),  implies
that, to avoid any kind of pathology associated with stability
subgroups of gauge transformations, one has to work with cotriads
belonging to a function space such that: i) there is no subgroup of
isometries in the action of $Diff\, \Sigma_{\tau}$ on the cotriads (no
cone over cone structure of singularities in the lower branch of the
diagram); ii) all the spin connections associated with the cotriads
are fully irreducible (no type of Gribov ambiguity in the upper branch
of the diagram). Both these requirements point towards the use of
special weighted Sobolev spaces like in Yang-Mills
theory\cite{lusa,moncr}.

It would be useful to make a systematic study of the relationships between
the stability groups ${\cal S}_R(\Sigma_{\tau},{}^3g) \supseteq {\cal S}
_{\Gamma}(\Sigma_{\tau},{}^3g) \supseteq Iso\, (\Sigma_{\tau},{}^3g)$ and the
stability groups ${\bar {\cal G}}^{\Omega}_{ROT} \supseteq {\bar {\cal G}}
^{\omega}_{ROT}$ and to show rigorously that the presence of isometries (Gribov
ambiguity) in the lower (upper) branch of the diagram implies the existence of
Gribov ambiguity (isometries) in the upper (lower) branch.

Let us make a  comment on the global Gribov ambiguity. Both in
Yang-Mills theory and in tetrad gravity  with angle-dependent boundary
conditions on noncompact spacelike Cauchy surfaces we cannot make an
one-point compactification of these surfaces: therefore, the relevant
principal fiber bundles remain trivial and have the global identity
cross section needed for the evaluation of Dirac's observables, but at
the price of having the local Gribov ambiguities described in this
Section. In the limit of angle-independent boundary conditions in
suitable weighted Sobolev spaces (only completely irreducible
connections; absence of local Gribov problem; well defined color
charges in Yang-Mills theory and absence of supertranslations in
tetrad gravity) we will {\it go on} to consider {\it trivial}
principal fiber bundle: even if now it would be possible to make the
one-point compactification, we are not doing it not to loose the
global identity cross section. Otherwise our construction of Dirac's
observables in the following Sections would become local.

In the more complex case in which magnetic monopoles or other objects
which require the use of a non-trivial principal bundle from the
beginning are assumed to exist , we remark that strictly speaking
action principles depending on gauge potentials on the base manifold
do not exist and one should reformulate the gauge theory starting from
an action principle defined on the principal bundle manifold and
depending on the connections on it \cite{marmo}. Then one would have
many more gauge degrees of freedom (the vertical vector fields) and
the search of Dirac's observables should be reformulated in this
framework.

See Ref.\cite{giul} for a treatment of large diffeomorphisms, the
analogous of the large gauge transformations (due to winding number)
of Yang-Mills theory \cite{lusa}, not connected to the identity.

The requirement of absence of isometries for every Riemann 3-manifold
$\Sigma_{\tau}$ (the Cauchy surfaces) in the foliation of the
spacetime $M^4$, associated with its allowed 3+1 splittings, should
not be an obstruction to the existence of 4-isometries of the
pseudo-Riemannian 4-manifold $M^4$. For instance Minkowski spacetime
has ten 4-isometries (the Killing vectors are associated to the
kinematical Poincar\'e group) and can be foliated with foliations
whose spacelike leaves admit no 3-isometry of their intrinsic
Riemannian structure.

\subsection{The Superhamiltonian Constraint as a Generator of
         Gauge Transformations.}

Let us now consider the gauge transformations generated by the
superhamiltonian constraint, whose meaning has never  been
completely clarified in the literature (see for instance
Refs.\cite{wa,anton}). Here we shall repeat what has been already
said in Ref.\cite{restmg} regarding metric gravity. Since in
tetrad gravity the superhamiltonian constraint is the same as in
metric gravity \cite{ru11}, only re-expressed in terms of the
cotriads and their momenta, the interpretation of the gauge
transformations generated by this constraint is the same in the
two theories.

In Ref.\cite{tei} the superhamiltonian and supermomentum
constraints of ADM metric  gravity are interpreted as the
generators of the change of the canonical data ${}^3g_{rs}$,
${}^3{\tilde \Pi}^{rs}$, under the normal and tangent deformations
of the spacelike hypersurface $\Sigma_{\tau}$ which generate
$\Sigma_{\tau +d\tau}$ \footnote{One thinks to $\Sigma_{\tau}$ as
determined by a cloud of observers, one per space point; the idea
of bifurcation and re-encounter of the observers is expressed by
saying that the data on $\Sigma_{\tau}$ (where the bifurcation
took place) are propagated to some final $\Sigma_{\tau + d\tau}$
(where the re-encounter arises) along different intermediate
paths, each path being a monoparametric family of surfaces that
fills the sandwich in between the two surfaces; embeddability of
$\Sigma_{\tau}$ in $M^4$ becomes the synonymous with path
independence; see also Ref.\cite{gr13} for the connection with the
theorema egregium of Gauss.}. Therefore, the algebra of
supermomentum and superhamiltonian constraints reflects the {\it
embeddability} of $\Sigma_{\tau}$ into $M^4$ (see also
Ref.\cite{anton}).

As a consequence of this geometrical property in the case of {\it
compact spacetimes without boundary} the superhamiltonian constraint
is interpreted as a {\it time-dependent Hamiltonian} for general
relativity in some  {\it internal} time variable defined in terms of
the canonical variables (see for instance Ref.\cite{beig} and the so
called {\it internal intrinsic many-fingered time} \cite{kku}).

The two main proposals for an {\it internal time} are:

i) The {\it intrinsic internal time} : it is the conformal factor
$q(\tau ,\vec \sigma )={1\over 6} ln\, det\, {}^3g_{rs}$ or $\phi
(\tau ,\vec
\sigma )= e^{{1\over 2}q(\tau ,\vec \sigma )}  =({}^3g)^{1/12}
> 0$ of the 3-metric. It is not a scalar and is
proportional to Misner's time $\Omega=-{1\over 3}\, ln\,
\sqrt{\hat \gamma}$ \cite{misner} for asymptotically flat spacetimes
(see Appendix C of II):  $q=-{1\over 2} \Omega$.

ii) York's {\it extrinsic internal time} ${\cal T}=-{{\epsilon
c^3}\over {12\pi G}}\, {}^3K={2\over {3\sqrt{\gamma}}}{}^3{\tilde
\Pi}$ \footnote{See Ref.\cite{beig} for a review of the known
results with York's extrinsic internal time, Ref.\cite{yoyo} for
York cosmic time versus proper time and Refs.\cite{ish,kuchar1}
for more general reviews about the problem of time in general
relativity.}.

There are two interpretations of the superhamiltonian constraint in
this framework

a) Either as a generator of time evolution (being a time-dependent
Hamiltonian) like in the commonly accepted viewpoint based on the
Klein-Gordon interpretation of the quantized superhamiltonian
constraint, i.e. the Wheeler-DeWitt equation \footnote{See Kuchar in
Ref.\cite{ku1} and Wheeler's evolution of 3-geometries in superspace
in Ref.\cite{dc11,mtw} ; see Ref.\cite{paren} for the cosmological
implications.}.

b) or as a quantum Hamilton-Jacobi  equation without any time (one
can introduce a concept of evolution, somehow connected with an
effective time, only in a WKB sense \cite{kiefer}).

 A related problem is the validity of the {\it full or thick sandwich conjecture}
\cite{dc11,mtw} \footnote{Given two nearby 3-metrics on Cauchy
surfaces $\Sigma_{\tau_1}$ and $\Sigma_{\tau_2}$, there is a unique
spacetime $M^4$, satisfying Einstein's equations, with these 3-metrics
on those Cauchy surfaces.} and of the {\it thin sandwich conjecture}
\footnote{Given ${}^3g$ and $\partial_{\tau}\, {}^3g$ on $\Sigma_{\tau}$, there
is a unique spacetime $M^4$ with these initial data satisfying
Einstein's equations; doing so the constraints become equations for
the lapse and shift functions against the logic of the Hamiltonian
presymplectic theory.}: see Ref. \cite{bafo} (and also
Ref.\cite{giusa}) for the non validity of the {\it full} case and for
the restricted validity (and its connection with constraint theory) of
the {\it thin} case.

Since the superhamiltonian  constraint is quadratic in the
momenta, one is naturally driven to make a comparison with the
free scalar relativistic particle described by the first class
constraint $p^2-\epsilon m^2\approx 0$. As shown in
Refs.\cite{lus1,dc1}, the constraint manifold in phase space has
1-dimensional gauge orbits (the two disjointed branches of the
mass-hyperboloid); the $\tau$-evolution generated by the Dirac
Hamiltonian $H_D=\lambda (\tau ) (p^2-\epsilon m^2)$ gives the
parametrized solution $x^{\mu}(\tau )$. Instead, if we go to the
reduced phase space by adding the non-covariant gauge-fixing
$x^o-\tau \approx 0$ and eliminating the pair of canonical
variables $x^o\approx \tau$, $p^o \approx \pm \sqrt{{\vec
p}^2+m^2}$, we get a frozen Jacobi description in terms of
independent Cauchy data, in which the same Minkowski trajectory of
the particle can be recovered in the non-covariant form $\vec
x(x^o)$ by introducing as Hamiltonian the energy generator $\pm
\sqrt{{\vec p}^2+m^2}$ of the Poincar\'e group \footnote{With the
variables of Ref.\cite{longhi}, one adds the covariant
gauge-fixing $p\cdot x/\sqrt{p^2} -\tau \approx 0$ and eliminates
the pair $T=p\cdot x/ \sqrt{p^2}$, $\epsilon =\eta \sqrt{p^2}
\approx \pm m$; now, since the invariant mass is constant, $\pm
m$, the non-covariant Jacobi data $\vec z= \epsilon (\vec x-\vec p
x^o/p^o)$, $\vec k=\vec p/\epsilon$ cannot be made to evolve.}.

This comparison would suggest to solve the superhamiltonian constraint
in one component of the ADM canonical momenta ${}^3{\tilde \Pi}^{rs}$,
namely in one component of the extrinsic curvature.

But, differently from the scalar particle, the solution of the
superhamiltonian constraint does not define the weak ADM energy,
which, instead, is connected with an integral over 3-space of that
part of the superhamiltonian constraint dictated by the associated
Gauss law, see Eqs.(5.4) of Ref.\cite{restmg}. Indeed, the
superhamiltonian constraint, being a secondary first class constraint
of a field theory, has an associated {\it Gauss law} like the
supermomentum constraints. In every Gauss law, the piece of the
secondary first class constraint corresponding to a divergence and
giving the {\it strong} form of the conserved charge (the strong ADM
energy in this case) as the flux through the surface at infinity of a
corresponding density depends on the variable which has to be
eliminated in the canonical reduction by using the constraint (the
conjugate variable is the gauge variable): once the constraint is
solved in this variable, it can be put inside the volume expression of
the {\it weak} form of the conserved charge to obtain its expression
in the reduced phase space; the strong ADM energy is the only known
charge, associated with a constraint bilinear in the momenta,
depending only on the coordinates and not on the momenta, so that this
implies that the superhamiltonian constraint has to be solved in one
of the components of the 3-metric.

This shows that  the right approach to the superhamiltonian
constraint is the one of Lichnerowicz\cite{conf} leading to the
conformal approach to the reduction of ADM metric gravity
\cite{york,cho,yoyo,ciuf} \footnote{See Appendix C of II for its
review and for some notions on mean extrinsic curvature slices,
for the TT ({\it transverse traceless})-decomposition and for more
comments about internal intrinsic and extrinsic times.}. In this
approach the superhamiltonian constraint supplemented with the
gauge fixing ${}^3K(\tau ,\vec \sigma )\approx 0$ (or $\approx
const.$; it is a condition on the internal extrinsic York time),
named {\it maximal slicing condition}, is considered as an
elliptic equation (the {\it Lichnerowicz equation}) to be solved
in the {\it conformal factor} $\phi (\tau ,\vec \sigma )=
e^{{1\over 2}q(\tau ,\vec \sigma )} > 0$ of the 3-metric
\footnote{Namely in its determinant ${}^3g = \phi^{12}$
(${}{}^3g=({}^3e)^2$ with ${}^3e=det ({}^3e_{(a)r})$ in tetrad
gravity), which can be extracted from it in a 3-covariant way.}
rather than in its conjugate momentum. Lichnerowicz has shown that
the superhamiltonian and supermomentum constraints plus the
maximal slicing condition of ADM metric gravity form a system of 5
elliptic differential equations which can be shown to have {\it
one and only one solution}; moreover, with this condition Schoen
and Yau \cite{p17} have shown that the ADM 4-momentum is timelike
(i.e. the ADM energy is positive or zero for Minkowski spacetime).
Moreover, Schoen-Yau have shown in their last proof of the
positivity of the ADM energy that one can relax the maximal
slicing condition. See the reviews\cite{cho,beig} with their rich
bibliography.

In the conformal approach one put ${}^3g_{rs}=\phi^4\,
{}^3\sigma_{rs}$ [$det\, {}^3\sigma_{rs} =1$] and ${}^3{\tilde
\Pi}^{rs}=\phi^{-10}\, {}^3{\tilde \Pi}^{rs}_A+{1\over 3} \, {}^3g^{rs}\,
{}^3\tilde \Pi$ [${}^3g_{rs}\, {}^3{\tilde \Pi}^{rs}_A=0$]. Then, one
makes the TT-decomposition ${}^3{\tilde \Pi}^{rs}_A={}^3{\tilde
\Pi}^{rs}_{TT} + {}^3{\tilde \Pi}^{rs}_L$ (the TT-part is the conformally rescaled
{\it distortion tensor}) with ${}^3{\tilde \Pi}^{rs}_L=(L
W_{\pi})^{rs}=W^{r|s}_{\pi} + W^{s|r}_{\pi} -{2\over 3}\, {}^3g^{rs}\,
W^u_{\pi \, |u}$, where $W^r_{\pi}$ is {\it York gravitomagnetic
vector potential}. The superhamiltonian and supermomentum constraints
are interpreted as coupled quasilinear elliptic equations for $\phi$
and $W^r_{\pi}$ (the four conjugate variables are free gauge
variables), which decouple with the maximal slicing condition
${}^3K=0$; the two physical degrees of freedom are hidden in
${}^3{\tilde \Pi}^{rs}_{TT}$ (and in two conjugate variables).

In Ref.\cite{yorkmap} it is shown that given the non-canonical
basis [see the end of Appendix C of II, in  particular its
Eq.(C7)] ${\cal T}=-{{\epsilon c^3}\over {12\pi G}} \,
{}^3K={2\over {3\sqrt{\gamma}}}{}^3{\tilde \Pi}$, ${\cal P}_{\cal
T}=-det\, {}^3g_{rs}=- \phi^{12}$,
${}^3\sigma_{rs}={}^3g_{rs}/(det\, {}^3g)^{1/3}$, ${}^3{\tilde
\Pi}^{rs}_A$ , there exists a canonical basis
 hidden in the variables ${}^3\sigma_{rs}$,
${}^3{\tilde \Pi}_A^{rs}$ (but it has {\it never} been found
explicitly) and that one can define the reduced phase space (the
{\it conformal superspace}) ${\tilde {\cal S}}$ \footnote{The {\it
conformal superspace} ${\tilde {\cal S}}$ may be defined as the
space of {\it conformal 3-geometries} on {\it closed} manifolds
and can be identified in a natural way with the space of {\it
conformal 3-metrics} (the quotient of superspace by the group
$Weyl\, \Sigma_{\tau}$ of {\it conformal Weyl rescalings}) modulo
space diffeomorphisms, or, equivalently, with the space of
Riemannian 3-metrics modulo space diffeomorphisms and conformal
transformations of the form ${}^3g_{rs}\mapsto \phi^4\,
{}^3g_{rs}$, $\phi > 0$. Instead, the {\it ordinary superspace}
${\cal S}$ is the space of {\it Lorentzian 4-metrics} modulo
spacetime diffeomorphisms. In this way a bridge is built towards
the phase superspace, which is mathematically connected with the
Moncrief splitting theorem\cite{mo,cho} valid for closed
$\Sigma_{\tau}$. See however Ref.\cite{cho} for what is known in
the asymptotically flat case by using weighted Sobolev spaces.},
in which one has gone to the quotient with respect to the space
diffeomorphisms and to the conformal rescalings. It is also shown
that one can define a {\it York map} from this reduced phase space
to the subset of the standard phase superspace \footnote{Quotient
of the ADM phase space with respect to the space diffeomorphisms
plus the gauge transformations generated by the superhamiltonian
constraint; it is the phase space of the superspace, the
configuration space obtained from the 3-metrics going to the
quotient with respect to the space- and {\it time}-
diffeomorphisms of the ADM formalism.} defined by the condition
${}^3K=const.$.

In the conformal approach one uses York's TT-variables \cite{york},
because most of the work on the Cauchy problem for Einstein's
equations in metric gravity is done by using spacelike hypersurfaces
$\Sigma$ of constant mean extrinsic curvature (CMC surfaces) in the
compact case (see Refs.\cite{cho,i1,i2}) and with the maximal slicing
condition ${\cal T}(\tau ,\vec \sigma )=0$. It may be extended to non
constant ${\cal T}$) in the asymptotically free case \footnote{See
also Ref.\cite{im} for recent work in the compact case with non
constant ${\cal T}$ and Ref.\cite{brill1} for solutions of Einstein's
equations in presence of matter which do not admit constant mean
extrinsic curvature slices.}.

Let us remark that in Minkowski spacetime ${}^3K(\tau ,\vec \sigma
)=0$ are the hyperplanes, while ${}^3K(\tau ,\vec \sigma )=const.$ are
the mass hyperboloids, corresponding to the instant and point form of
the dynamics according to Dirac\cite{dd} respectively (see
Refs.\cite{gaida} for other types of foliations).

Instead in  {\it asymptotically free} spacetimes  there exists a time
evolution in the mathematical time parametrizing the leaves
$\Sigma_{\tau}$ of the 3+1 splitting of $M^4$ {\it governed by the
weak ADM energy}\cite{fermi} as we have seen with the rest-frame
instant form of gravity. The superhamiltonian constraint is {\it not
connected with time evolution}: the strong and weak ADM energies are
only integrals of parts of this constraint. Instead it is the
generator of Hamiltonian gauge transformations.

As a constraint it determines the non-scalar conformal factor (the
determinant) of the 3-metric as a functional of ${}^3\sigma_{rs}$
and ${}^3{\tilde \Pi}^{rs}$ [of ${}^3e_{(a)r}$, ${}^3{\tilde
\pi}^r_{(a)}$ in tetrad gravity]. But this means that the
associated {\it gauge variable} is the {\it canonical momentum
conjugate to the conformal factor}. This variable, and not York
time, parametrizes the {\it normal deformation} of the embeddable
spacelike hypersurfaces $\Sigma_{\tau}$. Now, since different
$\Sigma_{\tau}$ corresponds to different 3+1 splittings of $M^4$,
in the class of the allowed ones going in an angle-independent way
to Minkowski spacelike hyperplanes, we get that the gauge
transformations generated by the superhamiltonian constraint
correspond to the transition from an allowed 3+1 splitting to
another one (this is the gauge orbit in the phase space over
superspace). Therefore {\it the theory is independent from the
choice of the 3+1 splitting like parametrized Minkowski theories}.

Since the solution of the Lichnerowicz equation gives the conformal
factor $\phi = e^{q/2}=({}^3g)^{1/12}$ as a function of its conjugate
momentum and of the remaining canonical variables as in the compact
case, also in the asymptotically free case {\it only the conformal
3-geometries contain the physical degrees of freedom, whose functional
form depends on the other gauge fixings, in particular on the choice
of the 3-coordinates}.

A gauge fixing to the superhamiltonian constraint is a choice of a
particular 3+1 splitting and this is done by fixing the momentum
conjugate to the conformal factor \footnote{A non-local information on
the extrinsic curvature of $\Sigma_{\tau}$, which becomes the York
time, or the maximal slicing condition, only with the special
canonical basis identified by the York map.}.

Therefore it is important to study the Shanmugadhasan canonical bases
of both metric and tetrad gravity, in which the conformal factor of
the 3-metric \footnote{Or better, if one succeeds in doing it, an
Abelianized form of the superhamiltonian constraint having zero
Poisson bracket with itself.} is one of the configurational canonical
variables (see Section VI). One of these bases should correspond to
the extension of the York map to asymptotically flat spacetimes: in it
the momentum conjugate to the conformal factor is just York time and
one can add the maximal slicing condition as a gauge fixing.

This leads to the conclusion that neither York's internal
extrinsic time nor Misner's internal intrinsic time are to be used
as time parameters: Misner's time (the conformal factor) is
determined by the Lichnerowicz equation while York's time (the
trace of the extrinsic curvature) by the gauge-fixing.  As said in
Section XI of Ref.\cite{restmg}, the rest-frame instant form of
metric and tetrad gravity uses a {\it mathematical time}
identified before quantization: the parameter $\tau \equiv
T_{(\infty )}$ labelling the WSW hypersurfaces and coinciding with
the rest-frame time of the {\it external} decoupled center of mass
of the universe considered as a {\it point particle clock}. We
refer to Ref.\cite{restmg} for further details and for the
connection to the either proper or coordinate time of physical
clocks.

\vfill\eject

\section{Multitemporal Equations and their Solution.}

In this Chapter we study the multitemporal equations\cite{lu1} (or
generalized Lie equations \cite{batalin}) associated with the gauge
transformations in ${\bar {\cal G}}_R$, to find a local
parametrization of the cotriads ${}^3e_{(a)r}(\tau ,\vec
\sigma )$ in terms of the parameters $\xi_r(\tau ,\vec \sigma )$ and
$\alpha_{(a)}(\tau ,\vec \sigma )$ of ${\bar {\cal G}}_R$. We shall
assume to have chosen a global coordinate system $\Xi$ on
$\Sigma_{\tau} \approx R^3$ to conform with the discussion of the
previous Section.

\subsection{The Multi-Temporal Equations for the Rotations.}

Let us start with the invariant subalgebra ${\bar g}_{ROT}$ (the
algebra of ${\bar {\cal G}}_{ROT}$) of rotations, whose generators are
the vector fields $X_{(a)}(\tau ,\vec \sigma )$ of Eqs.(\ref{III5}).
Since the group manifold of ${\bar {\cal G}}_{ROT}$ is a trivial
principal bundle $\Sigma_{\tau}^{(\Xi )}
\times SO(3)\approx L\Sigma_{\tau}$ over $\Sigma_{\tau}$, endowed with the
coordinate system $\Xi$, with structure group SO(3), we can use the
results of Ref.\cite{lusa} for the case of SO(3) Yang-Mills theory.

Let $\alpha_{(a)}$ be {\it canonical coordinates of first kind} on the
group manifold of SO(3). If $r^{(a)}$ are the generators of so(3),
$[r^{(a)},r^{(b)}]=\epsilon_{(a)(b)(c)}r^{(c)}$ \footnote{Instead
${\hat R}^{(a)}$ are the generators in the adjoint representation,
$({\hat R}^{(a)})_{(b)(c)}=\epsilon_{(a)(b)(c)}$.}, and if
$\gamma_{\alpha}(s)=exp_{SO(3)}\, (s \alpha_{(a)}r^{(a)})$ is a
one-parameter subgroup of SO(3) with tangent vector
$\alpha_{(a)}r^{(a)}$ at the identity $I\in SO(3)$, then the group
element $\gamma_{\alpha}(1)=exp_{SO(3)}(\alpha_{(a)}r^{(a)})
\in N_I\subset SO(3)$ \footnote{$N_I$ is a neighbourhood of the identity such that
$exp_{SO(3)}$ is a diffeomorphism from a neighbourhood of $0\in so(3)$
to $N_I$.} is given coordinates $\lbrace \alpha_{(a)} \rbrace$. If
${\tilde Y}
_{(a)}$ and ${\tilde \theta}_{(a)}$ are dual bases ($i_{{\tilde Y}_{(a)}}
{\tilde \theta}_{(b)}=\delta_{(a)(b)}$) of left invariant vector
fields and left invariant (or Maurer-Cartan) 1-forms on SO(3), we have
the standard Maurer-Cartan structure equations\footnote{$so(3)^{*}$ is the dual Lie
algebra; $TSO(3)\approx so(3)$, $T^{*}SO(3)
\approx so(3)^{*}$.}

\bea
&&[{\tilde Y}_{(a)},{\tilde Y}_{(b)}]=
\epsilon_{(a)(b)(c)}{\tilde Y}_{(c)}, \quad\quad
[{\tilde Y}_{(a)}{|}_I=r^{(a)}\in so(3)],\nonumber \\
 &&d{\tilde\theta}_{(a)}=-{1\over 2}\epsilon_{(a)(b)(c)} {\tilde \theta}
_{(b)}\wedge {\tilde \theta}_{(c)},\quad\quad
[{\tilde \theta}_{(a)}{|}_I=r_{(a)}\in so(3)^{*}].
\label{IV1}
\eea

\noindent  Then, from Lie theorems, on the
group manifold we have

\beq
{\tilde Y}_{(a)}=B_{(b)(a)}(\alpha ){{\partial}\over {\partial
\alpha_{(b)}}},\quad {\tilde \theta}_{(a)}=A_{(a)(b)}(\alpha )d\alpha
_{(b)},\quad  A(\alpha )=B^{-1}(\alpha ),\quad A(0)=B(0)=1,
\label{IV2}
\eeq

\noindent and the Maurer-Cartan equations become

\bea
&&{{\partial A_{(a)(c)}(\alpha )}\over {\partial \alpha_{(b)}}}
-{{\partial A_{(a)(b)}(\alpha )}\over {\partial \alpha_{(c)}}}=-\epsilon
_{(a)(u)(v)}A_{(u)(b)}(\alpha )A_{(v)(c)}(\alpha ),\nonumber \\
 &&{}\nonumber \\
 &&{\tilde Y}_{(b)} B_{(a)(c)}(\alpha )-{\tilde Y}_{(c)}B_{(a)(b)}(\alpha
)=B_{(u)(b)}(\alpha ) {{\partial B_{(a)(c)}(\alpha )}\over {\partial
\alpha_{(u)}}}-B_{(u)(c)} (\alpha ){{\partial B_{(a)(b)}(\alpha
)}\over {\partial \alpha_{(u)}}}=\nonumber \\
 &&= B_{(a)(u)}(\alpha )\epsilon_{(u)(b)(c)}.
\label{IV3}
\eea

By definition these coordinates are said canonical of first kind and
satisfy $A_{(a)(b)}(\alpha )\, \alpha_{(b)}=\alpha_{(a)}$, so that we
get $A(\alpha )=(e^{R\alpha}-1)/R\alpha$ with $(R\alpha
)_{(a)(b)}=({\hat R}^{(c)})_{(a)(b)}\alpha_{(c)}=\epsilon_{(a)(b)(c)}
\alpha_{(c)}$. The canonical 1-form on SO(3) is ${\tilde \omega}_{SO(3)}=
{\tilde \theta}_{(a)}r^{(a)}=A_{(a)(b)}(\alpha )d\alpha_{(b)}r^{(a)}$
\footnote{$= a^{-1}(\alpha )d_{SO(3)}\, a(\alpha )$, $a(\alpha )\in SO(3)$;
$d_{SO(3)}$ is the exterior derivative on SO(3).}. Due to the
Maurer-Cartan structure equations the 1-forms ${\tilde \theta}_{(a)}$
are not integrable on SO(3); however in the neighbourhood $N_I\subset
SO(3)$ we can integrate them along the {\it preferred defining line}
$\gamma_{\alpha}(s)$ defining the canonical coordinates of first kind
to get the phases

\beq
\Omega^{\gamma_{\alpha}}_{(a)}(\alpha (s))={}_{\gamma
_{\alpha}}\int_I^{\gamma_{\alpha}(s)}\, {\tilde \theta}_{(a)}{|}_{\gamma
_{\alpha}}={}_{\gamma_{\alpha}}\int_0^{\alpha (s)}\, A_{(a)(b)}(\bar \alpha )
d{\bar \alpha}_{(b)}.
\label{IV4}
\eeq

If $d_{\gamma_{\alpha}}=ds{{d\alpha_{(a)}(s)}\over {ds}}
{{\partial}\over {\partial \alpha_{(a)}}}{|}_{\alpha =\alpha
(s)}=d_{SO(3)} {|}_{\gamma_{\alpha}(s)}$ is the {\it directional
derivative} along $\gamma_{\alpha}$, on $\gamma_{\alpha}$ we have
$d_{\gamma_{\alpha}}\, {\Omega}^{\gamma
_{\alpha}}_{(a)}(\alpha (s))={\tilde \theta}_{(a)}(\alpha (s))$ and
$d_{\gamma_{\alpha}}\, {\tilde \theta}_{(a)}(\alpha (s))=0$ $\Rightarrow \,
d^2_{\gamma_{\alpha}}=0$. The analytic atlas ${\cal N}$ for the group manifold
of SO(3) is built by starting from the neighbourhood $N_I$ of the identity
with canonical coordinates of first kind by left multiplication by elements of
SO(3): ${\cal N}=\cup_{a\in SO(3)}\, \lbrace a \cdot N_I\rbrace$.

As shown in Ref.\cite{lusa} for $R^3\times SO(3)$, in a tubular
neighbourhood of the identity cross section $\sigma_I$ of the trivial
principal bundle $R^3\times SO(3)$, in which each fiber is a copy of
the SO(3) group manifold, we can define {\it generalized} canonical
coordinates of first kind on each fiber so to build a coordinatization
of $R^3\times SO(3)$. We now extend this construction from the flat
Riemannian manifold $(R^3,\delta
_{rs})$ to a Riemannian manifold $(\Sigma_{\tau},{}^3g_{rs})$ satisfying our
hypotheses, especially the Hadamard theorem, so
that the 3-manifold $\Sigma_{\tau}$, diffeomorphic to $R^3$, admits
global charts.

Let us consider the fiber SO(3) over a point $p\in \Sigma_{\tau}$,
chosen as origin $\vec \sigma =0$ of the global chart $\Xi$ on
$\Sigma_{\tau}$, with canonical coordinates of first kind
$\alpha_{(a)}=\alpha_{(a)}(\tau ,\vec 0)$ on it. For a given spin
connection ${}^3\omega_{(a)}$ on $\Sigma_{\tau}^{(\Xi )}\times SO(3)$
let us consider the ${}^3\omega$-{\it horizontal lifts} through each
point of the fiber SO(3) of the star of geodesics of the Riemann
3-manifold $(\Sigma_{\tau},{}^3g_{rs}={}^3e_{(a)r}\, {}^3e_{(a)s})$
emanating from $p\in \Sigma_{\tau}$ . If the spin connection
${}^3\omega_{(a)}$ is fully irreducible, $\Sigma_{\tau}\times SO(3)$
is in this way foliated by a connection-dependent family of global
cross sections defined by the ${}^3\omega$-horizontal lifts of the
star of geodesics \footnote{They are not ${}^3\omega$-horizontal cross
sections, as it was erroneously written in Ref.\cite{lusa}, since such
cross sections do not exist when the holonomy groups in each point of
$\Sigma_{\tau}\times SO(3)$ are not trivial.}. The canonical
coordinates of first kind on the reference SO(3) fiber may then be
parallel- (with respect to ${}^3\omega_{(a)}$) transported to all the
other fibers along these ${}^3\omega$-dependent global cross sections.
If $\tilde p=(p;\alpha_{(a)})=(\tau ,\vec 0;\alpha_{(a)}(\tau ,
\vec 0))$ is a point in $\Sigma_{\tau}\times SO(3)$ over $p\in \Sigma_{\tau}$,
if $\sigma_{(\tilde p)}:\Sigma_{\tau}\rightarrow \Sigma_{\tau}\times SO(3)$ is
the ${}^3\omega$-dependent cross section through $\tilde p$ and if ${}^3\omega
^{(\tilde p)}_{r(a)}(\tau ,\vec \sigma )d\sigma^r=\sigma^{*}_{(\tilde p)}\,
{}^3\omega_{(a)}$, then the coordinates of the point intersected by $\sigma
_{(\tilde p)}$ on the SO(3) fiber over the point $p^{'}$ of $\Sigma_{\tau}$ with
coordinates $(\tau ,\vec \sigma )$ are

\begin{eqnarray}
\alpha_{(a)}(\tau ,\vec \sigma )&=&\alpha_{(b)}(\tau ,\vec 0)\, \zeta^{(\omega
^{(\tilde p)}_{(c)})}_{(b)(a)}(\vec \sigma ,\vec 0;\tau )=\nonumber \\
&=&\alpha_{(b)}
(\tau ,\vec 0)\Big( P_{\gamma_{pp^{'}}}\,
e^{\int_{\vec 0}^{\vec \sigma}dz^r{\hat
R}^{(c)}\, {}^3\omega^{(\tilde p)}_{r(c)}(\tau,\vec z)}\Big)_{(b)(a)},
\label{IV5}
\end{eqnarray}

\noindent where $\zeta^{(\omega )}_{(b)(a)}(\vec \sigma ,\vec 0;\tau )$ is the
Wu-Yang non-integrable phase with the path ordering evaluated along
the geodesic $\gamma_{pp^{'}}$ from p to $p^{'}$. The infinitesimal
form is

\begin{eqnarray}
\alpha_{(a)}(\tau ,d\vec \sigma )&\approx& \alpha_{(a)}(\tau ,\vec 0)+{{\partial
\alpha_{(a)}(\tau ,\vec \sigma )}\over {\partial \sigma^r}}{|}_{\vec \sigma
=0} d\sigma^r \approx \nonumber \\
&\approx& \alpha_{(b)}(\tau ,\vec 0)
\Big[ \delta_{(b)(a)}+({\hat R}^{(c)})_{(b)(a)}
\, {}^3\omega_{r(c)}(\tau ,\vec 0)d\sigma^r\Big] ,
\label{IV6}
\end{eqnarray}

\noindent implying that the identity cross section $\sigma_I$ of $\Sigma_{\tau}
^{(\Xi )}\times SO(3)$
[$\alpha_{(a)}=\alpha_{(a)}(\tau ,\vec 0)=0$] is the origin for
all SO(3) fibers: $\alpha_{(a)}(\tau ,\vec \sigma ){|}_{\sigma_I}=0$. As shown
in Ref.\cite{lusa}, on $\sigma_I$ we also have $\partial_r \alpha_{(a)}(\tau ,
\vec \sigma ){|}_{\sigma_I}=0$. With this coordinatization the difference between
the coordinates $\alpha_{(a)}(\tau ,\vec 0)$ of a point $\tilde p$ in
the fiber over the reference point $p\in \Sigma_{\tau}$,
$\sigma_p^A=(\tau ,\vec 0)$, and the point ${\tilde p}^{'}$ on the
neighbouring fiber over $p^{'}\in \Sigma_{\tau}$,
$\sigma^A_{p^{'}}=(\tau ,d\vec \sigma )$, joined to $\tilde p$ by the
lift of the geodesic joining $p$ and $p^{'}$, is numerically equal to
the horizontal infinitesimal increment $\partial_r
\alpha_{(a)}(\tau ,\vec \sigma )d\sigma^r{|}_{\vec \sigma =\vec 0}$
in going from $\vec \sigma =\vec 0$ to $\vec 0 +d\vec \sigma$ in
$\Sigma_{\tau}$ of a function $\alpha_{(a)}(\tau ,\vec \sigma )$

\begin{equation}
d\alpha_{(a)}{|}_{\alpha =\alpha (\tau ,\vec \sigma )} = d\alpha_{(a)}(\tau ,
\vec \sigma )=\partial_r \alpha_{(a)}(\tau ,\vec \sigma ) d\sigma^r.
\label{IV7}
\end{equation}

\noindent These new coordinates of ${\tilde p}^{'}$ differ from the natural
canonical coordinates of first kind existing on the fiber to which
${\tilde p}^{'}$ belong by just this quantity, which then assumes the
meaning of a vertical infinitesimal increment added to the natural
coordinates.

With this coordinatization of $\Sigma_{\tau}^{(\Xi )}\times SO(3)$, in
the chosen global coordinate system $\Xi$ for $\Sigma_{\tau}$ in which
the identity cross section $\sigma_I$ is chosen as the origin of the
angles, as in Ref.\cite{lusa} we have the following realization for
the vector fields $X_{(a)}(\tau ,\vec \sigma )$ of Eqs.(\ref{III5})

\begin{equation}
X_{(a)}(\tau ,\vec \sigma )=B_{(b)(a)}(\alpha_{(e)}(\tau ,\vec \sigma ))
{{\tilde \delta}\over {\delta \alpha_{(b)}(\tau ,\vec \sigma )}}\,
\Rightarrow {{\tilde \delta}\over {\delta \alpha_{(a)}(\tau ,\vec \sigma )}}=
A_{(b)(a)}(\alpha_{(e)}(\tau ,\vec \sigma )) X_{(b)}(\tau ,\vec \sigma ),
\label{IV8}
\end{equation}

\noindent where the functional derivative is the {\it directional functional
derivative} along the path $\gamma_{\alpha (\tau ,\vec \sigma )}(s)$
in $\Sigma
_{\tau}^{(\Xi )}\times SO(3)$ originating at the identity cross section
$\sigma_I$ (the origin of all SO(3) fibers) in the SO(3) fiber over
the point $p\in \Sigma_{\tau}$ with coordinates $(\tau ,\vec \sigma
)$, corresponding in the above construction to the path
$\gamma_{\alpha}(s)$ defining the canonical coordinates of first kind
in the reference SO(3) fiber. It satisfies the commutator in
Eq.(\ref{III6}) due to the generalized Maurer-Cartan equations for
$\Sigma_{\tau}\times SO(3)$ [$A=B^{-1}$]

\begin{eqnarray}
B_{(u)(a)}(\alpha_{(e)}(\tau ,\vec \sigma ))&&{{\partial B_{(v)(b)}(\alpha
_{(e)})}\over {\partial \alpha_{(u)}}} {|}_{\alpha =\alpha (\tau ,\vec \sigma )}
-B_{(u)(b)}(\alpha_{(e)}(\tau ,\vec \sigma )){{\partial B_{(v)(a)}(\alpha_{(e)})
}\over {\partial \alpha_{(u)}}}
{|}_{\alpha =\alpha (\tau ,\vec \sigma )}=\nonumber \\
&=&B_{(v)(d)}(\alpha_{(e)}(\tau ,\vec \sigma )) \epsilon_{(d)(a)(b)},
\nonumber \\
{{\partial A_{(a)(c)}(\alpha_{(e)})}\over {\partial \alpha_{(b)}}}&{|}_{\alpha
=\alpha (\tau ,\vec \sigma )}&-{{\partial A_{(a)(b)}(\alpha_{(e)})}\over
{\partial \alpha_{(c)}}} {|}_{\alpha =\alpha (\tau ,\vec \sigma )}=\nonumber \\
&=&\epsilon_{(a)(u)(v)}A_{(u)(b)}(\alpha_{(e)}(\tau ,\vec \sigma ))A_{(v)(c)}
(\alpha_{(e)}(\tau ,\vec \sigma )),
\label{IV9}
\end{eqnarray}

\noindent holding pointwise on each fiber of $\Sigma_{\tau}^{(\Xi )}\times
SO(3)$ over $(\tau ,\vec \sigma )$ in a suitable tubular neighbourhood of the
identity cross section.

By defining a generalized canonical 1-form for ${\bar {\cal G}}_{ROT}$,

\bea
\tilde \omega &=& {\hat R}^{(a)}\, {\tilde \theta}_{(a)}(\tau ,\vec \sigma )=H_{(a)}
(\alpha_{(e)}(\tau ,\vec \sigma )) d\alpha_{(a)}(\tau ,\vec \sigma
),\nonumber \\
 &&{}\nonumber \\
 &&\text{where} \nonumber \\
 &&{}\nonumber \\
 {\tilde \theta}_{(a)}(\tau ,\vec \sigma ) &=& {\hat
\theta}_{(a)}(\alpha_{(e)} (\tau ,\vec \sigma
),\partial_r\alpha_{(e)}(\tau ,\vec \sigma ))={\tilde
\theta}_{(a)r}(\tau ,\vec \sigma )d\sigma^r=A_{(a)(b)}(\alpha_{(e)}(\tau ,
\vec \sigma )) d\alpha_{(b)}(\tau ,\vec \sigma ),\nonumber \\
 &&
\label{IV10}
\eea

\noindent are the generalized
Maurer-Cartan 1-forms on the Lie algebra ${\bar g}_{ROT}$ of
${\bar {\cal G}}_{ROT}$
and where we defined the matrices $H_{(a)}(\alpha_{(e)}(\tau ,\vec \sigma ))=
{\hat R}^{(b)}\, A_{(b)(a)}(\alpha_{(e)}(\tau ,\vec \sigma ))$, the previous
equations can be rewritten in the form of a zero curvature condition

\begin{equation}
{{\partial H_{(a)}(\alpha_{(e)})}\over {\partial \alpha_{(b)}}} {|}_{\alpha
=\alpha (\tau ,\vec \sigma )} - {{\partial H_{(b)}(\alpha_{(e)})}\over
{\partial \alpha_{(a)}}} {|}_{\alpha =\alpha (\tau ,\vec \sigma )} +
[H_{(a)}(\alpha_{(e)}(\tau ,\vec \sigma )),H_{(b)}(\alpha_{(e)}(\tau ,\vec
\sigma ))]=0.
\label{IV11}
\end{equation}

Eq.(3.19) of Ref.\cite{ru11} and Eqs.(\ref{III5})  give the following
multitemporal equations for the dependence of the cotriad
${}^3e_{(a)r}(\tau ,\vec \sigma )$ on the 3 gauge angles
$\alpha_{(a)}(\tau ,\vec \sigma )$

\begin{eqnarray}
X_{(b)}(\tau ,{\vec \sigma}^{'})\, {}^3e_{(a)r}(\tau ,\vec \sigma )&=&
B_{(c)(b)}(\alpha_{(e)}(\tau ,{\vec \sigma}^{'}) {{\tilde \delta \,
{}^3e_{(a)r}(\tau ,\vec \sigma )}\over {\delta \alpha_{(c)}(\tau ,{\vec
\sigma}^{'})}} =\nonumber \\
&=&-\lbrace {}^3e_{(a)r}(\tau ,\vec \sigma ),{}^3{\tilde M}_{(b)}(\tau ,{\vec
\sigma}^{'})\rbrace =-\epsilon_{(a)(b)(c)} \, {}^3e_{(c)r}(\tau ,\vec \sigma )
\delta^3(\vec \sigma ,{\vec \sigma}^{'}),\nonumber \\
&&{}\nonumber \\
\Rightarrow && {{\tilde \delta \, {}^3e_{(a)r}(\tau ,\vec \sigma )}\over
{\delta \alpha_{(b)}(\tau ,{\vec \sigma}^{'})}}=-\epsilon_{(a)(c)(d)}
A_{(c)(b)}(\alpha_{(e)}(\tau ,\vec \sigma ))\, {}^3e_{(a)r}(\tau ,\vec \sigma )
\delta^3(\vec \sigma ,{\vec \sigma}^{'})=\nonumber \\
&=&\Big[ {\hat R}^{(c)} A_{(c)(b)}(\alpha_{(e)}(\tau ,\vec \sigma ))
\Big]_{(a)(d)}\,
{}^3e_{(d)r}(\tau ,\vec \sigma ) \delta^3(\vec \sigma ,{\vec \sigma}^{'})=
\nonumber \\
&=& \Big[ H_{(b)}(\alpha_{(e)}(\tau ,\vec \sigma ))\Big]_{(a)(d)}\, {}^3e
_{(d)r}(\tau ,\vec \sigma ) \delta^3(\vec \sigma ,{\vec \sigma}^{'}).
\label{IV12}
\end{eqnarray}

\noindent These equations are a functional multitemporal generalization of the
matrix equation ${d\over {dt}} U(t,t_o)= h\, U(t,t_o)$ ,
$U(t_o,t_o)=1$, generating the concept of time-ordering. They are
integrable (i.e. their solution is path independent) due to
Eqs.(\ref{IV11}) and their solution is

\begin{equation}
{}^3e_{(a)r}(\tau ,\vec \sigma )={}^3R_{(a)(b)}(\alpha_{(e)}(\tau ,\vec
\sigma ))\, {}^3{\bar e}_{(b)r}(\tau ,\vec \sigma ),
\label{IV13}
\end{equation}

\noindent where \footnote{$l$ is an arbitrary path originating at the identity cross
section of $\Sigma_{\tau}^{(\Xi )}\times SO(3)$; due to the path
independence it can be replaced with the defining path $\gamma_{\alpha
(\tau ,\vec \sigma )}(s)=\hat \gamma (\tau ,\vec \sigma ;s)$.}

\begin{eqnarray}
{}^3R_{(a)(b)}(\alpha_{(e)}(\tau ,\vec \sigma ))&=& \Big( P\, e^{{}_{(l)}\int
_0^{\alpha_{(e)}(\tau ,\vec \sigma )}\, H_{(c)}({\bar \alpha}_{(e)}) {\cal
D}{\bar \alpha}_{(c)}} \Big)_{(a)(b)}=\nonumber \\
&=&\Big( P\, e^{{}_{(\hat \gamma )}\int_0^{\alpha_{(e)}(\tau ,\vec \sigma )}\,
H_{(c)}({\bar \alpha}_{(e)}) {\cal D}{\bar \alpha}_{(c)}} \Big)_{(a)(b)}=
\nonumber \\
&=&\Big( P\, e^{\Omega^{\hat \gamma}(\alpha_{(e)}(\tau ,\vec \sigma ))}
\Big)_{(a)(b)},
\label{IV14}
\end{eqnarray}

\noindent is a point dependent rotation matrix [${}^3R^T_{(a)(b)}(\alpha )=
{}^3R^{-1}_{(a)(b)}(\alpha )$ since ${\hat R}^{(a)\dagger}=-{\hat R}^{(a)}$].

In Eq.(\ref{IV14}) we introduced the generalized phase obtained by
functional integration along the defining path in $\Sigma_{\tau}^{(\Xi
)}\times SO(3)$ of the generalized Maurer-Cartan 1-forms

\begin{eqnarray}
\Omega^{\hat \gamma}(\alpha_{(e)}(\tau ,\vec \sigma ;s))&=& {}_{(\hat \gamma )}
\int_I^{\gamma_{\alpha (\tau ,\vec \sigma ;s)}}\, {\hat R}^{(a)} {\tilde \theta}
_{(a)} {|}_{\gamma_{\alpha (\tau ,\vec \sigma ;s)}}=\nonumber \\
&=&{}_{(\hat \gamma )}\int_0^{\alpha_{(e)}(\tau ,\vec \sigma ;s)}\, H_{(a)}
({\bar \alpha}_{(e)}) {\cal D}{\bar \alpha}_{(a)}=\nonumber \\
&=&{}_{(\hat \gamma )}\int_0^{\alpha_{(e)}(\tau ,\vec \sigma ;s)}\, {\hat R}
^{(a)} A_{(a)(b)}({\bar \alpha}_{(e)}) {\cal D} {\bar \alpha}_{(b)}.
\label{IV15}
\end{eqnarray}

As shown in Ref.\cite{lusa}, we have

\beq
d_{\hat \gamma}\, \Omega^{\hat \gamma} (\alpha_{(e)}(\tau ,\vec \sigma
;s))={\hat R}^{(a)} {\hat \theta}_{(a)} (\alpha_{(e)}(\tau ,\vec
\sigma ;s),\partial_r\alpha_{(e)}(\tau ,\vec \sigma ; s)),
\label{IV16}
\eeq

\noindent
where $d_{\hat \gamma}$ is the restriction of the {\it fiber or
vertical} derivative $d_V$ on $\Sigma_{\tau}^{(\Xi )}\times SO(3)$
(the BRST operator) to the defining path, satisfying $d^2_{\hat
\gamma}=0$ due to the generalized Maurer-Cartan equations.

In Eq.(\ref{IV13}), ${}^3{\bar e}_{(a)r}(\tau ,\vec \sigma )$ are the
cotriads evaluated at $\alpha_{(a)}(\tau ,\vec \sigma )=0$, i.e. on
the identity cross section. Being Cauchy data of Eqs.(\ref{IV12}),
they are independent from the angles $\alpha_{(a)}(\tau ,\vec \sigma
)$, satisfy $\lbrace {}^3{\bar e}
_{(a)r}(\tau ,\vec \sigma ),{}^3{\tilde M}_{(b)}(\tau ,{\vec \sigma}^{'})
\rbrace =0$ and depend only on 6 independent functions \footnote{The $\alpha_{(a)}(\tau
,\vec \sigma )$ are the 3 rotational gauge degrees of freedom hidden
in the 9 variables ${}^3e_{(a)r}(\tau ,\vec \sigma )$.}. We have not
found 3 specific conditions on cotriads implying their independency
from the angles $\alpha_{(a)}$.

Since Eq.(A24) of Ref.\cite{ru11} gives the law ${}^3\omega_r \mapsto
R\, {}^3\omega_r\, R^T -R \partial_r R^T$  for the spin connection
under local SO(3) rotations, under infinitesimal rotations we get
\footnote{${\hat D}^{(\omega )}_{(a)(b)r}(\tau ,\vec \sigma )$ is the SO(3)
covariant derivative in the adjoint representation.}

\begin{eqnarray}
X_{(b)}(\tau ,{\vec \sigma}^{'})\, {}^3\omega_{r(a)}(\tau ,\vec \sigma )&=&
B_{(c)(b)}(\alpha_{(e)}(\tau ,{\vec \sigma}^{'})) {{\tilde \delta \, {}^3\omega
_{r(a)}(\tau ,\vec \sigma )}\over {\delta \alpha_{(c)}(\tau ,{\vec \sigma}
^{'})}}  =\nonumber \\
&=&-\lbrace {}^3\omega_{r(a)}(\tau ,\vec \sigma ), {}^3{\tilde M}_{(b)}(\tau ,
{\vec \sigma}^{'})\rbrace =\nonumber \\
&=&\Big[ \delta_{(a)(b)}\partial_r+\epsilon_{(a)(c)(b)}
\, {}^3\omega_{r(c)}(\tau ,\vec \sigma )\Big] \delta^3(\vec \sigma ,{\vec
\sigma}^{'})=\nonumber \\
&=&\Big[ \delta_{(a)(b)}\partial_r-({\hat R}^{(c)}\, {}^3\omega_{r(c)}(\tau
,\vec \sigma ))_{(a)(b)}\Big] \delta^3(\vec \sigma ,{\vec \sigma}^{'})
={\hat D}^{(\omega
)}_{(a)(b)r}(\tau ,\vec \sigma ) \delta^3(\vec \sigma ,{\vec \sigma}^{'}),
\nonumber \\
&&{}
\label{IV17}
\end{eqnarray}

\noindent which is the same result as for the gauge potential of the SO(3)
Yang-Mills theory. We can use the results of Ref.\cite{lusa} to write
the solution of Eq.(\ref{IV17})

\begin{eqnarray}
{}^3\omega_{r(a)}(\tau ,\vec \sigma )&=&A_{(a)(b)}(\alpha_{(e)}(\tau ,\vec
\sigma )) \partial_r\alpha_{(b)}(\tau ,\vec \sigma )+{}^3\omega^{(T)}_{r(a)}
(\tau ,\vec \sigma ,\alpha_{(e)}(\tau ,\vec \sigma )),\nonumber \\
 &&{}\nonumber \\
 &&\text{with}\nonumber \\
 &&{}\nonumber \\
{{\partial \, {}^3\omega^{(T)}_{r(a)}(\tau ,\vec \sigma ,\alpha_{(e)})}\over
{\partial \alpha_{(b)}}}&&{|}_{\alpha =\alpha (\tau ,\vec \sigma )}=-\epsilon
_{(a)(d)(c)}A_{(d)(b)}(\alpha_{(e)}(\tau ,\vec \sigma ))\, {}^3\omega^{(T)}
_{r(c)}(\tau ,\vec \sigma ,\alpha_{(e)}(\tau ,\vec \sigma )).
\label{IV18}
\end{eqnarray}

In ${}^3\omega_{r(a)}(\tau ,\vec \sigma )d\sigma^r={\tilde \theta}_{(a)}(\tau ,
\vec \sigma )+{}^3\omega^{(T)}_{r(a)}(\tau ,\vec \sigma ,\alpha_{(e)}(\tau ,
\vec \sigma )) d\sigma^r$, the first term is a pure gauge spin connection
(the BRST {\it ghost}), while the second one is the {\it source} of
the field strength: ${}^3\Omega_{rs(a)}=\partial_r\,
{}^3\omega^{(T)}_{s(a)}-\partial_s\,
{}^3\omega^{(T)}_{r(a)}-\epsilon_{(a)(b)(c)}\,
{}^3\omega^{(T)}_{r(b)}\, {}^3\omega^{(T)}_{s(c)}$. Moreover, the
Hodge decomposition theorem (in the functional spaces where the spin
connections are fully irreducible) implies that ${}^3\omega^{(\perp
)}_{r(a)}(\tau ,\vec \sigma )\, {\buildrel {def}
\over =}\, {}^3\omega^{(T)}
_{r(a)}(\tau ,\vec \sigma ,\alpha_{(e)}(\tau ,\vec \sigma ))$ satisfies
${}^3\nabla^r\, {}^3\omega^{(\perp )}_{r(a)}=0$.

Since we have $X_{(b)}(\tau ,{\vec \sigma}^{'})\, {}^3\omega^{(\perp )}_{r(a)}
(\tau ,\vec \sigma )=-\epsilon_{(a)(c)(b)}\, {}^3\omega^{(\perp )}_{r(c)}(\tau
,\vec \sigma ) \delta^3(\vec \sigma ,{\vec \sigma}^{'})$, we get

\begin{eqnarray}
{{\tilde \delta \, {}^3\omega^{(\perp )}_{r(a)}(\tau ,\vec \sigma )}\over
{\delta \alpha_{(b)}(\tau ,{\vec \sigma}^{'})}}&=&[H_{(b)}(\alpha_{(e)}(\tau ,
\vec \sigma ))]_{(a)(c)}\, {}^3\omega^{(\perp )}_{r(c)}(\tau ,\vec \sigma )
\delta^3(\vec \sigma ,{\vec \sigma }^{'}),\nonumber \\
\Rightarrow && {}^3\omega^{(\perp )}_{r(a)}(\tau ,\vec \sigma )=\Big( P\,
e^{\Omega^{\hat \gamma}(\alpha_{(e)}(\tau ,\vec \sigma ))} \Big)_{(a)(b)}\,
{}^3{\bar \omega}^{(\perp )}_{r(b)}(\tau ,\vec \sigma ),\nonumber \\
&&{}^3\nabla^r\, {}^3{\bar \omega}^{(\perp )}_{r(a)}(\tau ,\vec \sigma )=0.
\label{IV19}
\end{eqnarray}

The transverse spin connection ${}^3{\bar \omega}^{(\perp )}_{r(a)}(\tau ,
\vec \sigma )$ is independent from the gauge angles $\alpha_{(a)}(\tau ,\vec
\sigma )$ and is the source of the field strength ${}^3{\bar \Omega}_{rs(a)}=
\partial_r\, {}^3{\bar \omega}^{(\perp )}_{s(a)}-\partial_s\, {}^3{\bar \omega}
^{(\perp )}_{r(a)}-\epsilon_{(a)(b)(c)}\, {}^3{\bar \omega}^{(\perp )}_{r(b)}
\, {}^3{\bar \omega}^{(\perp )}_{s(c)}$ invariant under the rotation gauge
transformations. Clearly, ${}^3{\bar \omega}_{r(a)}^{(\perp )}$ is built with
the reduced cotriads ${}^3{\bar e}_{(a)r}$.

Let us remark that for ${}^3\omega^F_{r(a)}(\tau , \vec \sigma )d\sigma^r=
{\tilde \theta}_{(a)}(\tau ,\vec \sigma )$ we get ${}^3\Omega_{rs(a)}(\tau ,
\vec \sigma )=0$ and then ${}^3R_{rsuv}=0$: in this case the Riemannian manifold
$(\Sigma_{\tau},{}^3g_{rs}={}^3e_{(a)r}\, {}^3e_{(a)s})$ becomes the
Euclidean manifold $(R^3,\, {}^3g^F_{rs})$ with ${}^3g^F_{rs}$ the
flat 3-metric in curvilinear coordinates. Since Eq.(\ref{IV13})
implies that ${}^3g_{rs}={}^3{\bar e}_{(a)r}\, {}^3{\bar e}_{(a)s}$
and since ${}^3g^F_{rs}(\tau ,\vec \sigma )
={{\partial {\tilde \sigma}^u}\over {\partial \sigma^r}}{{\partial {\tilde
\sigma}^v}\over {\partial \sigma^s}}\delta_{uv}={}^3g^F_{rs}(\vec \sigma )$, if
${\tilde \sigma}^u(\vec \sigma )$ are Cartesian coordinates, we get

\beq
{}^3\Gamma^{F\, u}_{rs}={{\partial \sigma^u}
\over {\partial {\tilde \sigma}^n}}{{\partial^2{\tilde \sigma}^n}\over {\partial
\sigma^r\partial \sigma^s}}={}^3e^u_{(a)}\partial_r\, {}^3e_{(a)s}=
{}^3\triangle^u_{rs},
\label{IV20}
\eeq

\noindent (see the remark after Eqs.(A19) of Ref.\cite{ru11} ). This implies that
for an arbitrary ${}^3g$ we have the decomposition
${}^3\Gamma^u_{rs}={}^3\triangle^u_{rs}+{}^3{\bar \Gamma}^u
_{rs}$ with ${}^3{\bar \Gamma}^u_{rs}={}^3e^u_{(a)}\, {}^3e_{(b)s}\, {}^3\omega
_{r(a)(b)}$ the source of the Riemann tensor. This implies that

\beq
{}^3{\bar e}^F
_{(a)r}(\tau ,\vec \sigma )={{\partial {\tilde \sigma}^u}\over {\partial
\sigma^r}}\, {}^3{\tilde {\bar e}}^F_{(a)u}(\tau ,{\vec {\tilde \sigma}})=
\delta_{(a)u} {{\partial {\tilde \sigma}^u(\vec \sigma )}\over {\partial
\sigma^r}}.
\label{IV21}
\eeq

 Therefore, a flat cotriad on $R^3$ has the form

\begin{equation}
{}^3e^F_{(a)r}(\tau ,\vec \sigma )={}^3R_{(a)(b)}(\alpha_{(e)}(\tau ,\vec
\sigma )) \delta_{(b)u} {{\partial {\tilde \sigma}^u(\vec \sigma )}\over
{\partial \sigma^r}}.
\label{IV22}
\end{equation}

Eqs.(3.19) of Ref.\cite{ru11}  imply the following multitemporal
equations for the momenta

\begin{eqnarray}
X_{(b)}(\tau ,{\vec \sigma }^{'})\, {}^3{\tilde \pi}^r_{(a)}(\tau ,\vec
\sigma )&=&B_{(c)(b)}(\alpha_{(e)}(\tau ,{\vec \sigma }^{'}) {{\tilde
\delta \, {}^3{\tilde \pi}^r_{(a)}(\tau ,\vec \sigma )}\over {\delta
\alpha_{(c)}(\tau ,{\vec \sigma }^{'})}}=\nonumber \\
&=&-\epsilon_{(a)(b)(c)}\, {}^3{\tilde \pi}^r_{(c)}(\tau ,\vec \sigma ) \delta
^3(\vec \sigma ,{\vec \sigma }^{'}),
\label{IV23}
\end{eqnarray}

\noindent whose solution is [${}^3{\bar {\tilde \pi}}^r_{(a)}(\tau ,\vec
\sigma )$ depends only on 6 independent functions]

\begin{equation}
{}^3{\tilde \pi}^r_{(a)}(\tau ,\vec \sigma )={}^3R_{(a)(b)}(\alpha_{(e)}(\tau ,
\vec \sigma ))\, {}^3{\bar {\tilde \pi}}^r_{(b)}(\tau ,\vec \sigma ).
\label{IV24}
\end{equation}

With the definition of SO(3) covariant derivative given in
Eq.(\ref{IV17}), the constraints ${\hat {\cal H}}_{(a)}(\tau ,\vec
\sigma )\approx 0$ of Eqs.(\ref{II11}), may be written as

\begin{eqnarray}
{\hat {\cal H}}_{(a)}(\tau ,\vec \sigma )&=&-{}^3e^r_{(a)}(\tau ,\vec \sigma )
\Big[ {}^3{\tilde \Theta}_r + {}^3\omega_{r(b)}\, {}^3{\tilde M}_{(b)}\Big]
(\tau ,\vec \sigma )=\nonumber \\
&=&{\hat D}^{(\omega )}_{(a)(b)r}(\tau ,\vec \sigma )\, {}^3{\tilde \pi}^r
_{(b)}(\tau ,\vec \sigma )\approx 0,
\label{IV25}
\end{eqnarray}

\noindent so that we have (${}^3{\tilde \pi}^{(T)r}_{(a)}(\tau ,\vec \sigma )$
is a field with zero SO(3) covariant divergence)

\begin{eqnarray}
{}^3{\tilde \pi}^r_{(a)}(\tau ,\vec \sigma )&=&{}^3{\tilde \pi}^{(T)r}_{(a)}
(\tau ,\vec \sigma )-\int d^3\sigma^{'}\, \zeta^{(\omega )r}_{(a)(b)}(\vec
\sigma ,{\vec \sigma }^{'};\tau )\, {\hat {\cal H}}_{(b)}(\tau ,{\vec \sigma }
^{'}),\nonumber \\
{}&&{\hat D}^{(\omega )}_{(a)(b)r}(\tau ,\vec \sigma )\, {}^3{\tilde \pi}
^{(T)r}_{(b)}(\tau ,\vec \sigma )\equiv 0,\nonumber \\
 &&{}\nonumber \\
 &&\text{with}\nonumber \\
 &&{}\nonumber \\
 {}^3{\tilde \pi}^{(T)r}_{(a)}(\tau ,\vec \sigma ) &=& \int d^3\sigma_1
 \Big[ \delta^r_s\delta_{(a)(b)}\delta^3(\vec \sigma ,{\vec \sigma}_1) +
 \zeta^{(\omega )r}_{(a)(c)}(\vec \sigma ,{\vec \sigma}_1;\tau )
 {\hat D}^{(\omega )}_{(c)(b)s}(\tau ,{\vec \sigma}_1)\Big]
 {}^3{\tilde \pi}^s_{(b)}(\tau ,{\vec \sigma}_1).
\label{IV26}
\end{eqnarray}

\noindent In this equation, we introduced the Green function of the SO(3)
covariant divergence, defined by

\begin{equation}
{\hat D}^{(\omega )}_{(a)(b)r}(\tau ,\vec \sigma )\, \zeta^{(\omega )r}_{(b)(c)}
(\vec \sigma ,{\vec \sigma }^{'};\tau )=-\delta_{(a)(c)} \delta^3(\vec \sigma ,
{\vec \sigma }^{'}).
\label{IV27}
\end{equation}

In Ref.\cite{lusa}, this Green function was evaluated for
$\Sigma_{\tau}=R^3$, the flat Euclidean space, by using the Green
function $\vec c(\vec \sigma - {\vec \sigma }^{'})$ of the flat
ordinary divergence [$\triangle ={\vec
\partial}^2_{\sigma}$] in Cartesian coordinates

\begin{eqnarray}
\vec c(\vec \sigma -{\vec \sigma }^{'})&=&{\vec \partial}_{\sigma}\, c(\vec
\sigma -{\vec \sigma }^{'})=-{{{\vec \partial}_{\sigma}}\over {\triangle}}
\delta^3(\vec \sigma -{\vec \sigma }^{'})={{\vec \sigma -{\vec \sigma }^{'}}
\over {4\pi |\, \vec \sigma -{\vec \sigma }^{'}|{}^3}}= {{\vec n(\vec \sigma -
{\vec \sigma }^{'})}\over {4\pi (\vec \sigma -{\vec \sigma }^{'})^2}},
\nonumber \\
&&{}\nonumber \\
{\vec \partial}_{\sigma}\cdot \vec c(\vec \sigma -{\vec \sigma }^{'})&=&
-\delta^3(\vec \sigma -{\vec \sigma }^{'}),
\label{IV28}
\end{eqnarray}

\noindent where $\vec n(\vec \sigma -{\vec \sigma }^{'})$ is the tangent to the
flat geodesic (straight line segment) joining the point of coordinates $\vec
\sigma $ and ${\vec \sigma }^{'}$, so that $\vec n(\vec \sigma -{\vec \sigma }
^{'})\cdot {\vec \partial}_{\sigma}$ is the directional derivative along the
flat geodesic.

With our special family of Riemannian 3-manifolds
$(\Sigma_{\tau},{}^3g)$, we would use Eq.(\ref{IV28}) in the special
global normal chart in which the star of geodesics originating from
the reference point p becomes a star of straight lines. In non normal
coordinates, the Green function $\vec c(\vec \sigma - {\vec \sigma
}^{'})$ will be replaced with the gradient of the {\it Synge world
function}\cite{synge} or {\it De Witt geodesic interval
bitensor}\cite{dew} $\sigma_{DW}(\vec \sigma ,{\vec \sigma }^{'})$
(giving the arc length of the geodesic from $\vec \sigma $ to ${\vec
\sigma }^{'}$) adapted from the Lorentzian spacetime $M^4$ to the Riemannian
3-manifold  $(\Sigma_{\tau}, {}^3g)$, i.e.

\beq
d^r_{\gamma_{pp^{'}}}(\vec \sigma , {\vec \sigma }^{'})={1\over
3}\sigma_{DW}^r(\vec \sigma ,{\vec \sigma }^{'})= {1\over 3}\,
{}^3\nabla^r_{\sigma}\, \sigma_{DW}(\vec \sigma ,{\vec \sigma }
^{'})={1\over 3} \partial^r_{\sigma}\, \sigma_{DW}(\vec \sigma ,{\vec \sigma }
^{'}),
\label{IV29}
\eeq

\noindent
giving in each point $\vec \sigma $ the tangent to the geodesic
$\gamma_{pp^{'}}$ joining the points p and $p^{'}$ of coordinates
$\vec \sigma $ and ${\vec \sigma }^{'}$ in the direction from $p^{'}$
to p. Therefore, the Green function is
\footnote{$\partial_rd^r_{\gamma_{pp^{'}}}(\vec \sigma ,{\vec
\sigma}^{'})=-\delta^3(\vec \sigma ,{\vec \sigma}^{'})$; $d^r_{\gamma_{pp^{'}}}
(\vec \sigma ,{\vec \sigma}^{'})\, \partial_r$ is the directional
derivative along the geodesic $\gamma_{pp^{'}}$ at $p$ of coordinates
$\vec \sigma$.}

\begin{equation}
\zeta^{(\omega )r}_{(a)(b)}(\vec \sigma ,{\vec \sigma }^{'};\tau )=d^r_{\gamma
_{pp^{'}}}(\vec \sigma ,{\vec \sigma }^{'})
\Big( P_{\gamma_{pp^{'}}}\, e^{\int^{\vec \sigma }_{{\vec \sigma }^{'}}
d\sigma_1^s\, {\hat R}^{(c)}\, {}^3\omega_{s(c)}(\tau ,{\vec \sigma }_1)}
\Big)_{(a)(b)},
\label{IV30}
\end{equation}

\noindent with the path ordering done
along the geodesic $\gamma_{pp^{'}}$. This path ordering ({\it Wu-Yang
non-integrable phase} or {\it geodesic Wilson line}) is defined on all
$\Sigma_{\tau}\times SO(3)$ only if the spin connection is fully
irreducible; it is just the parallel transporter of Eq.(\ref{IV5}).

Eqs.(\ref{IV13}) show the dependence of the cotriad on the 3 angles
$\alpha_{(a)} (\tau ,\vec \sigma )$, which therefore must be
expressible only in terms of the cotriad itself and satisfy $\lbrace
\alpha_{(a)}(\tau ,\vec \sigma ),\alpha
_{(b)}(\tau ,{\vec \sigma }^{'})\rbrace =0$. They are the rotational gauge
variables, canonically conjugate to  Abelianized rotation constraints
${\tilde \pi}^{\vec \alpha}_{(a)}(\tau ,\vec \sigma )\approx 0$. From
Eqs.(\ref{IV8}), since the functional derivatives commute, we see that
we have the following expression for the Abelianized
constraints\cite{lusa,hwang}

\begin{eqnarray}
{\tilde \pi}^{\vec \alpha}_{(a)}(\tau ,\vec \sigma )&=&-{}^3{\tilde
M}_{(b)} (\tau ,\vec \sigma ) A_{(b)(a)}(\alpha_{(e)}(\tau ,\vec
\sigma ))\approx 0,
\nonumber \\
&&{}\nonumber \\
\lbrace {\tilde \pi}^{\vec \alpha}_{(a)}(\tau ,\vec \sigma ),{\tilde \pi}
^{\vec \alpha}_{(b)}(\tau ,{\vec \sigma }^{'})\rbrace &=&0,\nonumber \\
\lbrace \alpha_{(a)}(\tau ,\vec \sigma ),{\tilde \pi}^{\vec \alpha}_{(b)}
(\tau ,{\vec \sigma }^{'})\rbrace &=&A_{(c)(b)}(\alpha_{(e)}(\tau
,{\vec \sigma }^{'})) X_{(c)}(\tau ,{\vec \sigma }^{'})
\alpha_{(a)}(\tau , \vec \sigma )=\nonumber \\
&=&\delta_{(a)(b)} \delta^3(\vec \sigma ,{\vec \sigma }^{'}).
\label{IV31}
\end{eqnarray}

The functional equation determining $\alpha_{(a)}(\tau ,\vec \sigma )$ in terms
of ${}^3e_{(a)r}(\tau ,\vec \sigma )$ is

\begin{eqnarray}
-\delta_{(a)(b)} \delta^3(\vec \sigma ,{\vec \sigma }^{'})&=&\lbrace \alpha
_{(a)}(\tau ,\vec \sigma ),{}^3{\tilde M}_{(c)}(\tau ,{\vec \sigma }^{'})
\rbrace A_{(c)(b)}(\alpha_{(e)}(\tau ,{\vec \sigma }^{'})=\nonumber \\
&=&\epsilon_{(c)(u)(v)}\, {}^3e_{(u)r}(\tau ,{\vec \sigma }^{'}) \lbrace
\alpha_{(a)}(\tau ,\vec \sigma ),{}^3{\tilde \pi}^r_{(v)}(\tau ,{\vec \sigma }
^{'})\rbrace A_{(c)(b)}(\alpha_{(e)}(\tau ,{\vec \sigma }^{'}))=\nonumber \\
&=&\epsilon_{(c)(u)(v)}\, A_{(c)(b)}(\alpha_{(e)}(\tau ,{\vec \sigma }^{'}))
\, {}^3e_{(u)r}(\tau ,{\vec \sigma }^{'}) {{\delta \alpha_{(a)}(\tau ,\vec
\sigma )}\over {\delta \, {}^3e_{(v)r}(\tau ,{\vec \sigma }^{'})}},\nonumber \\
&&{}\nonumber \\
\Rightarrow && \epsilon_{(b)(u)(v)}\, {}^3e_{(u)r}(\tau ,{\vec \sigma }^{'})
{{\delta \alpha_{(a)}(\tau ,\vec \sigma )}\over {\delta \, {}^3e_{(v)r}(\tau ,
{\vec \sigma }^{'})}}=-B_{(a)(b)}(\alpha_{(e)}(\tau ,\vec \sigma ))\delta^3
(\vec \sigma ,{\vec \sigma }^{'}),\nonumber \\
{}&&\epsilon_{(b)(u)(v)}\, {}^3e_{(u)r}(\tau ,{\vec \sigma }^{'})
\Big[ A_{(a)(c)}
(\alpha_{(e)}(\tau ,\vec \sigma )){{\delta \alpha_{(a)}(\tau ,\vec \sigma )}
\over {\delta \, {}^3e_{(v)r}(\tau ,{\vec \sigma }^{'})}}\Big] =-\delta_{(a)(b)}
\delta^3(\vec \sigma ,{\vec \sigma }^{'}),\nonumber \\
{}&&\epsilon_{(b)(u)(v)}\, {}^3e_{(u)r}(\tau ,{\vec \sigma }^{'}) {{\delta
\Omega^{\hat \gamma}_{(a)}(\alpha_{(e)}(\tau ,\vec \sigma ))}\over {\delta
\, {}^3e_{(v)r}(\tau ,{\vec \sigma }^{'})}}=-\delta_{(a)(b)}\delta^3(\vec
\sigma ,{\vec \sigma }^{'}),\nonumber \\
\Rightarrow && \epsilon_{(b)(u)(v)} {{\delta
\Omega^{\hat \gamma}_{(a)}(\alpha_{(e)}(\tau ,\vec \sigma ))}\over {\delta
\, {}^3e_{(v)r}(\tau ,{\vec \sigma }^{'})}}=-{1\over 3}\delta_{(a)(b)}
{}^3e^r_{(u)}(\tau ,{\vec \sigma}^{'}) \delta^3(\vec
\sigma ,{\vec \sigma }^{'}),\nonumber \\
{}&& {{\delta
\Omega^{\hat \gamma}_{(a)}(\alpha_{(e)}(\tau ,\vec \sigma ))}\over {\delta
\, {}^3e_{(b)r}(\tau ,{\vec \sigma }^{'})}}={1\over 6} ({\hat R}^{(u)})
_{(a)(b)}\, {}^3e^r_{(u)}(\tau ,\vec \sigma ) \delta^3(\vec
\sigma ,{\vec \sigma }^{'}).
\label{IV32}
\end{eqnarray}

This equation is not integrable like the corresponding one in the Yang-Mills
case \cite{lusa}. Having chosen a global coordinate system $\Xi$ on $\Sigma
_{\tau}$ as the conventional origin of pseudo-diffeomorphisms, the discussion in
the previous Section  allows to define the trivialization
$\Sigma^{(\Xi )}_{\tau}
\times SO(3)$ of the coframe bundle $L\Sigma_{\tau}$. If: \hfill\break
i) $\sigma^{(\Xi )}_I$ is the identity cross section of $\Sigma^{(\Xi
)}_{\tau} \times SO(3)$, corresponding to the coframe
${}^3\theta^I_{(a)}={}^3e^I_{(a)r} d\sigma^r$ in $L\Sigma_{\tau}$
($\sigma^r$ are the coordinate functions of $\Xi$);
\hfill\break
ii) $\sigma^{(\Xi )}$ is an arbitrary global cross section of $\Sigma^{(\Xi )}
_{\tau}\times SO(3)$, corresponding to a coframe ${}^3\theta_{(a)}={}^3e
_{(a)r} d\sigma^r$ in $L\Sigma_{\tau}$, in a tubular neighbourhood of the
identity cross section where the generalized canonical coordinates of first kind
 on the fibers of $\Sigma^{(\Xi )}_{\tau}\times SO(3)$
 are defined; \hfill\break iii) $\sigma^{(\Xi )}(s)$ is the family of
global cross sections of $\Sigma^{(\Xi )}_{\tau}\times SO(3)$
connecting $\Sigma^{(\Xi )}_I=\sigma^{(\Xi )}(s=0)$ and $\Sigma^{(\Xi
)}=\sigma^{(\Xi )} (s=1)$ so that on each fiber the point on
$\sigma^{(\Xi )}_I$ is connected with the point on $\Sigma^{(\Xi )}$
by the defining path $\hat \gamma$ of canonical coordinates of first
kind;\hfill\break then the formal solution of the previous equation is

\begin{equation}
\Omega^{\hat \gamma}_{(a)}(\alpha_{(e)}(\tau ,\vec \sigma
))={1\over 6} \,\,\, {}_{\hat \gamma}\int^{{}^3e_{(a)r}(\tau ,\vec
\sigma )} _{{}^3e^I_{(a)r}(\tau ,\vec \sigma )}\, ({\hat R}^{(u)}
)_{(a)(b)}\, {}^3e^r_{(u)}\, {\cal D}\, {}^3e_{(b)r},
 \label{IV33}
\end{equation}

\noindent where the path integral is made along the path of coframes connecting
${}^3\theta^I_{(a)}$ with ${}^3\theta_{(a)}$ just described. As in
Ref.\cite{lusa} , to get the angles $\alpha_{(a)}(\tau ,\vec \sigma )$ from
$\Omega^{\hat \gamma}_{(a)}(\alpha_{(e)}(\tau ,\vec \sigma ))$, we essentially
have to invert the equation $\Omega^{\hat \gamma}_{(a)}(\alpha_{(e)})=
{}_{\hat \gamma} \int^{\alpha_{(e)}}_{0}\, A_{(a)(b)}(\bar \alpha )
d{\bar \alpha}_{(b)}$ with $A=(e^{R\alpha}-1)/R\alpha$.

\subsection{The Multi-Temporal Equations for the Pseudo-Diffeomorphisms.}

Let us now study the multitemporal equations associated with
pseudo-diffeomorphisms to find the dependence of ${}^3e_{(a)r}(\tau
,\vec \sigma )$ on the parameters $\xi^r(\tau ,\vec \sigma )$.
Disregarding momentarily rotations, let us look for a realization of
vector fields ${\tilde Y}_r(\tau ,\vec \sigma )$ satisfying the last
line of Eqs.(\ref{III6}). If we put

\begin{equation}
{\tilde Y}_r(\tau ,\vec \sigma )=-{{\partial \xi^s(\tau ,\vec \sigma )}\over
{\partial \sigma^r}} {{\delta}\over {\delta \xi^s(\tau ,\vec \sigma )}},
\label{IV34}
\end{equation}

\noindent we find

\begin{eqnarray}
[{\tilde Y}_r(\tau ,\vec \sigma ),{\tilde Y}_s(\tau ,{\vec
\sigma }^{'})]&=&[{{\partial \xi^u(\tau ,\vec \sigma )}\over {\partial \sigma
^r}}{{\delta}\over {\delta \xi^u(\tau ,\vec \sigma )}},{{\partial \xi^v(\tau ,
{\vec \sigma}^{'})}\over {\partial \sigma^{{'}s}}}{{\delta}\over {\delta
\xi^v(\tau ,{\vec \sigma }^{'})}}]=\nonumber \\
&=&{{\partial \xi^u(\tau ,\vec \sigma )}\over {\partial \sigma^r}}{{\partial
\delta^3(\vec \sigma ,{\vec \sigma }^{'})}\over {\partial \sigma^{{'}s}}}
{{\delta}\over {\delta \xi^u(\tau ,{\vec \sigma }^{'})}}-{{\partial \xi^u(\tau
,{\vec \sigma }^{'})}\over {\partial \sigma^{{'}s}}}{{\partial \delta^3(\vec
\sigma ,{\vec \sigma }^{'})}\over {\partial \sigma^r}}{{\delta}\over {\delta
\xi^u(\tau ,\vec \sigma )}}=\nonumber \\
&=&-{{\partial \xi^u(\tau ,\vec \sigma )}\over {\partial \sigma^r}}{{\partial
\delta^3(\vec \sigma ,{\vec \sigma }^{'})}\over {\partial \sigma^{s}}}
{{\delta}\over {\delta \xi^u(\tau ,{\vec \sigma }^{'})}}+{{\partial \xi^u(\tau
,{\vec \sigma }^{'})}\over {\partial \sigma^{{'}s}}}{{\partial \delta^3(\vec
\sigma ,{\vec \sigma }^{'})}\over {\partial \sigma^{{'}r}}}{{\delta}\over
{\delta \xi^u(\tau ,\vec \sigma )}}=\nonumber \\
&=&\Big[ -{{\partial}\over {\partial \sigma^s}}\Big( {{\partial \xi^u(\tau ,\vec
\sigma )}\over {\partial \sigma^r}}\delta^3(\vec \sigma ,{\vec \sigma }^{'})
\Big) +{{\partial^2\xi^u(\tau ,\vec \sigma )}\over {\partial \sigma^r\partial
\sigma^s}}\delta^3(\vec \sigma ,{\vec \sigma }^{'})\Big] {{\delta}\over {\delta
\xi^u(\tau ,{\vec \sigma}^{'})}}+\nonumber \\
&+&\Big[ {{\partial}\over {\partial \sigma^{{'}r}}}\Big( {{\partial \xi^u(\tau
,{\vec \sigma}^{'})}\over {\partial \sigma^{{'}s}}}\delta^3(\vec \sigma ,{\vec
\sigma }^{'})\Big) -{{\partial^2\xi^u(\tau ,{\vec \sigma}^{'})}\over {\partial
\sigma^{{'}r}\partial \sigma^{{'}s}}}\delta^3(\vec \sigma ,{\vec \sigma }^{'})
\Big] {{\delta}\over {\delta \xi^u(\tau ,\vec \sigma )}}=\nonumber \\
&=&-{{\partial \delta^3(\vec \sigma ,{\vec \sigma }^{'})}\over
{\partial \sigma^s}}{{\partial \xi^u(\tau ,{\vec \sigma}^{'})}\over {\partial
\sigma^{{'}r}}}{{\delta}\over {\delta \xi^u(\tau ,{\vec \sigma}^{'})}}+
{{\partial \delta^3(\vec \sigma ,{\vec \sigma }^{'})}\over
{\partial \sigma^{{'}r}}}{{\partial \xi^u(\tau ,\vec \sigma )}\over {\partial
\sigma^s}}{{\delta}\over {\delta \xi^u(\tau ,\vec \sigma )}}=
\nonumber \\
&=&-{{\partial \delta^3(\vec \sigma ,{\vec \sigma }^{'})}\over
{\partial \sigma^{{'}s}}}{\tilde Y}_r(\tau ,{\vec \sigma }^{'})-{{\partial
\delta^3(\vec \sigma ,{\vec \sigma }^{'})}\over {\partial \sigma^{{'}r}}}
{\tilde Y}_s(\tau ,\vec \sigma ),
\label{IV35}
\end{eqnarray}

\noindent in accord with the last of Eqs.(\ref{III6}).
Therefore, the role of the Maurer-Cartan matrix B for rotations is
taken by minus the {\it Jacobian matrix} of the pseudo-diffeomorphism
$\vec \sigma \mapsto \vec \xi (\vec \sigma )$. To take into account
the noncommutativity of rotations and pseudo-diffeomorphisms [the
second line of Eqs.(\ref{III6})], we need the definition

\begin{equation}
Y_r(\tau ,\vec \sigma )=-\lbrace .,{}^3{\tilde \Theta}_r(\tau ,\vec \sigma )
\rbrace =-{{\partial \xi^s(\tau ,\vec \sigma )}\over {\partial \sigma^r}}
{{\delta}\over {\delta \xi^s(\tau ,\vec \sigma )}}-{{\partial \alpha_{(a)}
(\tau ,\vec \sigma )}\over {\partial \sigma^r}} {{\tilde \delta}\over
{\delta \alpha_{(a)}(\tau ,\vec \sigma )}}.
\label{IV36}
\end{equation}

Clearly the last line of Eqs.(\ref{III6}) is satisfied, while
regarding the second line we have consistently

\begin{eqnarray}
[X_{(a)}(\tau ,\vec \sigma ),Y_r(\tau ,{\vec \sigma }^{'})]&=&-[B_{(b)(a)}
(\alpha_{(e)}(\tau ,\vec \sigma )){{\tilde \delta}\over {\delta \alpha_{(b)}
(\tau ,\vec \sigma )}},{{\partial \alpha_{(c)}(\tau ,{\vec \sigma }^{'})}\over
{\partial \sigma^{{'}r}}}{{\tilde \delta}\over {\delta \alpha_{(c)}(\tau ,
{\vec \sigma }^{'})}}]=\nonumber \\
&=&-B_{(b)(a)}(\alpha_{(e)}(\tau ,\vec \sigma )){{\partial \delta^3(\vec \sigma
,{\vec \sigma }^{'})}\over {\partial \sigma^{{'}r}}}{{\tilde \delta}\over
{\delta \alpha_{(b)}(\tau ,{\vec \sigma }^{'})}}+\nonumber \\
&+&{{\partial \alpha_{(c)}(\tau
,\vec \sigma )}\over {\partial \sigma^r}}\delta^3(\vec \sigma ,{\vec \sigma }
^{'}){{\partial B_{(b)(a)}(\alpha_{(e)})}\over {\partial \alpha_{(c)}}}{|}
_{\alpha =\alpha (\tau ,\vec \sigma )}{{\tilde \delta}\over {\delta \alpha
_{(b)}(\tau ,\vec \sigma )}}=\nonumber \\
&=&-{{\partial \delta^3(\vec \sigma ,{\vec \sigma }^{'})}\over {\partial
\sigma^{{'}r}}} X_{(a)}(\tau ,{\vec \sigma }^{'}).
\label{IV37}
\end{eqnarray}

{}From Eqs.(\ref{IV36}) and (\ref{IV8}) we get

\begin{eqnarray}
{{\delta}\over {\delta \xi^r(\tau ,\vec \sigma )}}&=&-{{\partial \sigma^s(\vec
\xi )}\over {\partial \xi^r}}{|}_{\vec \xi =\vec \xi (\tau ,\vec \sigma )}
\Big[ Y_s(\tau ,\vec \sigma )+A_{(a)(b)}(\alpha_{(e)}(\tau ,\vec \sigma ))
{{\partial \alpha_{(b)}(\tau ,\vec \sigma )}\over {\partial \sigma^s}} X_{(a)}
(\tau ,\vec \sigma )\Big] =\nonumber \\
&=&{{\partial \sigma^s(\vec \xi )}\over {\partial \xi^r}}{|}_{\vec \xi =\vec
\xi (\tau ,\vec \sigma )}\Big[
\lbrace .,{}^3{\tilde \Theta}_s(\tau ,\vec \sigma )
\rbrace +{\tilde \theta}_{(a)s}(\tau ,\vec \sigma )\lbrace .,{}^3{\tilde M}
_{(a)}(\tau ,\vec \sigma )\rbrace \Big]\, {\buildrel {def}\over =}\nonumber \\
&{\buildrel {def} \over =}&\lbrace .,{\tilde \pi}^{\vec \xi}_r(\tau ,\vec
\sigma )\rbrace ,\nonumber \\
{}&&\nonumber \\
\Rightarrow && \lbrace \xi^r(\tau ,\vec \sigma ),{\tilde \pi}^{\vec \xi}_s(\tau
,{\vec \sigma }^{'})\rbrace =\delta^r_s \delta^3(\vec \sigma ,{\vec \sigma }
^{'}),\nonumber \\
{}&&\lbrace {\tilde \pi}^{\vec \xi}_r(\tau ,\vec \sigma ),{\tilde \pi}^{\vec
\xi}_s(\tau ,{\vec \sigma }^{'})\rbrace =0,
\label{IV38}
\end{eqnarray}

\noindent where ${\tilde \pi}^{\vec \xi}_r(\tau ,\vec \sigma )$ is the momentum
conjugate to the 3 gauge variables $\xi^r(\tau ,\vec \sigma )$, which
will be functions only of the cotriads. On the space of cotriads the
Abelianized form of the pseudo-diffeomorphism constraints is

\begin{eqnarray}
{\tilde \pi}^{\vec \xi}_r(\tau ,\vec \sigma )&=&{{\partial \sigma^s(\vec \xi )}
\over {\partial \xi^r}}{|}_{\vec \xi =\vec \xi (\tau ,\vec \sigma )} \Big[ {}^3
{\tilde \Theta}_s(\tau ,\vec \sigma )+{\hat {\tilde \theta}}_{(a)s}(\alpha_{(e)}
(\tau ,\vec \sigma ),\partial_u\alpha_{(e)}(\tau ,\vec \sigma ))\, {}^3{\tilde
M}_{(a)}(\tau ,\vec \sigma )\Big] =\nonumber \\
&=&{{\partial \sigma^s(\vec \xi )}
\over {\partial \xi^r}}{|}_{\vec \xi =\vec \xi (\tau ,\vec \sigma )} \Big[ {}^3
{\tilde \Theta}_s-{{\partial \alpha_{(a)}}\over {\partial \sigma^s}}
{\tilde
\pi}^{\vec \alpha}_{(a)}\Big] (\tau ,\vec \sigma )\approx 0,
\label{IV39}
\end{eqnarray}

\noindent and both $\xi^r(\tau ,\vec \sigma )$
and ${\tilde \pi}^{\vec \xi}_r(\tau ,\vec \sigma )$
have zero Poisson bracket with $\alpha_{(a)}(\tau ,\vec \sigma )$,
${\tilde \pi}^{\vec \alpha}_{(a)}(\tau ,\vec \sigma )$.

Therefore, the 6 gauge variables $\xi^r(\tau ,\vec \sigma )$ and
$\alpha_{(a)} (\tau ,\vec \sigma )$ and their conjugate momenta form 6
canonical pairs of a new canonical basis adapted to the rotation and
pseudo-diffeomorphisms constraints and replacing 6 of the 9 conjugate
pairs ${}^3e_{(a)r}(\tau ,\vec
\sigma )$, ${}^3{\tilde \pi}^r_{(a)}(\tau ,\vec \sigma )$.

{}From Eqs.(3.19) of Ref.\cite{ru11}  and from Eqs.(\ref{IV36}) and
(\ref{IV13}), we get

\begin{eqnarray}
Y_s(\tau ,{\vec \sigma }^{'})\, {}^3e_{(a)r}(\tau ,\vec \sigma )&=&
-\Big( {{\partial
\xi^u(\tau ,{\vec \sigma }^{'})}\over {\partial \sigma^{{'}s}}}{{\delta}
\over {\delta \xi^u(\tau ,{\vec \sigma }^{'})}}+{{\partial \alpha_{(c)}(\tau ,
{\vec \sigma }^{'})}\over {\partial \sigma^{{'}r}}}{{\tilde \delta}\over
{\delta \alpha_{(c)}(\tau ,{\vec \sigma }^{'})}}\Big) \cdot \nonumber \\
&\cdot & \Big[
{}^3R_{(a)(b)}(\alpha_{(e)}(\tau ,\vec \sigma ))\, {}^3{\bar e}_{(b)r}
(\tau ,\vec \sigma )\Big] =\nonumber \\
&=&-{}^3R_{(a)(b)}(\alpha_{(e)}(\tau ,\vec \sigma )){{\partial \xi^u(\tau ,
{\vec \sigma }^{'})}\over {\partial \sigma^{{'}s}}}{{\delta \, {}^3{\bar e}
_{(b)r}(\tau ,\vec \sigma )}\over {\delta \xi^u(\tau ,{\vec \sigma }^{'})}}-
\nonumber \\
&-&{{\partial \alpha_{(c)}(\tau ,{\vec \sigma }^{'})}\over {\partial \sigma
^{{'}s}}}{{\tilde \delta {}^3R_{(a)(b)}(\alpha_{(e)}(\tau ,\vec \sigma ))}\over
{\delta \alpha_{(c)}(\tau ,{\vec \sigma }^{'})}}\, {}^3{\bar e}_{(b)r}(\tau ,
\vec \sigma )=\nonumber \\
&=&-{}^3R_{(a)(b)}(\alpha_{(e)}(\tau ,\vec \sigma )){{\partial \xi^u(\tau ,
{\vec \sigma}^{'})}\over {\partial \sigma^{{'}s}}}{{\delta \, {}^3{\bar e}
_{(b)r}(\tau ,\vec \sigma )}\over {\delta \xi^u(\tau ,{\vec \sigma }^{'})}}-
\nonumber \\
&-&{{\partial \, {}^3R_{(a)(b)}(\alpha_{(e)}(\tau ,\vec \sigma ))}\over {\partial
\sigma^s}}\delta^3(\vec \sigma ,{\vec \sigma }^{'})\, {}^3{\bar e}_{(b)r}(\tau ,
\vec \sigma )=\nonumber \\
&&{}\nonumber \\
&=&-\lbrace {}^3e_{(a)r}(\tau ,\vec \sigma ),{}^3{\tilde \Theta}_s(\tau ,{\vec
\sigma }^{'})\rbrace =\nonumber \\
&=&-{{\partial \, {}^3e_{(a)r}(\tau ,\vec \sigma )}\over
{\partial \sigma^s}}\delta^3(\vec \sigma ,{\vec \sigma }^{'})+{}^3e_{(a)s}
(\tau ,\vec \sigma ){{\partial \delta^3(\vec \sigma ,{\vec \sigma }^{'})}\over
{\partial \sigma^r}}=\nonumber \\
&=&-{}^3R_{(a)(b)}(\alpha_{(e)}(\tau ,\vec \sigma )){{\partial \, {}^3{\bar e}
_{(b)r}(\tau ,\vec \sigma )}\over {\partial \sigma^s}}\delta^3(\vec \sigma ,
{\vec \sigma }^{'})-\nonumber \\
&-&{{\partial {}^3R_{(a)(b)}(\alpha_{(e)}(\tau ,\vec \sigma ))}
\over {\partial \sigma^s}}\, {}^3{\bar e}_{(b)r}(\tau ,\vec \sigma )\delta^3
(\vec \sigma ,{\vec \sigma }^{'})+\nonumber \\
&+&{}^3R_{(a)(b)}(\alpha_{(e)}(\tau ,\vec \sigma ))\, {}^3{\bar e}_{(b)r}
(\tau ,\vec \sigma ){{\partial \delta^3(\vec \sigma ,{\vec \sigma }^{'})}\over
{\partial \sigma^r}},
\label{IV40}
\end{eqnarray}

\noindent so that the pseudo-diffeomorphism multitemporal equations for
${}^3{\bar e}_{(a)r}(\tau ,\vec \sigma )$ are

\begin{eqnarray}
-{\tilde Y}_s(\tau ,{\vec \sigma }^{'})\, {}^3{\bar e}_{(a)r}(\tau ,\vec
\sigma )&=&{{\partial \xi^u(\tau ,{\vec \sigma }^{'})}\over {\partial \sigma
^{{'}s}}} {{\delta \, {}^3{\bar e}_{(a)r}(\tau ,\vec \sigma )}\over {\delta
\xi^u(\tau ,{\vec \sigma }^{'})}}=\nonumber \\
&=&{{\partial \, {}^3{\bar e}_{(a)r}(\tau ,\vec \sigma )}\over {\partial
\sigma^s}} \delta^3(\vec \sigma ,{\vec \sigma }^{'})-{}^3{\bar e}_{(a)s}(\tau ,
\vec \sigma ) {{\partial \delta^3(\vec \sigma ,{\vec \sigma }^{'})}\over
{\partial \sigma^{{'}r}}}.
\label{IV41}
\end{eqnarray}

Analogously, from Eqs.(3.19) of Ref.\cite{ru11} and Eqs.(\ref{IV36})
and (\ref{IV24}) we have

\begin{eqnarray}
Y_s(\tau ,{\vec \sigma }^{'})\, {}^3{\tilde \pi}^r_{(a)}(\tau ,\vec \sigma )&=&
-\Big( {{\partial
\xi^u(\tau ,{\vec \sigma }^{'})}\over {\partial \sigma^{{'}s}}}{{\delta}
\over {\delta \xi^u(\tau ,{\vec \sigma }^{'})}}+{{\partial \alpha_{(c)}(\tau ,
{\vec \sigma }^{'})}\over {\partial \sigma^{{'}r}}}{{\tilde \delta}\over
{\delta \alpha_{(c)}(\tau ,{\vec \sigma }^{'})}}\Big) \cdot \nonumber \\
&\cdot & \Big[ {}^3R_{(a)(b)}(\alpha_{(e)}(\tau ,\vec \sigma ))\, {}^3{\bar
{\tilde \pi}}^r_{(b)}(\tau ,\vec \sigma )\Big] =\nonumber \\
&=&-\Big[
{}^3R_{(a)(b)}(\alpha_{(e)}(\tau ,{\vec \sigma }^{'}))\, {}^3{\bar {\tilde
\pi}}^r_{(b)}(\tau ,{\vec \sigma }^{'})\Big] {{\partial \delta^3(\vec \sigma ,
{\vec \sigma }^{'})}\over {\partial \sigma^{{'}s}}}+\nonumber \\
&+&\delta^r_s {}^3R_{(a)(b)}
(\alpha_{(e)}(\tau ,\vec \sigma ))\, {}^3{\bar {\tilde \pi}}^u_{(b)}(\tau ,
\vec \sigma ){{\partial \delta^3(\vec \sigma ,{\vec \sigma }^{'})}
\over {\partial \sigma^{{'}u}}},
\label{IV42}
\end{eqnarray}

\noindent and we get the pseudo-diffeomorphism multitemporal equation for
${}^3{\bar {\tilde \pi}}^r_{(a)}(\tau ,\vec \sigma )$

\begin{eqnarray}
-{\tilde Y}_s(\tau ,{\vec \sigma }^{'})\, {}^3{\bar {\tilde \pi}}^r_{(a)}
(\tau ,\vec \sigma )&=&{{\partial \xi^u(\tau ,{\vec \sigma }^{'})}\over
{\partial \sigma^{{'}s}}}{{\delta \, {}^3{\bar {\tilde \pi}}^r_{(a)}(\tau ,
\vec \sigma )}\over {\delta \xi^u(\tau ,{\vec \sigma }^{'})}}=\nonumber \\
&=&-{}^3{\bar {\tilde \pi}}^r_{(a)}(\tau ,{\vec \sigma }^{'}){{\partial \delta
^3(\vec \sigma ,{\vec \sigma }^{'})}\over {\partial \sigma^{{'}s}}}-\delta^r_s
\, {}^3{\bar {\tilde \pi}}^u_{(a)}(\tau ,\vec \sigma ){{\partial \delta^3
(\vec \sigma ,{\vec \sigma }^{'})}\over {\partial \sigma^{{'}u}}}.
\label{IV43}
\end{eqnarray}

Let us remark that the Jacobian matrix satisfies an equation like
(\ref{IV41})

\begin{eqnarray}
-{\tilde Y}_s(\tau ,{\vec \sigma }^{'})&&{{\partial \xi^u(\tau ,\vec \sigma )}
\over {\partial \sigma^r}}={{\partial \xi^v(\tau ,{\vec \sigma }^{'})}\over
{\partial \sigma^{{'}s}}}{{\delta}\over {\delta \xi^v(\tau ,{\vec \sigma }^{'})
}}{{\partial \xi^u(\tau ,\vec \sigma )}\over {\partial \sigma^r}}=\nonumber \\
&=&{{\partial \xi^u(\tau ,{\vec \sigma }^{'})}\over {\partial \sigma^{{'}s}}}
{{\partial \delta^3(\vec \sigma ,{\vec \sigma }^{'})}\over {\partial \sigma^r}}
=-{{\partial^u(\tau ,{\vec \sigma }^{'})}\over {\partial \sigma^{{'}s}}}
{{\partial \delta^3(\vec \sigma ,{\vec \sigma }^{'})}\over {\partial
\sigma^{{'}r}}}=\nonumber \\
&=&-{{\partial \xi^u(\tau ,\vec \sigma )}\over {\partial \sigma^s}}
{{\partial \delta^3(\vec \sigma ,{\vec \sigma }^{'})}\over {\partial \sigma
^{{'}r}}}+{{\partial^2\xi^u(\tau ,\vec \sigma )}\over {\partial \sigma^r
\partial \sigma^s}}\delta^3(\vec \sigma ,{\vec \sigma }^{'})=\nonumber \\
&=&{{\partial}\over {\partial \sigma^s}}({{\partial \xi^u(\tau ,\vec \sigma )}
\over {\partial \sigma^r}})\delta^3(\vec \sigma ,{\vec \sigma }^{'})-{{\partial
\xi^u(\tau ,\vec \sigma )}\over {\partial \sigma^s}}{{\partial \delta^3
(\vec \sigma ,{\vec \sigma }^{'})}\over {\partial \sigma^{{'}r}}}.
\label{IV44}
\end{eqnarray}

\noindent so that the identity ${{\partial \xi^u(\tau ,{\vec \sigma }^{'})}
\over {\partial \sigma^{{'}s}}}{{\delta f(\tau ,\vec \xi (\tau ,\vec \sigma ))}
\over {\delta \xi^u(\tau ,{\vec \sigma }^{'})}}={{\partial f(\tau ,\vec \xi
(\tau ,\vec \sigma ))}\over {\partial \sigma^s}}\delta^3(\vec \sigma ,
{\vec \sigma }^{'})$, implies the following solutions of the
multitemporal equations \footnote{Again $\hat V(\vec \xi (\tau ,\vec
\sigma ))$ is the operator with the action $\hat V(\vec \xi (\tau
,\vec \sigma ))f(\tau ,\vec \sigma )= f(\tau ,\vec \xi (\tau ,\vec
\sigma ))$; and Eqs.(\ref{III7}) is used.}

\begin{eqnarray}
{}^3{\bar e}_{(a)r}(\tau ,\vec \sigma )&=&{{\partial \xi^s(\tau ,\vec \sigma )}
\over {\partial \sigma^r}}\, {}^3{\hat e}_{(a)s}(\tau ,\vec \xi (\tau ,
\vec \sigma ))={{\partial \xi^s(\tau ,\vec \sigma )}\over {\partial \sigma^r}}
\hat V(\vec \xi (\tau ,\vec \sigma ))\, {}^3{\hat e}_{(a)s}(\tau ,\vec \sigma )
,\nonumber \\
{}&&{{\delta \, {}^3{\hat e}_{(a)r}(\tau ,\vec \sigma )}\over {\delta \xi^s
(\tau ,{\vec \sigma }^{'})}}=0,\nonumber \\
{}^3e_{(a)r}(\tau,\vec \sigma )&=&{}^3R_{(a)(b)}(\alpha_{(e)}(\tau ,\vec
\sigma )) {{\partial \xi^s(\tau ,\vec \sigma )}\over {\partial \sigma^r}}\,
{}^3{\hat e}_{(b)s}(\tau ,\vec \xi (\tau ,\vec \sigma ))=\nonumber \\
&=&{{\partial \xi^s(\tau ,\vec \sigma )}\over {\partial \sigma^r}}\, {}^3R
_{(a)(b)}(\alpha^{'}_{(e)}(\tau ,\vec \xi (\tau ,\vec \sigma )))\, {}^3{\hat e}
_{(b)s}(\tau ,\vec \xi (\tau ,\vec \sigma )),\nonumber \\
&&{}\nonumber \\
{}^3g_{rs}(\tau ,\vec \sigma )&=&{{\partial \xi^u(\tau ,\vec \sigma )}\over
{\partial \sigma^r}} {{\partial \xi^v(\tau ,\vec \sigma )}\over {\partial
\sigma^s}}\, {}^3{\hat e}_{(a)u}(\tau ,\vec \xi (\tau ,\vec \sigma ))\,
{}^3{\hat e}_{(a)v}(\tau ,\vec \xi (\tau ,\vec \sigma )).
\label{IV45}
\end{eqnarray}

Here the cotriads ${}^3{\hat e}_{(a)r}(\tau ,\vec \sigma )$ depend
only on 3 degrees of freedom and are Dirac observables with respect to
both Abelianized rotations and pseudo-diffeomorphisms. Again, like in
the case of rotations, we have not found 3 specific conditions on the
cotriads implying this final reduction. This is due to the fact that,
even if one has a trivial coframe bundle, one does not know the group
manifold of $Diff\, \Sigma_{\tau}$ and there is no canonical identity
for pseudo-diffeomorphisms and therefore also for rotations inside the
gauge group ${\bar {\cal G}}_R$.

Eqs.(\ref{IV45}) are the counterpart in tetrad gravity of the
solutions of the 3 elliptic equations for the gravitomagnetic vector
potential ${\check W}^r$ of the conformal approach\cite{ciuf} (see the
end of Appendix C of II).

If ${{\partial \sigma^r(\vec \xi )}\over {\partial \xi^s}}{|}_{\vec \xi =\vec
\xi (\tau ,\vec \sigma )}$ is the inverse Jacobian matrix and $|\, {{\partial
\xi (\tau ,\vec \sigma )}\over {\partial \sigma}}\, |$ the determinant of the
Jacobian matrix, the following identities

\begin{eqnarray}
\delta^r_s&=&{{\partial \sigma^r(\vec \xi )}\over {\partial \xi^u}}{|}_{\vec \xi
=\vec \xi (\tau ,\vec \sigma )}\, {{\partial \xi^u(\tau ,\vec \sigma )}\over
{\partial \sigma^s}},\nonumber \\
\Rightarrow && {{\partial }\over {\partial \sigma^v}}{{\partial \sigma^r(\vec
\xi )}\over {\partial \xi^u}}{|}_{\vec \xi =\vec \xi (\tau ,\vec \sigma )}=-
{{\partial \sigma^s(\vec \xi )}\over {\partial \xi^u}}{|}_{\vec \xi =\vec \xi
(\tau ,\vec \sigma )}\, {{\partial \sigma^r(\vec \xi )}\over {\partial \xi^w}}
{|}_{\vec \xi =\vec \xi (\tau ,\vec \sigma )}\, {{\partial^2\xi^w(\tau ,\vec
\sigma )}\over {\partial \sigma^v\partial \sigma^s}},\nonumber \\
\Rightarrow && {{\delta}\over {\delta \xi^v(\tau ,{\vec \sigma }^{'})}}
{{\partial \sigma^r(\vec \xi )}\over {\partial \xi^u}}{|}_{\vec \xi =\vec \xi
(\tau ,\vec \sigma )}=-{{\partial \sigma^s(\vec \xi )}\over {\partial \xi^u}}
{|}_{\vec \xi =\vec \xi (\tau ,\vec \sigma )} {{\partial \sigma^r(\vec \xi )}
\over {\partial \xi^s}}{|}_{\vec \xi =\vec \xi (\tau ,\vec \sigma )}
{{\partial \delta^3(\vec \sigma ,{\vec \sigma }^{'})}\over {\partial \sigma^v}},
\nonumber \\
{}&&\downarrow \nonumber \\
-{\tilde Y}_s(\tau ,{\vec \sigma }^{'})&&{{\partial \sigma^r(\vec \xi )}\over
{\partial \xi^u}}{|}_{\vec \xi =\vec \xi (\tau ,\vec \sigma  )}=\nonumber \\
&=&{{\partial}\over {\partial \sigma^s}}
\Big( {{\partial \sigma^r(\vec \xi )}\over
{\partial \xi^u}}{|}_{\vec \xi =\vec \xi (\tau,\vec \sigma  )}\Big) \delta^3
(\vec \sigma ,{\vec \sigma }^{'})+\delta^r_s{{\partial \sigma^v(\vec \xi )}
\over {\partial \xi^u}}{|}_{\vec \xi -\vec \xi (\tau ,\vec \sigma )}\,
{{\partial \delta^3(\vec \sigma ,{\vec \sigma }^{'})}\over {\partial
\sigma^{{'}v}}},
\label{IV46}
\end{eqnarray}

\noindent and [use is done of $\delta ln\, det\, M=Tr\, (M^{-1}\delta M)$]

\begin{eqnarray}
{{\partial}\over {\partial \sigma^r}}&& |\, {{\partial \xi (\tau ,\vec \sigma )}
\over {\partial \sigma }}\, | = |\, {{\partial \xi (\tau ,\vec \sigma )}\over
{\partial \sigma}}\, |\, {{\partial \sigma^s(\vec \xi )}\over {\partial \xi^u}}
{|}_{\vec \xi =\vec \xi (\tau ,\vec \sigma )}\, {{\partial^2\xi^u(\tau ,\vec
\sigma )}\over {\partial \sigma^r\partial \sigma^s}},\nonumber \\
{{\delta}\over {\delta \xi^r(\tau ,{\vec \sigma }^{'})}} && |\, {{\partial \xi
(\tau ,\vec \sigma )}\over {\partial \sigma}}\, |=|\, {{\partial \xi (\tau ,
\vec \sigma )}\over {\partial \sigma}}\, |\, {{\partial \sigma^s(\vec \xi )}
\over {\partial \xi^r}}{|}_{\vec \xi =\vec \xi (\tau ,\vec \sigma )}\,
{{\partial \delta^3(\vec \sigma ,{\vec \sigma }^{'})}\over {\partial \sigma^s}},
\nonumber \\
{}&&\downarrow \nonumber \\
-{\tilde Y}_s(\tau ,{\vec \sigma }^{'})&& |\, {{\partial \xi (\tau ,\vec \sigma
)}\over {\partial \sigma}}\, |=-|\, {{\partial \xi (\tau ,\vec \sigma )}\over
{\partial \sigma }}\, |\, {{\partial \delta^3(\vec \sigma ,{\vec \sigma }^{'})}
\over {\partial \sigma^{{'}s}}},
\label{IV47}
\end{eqnarray}

\noindent allow to get

\begin{eqnarray}
{}^3{\tilde \pi}^r_{(a)}(\tau ,\vec \sigma )&=&{}^3R_{(a)(b)}(\alpha_{(a)}
(\tau ,\vec \sigma ))\, {}^3{\bar {\tilde \pi}}^r_{(b)}(\tau ,\vec \sigma )=
\nonumber \\
&=&{}^3R_{(a)(b)}(\alpha_{(e)}(\tau ,\vec \sigma ))\, |\, {{\partial \xi
(\tau ,\vec \sigma )}\over {\partial \sigma}}\, |\, {{\partial \sigma^r(\vec
\xi )}\over {\partial \xi^s}}{|}_{\vec \xi =\vec \xi (\tau ,\vec \sigma )}\,
{}^3{\hat {\tilde \pi}}^s_{(b)}(\tau ,\vec \xi (\tau ,\vec \sigma ))=
\nonumber \\
&=&{}^3R_{(a)(b)}(\alpha_{(e)}(\tau .\vec \sigma ))\, |\, {{\partial \xi (\tau ,
\vec \sigma )}\over {\partial \sigma}}\, |\, {{\partial \sigma^r(\vec \xi )}
\over {\partial \xi^s}}{|}_{\vec \xi =\vec \xi (\tau ,\vec \sigma )}\,
\hat V(\vec \xi (\tau ,\vec \sigma ))\, {}^3{\hat {\tilde \pi}}^s_{(b)}(\tau .
\vec \sigma ),
\label{IV48}
\end{eqnarray}

\noindent where ${}^3{\hat {\tilde \pi}}^r_{(a)}(\tau ,\vec \sigma )$ are Dirac
observables with respect to both Abelianized rotations and
pseudo-diffeomorphisms. In a similar way we get

\begin{equation}
{}^3e^r_{(a)}(\tau ,\vec \sigma )={}^3R_{(a)(b)}(\alpha_{(e)}(\tau ,\vec
\sigma )) {{\partial \sigma^r(\vec \xi )}\over {\partial \xi^s}}{|}_{\vec \xi =
\vec \xi (\tau ,\vec \sigma )}\, {}^3{\hat e}^r_{(b)}(\tau ,\vec \xi (\tau ,
\vec \sigma )),
\label{IV49}
\end{equation}

\noindent with ${}^3{\hat e}^r_{(a)}(\tau ,\vec \sigma )$ the Dirac observables
for triads dual to ${}^3e_{(a)r}(\tau ,\vec \sigma )$. The line element becomes

\begin{eqnarray}
ds^2&=&\epsilon \Big(
[N_{(as)}+n]^2 - [N_{(as) r}+n_r] {{\partial \sigma^r(\vec
\xi )}\over {\partial \xi^u}} {}^3{\hat e}^u_{(a)}(\vec \xi )\, {}^3{\hat e}^v
_{(a)}(\vec \xi ) {{\partial \sigma^s(\vec \xi )}\over {\partial \xi^v}}
[N_{(as) s}+n_s] \Big) (d\tau )^2-\nonumber \\
&-&2\epsilon [N_{(as) r}+n_r] d\tau d\sigma^r-\epsilon {{\partial \xi^u}\over
{\partial \sigma^r}}\, {}^3{\hat e}_{(a)u}(\vec \xi )\, {}^3{\hat e}
_{(a)v}(\vec \xi ) {{\partial \xi^v}\over {\partial \sigma^s}}
d\sigma^r d\sigma^s=\nonumber \\
&=&\epsilon \Big( [N_{(as)}+n]^2(d\tau )^2- [{}^3{\hat e}_{(a)u}(\vec \xi )
{{\partial \xi^u}\over {\partial \sigma^r}}d\sigma^r+{}^3{\hat e}^u_{(a)}(\vec
\xi ){{\partial \sigma^r(\vec \xi )}\over {\partial \xi^u}}(N_{(as)r}+n_r)
d\tau ]\nonumber \\
&&[{}^3{\hat e}_{(a)v}(\vec \xi )
{{\partial \xi^v}\over {\partial \sigma^s}}d\sigma^s+{}^3{\hat e}^v_{(a)}(\vec
\xi ){{\partial \sigma^s(\vec \xi )}\over {\partial \xi^v}}(N_{(as)s}+n_s)
d\tau ] \Big) .
\label{IV50}
\end{eqnarray}

To get $\xi^r(\tau ,\vec \sigma )$ in terms of the cotriads we have to
solve the equations \footnote{Use is done of Eq.(\ref{IV38}), of
(\ref{II11}) and of $\lbrace \xi^r(\tau ,\vec \sigma ),{}^3{\tilde
M}_{(a)}(\tau ,{\vec \sigma }^{'})\rbrace =0$.}

\begin{eqnarray}
\delta^r_s\delta^3(\vec \sigma ,{\vec \sigma }^{'})&=&\lbrace \xi^r(\tau ,\vec
\sigma ),{\tilde \pi}^{\vec \xi}_s(\tau ,{\vec \sigma }^{'})\rbrace =
\nonumber \\
&=&{{\partial \sigma^u(\vec \xi )}\over {\partial \xi^s}}{|}_{\vec \xi =\vec \xi
(\tau ,{\vec \sigma}^{'} )}\,
\lbrace \xi^r(\tau ,\vec \sigma ),{}^3{\tilde \Theta}_u
(\tau ,{\vec \sigma }^{'})\rbrace =\nonumber \\
&=&{{\partial \sigma^u(\vec \xi )}\over {\partial \xi^s}}{|}_{\vec \xi =\vec \xi
(\tau ,{\vec \sigma}^{'} )}
\Big[ \Big( {{\partial \, {}^3e_{(a)v}(\tau ,{\vec \sigma }
^{'})}\over {\partial \sigma^{{'}u}}}-{{\partial \, {}^3e_{(a)u}(\tau ,
{\vec \sigma }^{'})}\over {\partial \sigma^{{'}v}}}\Big) {{\delta \xi^r(\tau ,
\vec \sigma )}\over {\delta \, {}^3e_{(a)v}(\tau ,{\vec \sigma }^{'})}}-
\nonumber \\
&-&{}^3e_{(a)u}(\tau ,{\vec \sigma }^{'}){{\partial}\over {\partial
\sigma^{{'}v}}}\, {{\delta \xi^r(\tau ,\vec \sigma )}\over {\delta \, {}^3e
_{(a)v}(\tau ,{\vec \sigma }^{'})}}\Big] ,\nonumber \\
&&\Downarrow \nonumber \\
\Big( \Big[ \delta_{(a)(b)}\partial^{'}_v&-& {}^3e^u_{(a)} (\partial^{'}_u\,
{}^3e_{(b)v}-\partial^{'}_v\, {}^3e_{(b)u})\Big] (\tau ,{\vec \sigma}^{'})
{{\delta}\over {\delta \, {}^3e_{(b)v}(\tau ,{\vec \sigma}^{'})}}+\nonumber \\
&+&\delta^3(\vec \sigma ,{\vec \sigma }^{'})
{}^3e^u_{(a)}(\tau ,\vec \sigma ){{\partial}\over {\partial \sigma^u}}
\Big) \xi^r(\tau ,\vec \sigma ) =0.
\label{IV51}
\end{eqnarray}

We do not know how to solve these equations along some privileged path
in the group manifold of $Diff\, \Sigma_{\tau}$ after having chosen a
global coordinate system $\Xi$ as a conventional origin of
pseudo-diffeomorphisms \footnote{This identifies a conventional
identity cross section $\Sigma_{\tau}^{(\Xi )}$ in the proposed
description of $Diff\, \Sigma_{\tau}$ with the fibration
$\Sigma_{\tau}\times \Sigma_{\tau} \rightarrow \Sigma_{\tau}$ for the
case $\Sigma_{\tau}\approx R^3$.}, due to the poor understanding of
the geometry and differential structure of this group manifold.
Presumably, since the fibers of $\Sigma_{\tau}\times \Sigma_{\tau}$
are also copies of $\Sigma_{\tau}$, on each one of them one can try to
define an analogue of canonical coordinates of first kind by using the
geodesic exponential map:\hfill\break i) choose a reference fiber
$\Sigma_{\tau ,0}$ in $\Sigma_{\tau}\times \Sigma
_{\tau}$ over a point $p=(\tau ,\vec 0)$ chosen as origin in the base \footnote{Then
connected to all the points in base with geodesics; for
$\Sigma_{\tau}\approx R^3$ this is well defined; the global cross
sections corresponding to global coordinate systems should be
horizontal lifts of this geodesic star with respect to some notion of
connection on the fibration.};\hfill\break ii) if $q_o$ is the point
in $\Sigma_{\tau}\times \Sigma_{\tau}$ at the intersection of
$\Sigma_{\tau ,0}$ with the conventional identity cross section
$\Sigma_{\tau}^{(\Xi )}$ and $q_1$ the point where $\Sigma_{\tau ,0}$
intersects a nearby global cross section $\Sigma_{\tau}^{(\Xi^{'})}$
($\Xi^{'}$ is another global coordinate system on $\Sigma_{\tau}$), we
can consider the geodesic $\gamma_{q_oq_1}$ on $\Sigma_{\tau
,0}$;\hfill\break iii) use the geodesic exponential map along the
geodesic $\gamma_{q_oq_1}$ to define {\it pseudo-diffeomorphism
coordinates} $\vec \xi (\tau ,\vec 0)$ describing the transition from
the global coordinate system $\Xi$ to $\Xi^{'}$ over the base point
$p=(\tau ,\vec 0)$;\hfill\break iv) parallel transport these
coordinates on the fiber $\Sigma_{\tau ,0}$ to the other fibers along
the geodesics of the cross sections $\Sigma_{\tau}
^{(\Xi^{'})}$.

If this coordinatization of the group manifold of $Diff\, \Sigma_{\tau}$ for
$\Sigma_{\tau}\approx R^3$ can be justified, then one could try to solve the
previous equations.

Instead, we are able to give a formal expression for the operator $\hat V(\vec
\xi (\vec \sigma ))$ \footnote{For the sake of simplicity we do not consider the
$\tau$-dependence.}, whose action on functions $f(\vec \sigma )$ is
$\hat V (\vec \xi (\vec \sigma )) f(\vec \sigma )=f(\vec \xi (\vec
\sigma ))$. We have

\begin{equation}
\hat V(\vec \xi (\vec \sigma ))=P_{\gamma}\, e^{(\int_{\vec \sigma }^{\vec \xi
(\vec \sigma )} {{\partial \sigma^r(u)}\over {\partial u^s}} {\cal D}u^s)
{{\partial}\over {\partial \sigma^r}} },
\label{IV52}
\end{equation}

\noindent where the path ordering is along the geodesic $\gamma$ in $\Sigma
_{\tau}$ joining the points with coordinates $\vec \sigma $ and ${\vec
\sigma }^{'}=\vec \xi (\vec \sigma )$. For infinitesimal pseudo-diffeomorphisms
$\vec \sigma \mapsto {\vec \sigma }^{'}(\vec \sigma )=\vec \xi (\vec \sigma )=
\vec \sigma +\delta \vec \sigma (\vec \sigma )$ [with inverse ${\vec \sigma }
^{'}=\vec \xi \mapsto \vec \sigma (\vec \xi )=\vec \xi -\delta \vec \sigma
(\vec \xi )$], we have

\begin{eqnarray}
\hat V(\vec \sigma +\delta \vec \sigma )&\approx&
1+\Big[ \delta \sigma^s(\vec \sigma )
{{\partial \sigma^r(\vec \xi )}\over {\partial \xi^s}}{|}_{\vec \xi (\vec
\sigma )-\delta \vec \sigma (\vec \xi (\vec \sigma ))}\Big] {{\partial}\over
{\partial \sigma^r}}\approx 1+\delta \sigma^s(\vec \sigma ){{\partial}\over
{\partial \sigma^s}}:\nonumber \\
&:& f(\vec \sigma )\mapsto f(\vec \sigma )+\delta \sigma^s(\vec \sigma )
{{\partial f(\vec \sigma )}\over {\partial \sigma^s}}\approx f(\vec \sigma +
\delta \vec \sigma (\vec \sigma )).
\label{IV53}
\end{eqnarray}

Formally we have \footnote{If $\delta /\delta \xi^r(\vec \sigma )$ is
interpreted as the directional functional derivative along $\gamma$.}

\begin{eqnarray}
{{\delta}\over {\delta \xi^r({\vec \sigma }^{'})}} [\hat V(\vec \xi (\vec
\sigma )) f(\vec \sigma )]&=&\delta^3(\vec \sigma ,{\vec \sigma }^{'})
{{\partial \sigma^s(\vec \xi )}\over {\partial \xi^r}}{|}_{\vec \xi =\vec \xi
(\vec \sigma )}\, {{\partial}\over {\partial \sigma^s}}[\hat V(\vec \xi (\vec
\sigma ))f(\vec \sigma )]=\nonumber \\
&=&\delta^3(\vec \sigma ,{\vec \sigma }^{'}) {{\partial \sigma^s(\vec \xi )}
\over {\partial \xi^r}}{|}_{\vec \xi =\vec \xi (\vec \sigma )}{{\partial
f(\vec \xi (\vec \sigma ))}\over {\partial \sigma^s}}=\delta^3(\vec \sigma ,
{\vec \sigma }^{'}){{\partial f(\vec \xi )}\over {\partial \xi^r}}{|}_{\vec \xi
=\vec \xi (\vec \sigma )}=\nonumber \\
 &=&{{\delta f(\vec \xi (\vec \sigma ))}\over {\delta \xi^r({\vec \sigma }^{'})}}.
\label{IV54}
\end{eqnarray}

By using Eqs.(\ref{IV27}) and (3.19) of Ref.\cite{ru11}, we get

\begin{eqnarray}
{\hat D}^{(\omega )}_{(a)(b)r}(\tau ,\vec \sigma )&& \lbrace \zeta^{(\omega )r}
_{(b)(c)}(\vec \sigma ,{\vec \sigma}_1;\tau ),{}^3{\tilde M}_{(g)}(\tau ,
{\vec \sigma}_2)\rbrace =\nonumber \\
&=&-\epsilon_{(a)(d)(b)} \lbrace {}^3\omega_{s(d)}(\tau ,
\vec \sigma ),{}^3{\tilde M}_{(g)}(\tau ,{\vec \sigma}_2)\rbrace
\zeta^{(\omega )s}_{(b)(c)}(\vec \sigma ,{\vec \sigma}_1;\tau ),\nonumber \\
&&{}\nonumber \\
{\hat D}^{(\omega )}_{(a)(b)r}(\tau ,\vec \sigma )&& \lbrace \zeta^{(\omega )r}
_{(b)(c)}(\vec \sigma ,{\vec \sigma}_1;\tau ),{}^3{\tilde \Theta}_u(\tau ,{\vec
\sigma}_2)\rbrace =\nonumber \\
&=&-\epsilon_{(a)(d)(f)} \lbrace {}^3\omega_{s(d)}(\tau ,\vec
\sigma ),{}^3{\tilde \Theta}_u(\tau ,{\vec \sigma}_2)\rbrace \zeta^{(\omega )s}
_{(f)(c)}(\vec \sigma ,{\vec \sigma}_1;\tau ).
\label{IV55}
\end{eqnarray}

\noindent Then Eqs.(3.19) of Ref.\cite{ru11}, (\ref{IV17}) and (\ref{IV27}) imply the
following transformation properties under rotations and space
pseudo-diffeomorphisms of the Green function of the SO(3) covariant
divergence (which we do not know how to verify explicitly due to the
path-ordering contained in it)

\begin{eqnarray}
\lbrace \zeta^{(\omega )r}_{(a)(b)}(\vec \sigma ,{\vec \sigma}_1;\tau ),&&
{}^3{\tilde M}_{(g)}(\tau ,{\vec \sigma}_2)\rbrace ={{\partial}\over {\partial
\sigma_2^s}}\Big[
\zeta^{(\omega )r}_{(a)(e)}(\vec \sigma ,{\vec \sigma}_2;\tau )
\epsilon_{(e)(g)(f)} \zeta^{(\omega )s}_{(f)(b)}({\vec \sigma}_2,{\vec \sigma}
_1;\tau )\Big] +\nonumber \\
&+&\zeta^{(\omega )r}_{(a)(e)}(\vec \sigma ,{\vec \sigma}_2)\, {}^3\omega_{s(e)}
(\tau ,{\vec \sigma}_2)\, \zeta^{(\omega )s}_{(g)(b)}({\vec \sigma}_2,{\vec
\sigma}_1;\tau )-\nonumber \\
&-&\zeta^{(\omega )r}_{(a)(g)}(\vec \sigma ,{\vec \sigma}_2;\tau )
\, {}^3\omega_{s(f)}(\tau ,{\vec \sigma}_2)\, \zeta^{(\omega )s}_{(f)(b)}({\vec
\sigma}_2,{\vec \sigma}_1;\tau ),\nonumber \\
&&{}\nonumber \\
\lbrace \zeta^{(\omega )r}_{(a)(b)}(\vec \sigma ,{\vec \sigma}_1;\tau ),&&
{}^3{\tilde \Theta}_u(\tau ,{\vec \sigma}_2)\rbrace =\nonumber \\
&=&\int d^3\sigma_3 \zeta^{(\omega )r}
_{(a)(e)}(\vec \sigma ,{\vec \sigma}_3;\tau )\, \epsilon_{(e)(d)(f)}
\lbrace {}^3\omega_{s(d)}(\tau ,{\vec \sigma}_3), {}^3{\tilde \Theta}_u(\tau ,
{\vec \sigma}_2)\rbrace \zeta^{(\omega )s}_{(f)(b)}({\vec \sigma}_3,{\vec
\sigma}_1;\tau ).\nonumber \\
&&{}
\label{IV56}
\end{eqnarray}

Collecting all previous results, we obtain the following form for the
Dirac Hamiltonian  (\ref{II58}) of scenario b) on WSW hypersurfaces
with ${\tilde \lambda}_{AB}(\tau )=0$  \footnote{We have
$-n_{(a)}{\cal H}_{(a)}\approx n_{(a)}\,
 {}^3e^r_{(a)}\, {}^3{\tilde \Theta}_r\approx {\tilde n}^r {\tilde \pi}_r^{\vec \xi}$
 modulo ${}^3{\tilde M}_{(a)}\approx 0$, see Eqs.(\ref{II11}) and after
 Eq.(\ref{II58}).}

\begin{eqnarray}
{\hat H}_{(D)ADM}&=&\int d^3\sigma \Big[ n {\hat {\cal H}}+ n_{(a)}\,
{}^3e_{(a)}^r\, {}^3{\tilde \Theta}_r+\nonumber \\
 &+&\lambda_n\,
{\tilde \pi}^n+\lambda^{\vec n}_{(a)}{\tilde \pi}^{\vec n}_{(a)}
+\lambda_{(a)}^{\vec \varphi}{\tilde \pi}^{\vec \varphi}_{(a)}+{\hat
\mu}_{(a)}\, {}^3{\tilde M}_{(a)}\Big] (\tau ,\vec \sigma )-\nonumber \\
 &-& {\tilde \lambda}_{\tau}(\tau ) [\epsilon_{(\infty )}-{\hat P}^{\tau}_{ADM}]+
 {\tilde \lambda}_{\check r}(\tau ) {\hat P}^{\check r}_{ADM} =\nonumber \\
&=&\int d^3\sigma
\Big[ n {\hat {\cal H}}+ n_{(a)}\, {}^3e^r_{(a)} {{\partial
\xi^s}\over {\partial \sigma^r}}\, {\tilde \pi}_s^{\vec
\xi}+\lambda_n{\tilde \pi}^n
+\lambda^{\vec n}_{(a)}{\tilde \pi}^{\vec n}_{(a)}+\nonumber \\
 &+&\lambda_{(a)}^{\vec \varphi}{\tilde \pi}^{\vec
\varphi}_{(a)}-({\hat \mu}_{(b)} B_{(b)(a)}(\alpha_{(e)})+
n_{(b)}\, {}^3e^r_{(b)} {{\partial \alpha
_{(a)}}\over {\partial \sigma^r}}) {\tilde \pi}^{\vec \alpha}_{(a)}\Big] (\tau ,
\vec \sigma )-\nonumber \\
 &-& {\tilde \lambda}_{\tau}(\tau ) [\epsilon_{(\infty )}-{\hat P}^{\tau}_{ADM}]+
 {\tilde \lambda}_{\check r}(\tau ) {\hat P}^{\check r}_{ADM} =\nonumber \\
&=&\int d^3\sigma \Big[ n {\hat {\cal H}}+ n_{(a)}\, {}^3e^r_{(a)}
{{\partial \xi^s}\over {\partial \sigma^r}}\, {\tilde \pi}_s^{\vec
\xi}+
\nonumber \\
&+&\lambda_n{\tilde \pi}^n+\lambda^{\vec n}_{(a)}{\tilde \pi}^{\vec n}_{(a)}
+\lambda_{(a)}^{\vec \varphi}{\tilde \pi}^{\vec \varphi}_{(a)}+{\tilde \mu}
_{(a)} {\tilde \pi}^{\vec \alpha}_{(a)}\Big] (\tau ,\vec \sigma )-\nonumber \\
 &-& {\tilde \lambda}_{\tau}(\tau ) [\epsilon_{(\infty )}-{\hat P}^{\tau}_{ADM}]+
 {\tilde \lambda}_{\check r}(\tau ) {\hat P}^{\check r}_{ADM},
\label{IV57}
\end{eqnarray}

\noindent where ${\tilde \mu}_{(a)}$ are new Dirac multipliers.

The phase space action, which usually is incorrectly written without the primary
constraints, is

\begin{eqnarray}
\bar S&=&
\int d\tau d^3\sigma \Big[ {}^3{\tilde \pi}^r_{(a)} \partial_{\tau}\,
{}^3e_{(a)r}-n {\hat {\cal H}}+ n_{(a)}{\cal H}_{(a)}-\nonumber \\
&-&\lambda_n {\tilde \pi}^n-\lambda^{\vec n}_{(a)} {\tilde \pi}^{\vec n}_{(a)}-
\lambda^{\vec \varphi}_{(a)} {\tilde \pi}^{\vec \varphi}_{(a)}-\mu_{(a)}\,
{}^3{\tilde M}_{(a)}\Big] (\tau ,\vec \sigma )+\nonumber \\
 &+& {\tilde \lambda}_{\tau}(\tau ) [\epsilon_{(\infty )}-{\hat P}^{\tau}_{ADM}]-
 {\tilde \lambda}_{\check r}(\tau ) {\hat P}^{\check r}_{ADM} =\nonumber \\
 &=&\int d\tau d^3\sigma \Big[
{}^3{\tilde \pi}^r_{(a)}\partial_{\tau}\, {}^3e_{(a)r}- n {\hat {\cal
H}}- n_{(a)}\, {}^3e^r_{(a)}\, {}^3{\tilde
\Theta}_r-\nonumber \\
&-&\lambda_n {\tilde \pi}^n-\lambda^{\vec n}_{(a)} {\tilde \pi}^{\vec n}_{(a)}-
\lambda^{\vec \varphi}_{(a)} {\tilde \pi}^{\vec \varphi}_{(a)}-{\hat \mu}_{(a)}
\, {}^3{\tilde M}_{(a)}\Big] (\tau ,\vec \sigma )+\nonumber \\
 &+& {\tilde \lambda}_{\tau}(\tau ) [\epsilon_{(\infty )}-{\hat P}^{\tau}_{ADM}]-
 {\tilde \lambda}_{\check r}(\tau ) {\hat P}^{\check r}_{ADM} =\nonumber \\
 &=&\int d\tau d^3\sigma [{}^3{\tilde
\pi}^r_{(a)} \partial_{\tau}\, {}^3e
_{(a)r}-n {\hat {\cal H}}- n_{(a)}\, {}^3e^r_{(a)}
{{\partial \xi^s}\over {\partial \sigma^r}} {\tilde \pi}^{\vec \xi}_s-
\nonumber \\
&-&\lambda_n {\tilde \pi}^n-\lambda^{\vec n}_{(a)} {\tilde \pi}^{\vec n}_{(a)}-
\lambda^{\vec \varphi}_{(a)} {\tilde \pi}^{\vec \varphi}_{(a)}-{\tilde \mu}
_{(a)}\, {\tilde \pi}^{\vec \alpha}_{(a)}] (\tau ,\vec \sigma )+\nonumber \\
 &+& {\tilde \lambda}_{\tau}(\tau ) [\epsilon_{(\infty )}-{\hat P}^{\tau}_{ADM}]-
 {\tilde \lambda}_{\check r}(\tau ) {\hat P}^{\check r}_{ADM}.\nonumber \\
\label{IV58}
\end{eqnarray}

In conclusion the 18-dimensional phase space spanned by ${}^3e_{(a)r}$
and ${}^3{\tilde \pi}^r_{(a)}$ has a global (since
$\Sigma_{\tau}\approx R^3$) canonical basis, in which 12 variables are
$\alpha_{(a)}$, ${\tilde \pi}^{\vec \alpha}
_{(a)}\approx 0$, $\xi^r$, ${\tilde \pi}^{\vec \xi}_r\approx 0$. The remaining
6 variables, hidden in the reduced quantities ${}^3{\hat e}_{(a)r}$,
${}^3{\hat {\tilde \pi}}^r_{(a)}$, are 3 pairs of conjugate Dirac's
observables with respect to the gauge transformations in ${\bar {\cal
G}}_R$, namely they are invariant under Abelianized rotations and
space pseudo-diffeomorphisms \footnote{And, therefore, weakly
invariant under the original rotations and space
pseudo-diffeomorphisms.} connected with the identity and obtainable as
a succession of infinitesimal gauge transformations. However, since
space pseudo-diffeomorphisms connect different charts in the atlas of
$\Sigma_{\tau}$ and since $\xi^r(\tau ,\vec
\sigma )=\sigma^r$ means to choose as origin of space
pseudo-diffeomorphisms an arbitrary chart, the {\it functional form of
the Dirac's observables will depend on the chart chosen as origin}.
This will reflect itself in the {\it freedom of how to parametrize}
the reduced cotriad ${}^3{\hat e}_{(a)r}(\tau ,\vec \sigma )$ in terms
of only 3 independent functions: in each chart `c' they will be
denoted $Q^{(c)}_r(\tau ,\vec \sigma )$ and, if `c+dc' is a new chart
connected to `c' by an infinitesimal space pseudo-diffeomorphism of
parameters $\vec \xi (\tau ,\vec \sigma )$, then we will have
$Q^{(c+dc)}_r(\tau ,\vec \sigma )={{\partial \xi^s(\tau ,\vec
\sigma )}\over  {\partial \sigma^r}} Q^{(c)}_s(\tau ,\vec \xi (\tau ,\vec
\sigma ))$.

The invariants under pseudo-diffeomorphisms of a Riemannian 3-manifold
$(\Sigma_{\tau},{}^3g)$ (no explicit basis is known for them), can be
expressed in every chart `c' as functionals of the 3 independent
functions $Q^{(c)}_r(\tau ,\vec \sigma )$. Therefore, these 3
functions give a local coordinatization of the {\it space of
3-geometries} ({\it superspace or moduli space}) $Riem\,
\Sigma_{\tau}/Diff\, \Sigma_{\tau}$ \cite{witt,fis}.

By using Eqs.(\ref{IV45}) and (\ref{IV48}) in the Hamiltonian
expressions of the 4-tensors, we could get the most important
4-tensors on the pseudo-Riemannian 4-manifold $(M^4,\, {}^4g)$
expressed in terms of ${\tilde \lambda}_A$, $n$, ${\tilde
\pi}^n\approx 0$, $n_{(a)}$, ${\tilde \pi}^{\vec n}_{(a)}\approx 0$,
$\alpha_{(a)}$, ${\tilde \pi}^{\vec \alpha}_{(a)}\approx 0$, $\xi^r$,
${\tilde \pi}^{\vec \xi}_r\approx 0$, and of the (non canonically
conjugate) Dirac's observables with respect to the action of ${\bar
{\cal G}}_R$ , i.e. ${}^3{\hat e}_{(a)r}$, ${}^3{\hat {\tilde
\pi}}^r_{(a)}$ in the rest-frame instant form of tetrad gravity.
If we could extract from ${}^3{\hat e}_{(a)r}$,
${}^3{\hat {\tilde \pi}}^r_{(a)}$, the Dirac observables with respect
to the gauge transformations generated by the superhamiltonian
constraint ${\hat {\cal H}}(\tau ,\vec \sigma )\approx 0$, then we
could express all 4-tensors in terms of these final Dirac observables
(the independent Cauchy data of tetrad gravity), of the gauge
variables $n$, $n_{(a)}$, $\alpha_{(a)}$, $\xi^r$ and of the gauge
variable associated with ${\hat {\cal H}}(\tau ,\vec \sigma )\approx
0$, when all the constraints are satisfied. Therefore, we would get
not only a chart-dependent expression of the 4-metrics ${}^4g\in
Riem\, M^4$, but also of the {\it 4-geometries} in $Riem\, M^4/Diff\,
M^4$.

In the next Section we shall study the simplest charts of the atlas of
$\Sigma_{\tau}$, namely the 3-orthogonal ones.

\vfill\eject

\section{Quasi-Shanmugadhasan Canonical Transformation to the
         3-Orthogonal Gauges.}

In this Section we shall identify a Shanmugadhasan canonical
transformation to a canonical basis adapted to 13 of the 14 first
class constraints \footnote{In paper II there was a preliminary
attempt to find this canonical transformation, which was
subsequently realized to be wrong.}. This canonical basis will
then be specialized to 3-orthogonal coordinates for
$\Sigma_{\tau}$.

The quasi-Shanmugadhasan canonical transformation \cite{sha}
\footnote{{\it Quasi}- because we are not including the superhamiltonian
constraint ${\hat {\cal H}}(\tau ,\vec \sigma )
\approx 0$.} Abelianizing the rotation and pseudo-diffeomorphism constraints
${}^3{\tilde M}_{(a)}(\tau ,\vec \sigma )\approx 0$, ${}^3{\tilde \Theta}_r
(\tau ,\vec \sigma )\approx 0$, will send the canonical basis ${}^3e_{(a)r}
(\tau ,\vec \sigma )$, ${}^3{\tilde \pi}^r_{(a)}(\tau ,\vec \sigma )$, of
$T^{*}{\cal C}_e$ in a new basis whose conjugate pairs are $\Big( \alpha_{(a)}
(\tau ,\vec \sigma ), {\tilde \pi}^{\vec \alpha}_{(a)}(\tau ,\vec \sigma )
\approx 0\Big)$, $\Big( \xi^r(\tau ,\vec \sigma ),{\tilde \pi}^{\vec \xi}
_r(\tau ,\vec \sigma )\approx 0\Big)$ for the gauge sector and $\Big( Q_r(\tau
,\vec \sigma ),
{\tilde \Pi}^r(\tau ,\vec \sigma )\Big)$ for the sector of Dirac observables.

Therefore, we must parametrize the Dirac observables ${}^3{\hat e}_{(a)r}(\tau ,
\vec \sigma )$ in terms of three functions $Q_r(\tau ,\vec \sigma )$,
${}^3{\hat e}_{(a)r}(\tau ,\vec \sigma )={}^3{\hat e}_{(a)r}[\tau
,\vec \sigma ,Q_s(\tau ,\vec \sigma )]$, and then find how the Dirac
observables ${}^3{\hat {\tilde \pi}}^r_{(a)}(\tau ,\vec \sigma )$ are
expressible in terms of $Q_r(\tau ,\vec \sigma )$, ${\tilde
\Pi}^r(\tau ,\vec \sigma )$, ${\tilde \pi}^{\vec
\xi}_r(\tau ,\vec \sigma )$, ${\tilde \pi}^{\vec \alpha}_{(a)}(\tau ,
\vec \sigma )$ \footnote{They cannot depend on $\alpha_{(a)}(\tau ,\vec \sigma )$,
$\xi_r(\tau ,\vec \sigma )$, because they are Dirac observables.}.
Since from Eqs.(\ref{IV45}) we get

\begin{eqnarray}
{}^3g_{rs}(\tau ,\vec \sigma )&=&{}^3e_{(a)r}(\tau ,\vec \sigma )\, {}^3e_{(a)s}
(\tau ,\vec \sigma )={}^3{\bar e}_{(a)r}(\tau ,\vec \sigma )\, {}^3{\bar e}
_{(a)s}(\tau ,\vec \sigma )=\nonumber \\
&=&{{\partial \xi^u(\tau ,\vec \sigma )}\over {\partial \sigma^r}}{{\partial
\xi^v(\tau ,\vec \sigma )}\over {\partial \sigma^s}}\, {}^3{\hat e}_{(a)u}
[\tau ,\vec \xi (\tau ,\vec \sigma ), Q_w(\tau ,\vec \xi (\tau ,\vec
\sigma ))]\nonumber \\
 && {}^3{\hat e}_{(a)v}[\tau ,\vec \xi (\tau ,\vec \sigma ), Q_w(\tau ,
\vec \xi (\tau ,\vec \sigma ))]=\nonumber \\
&=&{{\partial \xi^u(\tau ,\vec \sigma )}\over {\partial \sigma^r}}{{\partial
\xi^v(\tau ,\vec \sigma )}\over {\partial \sigma^s}}\, {}^3{\hat g}_{uv}
[\tau ,\vec \xi (\tau ,\vec \sigma ), Q_w(\tau ,\vec \xi (\tau ,\vec
\sigma ))],
\label{V1}
\end{eqnarray}

\noindent the new metric ${}^3{\hat g}_{uv}(\tau ,\vec \xi )$ must depend only
on the functions $Q_w(\tau ,\vec \xi )$. This shows that the
parametrization of ${}^3{\hat e}_{(a)r}(\tau ,\vec \sigma )$ will
depend on the chosen system of coordinates, which will be {\it
declared} the origin $\vec \xi (\tau ,\vec \sigma )=\vec \sigma $ of
pseudo-diffeomorphisms from the given chart. Therefore, each Dirac
observable 3-metric ${}^3{\hat g}_{uv}$ is an element of {\it De Witt
superspace} \cite{dew} for Riemannian 3-manifolds: it defines a {\it
3-geometry} on $\Sigma_{\tau}$.

The simplest global systems of coordinates on
$\Sigma_{\tau}\approx R^3$, where to learn how to construct the
quasi-Shanmugadhasan canonical transformation, are the {\it
3-orthogonal} ones, in which ${}^3{\hat g}_{uv}$ is {\it
diagonal}. In it we have the parametrization

\begin{eqnarray}
{}^3{\hat e}_{(a)r}(\tau ,\vec \sigma )&=&
\delta_{(a)r} Q_r(\tau ,\vec \sigma ),\nonumber \\
 \Rightarrow&& {}^3{\hat e}^r_{(a)}(\tau ,\vec \sigma )= {{\delta^r_{(a)}}
\over {Q_r(\tau ,\vec \sigma )}}, \nonumber \\
\Rightarrow && {}^3{\hat g}_{rs}(\tau ,\vec \sigma )=\delta_{rs} Q^2_r(\tau ,
\vec \sigma ),\nonumber \\
ds^2&=&\epsilon \Big( [N_{(as)}+n]^2-
[N_{(as) r}+n_r] \sum_u{{\partial \sigma
^r(\vec \xi )}\over {\partial \xi^u}}{1\over {Q^2_u(\vec \xi )}}{{\partial
\sigma^s(\vec \xi )}\over {\partial \xi^u}}[N_{(as) s}+n_s] \Big) (d\tau )^2-
\nonumber \\
&-&2\epsilon [N_{(as) r}+n_r]d\tau d\sigma^r-\epsilon
\sum_u {{\partial \xi^u}\over
{\partial \sigma^r}} Q^2_u(\vec \xi ){{\partial \xi^u}\over {\partial
\sigma^s}} d\sigma^r d\sigma^s=\nonumber \\
&=&\epsilon \Big( [N_{(as)}+n]^2(d\tau )^2 -\delta_{uv}[Q_u{{\partial \xi^u}
\over {\partial \sigma^r}}d\sigma^r+{1\over {Q_u}}{{\partial \sigma^r(\vec
\xi )}\over {\partial \xi^u}}(N_{(as)r}+n_r)d\tau ]\nonumber \\
&&[Q_v{{\partial \xi^v}
\over {\partial \sigma^s}}d\sigma^s+{1\over {Q_v}}{{\partial \sigma^s(\vec
\xi )}\over {\partial \xi^v}}(N_{(as)s}+n_s)d\tau ],
\label{V2}
\end{eqnarray}

\noindent with $Q_r(\tau ,\vec \sigma )=1+h_r(\tau ,\vec \sigma ) > 0$ to avoid
singularities. The 3 functions $Q^2_r(\tau ,\vec \sigma )$ give a {\it
local parametrization of superspace}; the presence of singularities in
superspace depends on the boundary conditions for $Q_r(\tau ,\vec
\sigma )$, i.e. on the possible existence of stability subgroups
(isometries) of the group ${\bar {\cal G}}$ of gauge transformations,
which we assume to be absent if a suitable weighted Sobolev space is
chosen for cotriads.

The choice of the parametrization of ${}^3{\hat e}_{(a)r}$ is
equivalent to the {\it coordinate conditions} of
Refs.\cite{dirr,isha} \footnote{The 3-orthogonal gauges could be
implemented with three gauge-fixing constraints of the type
${}^3g_{rs}(\tau ,\vec \sigma )={}^3e_{(a)r}(\tau ,\vec \sigma )\,
{}^3e_{(a)s}(\tau ,\vec \sigma ) \approx 0$ for $r\not= s$. They
would replace our Shanmugadhasan-oriented constraints $\xi^r(\tau
,\vec \sigma )-\vec \sigma \approx 0$, but the information about
the parametrization would remain implicit.} \footnote{For instance
one would like to have a parametrization of the cotriads
${}^3{\hat e}_{(a)r}$ corresponding to {\it 3-normal} coordinates
around the point $\lbrace \tau ,\vec \sigma =0 \rbrace \in
\Sigma_{\tau}$. Cartan\cite{cart,spiv} showed that, given Riemann
normal coordinates $y^{\mu}$ at $p\in M^4$ [$y^{\mu}{|}_p=0$], one
can choose adapted orthonormal frames and coframes
${}^4E^{(N)}_{(\alpha )}={}^4E^{(N)\mu}_{(\alpha )}(y) \partial
/\partial y^{\mu}$, ${}^4\theta^{(N)(\alpha )}={}^4E^{(N)(\alpha
)}_{\mu}(y) dy^{\mu}$, obtained from ${}^4E^{(N)}_{(\alpha
)}{|}_p=\delta^{\mu}_{(\alpha )}
\partial /\partial y^{\mu}$, ${}^4\theta^{(N)(\alpha )}{|}_p=\delta^{(\alpha )}
_{\mu} dy^{\mu}$, by parallel transport along the geodesic arcs originating at
p. Then one has the following properties

\begin{eqnarray*}
{}^4E^{(N)(\alpha )}_{\mu}(y)\, y^{\mu} &=& \delta^{(\alpha )}_{\mu} y^{\mu}
\nonumber \\
{}^4\theta^{(N)(\alpha )}&=& \delta^{(\alpha )}_{\mu} [dy^{\mu}+
y^{\rho}y^{\sigma}\, N^{\mu}{}_{\rho\sigma\lambda}(y)\,
dy^{\lambda}],\nonumber \\
&&N_{\mu\rho\sigma\lambda}=-N_{\rho\mu\sigma\lambda}=-N_{\mu\rho\lambda\sigma}.
\nonumber \\
\end{eqnarray*}

These 4-coordinates are important for the {\it free fall observer}
description of phenomena, but are difficult to identify in the
canonical approach, which privileges 3-coordinates and rebuilds
4-coordinates only a posteriori after a complete fixation of the
Hamiltonian gauge and restriction to the solutions of Einstein
equations.

Since normal coordinates are the most natural from a differential
geometric point of view, let us look for a parametrization, in
such a system of 3-coordinates, of the Dirac observables
${}^3{\hat e}_{(a)r}(\tau ,\vec \sigma )$ on $\Sigma_{\tau}$ in
terms of 3 real functions ${\hat Q}_r(\tau ,\vec \sigma )$, whose
conjugate momenta will be denoted ${\hat {\tilde \Pi}}^r(\tau
,\vec \sigma )$.  The previous equation gives the Cartan
definition of orthonormal tetrads adapted to normal coordinates
for Lorentzian 4-manifolds. This suggests that for Riemannian
3-manifolds like $\Sigma_{\tau}$, to get 3-normal coordinates
geodesic at the point $\vec \sigma =0$ we have to parametrize the
reduced cotriads ${}^3{\hat e}_{(a)r}(\tau ,\vec \sigma )$  as
follows

\begin{eqnarray*}
{}^3{\hat e}_{(a)r}(\tau ,\vec \sigma )&=&\delta^s_{(a)}[\delta_{rs}+\sum_n
\epsilon_{run}\epsilon_{svn} \sigma^u\sigma^v {\hat Q}_n(\tau ,\vec \sigma )]
\nonumber \\
&\Rightarrow& {}^3{\hat e}_{(a)r}(\tau ,\vec \sigma )\, \sigma^r=\delta
_{(a)r}\, \sigma^r,\nonumber \\
\end{eqnarray*}

\noindent with $N_{surv}(\tau ,\vec \sigma )=\sum_n \epsilon_{sun}\epsilon
_{rvn}{\hat Q}_n(\tau ,\vec \sigma )=-N_{usrv}(\tau ,\vec \sigma )=-
N_{suvr}(\tau ,\vec \sigma )=N_{rvsu}(\tau ,\vec \sigma )$. Then one gets

\begin{eqnarray*}
{}^3{\hat g}_{rs}(\tau ,\vec \sigma )&=&{}^3{\hat e}_{(a)r}(\tau ,\vec \sigma )
\, {}^3{\hat e}_{(a)s}(\tau ,\vec \sigma )=\delta_{rs}+\nonumber \\
&+&\sigma^u\sigma^v [\sum_n\epsilon_{run}\epsilon_{svn}(2+{\vec \sigma}^2\,
{\hat Q}_n(\tau ,\vec \sigma )){\hat Q}_n(\tau ,\vec \sigma )-\sum_{nm}
\epsilon_{run}\epsilon_{svm}\sigma^n\sigma^m{\hat Q}_n(\tau ,\vec \sigma )
{\hat Q}_m(\tau ,\vec \sigma )].\nonumber \\
 &&{}\nonumber \\
\end{eqnarray*}
.}.

Eqs.(\ref{IV45}), rewritten in the form

\beq
{}^3e_{(a)r}(\tau ,\vec \sigma ) = {}^3R_{(a)(b)}(\alpha_{(e)}(\tau
,\vec \sigma )) {{\partial \xi^s(\tau ,\vec \sigma )}\over {\partial
\sigma^r}}\, {}^3{\hat e}_{(b)s}(\tau ,\vec \xi (\tau ,\vec \sigma ), Q_u(\tau
,\vec \xi (\tau ,\vec \sigma )],
\label{V3}
\eeq

\noindent define a {\it point} canonical trasformation, in which
the configuration variables transform in the following way

\beq
\begin{minipage}[t]{3cm}
\begin{tabular}{|l|} \hline
${}^3e_{(a)r}$ \\ \hline
\end{tabular}
\end{minipage} \ {\longrightarrow \hspace{.2cm}} \
\begin{minipage}[t]{4 cm}
\begin{tabular}{|l|l|l|} \hline
$\alpha_{(a)}$ & $\xi^r$ & $Q_r$\\ \hline
\end{tabular}
\end{minipage}
\label{V4}
\eeq

We have now to find the second half of this canonical transformation,
namely

\beq
\begin{minipage}[t]{2cm}
\begin{tabular}{|l|} \hline
 ${}^3{\tilde \pi}^r_{(a)}$ \\ \hline
\end{tabular}
\end{minipage} \ {\longrightarrow \hspace{.2cm}} \
\begin{minipage}[t]{2cm}
\begin{tabular}{|l|l|l|} \hline
 ${\tilde \pi}^{\vec \alpha}_{(a)}\, (\approx 0)$   & ${\tilde \pi}^{\vec \xi}_r
 \, (\approx 0)$ & ${\tilde \Pi}^r$  \\ \hline
\end{tabular}
\end{minipage}
\label{V5}
\eeq

For the cotriad canonical momentum we have [see Eq.(\ref{IV48})]

\bea
{}^3{\tilde \pi}^r_{(a)}(\tau ,\vec \sigma ) &=&
{}^3R_{(a)(b)}(\alpha_{(e)}(\tau ,\vec \sigma ))\, |{{\partial \xi
(\tau ,\vec \sigma )}\over {\partial \sigma }}|\, {{\partial
\sigma^r(\vec \xi )}\over {\partial \xi^s}}{|}_{\vec \xi =
\vec \xi (\tau ,\vec \sigma ))}\, {}^3{\hat {\tilde \pi}}^s_{(b)}(\tau
,\vec \xi (\tau ,\vec \sigma ))=\nonumber \\
 &=&{}^3{\tilde \pi}^r_{(a)}(\tau ,\vec \sigma |\alpha_{(e)},
 \xi^s, Q_s, {\tilde \pi}^{\vec \alpha}_{(e)}, {\tilde \pi}
 ^{\vec \xi}_s, {\tilde \Pi}^s] \equiv \nonumber \\
 &\equiv& {}^3{\check {\tilde \pi}}^r_{(a)}(\tau ,\vec \sigma |
 \alpha_{(e)}, \xi^s, Q_s, {\tilde \Pi}^s] +\nonumber \\
 &+& \int d^3\sigma_1\, {\tilde F}^r_{(a)(b)}(\vec \sigma ,{\vec \sigma}_1;\tau |
 \alpha_{(e)}, \xi^s, Q_s, {\tilde \Pi}^s]\, {\tilde \pi}^{\vec
 \alpha}_{(b)}(\tau ,{\vec \sigma}_1)+\nonumber \\
 &+& \sum_u \int d^3\sigma_1\, {\tilde G}^{ru}_{(a)}(\vec \sigma ,{\vec \sigma}_1;\tau |
 \alpha_{(e)}, \xi^s, Q_s, {\tilde \Pi}^s]\, {\tilde \pi}^{\vec
 \xi}_u(\tau ,{\vec \sigma}_1),
\label{V6}
\eea

\noindent with

\bea
{}^3{\check {\tilde \pi}}^r_{(a)}(\tau ,\vec \sigma ) &=& {}^3{\tilde
\pi}^r_{(a)}(\tau ,\vec \sigma ) {|}_{{\tilde \pi}^{\vec \alpha}_{(e)}=
{\tilde \pi}^{\vec \xi}_u=0},\nonumber \\
 {\tilde F}^r_{(a)(b)}(\vec \sigma ,{\vec \sigma}_1;\tau |
 \alpha_{(e)}, \xi^s, Q_s, {\tilde \Pi}^s] &=& {{\delta\, {}^3{\tilde \pi}
 ^r_{(a)}(\tau ,\vec \sigma )}\over {\delta {\tilde \pi}^{\vec \alpha}_{(b)}(\tau
 ,{\vec \sigma}_1)}} {|}_{{\tilde \pi}^{\vec \alpha}_{(e)}=
{\tilde \pi}^{\vec \xi}_u=0},\nonumber \\
 {\tilde G}^{ru}_{(a)}(\vec \sigma ,{\vec \sigma}_1;\tau |
 \alpha_{(e)}, \xi^s, Q_s, {\tilde \Pi}^s] &=& {{\delta\, {}^3{\tilde \pi}
 ^r_{(a)}(\tau ,\vec \sigma )}\over {\delta {\tilde \pi}^{\vec \xi}_u(\tau
 ,{\vec \sigma}_1)}} {|}_{{\tilde \pi}^{\vec \alpha}_{(e)}=
{\tilde \pi}^{\vec \xi}_u=0}.
\label{V7}
\eea

The last equality in Eq.(\ref{V6}) is a priori a strong equality
in the sense of Dirac: powers of the constraints are ineffective
near the constraint hypersurface. However, since the canonical
transformation is a {\it point} one, the old momenta depend {\it
linearly} upon the new ones, so that the strong equality sign may
be replaced with an ordinary equality sign. Indeed, if in a phase
space with canonical Darboux basis $q^i$, $p_i$, we make a {\it
point canonical transformation} $q^i, p_i \mapsto Q^i, P_i$ with
$q^i=q^i(Q)$ [whose inverse is $Q^i=Q^i(q)$] and $p_i=p_i(Q,P)$,
then the canonicity conditions $\delta^i_j= \{ q^i,p_j \} {}_{qp}
= \{ q^i(Q), p_j(Q,P) \} {}_{QP}= \sum_k {{\partial q^i(Q)}\over
{\partial Q^k}} {{\partial p_j(Q,P)}\over {\partial P_k}}$ imply
$p_j(Q,P)= \sum_k {{\partial Q^k(q)}\over {\partial
q^j}}{|}_{q=q(Q)}\,  P_k + F_j(Q)$; moreover, from $0=\{ p_i,p_j
\}_{q,p} = \{ p_i(Q,P),p_j(Q,P) \}_{Q,P}$ we get
$F_i(Q)={{\partial F(Q(q))}\over {\partial q^i}}{|}_{q=q(Q)}$. In
what follows we shall put $F(Q)=0$, since this corresponds to a so
called {\it trivial phase canonical transformation}.

Let us see what we can say about the dependence of the momenta
${}^3{\tilde \pi}^r_{(a)}(\tau ,\vec \sigma )$ upon the Abelianized
constraints ${\tilde \pi}^{\vec \alpha}_{(a)}(\tau ,\vec \sigma )
\approx 0$, ${\tilde \pi}^{\vec \xi}_r(\tau ,\vec \sigma ) \approx 0$.

Since the rotation constraints ${}^3{\tilde
M}_{(a)}=\epsilon_{(a)(b)(c)}\, {}^3e_{(b)r}\, {}^3{\tilde
\pi}^r_{(c)}={1\over 2}\epsilon_{(a)(b)(c)}\, {}^3{\tilde M}_{(b)(c)}$
may be written as

\bea
&&{}^3{\tilde M}_{(a)(b)}={}^3e
_{(a)r}\, {}^3{\tilde \pi}^r_{(b)}-{}^3e_{(b)r}\, {}^3{\tilde \pi}^r_{(a)}=
\epsilon_{(a)(b)(c)}\, {}^3{\tilde M}_{(c)}=-\epsilon_{(a)(b)(c)}\, {\tilde
\pi}^{\vec \alpha}_{(d)} B_{(d)(c)}(\alpha_{(e)})\approx 0,\nonumber \\
 &&
\label{V8}
\eea

\noindent due to Eqs.(\ref{IV31}), we may extract the following dependence
of ${}^3{\tilde \pi}^r_{(a)}(\tau ,\vec
\sigma )$ on ${\tilde \pi}^{\vec \alpha}_{(a)}(\tau ,\vec \sigma )$

\begin{eqnarray}
{}^3{\tilde \pi}^r_{(a)}&=&{}^3e^r_{(b)}\, {}^3e_{(b)s}\,
{}^3{\tilde \pi}^s _{(a)}=\nonumber \\
 &=&{1\over 2}\,
{}^3e^r_{(b)} \Big( {}^3e_{(b)s}\, {}^3{\tilde \pi}^s_{(a)}+{}^3e
_{(a)s}\, {}^3{\tilde \pi}^s_{(b)}\Big) +{1\over 2}\,
{}^3e^r_{(b)}\Big( {}^3e_{(b)s}\, {}^3{\tilde
\pi}^s_{(a)}-{}^3e_{(a)s}\, {}^3{\tilde \pi}^s_{(b)}\Big)
=\nonumber \\
 &=&{1\over 2}\, {}^3e^r_{(b)} \Big( {}^3e_{(b)s}\,
{}^3{\tilde \pi}^s_{(a)}+{}^3e _{(a)s}\, {}^3{\tilde
\pi}^s_{(b)}\Big) -\nonumber \\
 &-&{1\over 2}\, {}^3e^r_{(b)}\,
\epsilon_{(a)(b)(c)} {\tilde \pi} ^{\vec \alpha}_{(d)}
B_{(d)(c)}(\alpha_{(e)})\, {\buildrel {def} \over =}, \nonumber \\
 &{\buildrel {def} \over =}& {}^3{\tilde \pi}^{(M)r}_{(a)} -
{1\over 2}\, {}^3e^r_{(b)} \epsilon _{(a)(b)(c)} {\tilde
\pi}^{\vec \alpha}_{(d)} B_{(d)(c)}(\alpha_{(e)}), \nonumber \\
&&{}\nonumber \\
 && \text{with}  \nonumber \\
 &&{}\nonumber \\
 {}^3{\tilde \pi}^{(M)}_{(a)} &=& {1\over 2}\, Z_{(a)(b)} \, {}^3e^r_{(b)}=
 \sum_s P^{(M)r}_{(a)(b)s}\, {}^3{\tilde \pi}^s_{(b)},\nonumber \\
 &&{}\nonumber \\
Z_{(a)(b)}&=&Z_{(b)(a)}={}^3e_{(a)s}\, {}^3{\tilde
\pi}^s_{(b)}+{}^3e_{(b)s}\, {}^3{\tilde
\pi}^s_{(a)},\nonumber \\
P^{(M)r}_{(a)(b)s} &=& {1\over 2}\, {}^3e^r_{(c)}\Big(
{}^3e_{(a)s}\delta_{(c)(b)}+ {}^3e_{(c)s}\delta_{(a)(b)}\Big),\quad
P^{(M)r}_{(a)(c)u}\, P^{(M)u}_{(c)(b)s} = P^{(M)r}_{(a)(b)s}.
\label{V9}
\end{eqnarray}

Let us note that, due to the  projector $P^{(M)r}_{(a)(b)s}(\tau ,\vec
\sigma )$, ${}^3{\tilde \pi}^{(M)r}_{(a)}(\tau ,\vec
\sigma )$ is a solution of the rotation constraints ${}^3{\tilde
M}_{(a)}(\tau ,\vec \sigma )\approx 0$.  However, it is not known the
dependence of ${}^3{\tilde \pi}^{(M)r}_{(a)}(\tau ,\vec \sigma )$ upon
${\tilde \pi}^{\vec \xi}_r(\tau ,\vec \sigma )$.

To extract the dependence of ${}^3{\tilde \pi}^r_{(a)}(\tau ,\vec
\sigma )$ on ${\tilde \pi}^{\vec \xi}_r(\tau ,\vec \sigma )$, let us
recall Eqs.(\ref{II11}), (\ref{IV31}), (\ref{IV39})  and (\ref{IV26})

\begin{eqnarray}
{\hat {\cal H}}_{(a)}(\tau ,\vec \sigma )&=&{\hat D}^{(\omega )}_{(a)(b)r}
(\tau ,\vec \sigma )\, {}^3{\tilde \pi}^r_{(b)}(\tau ,\vec \sigma )=\nonumber \\
&=&-{}^3e^r_{(a)}(\tau ,\vec \sigma )
\Big[ {}^3{\tilde \Theta}_r+{}^3\omega_{r(b)}\,
{}^3{\tilde M}_{(b)}\Big] (\tau ,\vec \sigma )=\nonumber \\
&=&-{}^3e^r_{(a)}(\tau ,\vec \sigma )
\Big[ {{\partial \xi^s}\over {\partial \sigma^r}}
{\tilde \pi}^{\vec \xi}_s-\Big( B_{(b)(c)}(\alpha_{(e)})\,
{}^3\omega_{r(c)}- {{\partial \alpha_{(b)}}\over {\partial
\sigma^r}}\Big) \, {\tilde \pi}^{\vec
\alpha}_{(b)}\Big] (\tau ,\vec \sigma )\approx 0,\nonumber \\
 &&{}\nonumber \\
{}^3{\tilde \pi}^r_{(a)}(\tau ,\vec \sigma )&=&{}^3{\tilde
\pi}^{(T)r}_{(a)} (\tau ,\vec \sigma )-\int d^3\sigma_1\,
\zeta^{(\omega )r}_{(a)(b)}(\vec \sigma ,{\vec \sigma }_1;\tau )\,
{\hat {\cal H}}_{(b)}(\tau ,{\vec \sigma }_1) =\nonumber \\
 &=& {}^3{\tilde \pi}^{(T)r}_{(a)}(\tau ,\vec \sigma ) + \int d^3\sigma_1
 \zeta^{(\omega ) r}_{(a)(b)}(\vec \sigma ,{\vec \sigma}_1;\tau ) \nonumber \\
 &&{}^3e^s_{(b)}(\tau ,{\vec \sigma}_1) \Big[
{{\partial \xi^u}\over {\partial \sigma^s_1}} {\tilde \pi}^{\vec
\xi}_u-\Big( B_{(d)(c)}(\alpha_{(e)})\, {}^3\omega_{s(c)}- {{\partial
\alpha_{(d)}}\over {\partial \sigma^s_1}}\Big) \, {\tilde \pi}^{\vec
\alpha}_{(d)}\Big] (\tau ,{\vec \sigma}_1),\nonumber \\
 &&{}\nonumber \\
 &&{}\nonumber \\
 {}^3{\tilde \pi}^{(T)r}_{(a)} (\tau ,\vec \sigma ) &=& \int d^3\sigma_1\Big[
 \delta^r_s\delta_{(a)(b)}\delta^3(\vec \sigma ,{\vec \sigma}_1) +\zeta^{(\omega ) r}
 _{(a)(c)}(\vec \sigma ,{\vec \sigma}_1,\tau ) {\hat D}^{(\omega )}_{(c)(b)s}(\tau
 ,{\vec \sigma}_1)\Big] \, {}^3{\tilde \pi}^s_{(b)}(\tau ,{\vec \sigma}_1)=\nonumber \\
  &{\buildrel {def} \over =}& \int d^3\sigma_1 P^{(T)r}_{(a)(b) s}(\vec \sigma
  ,{\vec \sigma}_1,\tau )\, {}^3{\tilde \pi}^s_{(b)}(\tau ,{\vec \sigma}_1),\nonumber \\
   &&{}\nonumber \\
   &&{\hat D}^{(\omega )}_{(a)(b)r}(\tau ,\vec \sigma )\, {}^3{\tilde \pi}^{(T)r}_{(b)}(\tau
   ,\vec \sigma ) = 0.
\label{V10}
\end{eqnarray}

We have introduced the projector $P^{(T)r}_{(a)(b)s}(\vec \sigma
,{\vec \sigma}_1;\tau)$, satisfying\hfill\break $\int d^3\sigma_1
P^{(T)r}_{(a)(b)u}(\vec \sigma ,{\vec \sigma}_1;\tau )
P^{(T)u}_{(b)(c)s}({\vec \sigma}_1,{\vec \sigma}_2;\tau )=
P^{(T)r}_{(a)(c)s}(\vec \sigma ,{\vec \sigma}_2;\tau )$. In this
second presentation we have privileged the solution ${}^3{\tilde
\pi}^{(T)r}_{(a)}(\tau ,\vec \sigma )$ of the constraints
${\hat {\cal H}}_{(a)}(\tau ,\vec \sigma ) \approx 0$. However, it is
not known how ${}^3{\tilde \pi}^{(T)r}_{(a)}(\tau ,\vec \sigma )$
depends upon ${\tilde \pi}^{\vec \alpha}_{(a)}(\tau ,\vec \sigma )$.

Eqs.(\ref{V9}) and (\ref{V10}) show: i) that the dependence of the old
momenta upon the Abelianized constraints is linear; ii) but also that
the non-commutativity of the two projectors $P^{(M)r}_{(a)(b)s}(\tau
,\vec \sigma )$, $P^{(T)r}_{(a)(b)s}(\tau ,\vec \sigma )$ is an
obstruction to the determination of the kernels $F^r_{(a)(b)}$,
$G^{rs}_{(a)}$ of Eqs.(\ref{V6}) starting from these equations.

Therefore, let us come back to Eqs.(\ref{V7}) and let us use the point
nature of the canonical transformation

\bea
&&\begin{minipage}[t]{3cm}
\begin{tabular}{|l|} \hline
${}^3e_{(a)r}$ \\ \hline ${}^3{\tilde \pi}^r_{(a)}$ \\ \hline
\end{tabular}
\end{minipage} \ {\longrightarrow \hspace{.2cm}} \
\begin{minipage}[t]{4 cm}
\begin{tabular}{|l|l|l|} \hline
$\alpha_{(a)}$ & $\xi^r$ & $Q_r$\\ \hline
 ${\tilde \pi}^{\vec \alpha}_{(a)}$   & ${\tilde \pi}^{\vec \xi}_r$ &
 ${\tilde \Pi}^r$  \\ \hline
\end{tabular}
\end{minipage} ,\nonumber \\
 &&{}\nonumber \\
 &&\lbrace \alpha_{(a)}(\tau ,\vec
\sigma ),{\tilde \pi}^{\vec \alpha}_{(b)}(\tau ,{\vec \sigma }^{'})\rbrace
=\delta_{(a)(b)}\delta^3(\vec \sigma ,{\vec \sigma }^{'}), \nonumber \\
 &&\lbrace \xi^r
(\tau ,\vec \sigma ),{\tilde \pi}^{\vec \xi}_s(\tau ,{\vec \sigma }^{'})\rbrace
=\lbrace Q_s(\tau ,\vec \sigma ),{\tilde \Pi}^r(\tau ,{\vec \sigma }^{'})
\rbrace =\delta^r_s\delta^3(\vec \sigma ,{\vec \sigma }^{'}),\nonumber \\
 &&{}\nonumber \\
 &&\lbrace {}^3e_{(a)r}(\tau ,\vec \sigma ),\alpha_{(b)}(\tau ,{\vec \sigma }
^{'})\rbrace =\lbrace {}^3e_{(a)r}(\tau ,\vec \sigma ),\xi_s(\tau ,{\vec
\sigma }^{'})\rbrace =\nonumber \\
 &&=\lbrace {}^3e_{(a)r}(\tau ,\vec \sigma ),Q_s(\tau ,{\vec
\sigma }^{'})\rbrace =0,\nonumber \\
 &&{}\nonumber \\
 \delta^s_r\delta_{(a)(b)}\delta^3(\vec \sigma ,{\vec \sigma }^{'})&=&\lbrace
{}^3e_{(a)r}(\tau ,\vec \sigma ),{}^3{\tilde \pi}^s_{(b)}(\tau ,{\vec \sigma }
^{'})\rbrace =\nonumber \\
&=&\int d^3\sigma_1\Big[
\lbrace {}^3e_{(a)r}(\tau ,\vec \sigma ),{\tilde \pi}^{\vec
\alpha}_{(c)}(\tau ,{\vec \sigma }_1)\rbrace \lbrace \alpha_{(c)}(\tau ,
{\vec \sigma }_1),{}^3{\tilde \pi}^s_{(b)}(\tau ,{\vec \sigma }^{'})\rbrace +
\nonumber \\
&+&\lbrace {}^3e_{(a)r}(\tau ,\vec \sigma ),{\tilde \pi}^{\vec \xi}_u(\tau ,
{\vec \sigma }_1)\rbrace \lbrace \xi^u(\tau ,{\vec \sigma }_1),{}^3{\tilde \pi}
^s_{(b)}(\tau ,{\vec \sigma }^{'})\rbrace +\nonumber \\
&+&\lbrace {}^3e_{(a)r}(\tau ,\vec \sigma ),{\tilde \Pi}^u(\tau ,{\vec \sigma }
_1)\rbrace \lbrace Q_u(\tau ,{\vec \sigma }_1),{}^3{\tilde \pi}^s_{(b)}(\tau ,
{\vec \sigma }^{'})\rbrace \Big] =\nonumber \\
&=&\int d^3\sigma_1\Big[
{{\tilde \delta {}^3e_{(a)r}(\tau ,\vec \sigma )}\over
{\delta \alpha_{(c)}(\tau ,{\vec \sigma }_1)}}{{\delta {}^3{\tilde \pi}^s_{(b)}
(\tau ,{\vec \sigma }^{'})}\over {\delta {\tilde \pi}^{\vec \alpha}_{(c)}(\tau ,
{\vec \sigma }_1)}}+{{\delta {}^3e_{(a)r}(\tau ,\vec \sigma )}\over {\delta
\xi^u(\tau ,{\vec \sigma }_1)}}{{\delta {}^3{\tilde \pi}^s_{(b)}(\tau ,
{\vec \sigma }^{'})}\over {\delta {\tilde \pi}^{\vec \xi}_u(\tau ,{\vec \sigma
}_1)}}+\nonumber \\
&+&{{\delta {}^3e_{(a)r}(\tau ,\vec \sigma )}\over {\delta Q_u(\tau ,{\vec
\sigma }_1)}}{{\delta {}^3{\tilde \pi}^s_{(b)}(\tau ,{\vec \sigma }^{'})}\over
{\delta {\tilde \Pi}^u(\tau ,{\vec \sigma }_1)}}\Big] .
 \label{V11}
\eea

The two equations defining the kernels may be rewritten in the
following form

\bea
 {\tilde F}^r_{(a)(b)}(\vec \sigma ,{\vec \sigma}_1;\tau |
 \alpha_{(e)}, \xi^s, Q_s, {\tilde \Pi}^s] &=& {{\delta\, {}^3{\tilde \pi}
 ^r_{(a)}(\tau ,\vec \sigma )}\over {\delta {\tilde \pi}^{\vec \alpha}_{(b)}(\tau
 ,{\vec \sigma}_1)}} {|}_{{\tilde \pi}^{\vec \alpha}_{(e)}=
{\tilde \pi}^{\vec \xi}_u=0} =\nonumber \\
 &=& \{ \alpha_{(b)}(\tau ,{\vec \sigma}_1),\, {}^3{\tilde \pi}^r_{(a)}(\tau ,\vec \sigma )
 \} {|}_{{\tilde \pi}^{\vec \alpha}_{(e)}=
{\tilde \pi}^{\vec \xi}_u=0} = {{\delta \alpha_{(b)}(\tau ,{\vec
\sigma}_1)}\over {\delta\, {}^3e_{(a)r}(\tau ,\vec \sigma )}},\nonumber \\
 &&{}\nonumber \\
 {\tilde G}^{ru}_{(a)}(\vec \sigma ,{\vec \sigma}_1;\tau |
 \alpha_{(e)}, \xi^s, Q_s, {\tilde \Pi}^s] &=& {{\delta\, {}^3{\tilde \pi}
 ^r_{(a)}(\tau ,\vec \sigma )}\over {\delta {\tilde \pi}^{\vec \xi}_u(\tau
 ,{\vec \sigma}_1)}} {|}_{{\tilde \pi}^{\vec \alpha}_{(e)}=
{\tilde \pi}^{\vec \xi}_u=0} =\nonumber \\
 &=& \{ \xi^u(\tau ,{\vec \sigma}_1),\, {}^3{\tilde \pi}^r_{(a)}(\tau ,\vec \sigma )
 \}{|}_{{\tilde \pi}^{\vec \alpha}_{(e)}= {\tilde \pi}^{\vec \xi}_u=0}
 = {{\delta \xi^u(\tau ,{\vec
\sigma}_1)}\over {\delta\, {}^3e_{(a)r}(\tau ,\vec \sigma )}},
\label{V12}
\eea

\noindent where in both the final expressions there is no more the restriction
${\tilde \pi}^{\vec \alpha}_{(a)}(\tau ,\vec \sigma )={\tilde
\pi}^{\vec \xi}_r(\tau ,\vec \sigma )=0$ due to the point nature of the
canonical transformation. Therefore, Eq.(\ref{V6}) becomes

\bea
 {}^3{\tilde \pi}^r_{(a)}(\tau ,\vec \sigma ) &=& {}^3{\check {\tilde \pi}}
 ^r_{(a)}(\tau ,\vec \sigma | \alpha_{(e)},\xi^s,Q_s, {\tilde \Pi}^s] +\nonumber \\
 &+& \int d^3\sigma_1\, F^r_{(a)(b)}(\vec
\sigma ,{\vec \sigma}_1;\tau )\,
{\tilde \pi}^{\vec \alpha}_{(b)}(\tau ,{\vec \sigma}_1)+\nonumber \\
 &+& \sum_u \int d^3\sigma_1\, G^{ru}_{(a)}(\vec
\sigma ,{\vec \sigma}_1;\tau )\,
{\tilde \pi}^{\vec \xi}_u(\tau ,{\vec \sigma}_1),\nonumber \\
 &&{}\nonumber \\
 &&\text{with} \nonumber \\
 &&{}\nonumber \\
 F^r_{(a)(b)}(\vec \sigma ,{\vec \sigma}_1;\tau ) &=& {\tilde F}^r_{(a)(b)}(\vec
\sigma ,{\vec \sigma}_1;\tau | \alpha_{(e)}, \xi^s, Q_s],\nonumber \\
 &&{}\nonumber \\
 G^{ru}_{(a)}(\vec \sigma ,{\vec \sigma}_1;\tau ) &=& {\tilde G}^{ru}_{(a)}(\vec
\sigma ,{\vec \sigma}_1;\tau | \alpha_{(e)}, \xi^s, Q_s].
\label{V13}
\eea

This equation and the point nature of the canonical transformation
imply

\bea
&&{{\delta\, {}^3{\check {\tilde \pi}}^r_{(a)}(\tau ,\vec \sigma
)}\over {\delta {\tilde \Pi}^u(\tau ,{\vec \sigma}_1)}} = {{\delta\,
{}^3{\tilde \pi}^r_{(a)}(\tau ,\vec \sigma )}\over {\delta {\tilde
\Pi}^u(\tau ,{\vec \sigma}_1)}} =\nonumber \\
 &=& \{ Q_u(\tau ,{\vec \sigma}_1),\, {}^3{\tilde \pi}^r_{(a)}(\tau
 ,\vec \sigma ) \} = {{\delta Q_u(\tau ,{\vec \sigma}_1)}\over
 {\delta\, {}^3e_{(a)r}(\tau ,\vec \sigma )}}\,
 {\buildrel {def} \over =}\nonumber \\
 &&{}\nonumber \\
 &{\buildrel {def} \over =}& {\tilde {\cal K}}^r_{(a)u}(\vec \sigma ,{\vec \sigma}_1;\tau |
 \alpha_{(e)},\xi^s,Q_s] = {\cal K}^r_{(a)u}(\vec \sigma ,{\vec \sigma}_1;\tau ).
\label{V14}
\eea

In conclusion the quasi-Shanmugadhasan canonical transformation is
defined by

\bea
&&\begin{minipage}[t]{3cm}
\begin{tabular}{|l|} \hline
${}^3e_{(a)r}$ \\ \hline ${}^3{\tilde \pi}^r_{(a)}$ \\ \hline
\end{tabular}
\end{minipage} \ {\longrightarrow \hspace{.2cm}} \
\begin{minipage}[t]{4 cm}
\begin{tabular}{|l|l|l|} \hline
$\alpha_{(a)}$ & $\xi^r$ & $Q_r$\\ \hline
 ${\tilde \pi}^{\vec \alpha}_{(a)}$   & ${\tilde \pi}^{\vec \xi}_r$ &
 ${\tilde \Pi}^r$  \\ \hline
\end{tabular}
\end{minipage},\nonumber \\
 &&{}\nonumber \\
{}^3e_{(a)r}(\tau ,\vec \sigma ) &=& {}^3R_{(a)(b)}(\alpha_{(e)}(\tau
,\vec \sigma )) {{\partial \xi^s(\tau ,\vec \sigma )}\over {\partial
\sigma^r}}\, {}^3{\hat e}_{(b)s}(\tau ,\vec \xi (\tau ,\vec \sigma ), Q_u(\tau
,\vec \xi (\tau ,\vec \sigma )],\nonumber \\
 &&{}\nonumber \\
  {}^3{\tilde \pi}^r_{(a)}(\tau ,\vec \sigma ) &=& \sum_u \int d^3\sigma_1\,
  {\cal K}^r_{(a)u}(\vec \sigma ,{\vec \sigma}_1;\tau )\,
  {\tilde \Pi}^u(\tau ,{\vec \sigma}_1) +\nonumber \\
 &+& \int d^3\sigma_1\, F^r_{(a)(b)}(\vec
\sigma ,{\vec \sigma}_1;\tau )\,
{\tilde \pi}^{\vec \alpha}_{(b)}(\tau ,{\vec \sigma}_1)+\nonumber \\
 &+& \sum_u \int d^3\sigma_1\, G^{ru}_{(a)}(\vec
\sigma ,{\vec \sigma}_1;\tau )\,
{\tilde \pi}^{\vec \xi}_u(\tau ,{\vec \sigma}_1),
\label{V15}
\eea

\noindent where the kernels

\bea
F^r_{(a)(b)}({\vec \sigma} ,{\vec \sigma}_1;\tau ) &=& {\tilde
F}^r_{(a)(b)}({\vec \sigma} ,{\vec \sigma}_1;\tau |
\alpha_{(e)},\xi^s, Q_s] = {{\delta \alpha_{(b)}(\tau ,{\vec \sigma}_1)}\over
{\delta\, {}^3e_{(a)r}(\tau ,\vec \sigma )}},\nonumber \\
  G^{ru}_{(a)}({\vec \sigma} ,{\vec \sigma}_1;\tau ) &=&
 {\tilde G}^{ru}_{(a)}({\vec \sigma} ,{\vec \sigma}_1;\tau | \alpha_{(e)},
\xi^s, Q_s] = {{\delta \xi^u(\tau ,{\vec \sigma}_1)}\over
{\delta\, {}^3e_{(a)r}(\tau ,\vec \sigma )}},\nonumber \\
  {\cal K}^r_{(a)u}({\vec \sigma} ,{\vec \sigma}_1;\tau ) &=&
 {\tilde {\cal K}}^r_{(a)u}({\vec \sigma} ,{\vec \sigma}_1;\tau | \alpha_{(e)},
\xi^s, Q_s] = {{\delta Q_u(\tau ,{\vec \sigma}_1)}\over {\delta\,
{}^3e_{(a)r}(\tau ,\vec \sigma )}},
\label{V16}
\eea

\noindent are the matrix elements of the inverse of the Jacobian matrix of the point
canonical transformation

\bea
| {{\partial \Big( {}^3e_{(a)r} \Big) }\over {\partial \Big(
\alpha_{(b)}, \xi^s, Q_u \Big)}}| (\vec \sigma ,{\vec \sigma}_1;\tau ) &=&
\Big( {{\delta\, {}^3e_{(a)r}(\tau ,\vec \sigma )}\over {\delta \alpha_{(b)}(\tau
,{\vec \sigma}_1)}}, {{\delta\, {}^3e_{(a)r}(\tau ,\vec \sigma )}\over
{\delta \xi^s(\tau ,{\vec \sigma}_1)}}, {{\delta\, {}^3e_{(a)r}(\tau
,\vec \sigma )}\over {\delta Q_u(\tau ,{\vec \sigma}_1)}} \Big)
,\nonumber \\
  &&{}\nonumber \\
  &&{}\nonumber \\
  &&{}\nonumber \\
{{\tilde \delta {}^3e_{(a)r}(\tau ,\vec \sigma )}\over {\delta \alpha_{(c)}
(\tau ,{\vec \sigma }_1)}}&=&\delta^3(\vec \sigma ,{\vec \sigma }_1)\Big[
H_{(c)}(\alpha_{(e)}(\tau ,\vec \sigma )){}^3R(\alpha_{(e)}(\tau ,\vec \sigma ))
\Big]_{(a)(b)}\nonumber \\
&&\sum_u{{\partial \xi^u(\tau ,\vec \sigma )}\over {\partial \sigma^r}}
\delta_{(b)u}Q_u(\tau ,\vec \xi (\tau ,\vec \sigma ))=\nonumber \\
&=&\delta^3(\vec \sigma ,{\vec \sigma }_1)\epsilon_{(a)(n)(d)}A_{(d)(c)}
(\alpha_{(e)}(\tau ,\vec \sigma )){}^3R_{(n)(m)}(\alpha_{(e)}(\tau ,\vec
\sigma ))\cdot \nonumber \\
&\cdot& \sum_u{{\partial \xi^u(\tau ,\vec \sigma )}\over {\partial \sigma^r}}
\delta_{(m)u}Q_u(\tau ,\vec \xi (\tau ,\vec \sigma )),\nonumber \\
{{\delta {}^3e_{(a)r}(\tau ,\vec \sigma )}\over {\delta \xi^u(\tau ,{\vec
\sigma }_1)}}&=&{}^3R_{(a)(n)}(\alpha_{(e)}(\tau ,\vec \sigma ))  \nonumber \\
 &&\sum_v\delta_{(n)v}\Big[
{{\partial \xi^v(\tau ,\vec \sigma )}\over {\partial \sigma^r}}{{\partial
Q_v(\tau ,\vec \xi )}\over {\partial \xi^u}}{|}_{\vec \xi =\vec \xi (\tau ,
\vec \sigma )}\delta^3(\vec \sigma ,{\vec \sigma }_1)+\nonumber \\
&+&\delta^v_u
Q_v(\tau ,\vec \xi (\tau ,\vec \sigma )){{\partial \delta^3(\vec \sigma ,
{\vec \sigma }_1)}\over {\partial \sigma^r}}\Big] ,\nonumber \\
{{\delta {}^3e_{(a)r}(\tau ,\vec \sigma )}\over {\delta Q_u(\tau ,{\vec
\sigma }_1)}}&=&{}^3R_{(a)(n)}(\alpha_{(e)}(\tau ,\vec \sigma ))\sum_v
{{\partial \xi^v
(\tau ,\vec \sigma )}\over {\partial \sigma^r}}\delta_{(n)v}
\delta^u_v\delta^3(\vec \xi
(\tau ,\vec \sigma ),{\vec \sigma }_1),\nonumber \\
 &&{}\nonumber \\
 &&\text{with}\nonumber \\
 &&{}\nonumber \\
\delta^3(\vec \sigma ,{\vec \sigma}_1) \delta_{(a)(b)}\delta^s_r &=&
{{\delta\, {}^3e_{(a)r}(\tau ,\vec \sigma )}\over {\delta\,
{}^3e_{(b)s}(\tau ,{\vec \sigma}_1)}} =\nonumber \\
 &=& \int d^3\sigma_2\, \Big( {{\delta\, {}^3e_{(a)r}(\tau ,\vec \sigma )}\over
 {\delta \alpha_{(e)}(\tau ,{\vec \sigma}_2)}}\, F^s_{(b)(e)}({\vec \sigma}_1,
 {\vec \sigma}_2;\tau ) +\nonumber \\
 &+& {{\delta\, {}^3e_{(a)r}(\tau ,\vec \sigma )}\over {\delta \xi^u(\tau
,{\vec \sigma}_2)}} \, G^{su}_{(b)}({\vec \sigma}_1,{\vec
\sigma}_2;\tau ) + {{\delta\, {}^3e_{(a)r}(\tau ,\vec \sigma )}\over {\delta Q_v(\tau
,{\vec \sigma}_2)}}\, {\cal K}^s_{(b)v}({\vec \sigma}_1,{\vec
\sigma}_2;\tau ) \Big). \nonumber \\
 &&
\label{V17}
\eea

Even if Eqs.(\ref{V9}) and (\ref{V10}) give the solution of the
constraints ${}^3{\tilde M}_{(a)}(\tau ,\vec \sigma )\approx 0$ and
${\hat {\cal H}}_{(a)}(\tau ,\vec \sigma )\approx 0$ respectively,
their non-zero Poisson brackets \footnote{See Eqs.(\ref{II14}) for the
Poisson bracket of ${}^3{\tilde M}_{(a)}$ and ${}^3{\tilde
\Theta}_r$: from it we can deduce the quoted Poisson brackets.} imply
the necessity to solve these equations for the kernels to find the
connection of the old momenta with the Abelianized ones.

By comparing Eq.(\ref{V15}) with Eqs.(\ref{V8}), (\ref{V9}) and
(\ref{V10}) we get

\bea
{}^3{\check {\tilde \pi}}^r_{(a)}(\tau ,\vec \sigma ) &=& \sum_u \int
d^3\sigma_1\, {\cal K}^r_{(a)u}(\vec \sigma ,{\vec \sigma}_1;\tau )\,
  {\tilde \Pi}^u(\tau ,{\vec \sigma}_1) \approx \nonumber \\
  &\approx& {}^3{\tilde \pi}^{(M)r}_{(a)}(\tau ,\vec \sigma ) =
  P^{(M)r}_{(a)(b)s}(\tau ,\vec \sigma )\, {}^3{\tilde \pi}^s_{(b)}(\tau ,\vec \sigma )
  \approx \nonumber \\
  &\approx& {}^3{\tilde \pi}^{(T)r}_{(a)}(\tau ,\vec \sigma )=\int d^3\sigma_1
  P^{(T)r}_{(a)(b)s}({\vec \sigma} ,{\vec \sigma}_1;\tau )\,
  {}^3{\tilde \pi}^s_{(b)}(\tau ,{\vec \sigma}_1), \nonumber \\
  &&{}\nonumber \\
  &&\Downarrow \nonumber \\
  &&{}\nonumber \\
  \sum_u&&\int d^3\sigma_1 \Big[ {}^3e_{(a)r}(\tau ,\vec \sigma )\,
  {\cal K}^r_{(b)u} - {}^3e_{(b)r}(\tau ,\vec \sigma )\,
  {\cal K}^r_{(a)u}\Big] (\vec \sigma ,{\vec \sigma}_1;\tau )\,
  {\tilde \Pi}^u(\tau ,{\vec \sigma}_1) =\nonumber \\
  &=& -\epsilon_{(a)(b)(c)} \Big[ {\tilde \pi}^{\vec \alpha}_{(d)}\, B_{(d)(c)}(\alpha_{(e)})
  \Big] (\tau ,\vec \sigma ) -\nonumber \\
  &-& \int d^3\sigma_1 \Big[ {}^3e_{(a)r}(\tau ,\vec \sigma )\,
  F^r_{(b)(c)} - {}^3e_{(b)r}(\tau ,\vec \sigma )\,
  F^r_{(a)(c)}\Big] (\vec \sigma ,{\vec \sigma}_1;\tau )\,
  {\tilde \pi}^{\vec \alpha}_{(c)}(\tau ,{\vec \sigma}_1) -\nonumber \\
  &-& \sum_u \int d^3\sigma_1 \Big[ {}^3e_{(a)r}(\tau ,\vec \sigma )\,
  G^{ru}_{(b)} - {}^3e_{(b)r}(\tau ,\vec \sigma )\,
  G^{ru}_{(a)}\Big] (\vec \sigma ,{\vec \sigma}_1;\tau )\,
  {\tilde \pi}^{\vec \xi}_u(\tau ,{\vec \sigma}_1)\approx ,\nonumber \\
  &\approx& 0,\nonumber \\
  &&{}\nonumber \\
  \sum_u&& \int d^3\sigma_1 {\hat D}^{(\omega )}_{(a)(b)r}(\tau
  ,\vec \sigma ) {\cal K}^r_{(b)u}(\vec \sigma ,{\vec \sigma}_1;\tau )\,
  {\tilde \Pi}^u(\tau ,{\vec \sigma}_1) = \nonumber \\
  &=& -{}^3e^r_{(a)}(\tau ,\vec \sigma )\Big[ {{\partial \xi^s}\over {\partial \sigma^r}}
  {\tilde \pi}^{\vec \xi}_s -(B_{(b)(c)}(\alpha_{(e)})\, {}^3\omega_{r(c)} -{{\partial
  \alpha_{(b)}}\over {\partial \sigma^r}}) {\tilde \pi}^{\vec \alpha}_{(b)}
  \Big] (\tau ,\vec \sigma ) -\nonumber \\
  &-&\int d^3\sigma_1 {\hat D}^{(\omega )}_{(a)(b)r}(\tau ,\vec \sigma )\,
  F^r_{(b)(c)}(\vec \sigma ,{\vec \sigma}_1;\tau )\,
  {\tilde \pi}^{\vec \alpha}_{(c)}(\tau ,{\vec \sigma}_1) -\nonumber \\
  &-& \sum_u \int d^3\sigma_1 {\hat D}^{(\omega )}_{(a)(b)r}(\tau ,\vec \sigma )\,
  G^{ru}_{(b)}(\vec \sigma ,{\vec \sigma}_1;\tau )\, {\tilde \pi}^{\vec \xi}_u(\tau
  ,{\vec \sigma}_1) \approx 0,
\label{V18}
\eea

\noindent namely that the momenta ${}^3{\check {\tilde
\pi}}^r_{(a)}(\tau ,\vec \sigma )$ are {\it simultaneously weak
solutions} of the constraints ${}^3{\tilde M}_{(a)}(\tau ,\vec
\sigma )\approx 0$ and ${\hat {\cal H}}_{(a)}(\tau ,\vec \sigma
)\approx 0$ and, then, also of ${}^3{\tilde \Theta}_r(\tau ,\vec
\sigma )\approx 0$. Therefore, the Einstein equations ${}^4{\bar
G}_{lr}(\tau ,\vec \sigma )\, {\buildrel \circ \over =}\, 0$ (see
after Eq.(A10) of Ref.\cite{ru11}) are satisfied, if the kernels
${\cal K}^r_{(a)u}$, solutions of the last line of
Eqs.(\ref{V17}), also satisfy these equations. Since in every
gauge (like the 3-orthogonal ones) Eqs.(\ref{V17}) will turn out
to be linear homogeneous and inhomogeneous partial differential
equations (see Eqs.(\ref{V23}) for the 3-orthogonal gauges), their
solution will depend on arbitrary functions: the solutions of the
homogeneous equations associated to the inhomogeneous ones.
Eqs.(\ref{V18}) are restrictions on these arbitrary functions.

The class of 3-orthogonal gauges is defined by putting ${\tilde
\pi}^{\vec
\alpha}_{(a)}(\tau ,\vec \sigma )= {\tilde \pi}^{\vec \xi}_r(\tau ,\vec \sigma )=0$,
by adding the gauge fixings $\xi^r(\tau ,\vec \sigma )- \sigma^r
=0$, $\alpha_{(a)}(\tau ,\vec \sigma ) =0$ and by parametrizing the reduced cotriad
${}^3{\hat e}_{(a)r}(\tau ,\vec \sigma |Q_u]$ as in Eq.(\ref{V2}).

In these gauges Eq.(\ref{V17}) becomes

\bea
{{\delta\, {}^3e_{(a)r}(\tau ,\vec \sigma )}\over {\delta
\alpha_{(b)}(\tau,{\vec \sigma}_1)}}{|}_{3-0} &=& \epsilon_{(a)(c)(b)} \delta_{(c)r}
Q_r(\tau ,\vec \sigma ) \delta^3(\vec \sigma ,{\vec
\sigma}_1),\nonumber \\
 {{\delta\, {}^3e_{(a)r}(\tau ,\vec \sigma )}\over {\delta
\xi^u(\tau,{\vec \sigma}_1)}}{|}_{3-0} &=&  \delta_{(a)r} {{\partial Q_r(\tau
,\vec \sigma )}\over {\partial \sigma^u}} \delta^3(\vec \sigma ,{\vec
\sigma}_1)+ \delta_{(a)u} Q_u(\tau ,\vec \sigma ) {{\partial
\delta^3(\vec \sigma ,{\vec \sigma}_1)}\over {\partial \sigma^r}},\nonumber \\
 {{\delta\, {}^3e_{(a)r}(\tau ,\vec \sigma )}\over {\delta
Q_u(\tau,{\vec \sigma}_1)}}{|}_{3-0} &=&  \delta_{(a)r} \delta_{ru}
\delta^3(\vec \sigma ,{\vec \sigma}_1),
 \label{V19}
\eea

\noindent and we have to solve the following equations for the kernels
restricted to these gauges

\bea
\delta^3(\vec \sigma ,{\vec \sigma}_1) \delta_{(a)(b)}\delta^s_r &=&
\epsilon_{(a)(d)(c)} \delta_{(d)r} Q_r(\tau ,\vec \sigma ) F^s_{(b)(c)}({\vec
\sigma}_1 ,{\vec \sigma};\tau ) +\nonumber \\
 &+&\delta_{(a)r}{\cal K}^s_{(b)r}({\vec \sigma}_1 ,{\vec \sigma};\tau )+
 \nonumber \\
 &+& \delta_{(a)r} \sum_u{{\partial Q_r(\tau ,\vec \sigma )}\over {\partial
 \sigma^u}} G^{su}_{(b)}({\vec \sigma}_1 ,{\vec \sigma};\tau ) +\sum_u
 \delta_{(a)u} Q_u(\tau ,\vec \sigma ) {{\partial}\over {\partial
 \sigma^r}} G^{su}_{(b)}({\vec \sigma}_1 ,{\vec \sigma};\tau ),\nonumber \\
 &&{}\nonumber \\
 F^r_{(a)(b)}({\vec \sigma} ,{\vec \sigma}_1;\tau ) &=&
 {\tilde F}^r_{(a)(b)3-0}({\vec \sigma} ,{\vec \sigma}_1;\tau |Q_v],\nonumber \\
 G^{ru}_{(a)}({\vec \sigma} ,{\vec \sigma}_1;\tau) &=&
 {\tilde G}^{ru}_{(a)3-0}({\vec \sigma} ,{\vec \sigma}_1;\tau |Q_v],\nonumber \\
 {\cal K}^r_{(a)u}({\vec \sigma} ,{\vec \sigma}_1;\tau) &=&
 {\tilde {\cal K}}^r_{(a)u3-0}({\vec \sigma} ,{\vec \sigma}_1;\tau |Q_v].
 \label{V20}
 \eea

In Eqs.(\ref{V20}) we must separate the cases $a=r$ and $a\not= r$.

To select $a=r$ let us multiply Eqs.(\ref{V20}) by $\delta_{(a)r}$ and
then let us sum over $(a)$. If we make the substitutions $s
\mapsto r$, $r \mapsto u$, $b \mapsto a$ in the final expression, we get

\bea {\cal K}^r_{(a)u}({\vec \sigma}_1 ,{\vec \sigma};\tau ) &=&
\delta_{(a)}^r\delta_{(a)u} \delta^3({\vec \sigma}_1 ,{\vec
\sigma}) -\nonumber \\
 &-& Q_u(\tau ,\vec \sigma ) {{\partial G^{ru}_{(a)}({\vec \sigma}_1 ,{\vec \sigma};\tau )}
 \over {\partial \sigma^u}} -\sum_v {{\partial Q_u(\tau ,\vec \sigma )}\over
 {\partial \sigma^v}}\, G^{rv}_{(a)}({\vec \sigma}_1 ,{\vec \sigma};\tau ).
 \label{V21}
 \eea

To select $a\not= r$, with the antisymmetry $a \leftrightarrow r$,
let us multiply Eqs.(\ref{V20}) by
$\sum_{(m)}\epsilon_{(a)(m)(d)}\delta_{(m)r}$ and then let us sum over
$(a)$. If we make the substitutions $b \mapsto a$, $d \mapsto b$ at
the end of the calculations, we get the result that each
$F^s_{(a)(b)}$ can be expressed in two different ways in terms of the
$G$'s \footnote{For $b=r$ the following equations give $0=0$.}

\bea
(1-\delta_{(b)r})\, F^s_{(a)(b)}({\vec \sigma}_1 ,{\vec \sigma};\tau )
 &=& \epsilon_{(a)(r)(b)} \delta^s_r {{\delta^3({\vec \sigma}_1,{\vec
\sigma})}\over {Q_r(\tau ,\vec \sigma )}}-\nonumber \\
 &-& \sum_u \epsilon_{(u)(r)(b)} {{Q_u(\tau ,\vec \sigma )}\over
 {Q_r(\tau ,\vec \sigma )}} {{\partial G^{su}_{(a)}({\vec \sigma}_1
 ,{\vec \sigma};\tau )}\over {\partial \sigma^r}},\nonumber \\
 &&{}\nonumber \\
 &&\Downarrow \qquad \sum_r,\nonumber \\
 &&{}\nonumber \\
 2 F^s_{(a)(b)}({\vec \sigma}_1 ,{\vec \sigma};\tau )
 &=& \epsilon_{(a)(s)(b)}  {{\delta^3({\vec \sigma}_1 ,{\vec
\sigma})}\over {Q_s(\tau ,\vec \sigma )}}-\nonumber \\
 &-& \sum_{u,r} \epsilon_{(u)(r)(b)} {{Q_u(\tau ,\vec \sigma )}\over
 {Q_r(\tau ,\vec \sigma )}} {{\partial G^{su}_{(a)}({\vec \sigma}_1
 ,{\vec \sigma};\tau )}\over {\partial \sigma^r}}.
\label{V22}
\eea

As a consequence the $G^{ru}_{(a)}$'s are determined by the following
linear partial differential equations  \footnote{$r_1, r_2
\not= b$, $r_1 \not= r_2$; $\epsilon_{(a)(r_1)(b)}=-\delta_{(a)r_2}\epsilon_{(r_1)(r_2)(b)}$,
$\epsilon_{(a)(r_2)(b)}=\delta_{(a)r_1}\epsilon_{(r_1)(r_2)(b)}$,
$\epsilon_{(u)(r_1)(b)}=-\delta_{(u)r_2}\epsilon_{(r_1)(r_2)(b)}$,
$\epsilon_{(u)(r_2)(b)}=
\delta_{(u)r_1}\epsilon_{(r_1)(r_2)(b)}$; $Q_r=\sqrt{{}^3{\hat g}_{rr}}$.}

\bea
 &&\epsilon_{(a)(r_1)(b)} \delta^s_{r_1} {{\delta^3({\vec
\sigma}_1,{\vec \sigma})}\over {Q_{r_1}(\tau ,\vec \sigma )}}
-\sum_u \epsilon_{(u)(r_1)(b)} {{Q_u(\tau ,\vec \sigma )}\over
{Q_{r_1}(\tau ,\vec \sigma )}} {{\partial G^{su}_{(a)}({\vec
\sigma}_1 ,{\vec \sigma};\tau )}\over {\partial
\sigma^{r_1}}}=\nonumber \\
 =&& \epsilon_{(a)(r_2)(b)} \delta^s_{r_2} {{\delta^3({\vec \sigma}_1,{\vec
\sigma})}\over {Q_{r_2}(\tau ,\vec \sigma )}} -\sum_u
\epsilon_{(u)(r_2)(b)} {{Q_u(\tau ,\vec \sigma )}\over
{Q_{r_2}(\tau ,\vec \sigma )}} {{\partial G^{su}_{(a)}({\vec
\sigma}_1
 ,{\vec \sigma};\tau )}\over {\partial \sigma^{r_2}}},\nonumber \\
 &&{}\nonumber \\
 &&\Downarrow \nonumber \\
 &&{}\nonumber \\
 &&{1\over {Q_{r_1}^2(\tau ,\vec \sigma )}}\, {{\partial G^{sr_2}_{(a)}
 ({\vec \sigma}_1 ,{\vec \sigma};\tau )}\over {\partial \sigma^{r_1}}}+
 {1\over {Q_{r_2}^2(\tau ,\vec \sigma )}}\, {{\partial G^{sr_1}_{(a)}
 ({\vec \sigma}_1 ,{\vec \sigma};\tau )}\over {\partial \sigma^{r_2}}}
   =\nonumber \\
   =&& \Big[ {{\delta_{(a)r_1}\delta^s_{r_2}}\over {Q_{r_1}(\tau ,\vec \sigma )
   Q^2_{r_2}(\tau ,\vec \sigma )}}+
 {{\delta_{(a)r_2}\delta^s_{r_1}}\over {Q^2_{r_1}(\tau ,\vec \sigma )
   Q_{r_2}(\tau ,\vec \sigma )}}\Big]   \delta^3({\vec \sigma}_1,{\vec \sigma}),
   \nonumber \\
   &&{}\nonumber \\
 &&\Downarrow \nonumber \\
 &&{}\nonumber \\
 1)&&\,\, s=a\quad \text{ homogeneous equations}:\nonumber \\
 &&{}\nonumber \\
 &&{1\over {Q_1^2(\tau ,\vec \sigma )}}\, {{\partial G^{a2}_{(a)}
 ({\vec \sigma}_1 ,{\vec \sigma};\tau )}\over {\partial \sigma^1}} +
 {1\over {Q_2^2(\tau ,\vec \sigma )}}\, {{\partial G^{a1}_{(a)}
 ({\vec \sigma}_1 ,{\vec \sigma};\tau )}\over {\partial \sigma^2}} =\nonumber \\
 &&= {1\over {Q_2^2(\tau ,\vec \sigma )}}\, {{\partial G^{a3}_{(a)}
 ({\vec \sigma}_1 ,{\vec \sigma};\tau )}\over {\partial \sigma^2}} +
 {1\over {Q_3^2(\tau ,\vec \sigma )}}\, {{\partial G^{a2}_{(a)}
 ({\vec \sigma}_1 ,{\vec \sigma};\tau )}\over {\partial \sigma^3}} =\nonumber \\
 &&= {1\over {Q_3^2(\tau ,\vec \sigma )}}\, {{\partial G^{a1}_{(a)}
 ({\vec \sigma}_1 ,{\vec \sigma};\tau )}\over {\partial \sigma^3}} +
 {1\over {Q_1^2(\tau ,\vec \sigma )}}\, {{\partial G^{a3}_{(a)}
 ({\vec \sigma}_1 ,{\vec \sigma};\tau )}\over {\partial \sigma^1}} = 0,
 \qquad a=1,2,3;\nonumber \\
 &&{}\nonumber \\
 2)&&\,\, s\not= a\,\, [s\not= r,\, r\not= a]\quad \text{in-homogeneous equations}:\nonumber \\
 &&{}\nonumber \\
 &&{1\over {Q_s^2(\tau ,\vec \sigma )}}\, {{\partial G^{sr}_{(a)}
 ({\vec \sigma}_1 ,{\vec \sigma};\tau )}\over {\partial \sigma^s}} +
 {1\over {Q_r^2(\tau ,\vec \sigma )}}\, {{\partial G^{ss}_{(a)}
 ({\vec \sigma}_1 ,{\vec \sigma};\tau )}\over {\partial \sigma^r}} =\nonumber \\
  &&={1\over {Q_r^2(\tau ,\vec \sigma )}}\, {{\partial G^{sa}_{(a)}
 ({\vec \sigma}_1 ,{\vec \sigma};\tau )}\over {\partial \sigma^r}} +
 {1\over {Q_a^2(\tau ,\vec \sigma )}}\, {{\partial G^{sr}_{(a)}
 ({\vec \sigma}_1 ,{\vec \sigma};\tau )}\over {\partial \sigma^a}} =0,\nonumber \\
 &&{1\over {Q_a^2(\tau ,\vec \sigma )}}\, {{\partial G^{ss}_{(a)}
 ({\vec \sigma}_1 ,{\vec \sigma};\tau )}\over {\partial \sigma^a}} +
 {1\over {Q_s^2(\tau ,\vec \sigma )}}\, {{\partial G^{sa}_{(a)}
 ({\vec \sigma}_1 ,{\vec \sigma};\tau )}\over {\partial \sigma^s}}
 = {{\delta^3({\vec \sigma}_1 ;{\vec \sigma})}\over {Q_a(\tau ,\vec \sigma )
 Q_s^2(\tau ,\vec \sigma )}}.
 \label{V23}
 \eea

Each set of homogeneous equations, considered as equations for
functions of $\vec \sigma$ and disregarding the $\tau$-dependence,
is of the form $a_2(\vec \sigma ) \partial_2\, u_{(1)}(\vec \sigma
)+a_1(\vec \sigma ) \partial_1\, u_{(2)}(\vec \sigma ) = a_3(\vec
\sigma )\partial_3\, u_{(2)}(\vec \sigma )+a_2(\vec \sigma )
\partial_2\, u_{(3)}(\vec \sigma ) = a_1(\vec \sigma )
\partial_1\, u_{(3)}(\vec \sigma ) + a_3(\vec \sigma )
\partial_3\, u_{(1)}(\vec \sigma ) =0$ [$a_i=Q_i^{-2}$]. This is a
system of three linear partial differential equations for the
three unknown functions $u_{(i)}(\vec \sigma )$ of {\it elliptic}
type, since the determinant of its characteristic matrix
\cite{chester} is $2a_1(\vec \sigma )a_2(\vec \sigma )a_3(\vec
\sigma ) \xi_1\xi_2\xi_3 \not= 0$. Moreover it is integrable,
since $u_{(r)}(\vec \sigma ) = f_{(r)}(\sigma^r)$ with arbitrary
$f_{(r)}$ are solutions of the system. We do not know whether they
exhaust all the possible solutions. Therefore, $G^{ar}_{(a)}({\vec
\sigma}_1,\vec \sigma ;\tau )=h^{ar}_{(a)}({\vec
\sigma}_1,\sigma^r;\tau )$, with $h^{ar}_{(a)}$ arbitrary
functions, are solutions of the homogeneous equations.

As a consequence, if ${\bar G}^{ru}_{(a)}({\vec \sigma}_1,\vec
\sigma ;\tau )$, $r\not= a$, is a particular solution of each set
of inhomogeneous equations (\ref{V23}), then the general solution
is $G^{ru}_{(a)}({\vec \sigma}_1,\vec \sigma ;\tau )={\bar
G}^{ru}_{(a)}({\vec \sigma}_1,\vec \sigma ;\tau
)+g^{ru}_{(a)}({\vec \sigma}_1,\vec \sigma ;\tau )$, $r\not= a$,
with the $g^{ru}_{(a)}$'s arbitrary homogeneous solutions (again
with $g^{ru}_{(a)}({\vec \sigma}_1,\sigma^r;\tau )$, if this is
the most general solution of the associated homogeneous
equations). A way to find a particular solution may be to define
$G^{ru}_{f(a)}(\vec \sigma ;\tau )=\int d^3\sigma_1\, f({\vec
\sigma}_1) G^{ru}_{(a)}({\vec \sigma}_1,\vec \sigma ;\tau )$:
then, disregarding the $\tau$-dependence, we get the system of
elliptic linear partial differential equations [$a_s=Q_s^{-2}$,
$s\not= r,a$, $r\not= a$] $\,\,\, a_s(\vec \sigma ) \partial_s\,
G^{sr}_{f(a)}(\vec \sigma )+a_r(\vec \sigma )\partial_r\,
G^{ss}_{f(a)}(\vec \sigma ) = a_r(\vec \sigma )\partial_r\,
G^{sa}_{f(a)}(\vec \sigma )+a_a(\vec \sigma )\partial_a\,
G^{sr}_{f(a)}(\vec \sigma )=0$ , $a_a(\vec \sigma )\partial_a\,
G^{ss}_{f(a)}(\vec \sigma ) +a_s(\vec \sigma )\partial_s\,
G^{sa}_{f(a)}(\vec \sigma ) = f(\vec \sigma )/\sqrt{a_a(\vec
\sigma )} a_s(\vec \sigma )$. Each particular solution of this
system which is a functional linear in $f(\vec \sigma )$ will
allow to find a particular solution ${\bar G}^{ru}_{(a)}$.

Therefore the kernels $G^{ru}_{(a)}({\vec \sigma}_1,\vec \sigma
;\tau )$ solutions of Eqs.(\ref{V23}) can be written in the
following form

\beq
 G^{ru}_{(a)}({\vec \sigma}_1,\vec \sigma ;\tau ) =
\delta^r_{(a)} h^{au}_{(a)}({\vec \sigma}_1,\vec \sigma ;\tau )+
(1-\delta^r_{(a)})[{\bar G}^{ru}_{(a)}({\vec \sigma}_1,\vec \sigma
;\tau )+g^{ru}_{(a)}({\vec \sigma}_1,\vec \sigma ;\tau )],
 \label{V23a}
 \eeq

\noindent with arbitrary $h^{au}_{(a)}$'s and $g^{ru}_{(a)}$'s.
Then, Eq(\ref{V21}) gives the following expression for the kernels
${\cal K}^r_{(a)u}$'s

\bea
 &&{\cal K}^r_{(a)u}({\vec \sigma}_1,\vec \sigma )=\delta^r_{(a)}
 \delta_{(a)u} \delta^3({\vec \sigma}_1,\vec \sigma )-\nonumber \\
 &-& Q_u(\tau ,\vec \sigma ) \Big( \delta^r_{(a)} {{\partial
 h^{au}_{(a)}({\vec \sigma}_1,\vec \sigma ;\tau )}\over {\partial
 \sigma^u}} +(1-\delta^r_{(a)})[ {{\partial
 {\bar G}^{ru}_{(a)}({\vec \sigma}_1,\vec \sigma ;\tau )}\over {\partial
 \sigma^u}}+{{\partial
 g^{ru}_{(a)}({\vec \sigma}_1,\vec \sigma ;\tau )}\over {\partial
 \sigma^u}}]\Big) -\nonumber \\
 &-&\sum_v {{\partial Q_u(\tau ,\vec \sigma )}\over {\partial
 \sigma^v}} \Big( \delta^r_{(a)} h^{av}_{(a)}({\vec \sigma}_1,\vec
 \sigma ;\tau ) +(1-\delta^r_{(a)})[{\bar G}^{rv}_{(a)}({\vec
 \sigma}_1,\vec \sigma ;\tau ) +g^{rv}_{(a)}({\vec \sigma}_1,\vec
 \sigma ;\tau )]\Big).
 \label{V23b}
 \eea

The solutions of Eqs.(\ref{V23}) for the $G^{ru}_{(a)}$'s are
restricted by the requirement that the ${\cal K}^r_{(a)u}$'s of
Eqs.(\ref{V21}) satisfy Eqs.(\ref{V18}), which in the 3-orthogonal
gauges become \footnote{In the last line we give ${}^3{\hat
\omega}_{r(a)}(\tau ,\vec \sigma )$ in the 3-orthogonal gauges, see
Eq.(\ref{VI14}) of next Section.}

\bea
&&\sum_r Q_r(\tau ,\vec \sigma ) \Big[ \delta_{(a)r} {\cal K}^r_{(b)u}
- \delta_{(b)r} {\cal K}^r_{(a)u} \Big] (\vec \sigma ,{\vec \sigma}_1;\tau )
=\nonumber \\
 &&= Q_u(\tau ,{\vec \sigma}_1 ) \Big[ Q_a(\tau ,\vec \sigma ){{\partial G^{au}_{(b)}
(\vec \sigma ,{\vec \sigma}_1;\tau )}\over {\partial
\sigma^u_1}}-Q_b(\tau ,\vec \sigma ){{\partial G^{bu}_{(a)}
(\vec \sigma ,{\vec \sigma}_1;\tau )}\over {\partial
\sigma^u_1}} \Big] +\nonumber \\
 &&+\sum_v {{\partial Q_u(\tau ,{\vec \sigma}_1 )}\over {\partial \sigma^v_1}}
\Big[ Q_a(\tau ,\vec \sigma ) G^{av}_{(b)}(\vec \sigma ,{\vec \sigma}_1;\tau )-
Q_b(\tau ,\vec \sigma ) G^{bv}_{(a)}(\vec \sigma ,{\vec \sigma}_1;\tau
)\Big] = \nonumber \\
 &&= 0,\qquad a\not= b,\nonumber \\
 &&{}\nonumber \\
 &&{\hat D}^{(\hat \omega )}_{(a)(b)r}(\tau ,\vec \sigma ) {\cal K}^r_{(b)u}
 (\vec \sigma ,{\vec \sigma}_1;\tau ) =\nonumber \\
 &&=\Big( \delta_{(a)(b)}\partial_r + \epsilon_{(a)(b)(c)}\, {}^3{\hat \omega}_{r(c)}
 (\tau ,\vec \sigma )\Big) {\cal K}^r_{(b)u}
 (\vec \sigma ,{\vec \sigma}_1;\tau ) =\nonumber \\
 &&= \Big( \delta_{(a)(b)} \partial_r + \sum_u(\delta_{(a)r}\delta_{(b)u}-
 \delta_{(a)u}\delta_{(b)r}) {{\partial_uQ_r(\tau ,\vec \sigma )}\over
 {Q_u(\tau ,\vec \sigma )}}\Big)
\Big[ \delta_{(b)}^r\delta_{(b)u} \delta^3(\vec \sigma ,{\vec \sigma}_1) -\nonumber \\
 &-& Q_u(\tau ,{\vec \sigma}_1 ) {{\partial G^{ru}_{(b)}(\vec \sigma ,{\vec \sigma}_1;\tau )}
 \over {\partial \sigma^u_1}} -\sum_v {{\partial Q_u(\tau ,{\vec \sigma}_1 )}\over
 {\partial \sigma^v_1}}\, G^{rv}_{(b)}(\vec \sigma ,{\vec \sigma}_1;\tau ) \Big]
 = 0,\nonumber \\
 &&{}\nonumber \\
 &&{}^3{\hat \omega}_{r(a)}(\tau ,\vec \sigma ) =\sum_u
 \epsilon_{(a)(m)(n)}\delta_{(m)r}\delta_{(n)u}{{\partial_uQ_r(\tau ,\vec \sigma )}\over
 {Q_u(\tau ,\vec \sigma )}}
\label{V24}
\eea

The first set of Eqs.(\ref{V24}) becomes the following set of
three linear partial differential equations to get the
$g^{ru}_{(a)}$'s with $a\not= b$ in terms of the ${\bar
G}^{ru}_{(a)}$'s

\bea
 &&Q_a(\tau ,{\vec \sigma}_1) {{\partial
 g^{au}_{(b)}({\vec \sigma}_1,\vec \sigma ;\tau )}\over {\partial
 \sigma^u}} -Q_b(\tau ,{\vec \sigma}_1) {{\partial
 g^{bu}_{(a)}({\vec \sigma}_1,\vec \sigma ;\tau )}\over {\partial
 \sigma^u}}+\nonumber \\
 &+&\sum_v {{\partial ln\, Q_u(\tau ,\vec \sigma )}\over {\partial
 \sigma^v}} [Q_a(\tau ,{\vec \sigma}_1) g^{av}_{(b)}({\vec
 \sigma}_1,\vec \sigma ;\tau )-Q_b(\tau ,{\vec \sigma}_1)
 g^{bv}_{(a)}({\vec \sigma}_1,\vec \sigma ;\tau )]=\nonumber \\
 &=&-\Big[ Q_a(\tau ,{\vec \sigma}_1) {{\partial
 {\bar G}^{au}_{(b)}({\vec \sigma}_1,\vec \sigma ;\tau )}\over {\partial
 \sigma^u}} -Q_b(\tau ,{\vec \sigma}_1) {{\partial
 {\bar G}^{bu}_{(a)}({\vec \sigma}_1,\vec \sigma ;\tau )}\over {\partial
 \sigma^u}}+\nonumber \\
 &+&\sum_v {{\partial ln\, Q_u(\tau ,\vec \sigma )}\over {\partial
 \sigma^v}} [Q_a(\tau ,{\vec \sigma}_1) {\bar G}^{av}_{(b)}({\vec
 \sigma}_1,\vec \sigma ;\tau )-Q_b(\tau ,{\vec \sigma}_1)
 {\bar G}^{bv}_{(a)}({\vec \sigma}_1,\vec \sigma ;\tau )] \Big]
 =\nonumber \\
 &{\buildrel {def} \over =}& m^u_{ab}({\vec \sigma}_1,\vec \sigma
 ;\tau ),\nonumber \\
 &&{}\nonumber \\
 &&\text{or}\nonumber \\
 &&{}\nonumber \\
 &&{{\partial f^u_{ab}({\vec \sigma}_1,\vec \sigma ;\tau )}\over
 {\partial \sigma^u}}+ \sum_v{{\partial ln\, Q_u(\tau ,\vec \sigma
 )}\over {\partial \sigma^v}} f^v_{ab}({\vec \sigma}_1,\vec \sigma
 ;\tau ) = m^u_{ab}({\vec \sigma}_1,\vec \sigma ;\tau ),\nonumber
 \\
 &&{}\nonumber \\
 &&\text{with}\nonumber \\
 &&{}\nonumber \\
 &&f^u_{ab}({\vec \sigma}_1,\vec \sigma ;\tau )= Q_a(\tau ,{\vec
 \sigma}_1) g^{au}_{(b)}({\vec \sigma}_1,\vec \sigma ;\tau
 )-Q_b(\tau ,{\vec \sigma}_1) g^{bu}_{(a)}({\vec \sigma}_1,\vec
 \sigma ;\tau ).
 \label{V24a}
 \eea

 \noindent For each pair $a\not= b$, this is a system of three
 elliptic linear partial differential equations for the
 $f^u_{ab}$. Each choice of the $g^{au}_{(b)}$'s, $a\not= b$,
 which gives a solution of this system, implies that the
 associated kernels ${\cal K}^r_{(a)u}$'s satisfy the rotation
 constraints.

 Having found a solution for the $g^{au}_{(b)}$'s, $a\not= b$, the
 second set of Eqs.(\ref{V24}) becomes the following set of
 equations for the $h^{au}_{(a)}$'s in terms of the ${\bar
 G}^{ru}_{(a)}$'s and $g^{ru}_{(a)}$'s

 \bea
  &&\sum_{r,(b)} {\hat D}^{(\omega )}_{(a)(b)r}(\tau ,{\vec
  \sigma}_1) \delta^r_{(b)}\Big[ q_u(\tau ,\vec \sigma )
  {{\partial h^{bu}_{(b)}({\vec \sigma}_1,\vec \sigma ;\tau
  )}\over {\partial \sigma^u}}-\sum_v{{\partial  Q_u(\tau ,\vec
  \sigma )}\over {\partial \sigma^v}} h^{bv}_{(b)}({\vec
  \sigma}_1,\vec \sigma ;\tau )\Big] =\nonumber \\
  &&{}\nonumber \\
  &&=\sum_{r,(b)} {\hat D}^{(\omega )}_{(a)(b)r}(\tau ,{\vec
  \sigma}_1) \Big[ \delta^r_{(b)}\delta_{(b)u} \delta^3({\vec
  \sigma}_1,\vec \sigma )-\nonumber \\
  &&-(1-\delta^r_{(b)})\Big( q_u(\tau ,\vec \sigma )[ {{\partial
  {\bar G}^{ru}_{(b)}({\vec \sigma}_1,\vec \sigma ;\tau
  )}\over {\partial \sigma^u}}+ {{\partial g^{ru}_{(b)}({\vec \sigma}_1,\vec
  \sigma ;\tau )}\over {\partial \sigma^u}}]+\nonumber \\
  &&+\sum_v {{\partial Q_u(\tau ,\vec \sigma )}\over {\partial
  \sigma^v}}[{\bar G}^{rv}_{(b)}({\vec \sigma}_1,\vec \sigma ;\tau
  )+g^{rv}_{(b)}({\vec \sigma}_1,\vec \sigma ;\tau )]\Big)\Big] .
  \label{V24b}
  \eea

  By using the Green function of Eqs.(\ref{IV27}), (\ref{IV30}),
  we get ($f^{su}_{(T)}$ are solutions of the homogeneous equation)

  \bea
  &&{{\partial h^{su}_{(s)}({\vec \sigma}_1,\vec \sigma ;\tau
  )}\over {\partial \sigma^u}} +\sum_v {{\partial ln\, Q_u(\tau
  ,\vec \sigma )}\over {\partial \sigma^v}} h^{sv}_{(s)}({\vec
  \sigma}_1,\vec \sigma ;\tau )=\nonumber \\
  &&{}\nonumber \\
 &&= f^{su}_{(T)}({\vec \sigma}_1,\vec \sigma ;\tau )-\nonumber \\
  &&-\int d^3\sigma_2 \sum_{(a)}\zeta^{(\omega )s}_{(s)(a)}({\vec \sigma}_1,{\vec
  \sigma}_2;\tau )\sum_{r,(b)} {\hat D}^{(\omega )}_{(a)(b)r}(\tau ,{\vec
  \sigma}_2) \Big[ {{\delta^r_{(b)}\delta_{(b)u}}\over {Q_u(\tau ,\vec \sigma )}}
   \delta^3({\vec \sigma}_2,\vec \sigma )-\nonumber \\
  &&-(1-\delta^r_{(b)})\Big( Q_u(\tau ,\vec \sigma )[ {{\partial
  {\bar G}^{ru}_{(b)}({\vec \sigma}_2,\vec \sigma ;\tau
  )}\over {\partial \sigma^u}}+ {{\partial g^{ru}_{(b)}({\vec \sigma}_2,\vec
  \sigma ;\tau )}\over {\partial \sigma^u}}]+\nonumber \\
  &&+\sum_v {{\partial ln\, Q_u(\tau ,\vec \sigma )}\over {\partial
  \sigma^v}}[{\bar G}^{rv}_{(b)}({\vec \sigma}_2,\vec \sigma ;\tau
  )+g^{rv}_{(b)}({\vec \sigma}_2,\vec \sigma ;\tau )]\Big)\Big] .
  \label{V24c}
  \eea

  \noindent Again this is a system of elliptic linear partial differential
  equations for the $h^{au}_{(a)}$'s with fixed $a$.

Finally, we have to find the conditions imposed on the kernels by
the vanishing of the Poisson brackets of the old momenta thought
as functions of the new variables through Eq.(\ref{V15}): $\{
{}^3{\tilde \pi}^r_{(a)}(\tau ,\vec \sigma ), {}^3{\tilde
\pi}^s_{(c)}(\tau ,{\vec \sigma}^{'}) \} =0$. It turns out that we
get the following quasi-linear partial differential equations for
the remaining arbitrariness in the homogeneous solutions [the
kernels $F^r_{(a)(b)}$'s are given in Eqs.(\ref{V22})]:

\bea
 &&\sum_v\Big( {\cal K}^s_{(c)v}({\vec
 \sigma}^{'},{\vec \sigma}_2;\tau ){{\delta {\cal K}^r_{(a)u}(\vec
 \sigma ,{\vec \sigma}_1;\tau )}\over {\delta Q_v(\tau ,{\vec
 \sigma}_2)}}- {\cal K}^r_{(a)v}({\vec
 \sigma},{\vec \sigma}_2;\tau ){{\delta {\cal K}^s_{(c)u}({\vec
 \sigma}^{'} ,{\vec \sigma}_1;\tau )}\over {\delta Q_v(\tau ,{\vec
 \sigma}_2)}}\Big) +\nonumber \\
 &&+\Big( F^s_{(c)(d)}({\vec \sigma}^{'}, {\vec \sigma}_2;\tau
 ){{\delta {\cal K}^r_{(a)u}(\vec \sigma ,{\vec \sigma}_1)}\over
 {\delta \alpha_{(d)}(\tau ,{\vec \sigma}_2)}}-
 F^r_{(a)(d)}({\vec \sigma}, {\vec \sigma}_2;\tau
 ){{\delta {\cal K}^s_{(c)u}({\vec \sigma}^{'} ,{\vec \sigma}_1)}\over
 {\delta \alpha_{(d)}(\tau ,{\vec \sigma}_2)}}\Big) +\nonumber \\
 &&+\sum_v\Big( G^{sv}_{(c)}({\vec \sigma}^{'},{\vec
 \sigma}_2;\tau ) {{\delta {\cal K}^r_{(a)u}({\vec \sigma},{\vec
 \sigma}_1;\tau )}\over {\delta \xi^v(\tau ,{\vec \sigma}_2)}}-
 G^{rv}_{(a)}({\vec \sigma},{\vec
 \sigma}_2;\tau ) {{\delta {\cal K}^s_{(c)u}({\vec \sigma}^{'},{\vec
 \sigma}_1;\tau )}\over {\delta \xi^v(\tau ,{\vec
 \sigma}_2)}}\Big)=0,\nonumber \\
 &&{}\nonumber \\
 && \sum_v\Big( {\cal K}^s_{(c)v}({\vec \sigma}^{'},{\vec
 \sigma}_2;\tau ) {{\delta F^r_{(a)(b)}({\vec \sigma},{\vec
 \sigma}_1;\tau )}\over {\delta Q_v(\tau ,{\vec \sigma}_2)}}-
 {\cal K}^r_{(a)v}({\vec \sigma},{\vec
 \sigma}_2;\tau ) {{\delta F^s_{(c)(b)}({\vec \sigma}^{'},{\vec
 \sigma}_1;\tau )}\over {\delta Q_v(\tau ,{\vec \sigma}_2)}}\Big)
 +\nonumber \\
 &&+\Big( F^s_{(c)(d)}({\vec \sigma}^{'},{\vec \sigma}_2;\tau )
 {{\delta F^r_{(a)(b)}({\vec \sigma},{\vec \sigma}_1;\tau )}\over
 {\delta \alpha_{(d)}(\tau ,{\vec \sigma}_2)}}-
 F^r_{(a)(d)}({\vec \sigma},{\vec \sigma}_2;\tau )
 {{\delta F^s_{(c)(b)}({\vec \sigma}^{'},{\vec \sigma}_1;\tau )}\over
 {\delta \alpha_{(d)}(\tau ,{\vec \sigma}_2)}}\Big) +\nonumber \\
 &&+\sum_v\Big( G^{sv}_{(c)}({\vec \sigma}^{'},{\vec \sigma}_2;\tau
 ) {{\delta F^r_{(a)(b)}({\vec \sigma},{\vec \sigma}_1;\tau
 )}\over {\delta \xi^v(\tau ,{\vec \sigma}_2)}}-
 G^{rv}_{(a)}({\vec \sigma},{\vec \sigma}_2;\tau
 ) {{\delta F^s_{(c)(b)}({\vec \sigma}^{'},{\vec \sigma}_1;\tau
 )}\over {\delta \xi^v(\tau ,{\vec \sigma}_2)}}\Big) =0,\nonumber \\
 &&{}\nonumber \\
 && \sum_v\Big( {\cal K}^s_{(c)v}({\vec \sigma}^{'},{\vec
 \sigma}_2;\tau ) {{\delta G^{ru}_{(a)}({\vec \sigma},{\vec
 \sigma}_1;\tau )}\over {\delta Q_v(\tau ,{\vec \sigma}_2)}}-
{\cal K}^r_{(a)v}({\vec \sigma},{\vec
 \sigma}_2;\tau ) {{\delta G^{su}_{(c)}({\vec \sigma}^{'},{\vec
 \sigma}_1;\tau )}\over {\delta Q_v(\tau ,{\vec \sigma}_2)}}\Big)
 +\nonumber \\
 &&+\Big( F^s_{(c)(d)}({\vec \sigma}^{'},{\vec \sigma}_2;\tau )
 {{\delta G^{ru}_{(a)}({\vec \sigma},{\vec \sigma}_1;\tau )}\over
 {\delta \alpha_{(d)}(\tau ,{\vec \sigma}_2)}}-
 F^r_{(a)(d)}({\vec \sigma},{\vec \sigma}_2;\tau )
 {{\delta G^{su}_{(c)}({\vec \sigma}^{'},{\vec \sigma}_1;\tau )}\over
 {\delta \alpha_{(d)}(\tau ,{\vec \sigma}_2)}}\Big) +\nonumber \\
 &&+\sum_v\Big( G^{sv}_{(c)}({\vec \sigma}^{'},{\vec
 \sigma}_2;\tau ) {{\delta G^{ru}_{(a)}({\vec \sigma},{\vec
 \sigma}_1;\tau )}\over {\delta \xi^v(\tau ,{\vec \sigma}_2)}}-
 G^{rv}_{(a)}({\vec \sigma},{\vec
 \sigma}_2;\tau ) {{\delta G^{su}_{(c)}({\vec \sigma}^{'},{\vec
 \sigma}_1;\tau )}\over {\delta \xi^v(\tau ,{\vec
 \sigma}_2)}}\Big) =0,\nonumber \\
 &&{}\nonumber \\
 &&\Downarrow \text{in 3-orthogonal gauges}\nonumber \\
 &&{}\nonumber \\
 &&\sum_v\Big( {\cal K}^s_{(c)v}({\vec
 \sigma}^{'},{\vec \sigma}_2;\tau ){{\delta {\cal K}^r_{(a)u}(\vec
 \sigma ,{\vec \sigma}_1;\tau )}\over {\delta Q_v(\tau ,{\vec
 \sigma}_2)}}- {\cal K}^r_{(a)v}({\vec
 \sigma},{\vec \sigma}_2;\tau ){{\delta {\cal K}^s_{(c)u}({\vec
 \sigma}^{'} ,{\vec \sigma}_1;\tau )}\over {\delta Q_v(\tau ,{\vec
 \sigma}_2)}}\Big) =0,\nonumber \\
  && \sum_v\Big( {\cal K}^s_{(c)v}({\vec \sigma}^{'},{\vec
 \sigma}_2;\tau ) {{\delta F^r_{(a)(b)}({\vec \sigma},{\vec
 \sigma}_1;\tau )}\over {\delta Q_v(\tau ,{\vec \sigma}_2)}}-
 {\cal K}^r_{(a)v}({\vec \sigma},{\vec
 \sigma}_2;\tau ) {{\delta F^s_{(c)(b)}({\vec \sigma}^{'},{\vec
 \sigma}_1;\tau )}\over {\delta Q_v(\tau ,{\vec \sigma}_2)}}\Big)
 =0,\nonumber \\
  && \sum_v\Big( {\cal K}^s_{(c)v}({\vec \sigma}^{'},{\vec
 \sigma}_2;\tau ) {{\delta G^{ru}_{(a)}({\vec \sigma},{\vec
 \sigma}_1;\tau )}\over {\delta Q_v(\tau ,{\vec \sigma}_2)}}-
{\cal K}^r_{(a)v}({\vec \sigma},{\vec
 \sigma}_2;\tau ) {{\delta G^{su}_{(c)}({\vec \sigma}^{'},{\vec
 \sigma}_1;\tau )}\over {\delta Q_v(\tau ,{\vec
 \sigma}_2)}}\Big)=0.
 \label{V24d}
 \eea

After having found the general solutions for the kernels
$G^{ru}_{(a)}$'s, ${\cal K}^r_{(a)u}$'s and $F^r_{(a)(b)}$'s,
every remaining arbitrariness will be fixed by the boundary
conditions at spatial infinity, which must be the ones given in
Eqs.(\ref{II25}) for the momenta ${}^3{\tilde \pi}^r_{(a)}(\tau
,\vec \sigma )$, as implied by Eqs.(\ref{V15}). The final
solutions are equivalent not only to the solution of the rotation
and diffeomorphisms constraints, but also to their Abelianization
in the 3-orthogonal gauges with $\alpha_{(a)}(\tau ,\vec \sigma
)=0$.

Even if we do not know explicitly the kernels ${\cal K}^r_{(a)u}$
\footnote{The solution of Eqs.(\ref{V23}), (\ref{V24}) is
equivalent to solve the three Einstein equations corresponding to
the supermomentum constraints of metric gravity; the separate
solution of the rotation and diffeomorphism constraints of tetrad
gravity is not enough to find their simultaneous Abelianization:
this implies Eqs.(\ref{V24}).}, we are able to get the following
decomposition of the original variables in the 3-orthogonal gauges

\bea
{}^3e_{(a)r}(\tau ,\vec \sigma ){|}_{3-0} &=& \delta_{(a)r} Q_r(\tau
,\vec \sigma ),\nonumber \\
 {}^3{\tilde \pi}^r_{(a)}(\tau ,\vec \sigma ){|}_{3-0} &=& \sum_u
 \int d^3\sigma_1\, {\tilde {\cal K}}^r_{(a)u}(\vec \sigma ,{\vec \sigma}_1;
 \tau |Q_v]\, {\tilde \Pi}^u(\tau ,{\vec \sigma}_1)=\nonumber \\
 &=&\sum_u \int d^3\sigma_1 {\cal K}^r_{(a)u}(\vec \sigma ,{\vec
 \sigma}_1;\tau ) {\tilde \Pi}^u(\tau ,{\vec \sigma}_1).
\label{V25}
\eea

From Eqs.(\ref{II12})  we get  [${}^3\hat e =\sqrt{\hat \gamma}
=Q_1Q_2Q_3$]

\begin{eqnarray}
{}^3{\hat K}_{rs}&=&\sum_u {{\epsilon \, 4\pi G\,  Q_rQ_sQ_u}\over
{c^3\, Q_1Q_2Q_3}} \, {}^3G_{o(a)(b)(c)(d)}\, \delta_{(\bar a)r}
\delta_{(\bar b)s} \delta_{(\bar c)u}\, {}^3{\tilde \pi}^u_{(\bar
d)}{|}_{3-O},\nonumber \\
 &&{}\nonumber \\
  {}^3\hat K&=&-{{\epsilon 8\pi G}\over {c^3\, Q_1Q_2Q_3}}\, {}^3{\hat {\tilde
\Pi}}=-{{\epsilon \, 4\pi G}\over {c^3\,
Q_1Q_2Q_3}}\sum_r\delta_{(a)r}Q_r\, {}^3{\tilde
\pi}^r_{(a)}{|}_{3-O},\nonumber \\
 &&{}\nonumber \\
  {}^3{\hat {\tilde \Pi}}^{rs}&=&{{\epsilon \, c^3}\over {16\pi G}} Q_1Q_2Q_3
  ({}^3{\hat K}^{rs}-Q_r^2\delta^{rs}\, {}^3\hat K)
={1\over 4}\Big[ {{\delta^r_{(a)}\, {}^3{\tilde \pi}^s_{(a)}{|}_{3-O}}\over {Q_r}}+
{{\delta^s_{(a)}\, {}^3{\tilde \pi}^r_{(a)}{|}_{3-O}}\over
{Q_s}}\Big].
\label{V26}
\end{eqnarray}

The first line allows to get $\partial_{\tau}\, {}^3{\hat g}_{rs}(\tau
,\vec \sigma )$ in terms of lapse, shift functions and the final
variables by using Eq.(\ref{II12}).

With the canonical transformation (\ref{V11}) the functions
${}^3\Gamma^u_{rs}$, ${}^3R_{rsuv}$, ${}^3\omega_{r(a)}$,
${}^3\Omega_{rs(a)}$ and, by using Eqs.(\ref{II12}), ${}^3K_{rs}$ (and
also the metric ADM momentum ${}^3{\tilde \Pi}^{rs}$ of
Eq.(\ref{II12})) may now be expressed in terms of $\alpha_{(a)}$,
${\tilde \pi}^{\vec
\alpha}_{(a)}$, $\xi^r$, ${\tilde \pi}^{\vec \xi}_r$, $Q_r$, ${\tilde
\Pi}^r$.

 Instead the inverse canonical transformation cannot be computed
explicitly till when one does not understand how to solve
Eqs.(\ref{IV51}). If the solution $\xi^r[{}^3e_{(c)s}]$ of this
equation would be known, then with $\alpha_{(a)}[{}^3e_{(c)s}]$, given
by Eq.(\ref{IV33}), we implicitly would get from Eqs.(\ref{V3}),
(\ref{V2}) with the 3-orthogonal parametrization  of the reduced
cotriad \footnote{$\vec \sigma (\tau ,\xi^u )$ is the inverse of
$\xi^r=\xi^r(\tau ,\vec \sigma )$.}

\beq
Q_r(\tau ,\vec \sigma ) = \Big( ({}^3R^{-1})_{(b)(a)}(\alpha_{(e)})
\Big[ ({{\partial \xi}\over {\partial \sigma}})^{-1}\Big]^s_r\, {}^3e_{(a)s}
\Big) (\tau ,\vec \sigma (\tau ,\xi^u)).
\label{V27}
\eeq

Then  Eqs.(\ref{V15}) could be inverted to get ${\tilde
\Pi}^u[{}^3e_{(a)r}, {}^3{\tilde \pi}^r_{(a)}]$.

Let us remark that if instead of the parametrization (\ref{V2})
corresponding to the choice of 3-orthogonal coordinates for
$\Sigma^{(WSW)}_{\tau}$ we had made a different choice of
3-coordinates (for instance the 3-normal ones), we would have obtained
a different quasi-Shanmugadhasan canonical transformation

\beq
\begin{minipage}[t]{3cm}
\begin{tabular}{|l|} \hline
${}^3e_{(a)r}$ \\ \hline ${}^3{\tilde \pi}^r_{(a)}$ \\ \hline
\end{tabular}
\end{minipage} \ {\longrightarrow \hspace{.2cm}} \
\begin{minipage}[t]{4 cm}
\begin{tabular}{|l|l|l|} \hline
$\alpha_{(a)}^{'}$ & $\xi^{{'}\, r}$ & $Q^{'}_r$\\ \hline
 ${\tilde \pi}^{\vec \alpha}_{(a)}$   & ${\tilde \pi}^{\vec \xi}_r$ &
 ${\tilde \Pi}^{{'}\, r}$  \\ \hline
\end{tabular}
\end{minipage}
\label{V28}
\eeq

\noindent with  ${}^3{\tilde \pi}^r_{(a)}$ given by Eq.(\ref{V15})
with different kernels.

Since there is no canonical identity for pseudo-diffeomorphisms and
since the rotations do not have zero Poisson brackets with them, in
general the point canonical transformation connecting Eqs.(\ref{V11})
and (\ref{V28}) will require a redefinition of both the parameters of
pseudo-diffeomorphisms and of the angles:

\bea
\alpha^{'}_{(a)} &=& \alpha^{'}_{(a)}[\alpha_{(b)}, \xi^s, Q^s],\nonumber \\
 \xi^{{'}\, r} &=& \xi^{{'}\, r}[\alpha_{(b)}, \xi^s, Q^s],\nonumber \\
 Q^{{'}\, r} &=& Q^{{'}\, r}[\alpha_{(b)}, \xi^s, Q^s].
\label{V29}
\eea

{}From Eqs.(\ref{II55}) and (\ref{V2}) the asymptotic behaviour at
spatial infinity of the canonical variables $Q_r$, ${\tilde \Pi}^r$
parametrizing the phase space over superspace (the space of
3-geometries) is

\bea
Q_r(\tau ,\vec \sigma )\, &{\rightarrow}_{r\, \rightarrow \infty}&\,
1+ {M\over {2r}} + O(r^{-3/2}),\nonumber \\
 {\tilde \Pi}^r(\tau ,\vec \sigma )\, &{\rightarrow}_{r\, \rightarrow \infty}&\, O(r^{-2}),
 \label{V30}
 \eea

\noindent since $\alpha_{(a)}(\tau ,\vec \sigma )\, {\rightarrow}_{r \rightarrow \infty}\,
 O(r^{-(1+\epsilon )})$.

The only non-Abelianized constraint is now the superhamiltonian one
given in Eq.(\ref{II11}). Its expression in the 3-orthogonal gauges is

\bea {\hat {\cal H}}(\tau ,\vec \sigma ) &=& \epsilon \Big[
{{c^3}\over {16\pi G}}\, {}^3e\, {}^3R -{{2\pi G }\over {c^3 \,
{}^3e}}\, {}^3G_{o(a)(b)(c)(d)}\, {}^3e_{(a)r}\, {}^3{\tilde
\pi}^r_{(b)}\, {}^3e_{(c)s}\, {}^3{\tilde \pi}^s_{(d)}\Big] (\tau
,\vec \sigma ) =\nonumber \\
 &&{}\nonumber \\
 &=& {{\epsilon \, c^3}\over {16\pi G}}
  \Big[ Q_1Q_2Q_3\, {}^3R\Big] (\tau ,\vec \sigma ) -{{\epsilon \, 2\pi G}
 \over {c^3\, [Q_1Q_2Q_3](\tau ,\vec \sigma )}}
  \sum_{rs}[Q_rQ_s](\tau ,\vec \sigma )\nonumber \\
 &&\delta_{(a)r}\Big[ \sum_u \int d^3\sigma_1 {\cal K}^r_{(b)u}(\vec \sigma
 ,{\vec \sigma}_1;\tau )\, {\tilde \Pi}^u(\tau ,{\vec \sigma}_1)+\nonumber \\
  &+&\int d^3\sigma_1
 F^r_{(b)(m)}(\vec \sigma ,{\vec \sigma}_1;\tau ) {\tilde \pi}^{\vec \alpha}_{(m)}(\tau
 ,{\vec \sigma}_1)+\nonumber \\
 &+&\sum_u \int d^3\sigma_1 G^{ru}_{(b)}(\vec \sigma ,{\vec \sigma}_1;\tau )\,
 {\tilde \pi}^{\vec \xi}_u(\tau ,{\vec \sigma}_1) \Big] \nonumber \\
 &&\delta_{(c)s}\Big[ \sum_v\int d^3\sigma_2 {\cal K}^s_{(d)v}(\vec \sigma
 ,{\vec \sigma}_2;\tau )\, {\tilde \Pi}^u(\tau ,{\vec \sigma}_2)+\nonumber \\
  &+&\int d^3\sigma_2
 F^s_{(d)(n)}(\vec \sigma ,{\vec \sigma}_2;\tau ) {\tilde \pi}^{\vec \alpha}_{(n)}(\tau
 ,{\vec \sigma}_2)+\nonumber \\
 &+&\sum_v \int d^3\sigma_2 G^{sv}_{(d)}(\vec \sigma ,{\vec \sigma}_2;\tau )\,
 {\tilde \pi}^{\vec \xi}_v(\tau ,{\vec \sigma}_2) \Big] \nonumber \\
 &&{}\nonumber \\
 &{\buildrel {3-orth.gauge} \over \rightarrow}&
  {\hat {\cal H}}_R(\tau ,\vec \sigma )=\nonumber \\
 &=&{{\epsilon \, c^3}\over {16\pi G}}
   \Big[ Q_1Q_2Q_3\, {}^3\hat R\Big] (\tau ,\vec \sigma ) -{{\epsilon \, 2\pi G}
 \over {c^3\, [Q_1Q_2Q_3](\tau ,\vec \sigma )}}\, {}^3G_{o(a)(b)(c)(d)}
 \sum_{rs} [Q_rQ_s](\tau ,\vec \sigma )\nonumber \\
 && \sum_{uv}\delta_{(a)r}\delta_{(c)s}
 \int d^3\sigma_1 {\cal K}^r_{(b)u}(\vec \sigma ,{\vec \sigma}_1;\tau )\,
 {\tilde \Pi}^u(\tau,{\vec \sigma}_1)\, \int d^3\sigma_2
  {\cal K}^s_{(d)v}(\vec \sigma ,{\vec \sigma}_2;\tau )\,
 {\tilde \Pi}^v(\tau,{\vec \sigma}_2).\nonumber \\
 &&{}
\label{V31}
\eea

\vfill\eject

\section{A New Canonical Basis and the Superhamiltonian constraint
as the Lichnerowicz Equation in the 3-Orthogonal Gauges.}

We shall now look at tetrad gravity in the quasi-Shanmugadhasan
canonical basis of Eq.(\ref{V11}) associated  with 3-orthogonal
coordinates on the WSW hypersurfaces $\Sigma^{(WSW)}_{\tau}$, namely
in the 3-orthogonal gauges of Eq.(\ref{V2}). We shall study in more
detail the six gauge fixings on the gauge variables $\alpha_{(a)}(\tau
,\vec \sigma )$, $\xi^r(\tau ,\vec \sigma )$ needed to get this gauge.
Then we shall give a more convenient canonical basis for the
superspace sector and finally we shall show how the superhamiltonian
constraint becomes the reduced Lichnerowicz equation. Also the gauge
fixing on the last gauge variable, replacing the maximal slicing
condition of the conformal approach, will be studied.

\subsection{The Gauge-Fixings for the 3-Orthogonal Gauges.}

Let us study the  phase space spanned by the canonical coordinates
$n$, ${\tilde
\pi}^n\approx 0$, $n_{(a)}$, ${\tilde \pi}^{\vec n}
_{(a)}\approx 0$, $\varphi_{(a)}$, ${\tilde \pi}^{\vec \varphi}_{(a)}\approx 0$
(for the spacetime description), and $Q_r$, ${\tilde \Pi}^r$ (for the
superspace of 3-geometries).

Let us add  the gauge-fixing constraints  on the boost parameters

\bea
\varphi_{(a)}(\tau ,\vec \sigma ) &\approx& 0,\nonumber \\
 \partial_{\tau}\varphi_{(a)}(\tau ,\vec \sigma )\, &{\buildrel \circ \over =}\,&
 \{ \varphi_{(a)}(\tau ,\vec \sigma ), {\hat H}_{(D)ADM} \} =
 \lambda^{\vec \varphi}_{(a)}(\tau ,\vec \sigma ) \approx 0,
 \label{VI1}
 \eea

\noindent namely let us restrict to
the {\it surface-forming timelike congruence  of the Eulerian
observers} at rest on $\Sigma^{(WSW)}_{\tau}$.

Since there is no canonical origin in the group of
pseudo-diffeomorphisms and since rotations do not have zero Poisson
bracket with them, we shall make the {\it convention} that in the
canonical basis (\ref{V11}) the 3-orthogonal gauges corresponding  to
Eqs.(\ref{V2}) are identified by the gauge fixings

\bea
\xi^r(\tau ,\vec \sigma ) - \sigma^r &\approx& 0,\nonumber \\
 \alpha_{(a)}(\tau ,\vec \sigma )  &\approx& 0.
 \label{VI2}
 \eea

The gauge fixings $\xi^r(\tau ,\vec \sigma ) - \sigma^r \approx 0$
are equivalent to the statement  that a 3-orthogonal coordinate
system $\vec \sigma$ is chosen as {\it reference origin} for the
pseudo-diffeomorphisms parametrized by the $\xi^r$'s, while the
gauge fixings $\alpha_{(a)}(\tau ,\vec \sigma ) \approx 0$ say
that in the associated SO(3)-principal frame bundle over
$\Sigma^{(WSW)}_{\tau}$ with these coordinates we choose the
identity cross-section as an {\it origin for the rotations}
parametrized by the $\alpha_{(a)}$'s \footnote{Definition of the
standard of {\it non rotation} of the coordinates in each point.}.
A different 3-orthogonal gauge can be defined by changing the
convention (\ref{VI2}) to $\xi^r(\tau ,\vec \sigma )-f^r(\tau
,\vec \sigma )\approx 0$, $\alpha_{(a)}(\tau ,\vec \sigma
)-g_{(a)}(\tau ,\vec \sigma ) \approx 0$ for given functions $f^r$
and $g_{(a)}$. But this would only correspond to make a point
canonical transformation from the basis (\ref{V11}) to a new basis
with the same $Q_r$ and with $\alpha_{(a)} \mapsto
\alpha^{'}_{(a)}=\alpha_{(a)}-g_{(a)}$, $\xi^r \mapsto \xi^{{'}\,
r} =\xi^r - f^r$ (in this way one could obtain {\it rotating}
3-orthogonal gauges).

The meaning of the previous gauge fixings is to restrict the
Cauchy data of cotriads on $\Sigma^{(WSW)}_{\tau}$ by eliminating
the gauge degrees of freedom of boosts, rotations and space
pseudo-diffeomorphisms, i.e. by restricting ourselves to
3-orthogonal coordinates on $\Sigma^{(WSW)}_{\tau}$ and by having
made the choice of the $\Sigma^{(WSW)}_{\tau}$-adapted tetrads
${}^4_{(\Sigma_{\tau})}{\check {\tilde E}}^A_{(\alpha )}$
\footnote{See Eqs.(\ref{II1}), (\ref{II2}) rewritten in terms of
the Dirac observables ${}^3{\hat e}^r_{(a)}$ dual to ${}^3{\hat
e}_{(a)r}$.} as the {\it reference nongeodesic congruence of
timelike Eulerian nonrotating observers} with 4-velocity field
$l^A(\tau ,\vec \sigma )$.

By remembering Eqs.(\ref{IV31}), (\ref{IV38}) and (\ref{V11}), the
Dirac brackets are strongly equal to

\begin{eqnarray}
\lbrace A(\tau ,\vec \sigma )&,&B(\tau ,{\vec \sigma}^{'})\rbrace {}^{*}=
\lbrace A(\tau ,\vec \sigma ),B(\tau ,{\vec \sigma}^{'})\rbrace +\nonumber \\
&+&\int d^3\sigma_1\Big[ \lbrace A(\tau ,\vec \sigma ),\alpha_{(a)}(\tau ,{\vec
\sigma}_1)\rbrace \lbrace {\tilde \pi}^{\vec \alpha}_{(a)}(\tau ,{\vec
\sigma}_1),B(\tau ,{\vec \sigma}^{'})\rbrace -\nonumber \\
 &-&\lbrace A(\tau ,\vec \sigma ), {\tilde \pi}^{\vec
\alpha}_{(a)}(\tau ,{\vec \sigma}_1)\rbrace
 \lbrace \alpha_{(a)}(\tau ,{\vec \sigma}_1)
 ,B(\tau ,{\vec \sigma}^{'})\rbrace +\nonumber \\
  &+&\lbrace A(\tau
,\vec \sigma ),\xi^r(\tau ,{\vec \sigma}_1)\rbrace \lbrace {\tilde
\pi}^{\vec \xi}_r(\tau ,{\vec \sigma}_1),B(\tau ,{\vec \sigma}^{'})
\rbrace -\nonumber \\
 &-&\lbrace A(\tau ,\vec \sigma ),{\tilde \pi}^{\vec \xi}_r(\tau ,{\vec
\sigma}_1)\rbrace \lbrace\xi^r(\tau ,{\vec \sigma}_1),B(\tau ,{\vec \sigma}^{'})
\rbrace \Big] \equiv \nonumber \\
 &\equiv& \lbrace A(\tau ,\vec \sigma ),B(\tau ,{\vec \sigma}^{'})\rbrace +
\nonumber \\
 &+& \int d^3\sigma_1 \Big( \, \Big[\, \lbrace A(\tau ,\vec \sigma ),\alpha
_{(a)}(\tau ,{\vec \sigma}_1)\rbrace
 \lbrace {}^3{\tilde M}_{(b)}(\tau ,{\vec
\sigma}_1),B(\tau ,{\vec \sigma}^{'})\rbrace -\nonumber \\
 &-&\lbrace A(\tau ,\vec \sigma ),{}^3{\tilde M}
_{(b)}(\tau ,{\vec \sigma}_1)\rbrace
 \lbrace \alpha_{(a)}(\tau ,{\vec \sigma}_1)
 ,B(\tau ,{\vec \sigma}^{'})\rbrace \Big] \cdot \nonumber \\
 &\cdot& A_{(b)(a)}(\alpha_{(e)}(\tau ,{\vec \sigma}_1)) + {{\partial
\sigma^s_1(\vec \xi )}\over {\partial \xi^r}}{|}_{\vec \xi =\vec \xi (\tau ,
{\vec \sigma}_1)} \cdot \nonumber \\
&\cdot& \Big[ \lbrace A(\tau ,\vec \sigma ),\xi^r(\tau ,{\vec \sigma}_1)\rbrace
\lbrace {}^3{\tilde \Theta}_s(\tau ,{\vec \sigma}_1),B(\tau ,{\vec \sigma}^{'})
\rbrace -\nonumber \\
&-&\lbrace A(\tau ,\vec \sigma ),{}^3{\tilde \Theta}_s(\tau ,{\vec
\sigma}_1)\rbrace \lbrace \xi^r(\tau ,{\vec \sigma}_1),B(\tau ,{\vec
\sigma}^{'}) \rbrace \Big] +\nonumber \\
&+&{{\partial \sigma^s_1(\vec \xi )}\over {\partial \xi^r}}{|}_{\vec \xi =\vec
\xi (\tau ,{\vec \sigma}_1)} A_{(b)(a)}(\alpha_{(e)}(\tau ,{\vec \sigma}_1))
{{\partial \alpha_{(a)}(\tau ,{\vec \sigma}_1)}\over {\partial \sigma^s_1}}
\cdot \nonumber \\
&\cdot& \Big[ \lbrace A(\tau ,\vec \sigma ),\xi^r(\tau ,{\vec \sigma}_1)\rbrace
\lbrace {}^3{\tilde M}_{(b)}(\tau ,{\vec \sigma}_1),B(\tau ,{\vec \sigma}^{'})
\rbrace -\nonumber \\
&-&\lbrace A(\tau ,\vec \sigma ),{}^3{\tilde M}_{(b)}(\tau ,{\vec
\sigma}_1)\rbrace \lbrace \xi^r(\tau ,{\vec \sigma}_1),B(\tau ,{\vec \sigma}
^{'})\rbrace \Big]\, \Big) .
\label{VI3}
\end{eqnarray}

\noindent Since the variables $\alpha_{(a)}(\tau ,\vec \sigma )$, $\xi^r(\tau ,
\vec \sigma )$, are not known as explicit functions of the cotriads, these
Dirac brackets can be used only implicitly. We must have
$\alpha_{(a)}(\tau ,\vec \sigma )\,
\rightarrow \, O(r^{-(1+\epsilon )})$ and $\xi^r(\tau ,\vec \sigma )\,
\rightarrow \, \sigma^r+O(r^{-\epsilon})$ for $r\, \rightarrow \infty$
due to  Eqs.(\ref{II55}).

We have seen in Section IV that the differential geometric description
for rotations already showed that the restriction to the identity
cross section $\alpha_{(a)}(\tau ,
\vec \sigma )=0$ implied also $\partial_r\alpha_{(a)}(\tau ,\vec \sigma )=0$;
we also have $A_{(a)(b)}(\alpha_{(e)}(\tau ,\vec \sigma )){|}_{\alpha
=0}=0$. When we add the gauge-fixings $\alpha_{(a)}(\tau ,\vec \sigma
)\approx 0$, the derivatives of all orders of $\alpha_{(a)}(\tau ,\vec
\sigma )$ weakly vanish at $\alpha_{(a)}(\tau ,\vec \sigma )=0$.
Similarly, it can be shown that, if we have the pseudo-diffeomorphism
$\vec \xi (\tau ,\vec
\sigma )=\vec \sigma + {\hat {\vec \xi}}(\tau ,\vec \sigma )$, so that
for $\vec \xi (\tau ,\vec \sigma )\rightarrow \vec \sigma $ we have
${\hat {\vec \xi}}(\tau ,\vec \sigma )
\rightarrow \delta \vec \sigma  (\tau ,\vec \sigma )$, then the quantities
${}^3e_{(a)r}(\tau ,\vec \sigma )$, $\partial_r\, {}^3e_{(a)s}(\tau ,\vec
\sigma )$, ${}^3\omega_{r(a)}(\tau ,\vec \sigma )$, ${}^3\Omega_{rs(a)}(\tau ,
\vec \sigma )$, become functions only of $Q_r(\tau ,\vec \sigma )$ and
${\tilde \Pi}^r(\tau ,\vec \sigma )$ for $\vec \xi (\tau ,\vec \sigma
)\rightarrow \vec \sigma $ and $\alpha_{(a)} (\tau ,\vec \sigma
)\rightarrow 0$ only if we have the following behaviour of the
parameters $\xi^r(\tau ,\vec \sigma )$

\begin{eqnarray}
{{\partial \delta \sigma^r(\tau ,\vec \sigma )}\over {\partial \sigma^s}}
{|}_{\vec \xi =\vec \sigma}=0 &\Rightarrow& {{\partial \xi^r(\tau ,\vec \sigma
)}\over {\partial \sigma^s}}{|}_{\vec \xi =\vec \sigma }=\delta^r_s,\quad
{{\partial^2 \xi^r(\tau ,\vec \sigma )}\over {\partial \sigma^s\partial
\sigma^u}}{|}_{\vec \xi =\vec \sigma}=0,\nonumber \\
{{\partial^2\delta \sigma^r(\tau ,\vec \sigma )}\over {\partial \sigma^u\partial
\sigma^v}}{|}_{\vec \xi =\vec \sigma }=0
&\Rightarrow& [{{\partial}\over {\partial
\sigma^u}}{{\partial \sigma^r(\tau ,\vec \sigma )}\over {\partial \xi^v}}
{|}_{\vec \xi =\vec \xi (\tau ,\vec \sigma )}]{|}_{\vec \xi =\vec \sigma }=0,
\nonumber \\
{{\partial^3\delta \sigma^r(\tau ,\vec \sigma )}\over {\partial \sigma^s
\partial \sigma^u\partial \sigma^v}}{|}_{\vec \xi =\vec \sigma }=0
&\Rightarrow& [{{\partial^2}\over
{\partial \sigma^u\partial \sigma^v}}{{\partial \sigma^r(\vec \xi)}\over
{\partial \xi^s}}{|}_{\vec \xi =\vec \xi (\tau ,\vec \sigma )}]{|}_{\vec \xi =
\vec \sigma }=0.
\label{VI4}
\end{eqnarray}

\noindent These conditions should be satisfied by the parameters of
pseudo-diffeomorphisms near the identity, i.e. near the chart chosen
as reference chart (the 3-orthogonal one in this case). With the
gauge-fixings $\xi^r(\tau ,\vec \sigma )\approx \vec \sigma$ all these
properties are satisfied.

Let us remember the Dirac Hamiltonian (\ref{IV57}) valid on WSW
hypersurfaces $\Sigma^{(WSW)}_{\tau}$ before going to the rest-frame
instant form with ${\tilde \lambda}_{\tau}(\tau )=\epsilon$, ${\tilde
\lambda}_r(\tau )=0$

\bea {\hat H}_{(D)ADM}&=&\int d^3\sigma [n {\hat {\cal H}}+{\tilde
n}^r{\tilde \pi}^{\vec \xi}_r+\lambda_n{\tilde \pi}
^n+\lambda^{\vec n}_{(a)}{\tilde \pi}^{\vec n}_{(a)}+
\lambda^{\vec \varphi}_{(a)}{\tilde \pi}^{\vec \varphi}_{(a)}
+{\tilde \mu}_{(a)}{\tilde \pi}^{\vec \alpha}_{(a)}](\tau ,\vec
\sigma )-\nonumber \\
 &-&{\tilde \lambda}_{\tau}(\tau ) [\epsilon_{(\infty )} -
 {\hat P}^{\tau}_{ADM}] +{\tilde \lambda}_r(\tau ) {\hat P}^r_{ADM},
 \label{VI5}
 \eea

\noindent
with ${\tilde n}^r=n_{(a)}\, {}^3e^s_{(a)}{{\partial \xi^r}
\over {\partial \sigma^s}}=n_u\, {}^3g^{uv} {{\partial \xi^r}\over {\partial
\sigma^v}}=n^v {{\partial \xi^r}\over {\partial \sigma^v}}$.

The time constancy of the gauge fixings $\alpha_{(a)}(\tau ,\vec
\sigma )\approx 0$, $\xi^r(\tau ,\vec \sigma )-\sigma^r \approx 0$
[implying ${\tilde n}^r\approx n^r$] gives\footnote{As shown in
Ref.\cite{ber1}, when one has a chain of a primary and a secondary
first class constraint, both of them appear in the Dirac
Hamiltonin: the primary constraint with in front the arbitrary
Dirac multiplier and the secondary one with in front some function
of the canonical variables not determined by the Hamilton
equations. To make the gauge fixing one has to follow
Ref.\cite{giap}: i) add a gauge fixing $\chi_1\approx 0$ to the
secondary constraint; ii) find the implied gauge fixing
$\chi_2\approx 0$ for the primary constraint by requiring the time
constancy of $\chi_1\approx 0$; iii) determine the Dirac
multiplier by requiring the time constancy of $\chi_2\approx 0$.}

\bea
\partial_{\tau} \alpha_{(a)}(\tau ,\vec \sigma )
\, &{\buildrel \circ \over =}\,&
\{ \alpha_{(a)}(\tau ,\vec \sigma ), {\hat H}_{(D)ADM} \} =
 {\tilde \mu}_{(a)}(\tau ,\vec \sigma )  \approx 0,  \nonumber \\
 &&{}\nonumber \\
 \rightarrow && {\tilde \mu}_{(a)}(\tau ,\vec \sigma )\quad \text{determined},
\label{VI6}
\eea

\bea
\partial_{\tau} [\xi^r(\tau ,\vec \sigma )&-&\sigma^r]\, {\buildrel \circ \over
=}\, \{ \xi^r(\tau ,\vec \sigma ), {\hat H}_{(D)ADM} \} \approx n^{s}(\tau ,\vec
\sigma )+\nonumber \\
&+&\int d^3\sigma_1 n(\tau ,{\vec \sigma}_1) \{ \xi^r(\tau ,\vec
\sigma ),{\hat {\cal H}}(\tau ,{\vec \sigma}_1)\}
-{\tilde \lambda}_A(\tau ) \{ \xi^r(\tau ,\vec \sigma ),{\hat
P}^A_{ADM} \} \approx 0,\nonumber \\
 &&{}\nonumber \\
 \Rightarrow&& \quad n_r(\tau ,\vec \sigma )-{\hat n}_r(\tau ,\vec \sigma |r
_{\bar a}, \pi_{\bar a}, {\tilde \lambda}_A] \approx 0,\nonumber \\
 &&{}\nonumber \\
 &&{\hat n}_r(\tau ,\vec \sigma |r
_{\bar a}, \pi_{\bar a}, {\tilde \lambda}_A] ={}^3g_{rs}
(\tau ,\vec \sigma ) \Big[ {\tilde \lambda}_A(\tau ) {{\delta {\hat
P}^A_{ADM}}\over {\delta {\tilde \pi}_s^{\vec \xi}(\tau ,\vec \sigma
 )}}- \nonumber \\
 &-&\int d^3\sigma_1 n(\tau ,{\vec \sigma}_1) {{\delta {\hat {\cal H}}(\tau
 ,{\vec \sigma}_1)}\over {\delta {\tilde \pi}_s^{\vec \xi}(\tau ,\vec \sigma )}}\Big],
 \nonumber \\
 &&\Downarrow \nonumber \\
  \partial_{\tau}
\Big[ n_r(\tau ,\vec \sigma )&&-{\hat n}_r(\tau ,\vec \sigma |r
_{\bar a}, \pi_{\bar a}, {\tilde \lambda}_A] \Big] \approx
\nonumber \\ &&\approx [\lambda^{\vec n}_{(a)}\, {}^3{\hat
e}_{(a)r}](\tau ,\vec \sigma )-\{ {\hat n}_r(\tau ,\vec \sigma |r
_{\bar a}, \pi_{\bar a}, {\tilde \lambda}_A] , {\hat H}_{(D)ADM}
\} \approx 0,\nonumber \\ &&{}\nonumber \\ \Rightarrow&& \quad
\lambda^{\vec n}_{(a)}(\tau ,\vec \sigma )\quad \text{determined}.
\label{VI7} \eea

Therefore, the {\it shift functions do not vanish} in the 3-orthogonal
gauges avoiding the {\it synchronous} coordinates of $M^4$ with their
tendency to develop coordinate singularities in short times
\cite{misner,lifsh}. This shows the presence of gravitomagnetism and
the non-validity of Einstein simultaneity convention in these gauges,
since ${}^4g_{\tau r}\not= 0$. See Ref.\cite{yoyo} for the problems of
the fixation of the lapse and shift functions in ADM metric gravity
(coordinate conditions to rebuild spacetime) and for the origin of the
coordinates used in numerical gravity (see Ref.\cite{nume} for a
recent review of it). Let us remark that in Refs.\cite{york1} the
Einstein's equations corresponding to the supermomentum constraints of
metric gravity are thought as elliptic equations for the shift
functions in the framework either of a new conformal thin sandwich
formulation or of an enlargement of Einstein's equations to get a
manifestly hyperbolic system of equations: this approach, even if
mathematically legitimate, is completely innatural from the
Hamiltonian point of view with its interpretation based on constraint
theory.

In the rest-frame instant form of tetrad gravity with ${\tilde
\lambda}_A(\tau )=(\epsilon ;\vec 0)$ we get
\footnote{From Eqs.(\ref{II25}), (\ref{II11}) and from ${}^3{\tilde
\Pi}^{rs}={1\over 4}({}^3e^r_{(a)}\, {}^3{\tilde \pi}^s_{(a)}+{}^3e^s_{(a)}\,
{}^3{\tilde \pi}^r_{(a)})$ we get ${{\delta {\hat
P}^{\tau}_{ADM}}\over {\delta {\tilde \pi}_s^{\vec \xi}(\tau ,\vec
\sigma )}}=\int d^3\sigma_1 {{\delta {\hat {\cal H}}(\tau ,{\vec \sigma}_1)}\over {\delta
{\tilde \pi}_s^{\vec \xi}(\tau ,\vec \sigma )}}$.}

\bea
 &&n_r(\tau ,\vec \sigma ) \approx  {}^3e_{(a)r}(\tau ,\vec
\sigma )\, {}^3e_{(a)s}(\tau ,\vec \sigma ) \int d^3\sigma_1
[\epsilon -n(\tau ,{\vec \sigma}_1)] {{\delta {\hat {\cal H}}(\tau
,{\vec \sigma}_1)}\over {\delta {\tilde \pi}_s^{\vec \xi}(\tau
,\vec \sigma )}}\nonumber \\
  &{\buildrel {3-orth\, gauge} \over \rightarrow}& Q^2_r(\tau
  ,\vec \sigma ) \int d^3\sigma_1
[\epsilon -n(\tau ,{\vec \sigma}_1)] {{\delta {\hat {\cal H}}(\tau
,{\vec \sigma}_1)}\over {\delta {\tilde \pi}_r^{\vec \xi}(\tau
,\vec \sigma )}}{|}_{3-0},\nonumber \\
 &&{}\nonumber \\
 &&\text{with}\nonumber \\
 &&{}\nonumber \\
  &&{{\delta {\hat {\cal H}}(\tau ,{\vec \sigma}_1)}\over {\delta
{\tilde \pi}_s^{\vec \xi}(\tau ,\vec \sigma )}}=\int d^3\sigma_2
{{\delta {\hat {\cal H}}(\tau ,{\vec \sigma}_1)}\over {\delta {\tilde
\pi}_{(d)}^u(\tau ,{\vec \sigma}_2)}}
{{\delta {\tilde \pi}^u_{(d)}(\tau ,{\vec \sigma}_2)}\over {\delta
{\tilde \pi}_s^{\vec \xi}(\tau ,\vec \sigma )}}=\nonumber \\
 &=&-{{\epsilon \, 4\pi G}\over {c^3\, {}^3e(\tau ,{\vec \sigma}_1)}}\,
 {}^3G_{o(a)(b)(c)(d)}\, [{}^3e_{(a)v}\,
 {}^3{\tilde \pi}^v_{(b)}\, {}^3e_{(c)u}](\tau ,{\vec \sigma}_1)\,
 G^{us}_{(d)}({\vec \sigma}_1, \vec \sigma ;\tau ) \nonumber \\
  &{\buildrel {3-orth\, gauge} \over \rightarrow}&
 -{{\epsilon \, 2\pi G}\over {c^3\, [Q_1Q_2Q_3](\tau ,{\vec \sigma}_1)}}\,
 {}^3G_{o(a)(b)(c)(d)}\nonumber \\
 &&\sum_{wu} [Q_wQ_u](\tau ,{\vec \sigma}_1) \delta_{(a)w}\sum_v\int d^3\sigma_2
 {\cal K}^w_{(b)v}({\vec \sigma}_1,{\vec \sigma}_2;\tau )
 {\tilde \Pi}^v(\tau ,{\vec \sigma}_2)\, \delta_{(c)u} G^{us}_{(d)}({\vec
 \sigma}_1, \vec \sigma ;\tau ),
 \label{VI8}
\eea

\noindent where we used Eqs.(\ref{II11}) and (\ref{V12}). We see
that the shift functions in the 3-orthogonal gauges {\it depend
from the not yet determined lapse function $n(\tau ,\vec \sigma
)$}. Only after having added the gauge fixing to the
superhamiltonian constraint, namely only after having selected a
completely fixed 3-orthogonal gauge, the lapse and shift functions
will be determined. This is a feature common to all the completely
fixed gauges.

The Dirac Hamiltonian in the 3-orthogonal gauges, but not yet in the
rest-frame instant form, is

\beq
{\hat H}_{(D)ADM,R}=\int d^3\sigma [n {\hat {\cal H}}_R+\lambda_n{\tilde \pi}
^n](\tau ,\vec \sigma )-{\tilde \lambda}_{\tau}(\tau ) [\epsilon_{(\infty )} -
 {\hat P}^{\tau}_{ADM}] +{\tilde \lambda}_r(\tau ) {\hat P}^r_{ADM},
 \label{VI9}
 \eeq

\noindent
where ${\hat {\cal H}}_R$ is the reduced superhamiltonian constraint.

\subsection{A New Canonical Basis for Superspace in the
         3-Orthogonal Gauges.}

Let us now consider a new canonical transformation from the basis
$Q_u(\tau ,\vec \sigma )$, ${\tilde \Pi}^u(\tau ,\vec \sigma )$
\footnote{Where ${}^3{\hat e}_{(a)r}(\tau ,\vec \sigma
)=\delta_{(a)r}Q_r(\tau ,\vec \sigma )$, see Eq.(\ref{V2}).} to a
new basis $q_u(\tau ,\vec \sigma )$, $\rho_u(\tau ,\vec \sigma )$
defined in the following way

\begin{eqnarray}
q_u(\tau ,\vec \sigma )&=&ln\, Q_u(\tau ,\vec \sigma ),\qquad Q_u(\tau
 ,\vec \sigma )=e^{q_u(\tau ,\vec \sigma )},\nonumber \\
 \rho_u(\tau ,\vec \sigma )&=&Q_u(\tau ,\vec \sigma )\, {\tilde \Pi}^u(\tau ,
\vec \sigma ),\quad\quad {\tilde \Pi}^u(\tau ,\vec \sigma )=e^{-q_u(\tau ,\vec
\sigma )}\, \rho_u(\tau ,\vec \sigma ),\nonumber \\
 &&{}\nonumber \\
&&\lbrace q_u(\tau ,\vec \sigma ),\rho_v(\tau ,{\vec \sigma }^{'})\rbrace =
\delta_{uv}\delta^3(\vec \sigma ,{\vec \sigma }^{'}).
\label{VI10}
\end{eqnarray}

It is convenient to make one more canonical transformation, like for
the determination of the center of mass of a particle
system\cite{lus1}, to the following new canonical basis

\bea
&&q(\tau ,\vec \sigma )={1\over 3}\sum_u q_u(\tau ,\vec \sigma )=
{1\over 3} \sum_u ln\, Q_u(\tau ,\vec \sigma )\, {\rightarrow}_{r\,
\rightarrow \infty}\, {M\over {2r}} + O(r^{-3/2}) ,\nonumber \\
&&r_{\bar a}(\tau ,\vec \sigma )=\sqrt{3} \sum_u \gamma_{\bar au}
q_u(\tau ,\vec \sigma )=\sqrt{3}\sum_u\gamma_{\bar au} ln\, Q_u(\tau
,\vec \sigma )\, {\rightarrow}_{r\, \rightarrow \infty}\,
O(r^{-3/2}),\nonumber \\
 &&\quad\quad \bar a=1,2,\nonumber \\
 &&\rho (\tau ,\vec
\sigma )=\sum_u \rho_u(\tau ,\vec \sigma )=\sum_u [Q_u{\tilde
\Pi}^u](\tau ,\vec \sigma )\, {\rightarrow}_{r\, \rightarrow \infty}\, O(r^{-2}),\nonumber \\
 &&\pi_{\bar a}(\tau ,\vec
\sigma )={1\over {\sqrt{3}}} \sum_u \gamma_{\bar au}
\rho_u(\tau ,\vec \sigma )={1\over {\sqrt{3}}}\sum_u\gamma_{\bar au}
[Q_u{\tilde \Pi}^u](\tau ,\vec \sigma )\, {\rightarrow}_{r\,
\rightarrow \infty}\, O(r^{-2}),\nonumber \\
 &&\quad\quad \bar a=1,2,\nonumber \\
 &&{}\nonumber \\
 &&\lbrace q(\tau ,\vec \sigma ),\rho (\tau ,{\vec \sigma }^{'})\rbrace =
\delta^3(\vec \sigma ,{\vec \sigma }^{'}),\quad\quad
\lbrace r_{\bar a}(\tau ,\vec \sigma ),\pi_{\bar b}(\tau ,{\vec \sigma }^{'})
\rbrace =\delta_{\bar a\bar b}\delta^3(\vec \sigma ,{\vec \sigma }^{'}),
\nonumber \\
 &&{}\nonumber \\
 &&{}\nonumber \\
&&q_u(\tau ,\vec \sigma )=q(\tau ,\vec \sigma )+{1\over {\sqrt{3}}} \sum_{\bar
a}\gamma_{\bar au} r_{\bar a}(\tau ,\vec \sigma ),\quad\quad
Q_u(\tau ,\vec \sigma )=e^{q_u(\tau ,\vec \sigma )},\nonumber \\
&&\rho_u(\tau ,\vec \sigma )={1\over 3} \rho(\tau ,\vec \sigma )+\sqrt{3}
\sum_{\bar a} \gamma_{\bar au} \pi_{\bar a}(\tau ,\vec \sigma ),\quad\quad
{\tilde \Pi}^u(\tau ,\vec \sigma )=[e^{-q_u} \rho_u](\tau ,\vec \sigma ),
\nonumber \\
 &&{}\nonumber \\
 &&{}\nonumber \\
&&\begin{minipage}[t]{2cm}
\begin{tabular}{|l|} \hline
${}^3e_{(a)r}$ \\ \hline ${}^3{\tilde \pi}^r_{(a)}$ \\ \hline
\end{tabular}
\end{minipage} \ {\longrightarrow \hspace{.2cm}} \
\begin{minipage}[t]{4 cm}
\begin{tabular}{|l|l|l|l|} \hline
$\alpha_{(a)}$ & $\xi^{r}$ & $q$ & $r_{\bar a}$\\ \hline
 ${\tilde \pi}^{\vec \alpha}_{(a)}$   & ${\tilde \pi}^{\vec \xi}_r$ &
 $\rho$ & $\pi_{\bar a}$ \\ \hline
\end{tabular}
\end{minipage} \ {\longrightarrow \hspace{.2cm}} \
\begin{minipage}[t]{4 cm}
\begin{tabular}{|l|l|l|l|} \hline
$\alpha_{(a)}$ & $\xi^{r}$ & $\phi$ & $r_{\bar a}$\\ \hline
 ${\tilde \pi}^{\vec \alpha}_{(a)}$   & ${\tilde \pi}^{\vec \xi}_r$ &
 $\pi_{\phi}$ & $\pi_{\bar a}$ \\ \hline
\end{tabular}
\end{minipage} ,\nonumber \\
 &&{}\nonumber \\
 &&{}\nonumber \\
\phi (\tau ,\vec \sigma )&=& e^{{1\over 2}q(\tau ,\vec \sigma )}\,
{\rightarrow}_{r\, \rightarrow \infty}\, 1 + {M\over {4r}}
+O(r^{-3/2}),\quad\quad \pi_{\phi}(\tau ,\vec \sigma )= 2 e^{{1\over
2}q(\tau ,\vec \sigma )} \rho (\tau ,\vec \sigma ),\nonumber \\
 &&\{ \phi (\tau ,\vec \sigma ), \pi_{\phi}(\tau ,{\vec \sigma}^{'}) \}
 = \delta^3(\vec \sigma ,{\vec \sigma}^{'}),\nonumber \\
 &&{}\nonumber \\
 &&{}\nonumber \\
 Q_u(\tau ,\vec \sigma ) &=& \Big( e^{q+{1\over {\sqrt{3}}}
 \sum_{\bar a}\gamma_{\bar au}r_{\bar a}}\Big) (\tau ,\vec \sigma ) =
 \phi^2(\tau ,\vec \sigma ) e^{{1\over {\sqrt{3}}}\sum_{\bar a}
 \gamma_{\bar au}r_{\bar a}(\tau ,\vec \sigma )},\nonumber \\
 {\tilde \Pi}^u(\tau ,\vec \sigma ) &=& e^{-q(\tau ,\vec \sigma )}e^{-{1\over
 {\sqrt{3}}}\sum_{\bar a}\gamma_{\bar au}r_{\bar a}(\tau ,\vec \sigma )}
 [{1\over 3}\rho +\sqrt{3}\sum_{\bar b}\gamma_{\bar bu}\pi_{\bar b}](\tau ,\vec \sigma )=
 \nonumber \\
 &=&\phi^{-2}(\tau ,\vec \sigma )e^{-{1\over
 {\sqrt{3}}}\sum_{\bar a}\gamma_{\bar au}r_{\bar a}(\tau ,\vec \sigma )}
 [{1\over 6}\phi^{-1}\pi_{\phi} +
 \sqrt{3}\sum_{\bar b}\gamma_{\bar bu}\pi_{\bar b}](\tau ,\vec \sigma ). \nonumber \\
 &&
\label{VI11}
\eea

\noindent where $\gamma_{\bar au}$ are numerical constants satisfying\cite{lus1}

\begin{equation}
\sum_u \gamma_{\bar au}=0,\quad\quad \sum_u\, \gamma_{\bar au}\gamma_{\bar bu}
=\delta_{\bar a\bar b},\quad\quad \sum_{\bar a}\, \gamma_{\bar au}\gamma
_{\bar av}=\delta_{uv}-{1\over 3}.
\label{VI12}
\end{equation}

Therefore, Eqs.(\ref{VI11}) define a one-parameter family of canonical
transformations, one for each solution of Eqs.(\ref{VI12}).

In Eq.(\ref{VI10}) we have also shown the canonical transformation
from the canonical pair $q$, $\rho$ to the canonical pair $\phi$,
$\pi_{\phi}$ consisting in the conformal factor of the 3-metric and
its conjugate momentum.

In terms of these variables we have [$N_{(as)}=-{\tilde
\lambda}_{\tau}(\tau )$, $N_{(as)r}=-{\tilde \lambda}_r(\tau )$]

\begin{eqnarray}
{}^3{\hat g}_{rs}&=&e^{2q}\, \left( \begin{array}{ccc} e^{ {2\over {\sqrt{3}}}
\sum_{\bar a}\gamma_{\bar a1}r_{\bar a}}&0&0\\ 0& e^{ {2\over {\sqrt{3}}}
\sum_{\bar a}\gamma_{\bar a2}r_{\bar a}}&0\\ 0&0& e^{ {2\over {\sqrt{3}}}
\sum_{\bar a}\gamma_{\bar a3}r_{\bar a}} \end{array} \right) =
\phi^4\,\, {}^3{\hat g}^{diag}_{rs},
\nonumber \\
&&{}\nonumber \\
\hat \gamma &=&{}^3\hat g= {}^3{\hat e}^2=e^{6q}= \phi^{12},\quad\quad
det\, |{\hat g}^{diag}_{rs}| =1,\nonumber \\
 &&{}\nonumber \\
  d{\hat s}^2&=&\epsilon \Big( [N_{(as)}+n]^2-e^{-2q} \sum_u e^{ -{2\over
{\sqrt{3}}}\sum_{\bar a} \gamma_{\bar au} r_{\bar a}} [N_{(as)
u}+n_u]^2\Big) (d\tau )^2 -\nonumber \\
 &&-2\epsilon
[N_{(as)r}+n_r]d\tau d\sigma^r-\epsilon e^{2q} \sum_u e^{ {2\over
{\sqrt{3}}}\sum_{\bar a} \gamma_{\bar au} r_{\bar a}}
(d\sigma^u)^2=\nonumber \\
 &&=\epsilon \Big( [N_{(as)}+n]^2(d\tau )^2-\nonumber \\
 &&-\delta_{uv}[\phi^2 e^{ {2\over {\sqrt{3}}}\sum_{\bar
a} \gamma_{\bar au} r_{\bar a}} d\sigma^u+\phi^{-2} e^{- {2\over
{\sqrt{3}}}\sum_{\bar a} \gamma_{\bar au} r_{\bar a}}(N_{(as)u}+n_u)
d\tau ]\nonumber \\
 &&[\phi^2 e^{ {2\over {\sqrt{3}}}\sum_{\bar a}
\gamma_{\bar av} r_{\bar a}} d\sigma^v+\phi^{-2} e^{- {2\over
{\sqrt{3}}}\sum_{\bar a} \gamma_{\bar av} r_{\bar a}}(N_{(as)v}+n_v)
d\tau ] \Big),\nonumber \\
 &&{}\nonumber \\
 q&=& 2 ln\, \phi = {1\over 6}\,
ln\, {}^3\hat g,\nonumber \\ r_{\bar a}&=&{{\sqrt{3}}\over 2} \sum_r
\gamma_{\bar ar}\, ln\, {{{}^3{\hat g}
_{rr}}\over {{}^3\hat g}}.
\label{VI13}
\end{eqnarray}

We see that the freedom in the choice of the solutions of
Eqs.(\ref{VI12}) allows to put two of the diagonal elements of
${}^3{\hat g}^{diag}_{rs}$ equal but only in one point $\vec
\sigma$. To select which diagonal elements is convenient to make
equal, we should need an {\it intrinsic 2+1 splitting of the WSW
hypersurfaces}. In Ref.\cite{ckl} there is the statement that the
independent degrees of freedom of the gravitational field are
described by symmetric trace-free 2-tensors on 2-planes. On the
other hand, the tangent plane in each point of the WSW
hypersurfaces is naturally decomposed in the {\it gauge direction}
identified by the shift functions $n_r(\tau ,\vec \sigma )$ in
that point and in the orthogonal 2-plane. However, this has to be
done after the addition of the gauge fixing to the
superhamiltonian constraint, so that the lapse and shift functions
are determined. Moreover, Eqs.(\ref{VI11}) and (\ref{VI12}) should
be generalized so that the $\gamma_{\bar au}$'s become point
dependent. This will be studied in a future paper, because the
accomplishment of this result would identify the tensoriality
quoted in Ref.\cite{ckl} for the Dirac observables $r_{\bar
a}(\tau ,\vec \sigma )$, which, in each point, would be functions
only of the physical 2-plane in that point. Let us remark that
this type of $3+1 \rightarrow 2+1+1$ splitting is reminiscent of
the 2+2 splittings of Refs.\cite{inverno}.

Another possible canonical basis could be obtained by the point
canonical transformation $\phi , r_{\bar a}\, \mapsto \, \phi ,
\lambda_i^{CY}$ ($i=1,2$), where the $\lambda_i^{CY}$ are two
independent eigenvalues of the Cotton-York 3-tensor defined in
Appendix A (this tensor has only two independent components and
vanishes for $r_{\bar a} \rightarrow \infty$, namely when the
hypersurfaces are 3-conformally flat).

In terms of the variables $q=2 ln\, \phi, r_{\bar a}$, we have the
following results \footnote{Use is done of Eqs.(A25) of
Ref.\cite{ru11}; see also Appendix A.} \footnote{$\gamma_{PP_1}$
is the geodesic between P and $P_1$ for the 3-metric; $\phi
=e^{q/2}$, $\rho = \pi_{\phi}/2 \phi$.}

\begin{eqnarray}
{}^3{\hat e}_{(a)r}&=& \delta_{(a)r}e^{q_r}=\delta_{(a)r} e^{q+
{1\over {\sqrt{3}}}\sum_{\bar a}\gamma_{\bar ar}r_{\bar a}}=
 \delta_{(a)r} \phi^2 e^{{1\over {\sqrt{3}}}\sum_{\bar a}
\gamma_{\bar ar}r_{\bar a}}\nonumber \\
 &&{\rightarrow}_{r_{\bar a}\rightarrow 0}\,
 \delta_{(a)r} e^q =\delta_{(a)r} \phi^2 \nonumber \\
 &&{\rightarrow}_{q,r_{\bar a}\, \rightarrow 0}\,  \delta_{(a)r},\nonumber \\
 &&{\rightarrow}_{q\, \rightarrow 0}\, \delta_{(a)r} e^{{1\over
{\sqrt{3}}}\sum_{\bar a}\gamma_{\bar ar}r_{\bar a}},\nonumber \\
 &&{}\nonumber \\
 {}^3{\hat e}^r_{(a)}&=&\delta_{(a)}^r e^{-q_r}=
 \delta_{(a)}^r e^{-q-{1\over {\sqrt{3}}}\sum_{\bar a}
\gamma_{\bar ar}r_{\bar a}} =
 \delta^r_{(a)} \phi^{-2} e^{-{1\over {\sqrt{3}}}\sum_{\bar a}
\gamma_{\bar ar}r_{\bar a}}\nonumber \\
 &&{\rightarrow}_{r_{\bar a}\rightarrow 0}\,
  \delta_{(a)}^r e^{-q}= \delta^r_{(a)}\phi^{-2}\nonumber \\
 &&{\rightarrow}_{q,r_{\bar a}\, \rightarrow 0}\,
 \delta^r_{(a)},\nonumber \\
 &&{\rightarrow}_{q\, \rightarrow 0}\,
 \delta^r_{(a)} e^{-{1\over {\sqrt{3}}}\sum
_{\bar a}\gamma_{\bar ar}r_{\bar a}},\nonumber \\
&&{}\nonumber \\
 {}^3{\hat g}_{rs}&=&\delta_{rs} e^{2q_r}=\delta_{rs}
e^{2q+{2\over {\sqrt{3}}}
\sum_{\bar a}\gamma_{\bar ar}r_{\bar a}} = \delta_{rs} \phi^4
e^{{2\over {\sqrt{3}}}\sum_{\bar a}\gamma_{\bar ar}r_{\bar
a}}\nonumber \\
 &&{\rightarrow}_{r_{\bar a}\rightarrow 0}\,
\delta_{rs}e^{2q}=\delta_{rs} \phi^4\, {\rightarrow}
_{q\, \rightarrow 0}\, \delta_{rs},\quad {\rightarrow}_{q\, \rightarrow 0}\,
\delta_{rs} e^{{2\over {\sqrt{3}}}\sum_{\bar a}\gamma_{\bar ar}r_{\bar a}}
,\nonumber \\
 {}^3{\hat g}^{rs}&=&\delta^{rs} e^{-2q_r}=\delta^{rs}
e^{-2q-{2\over {\sqrt{3}}}
\sum_{\bar a}\gamma_{\bar ar}r_{\bar a}} =\delta^{rs} \phi^{-4}
e^{- {2\over {\sqrt{3}}}\sum_{\bar a}\gamma_{\bar ar}r_{\bar
a}}\nonumber \\
 &&{\rightarrow}_{r_{\bar a}\rightarrow 0}\,
\delta^{rs}e^{-2q}=\delta^{rs} \phi^{-4}\, {\rightarrow}
_{q\, \rightarrow 0}\, \delta^{rs},\quad {\rightarrow}_{q\, \rightarrow 0}\,
\delta^{rs} e^{-{2\over {\sqrt{3}}}\sum_{\bar a}\gamma_{\bar ar}r_{\bar a}}
,\nonumber \\
 &&{}\nonumber \\
  {}^3\hat e &=& \sqrt{\hat
\gamma}=e^{\sum_rq_r}=e^{3q}=\phi^6\, {\rightarrow}
_{q\, \rightarrow 0}\, 1,\nonumber \\
 &&{}\nonumber \\
 {}^3{\hat \Gamma}^r_{uv}&=& -\delta_{uv}
\sum_s \delta^r_s e^{2(q_u-q_s)} \partial_sq_u
+\delta^r_u \partial_vq_u +\delta^r_v \partial_uq_v=\nonumber \\
&&=-\delta_{uv}\sum_s\delta^r_s e^{{2\over {\sqrt{3}}}\sum_{\bar a}(\gamma
_{\bar au}-\gamma_{\bar as})r_{\bar a}} \Big[ 2\partial_sln\, \phi +{1\over {\sqrt{3}}}
\sum_{\bar b}\gamma_{\bar bu}\partial_sr_{\bar b}\Big] +\nonumber \\
&&+\delta^r_u\Big[ 2\partial_vln\, \phi +{1\over {\sqrt{3}}}\sum_{\bar
a}\gamma_{\bar au}
\partial_vr_{\bar a}\Big] +\delta^r_v\Big[ 2\partial_uln\, \phi +{1\over {\sqrt{3}}}
\sum_{\bar a}\gamma_{\bar av}\partial_ur_{\bar a}\Big] \nonumber \\
&&{\rightarrow}_{r_{\bar a}\rightarrow 0}\,
2[-\delta_{uv}\sum_s\delta^r_s
\partial_sln\, \phi +\delta^r_u \partial_vln\, \phi
 +\delta^r_v \partial_uln\, \phi \, {\rightarrow}_{q\,
\rightarrow 0}\, 0,\nonumber \\
&&{\rightarrow}_{q\, \rightarrow 0}\, {1\over {\sqrt{3}}}
\Big( -\delta_{uv}\sum_s
\delta^r_s e^{{2\over {\sqrt{3}}}\sum_{\bar a}(\gamma_{\bar au}-\gamma
_{\bar as})r_{\bar a}}\sum_{\bar b}\gamma_{\bar bu}\partial_sr_{\bar b}+\nonumber \\
 &+&\sum
_{\bar a}\Big[ \delta^r_u\gamma_{\bar au}\partial_vr_{\bar a}+\delta^r_v\gamma
_{\bar av}\partial_ur_{\bar a}\Big] \Big), \nonumber \\
 &&{}\nonumber \\
 &&\sum_u\, {}^3{\hat \Gamma}^u_{uv} = \partial_v \sum_u q_u =
3\partial_v q= 6 \partial_v ln\, \phi,
\nonumber \\
 &&{}\nonumber \\
 {}^3{\hat \omega}_{r(a)} &=& \epsilon_{(a)(b)(c)} \,
 \delta_{(b)r}\delta_{(c)u} \, e^{q_r-q_u} \partial_uq_r=\nonumber \\
 &=& \epsilon_{(a)(b)(c)} \,
 \delta_{(b)r}\delta_{(c)u} \,  e^{{1\over {\sqrt{3}}}\sum
_{\bar a}(\gamma_{\bar ar}-\gamma_{\bar au})r_{\bar a}}\Big[ 2\partial_uln\, \phi +{1
\over {\sqrt{3}}}\sum_{\bar b}\gamma_{\bar br} \partial_ur_{\bar b}\Big]  \Big]
\nonumber \\
&&{\rightarrow}_{r_{\bar a}\rightarrow 0}\, \epsilon_{(a)(b)(c)}
\, \delta_{(b)r}\delta_{(c)u}\, 2\partial_uln\, \phi  \,
 {\rightarrow}_{q\, \rightarrow 0}\, 0,\nonumber \\
&&{\rightarrow}_{q\, \rightarrow 0}\, {1\over {\sqrt{3}}}
\epsilon_{(a)(b)(c)} \sum_u \delta_{(b)r}\delta_{(c)u}\,
e^{{1\over {\sqrt{3}}}\sum_{\bar a} (\gamma_{\bar ar}-\gamma_{\bar
au})r_{\bar a}}\sum_{\bar b}\gamma_{\bar br}
\partial_ur_{\bar b},\nonumber \\
 &&{}\nonumber \\
\zeta^{(\hat \omega ) r}_{(a)(b)}(\vec \sigma ,{\vec \sigma}_1;\tau )
 &=& d^r_{\gamma_{PP_1}}(\vec
\sigma ,{\vec \sigma}_1) \Big( P_{\gamma_{PP_1}}\, e^{\int
^{\vec \sigma}_{{\vec \sigma}_1}d\sigma_2^w\, {}^3{\hat \omega}_{w(c)}(\tau ,
{\vec \sigma}_2){\hat R}^{(c)} }\, \Big)_{(a)(b)}\nonumber \\
 {\rightarrow}_{q,r_{\bar a} \rightarrow 0}&& \zeta^{(o)r}_{(a)(b)}(\vec \sigma ,{\vec \sigma}_1)
 =-{{\sigma^r-\sigma_1^r}\over {4\pi |\vec
 \sigma -{\vec \sigma}_1|^3}} \delta_{(a)(b)}=-\delta_{(a)(b)}
 {{\partial}\over {\partial \sigma_1^r}}{1\over {4\pi |\vec \sigma
 -{\vec \sigma}_1|}}.
 \label{VI14}
\end{eqnarray}

See Appendix A for the expression of ${}^3\Omega_{rs(a)}$,
${}^3R_{rsuv}$, ${}^3R_{rs}$, ${}^3R$  \footnote{We find that
${}^3\omega_{r(a)}$ goes  as $O(r^{-2})$, while ${}^3{\hat
R}_{rsuv}(\tau ,\vec \sigma )$ and ${}^3{\hat
\Omega}_{rs(a)}(\tau ,\vec \sigma )$ go as $O(r^{-3})$ for $r\,
\rightarrow \infty$.} and of other 3-tensors.

Moreover from Eqs.(\ref{V15})-(\ref{V20}), by choosing the
representation of the cotriad momentum satisfying ${\hat {\cal
H}}_{(a)}(\tau ,\vec \sigma ) =0$, we find

\begin{eqnarray}
{}^3{\hat {\tilde \pi}}^r_{(a)}(\tau ,\vec \sigma )&=& {}^3{\tilde
\pi}^r_{(a)}(\tau ,\vec \sigma ){|}_{\alpha_{(a)}=
 0, \xi^r=\sigma^r, {\tilde
\pi}^{\vec \alpha}_{(a)}= {\tilde \pi}^{\vec \xi}_r=0}=
\nonumber \\
 &=& {}^3{\check {\tilde \pi}}^r_{(a)}(\tau ,\vec \sigma ){|}_{\alpha_{(a)}=
 0, \xi^r=\sigma^r}=  \nonumber \\
 &=& \sum_s \int d^3\sigma_1 {\cal K}^r_{(a)s}(\vec \sigma ,{\vec \sigma}_1;\tau )
 \, {\tilde \Pi}^s(\tau ,{\vec \sigma}_1) =\nonumber \\
 &=&\sum_s \int
d^3\sigma_1 {\cal K}^r_{(a)s}(\vec \sigma ,{\vec \sigma}_1;\tau )
(\phi^{-2}e^{-{1\over {\sqrt{3}}}\sum_{\bar a}\gamma_{\bar as}r_{\bar
a}}) (\tau ,{\vec \sigma}_1) \nonumber \\ &&\Big[ {1\over 3}\rho
+\sqrt{3} \sum_{\bar b}
\gamma_{\bar bs} \pi_{\bar b}\Big] (\tau ,{\vec \sigma}_1)\qquad
\rho =\pi_{\phi}/2\phi\nonumber \\
{\rightarrow}_{\rho \rightarrow 0}&& \sqrt{3} \sum_{s,\bar b}
\gamma_{\bar bs}\int d^3\sigma_1 {\cal K}^r_{(a)s}(\vec \sigma ,{\vec
\sigma}_1;\tau |\phi ,r_{\bar a}] (\phi^{-2}e^{-{1\over
{\sqrt{3}}}\sum_{\bar a}\gamma_{\bar as}r_{\bar a}}
\pi_{\bar b})(\tau ,{\vec \sigma}_1),\nonumber \\
 &&{}\nonumber \\
 &&\text{with the kernel} \nonumber \\
 &&{}\nonumber \\
 {\cal K}^r_{(a)s}(\vec \sigma ,{\vec \sigma}_1,\tau ) &=&
 {\tilde {\cal K}}^r_{(a)s}(\vec \sigma ,{\vec \sigma}_1,\tau
|\phi ,r_{\bar a}]\, {\buildrel {def} \over =}\,
\delta^r_{(a)}\delta^r_s\delta^3(\vec \sigma ,{\vec \sigma}_1)+{\cal T}^r
_{(a)s}(\vec \sigma ,{\vec \sigma}_1,\tau ),\nonumber \\
 &&{}\nonumber \\
  {\cal T}^r_{(a)s}(\vec \sigma ,{\vec \sigma}_1,\tau ) &=&
 {\tilde {\cal T}}^r_{(a)s}(\vec \sigma ,{\vec \sigma}_1,\tau |\phi ,r_{\bar a}]=\nonumber \\
 &=&-Q_s(\tau ,{\vec \sigma}_1){{\partial G^{rs}_{(a)}(\vec \sigma ,{\vec \sigma}_1;\tau |
 \phi ,r_{\bar a}]}\over {\partial \sigma_1^u}}-\sum_v {{\partial Q_s(\tau
 ,{\vec \sigma}_1)}\over {\partial \sigma_1^v}} G^{rv}_{(a)}(\vec \sigma
 ,{\vec \sigma}_1;\tau |\phi ,r_{\bar a}], \nonumber \\
 &&
 \label{VI15}
 \eea

\noindent where the splitting of  ${\cal K}^r_{(a)u}$ follows from Eqs.(\ref{V21}).

 From Eqs.(3.12) and (4.2)of Ref.\cite{ru11} in the 3-orthogonal gauges
  the extrinsic curvature of $\Sigma^{(WSW)}_{\tau}$,
 the ADM momentum and the ADM Wheeler-DeWitt supermetric become

\bea
 {}^3{\hat K}_{rs}(\tau ,\vec \sigma ) &=&{{\epsilon \, 4\pi G}\over {c^3}} [e^{
{1\over {\sqrt{3}}}\sum_{\bar c} (\gamma_{\bar cr}+\gamma_{\bar
cs})r_{\bar c} } \sum_u(\delta_{ru}\delta_{(a)s}
+\delta_{su}\delta_{(a)r}-\delta_{rs}\delta_{(a)u})\nonumber \\
 &&e^{{1\over {\sqrt{3}}} \sum_{\bar c}\gamma_{\bar cu}r_{\bar c}}\,
{}^3{\hat {\tilde \pi}}^u_{(a)}](\tau ,\vec \sigma ),\nonumber \\
 &&{}\nonumber \\
{}^3{\hat K}(\tau ,\vec \sigma ) &=&-{{\epsilon \, 4\pi G}\over
{c^3}} [\phi^{-4} \sum_u \delta_{(a)u} e^{ {1\over
{\sqrt{3}}}\sum_{\bar c}\gamma_{\bar cu}r_{\bar c}}\, {}^3{\hat
{\tilde \pi}} ^u_{(a)}](\tau ,\vec \sigma )=\qquad [\rho
=\pi_{\phi}/2\phi ]\nonumber \\
 &=&-{{\epsilon \, 4\pi G}\over
{c^3}}\phi^{-4}(\tau ,\vec \sigma ) \sum_u\delta_{(a)u} e^{
{1\over {\sqrt{3}}}\sum_{\bar c}\gamma_{\bar cu}r_{\bar c}(\tau
,\vec \sigma )}\nonumber \\
 &&\sum_s \int
d^3\sigma_1 {\cal K}^u_{(a)s}(\vec \sigma ,{\vec \sigma}_1;\tau
|\phi ,r_{\bar a}] (\phi^{-2}e^{-{1\over {\sqrt{3}}}\sum_{\bar
a}\gamma_{\bar as}r_{\bar a}}) (\tau ,{\vec \sigma}_1) \nonumber
\\ &&\Big[ {1\over 3}\rho +\sqrt{3} \sum_{\bar b} \gamma_{\bar bs}
\pi_{\bar b}\Big] (\tau ,{\vec \sigma}_1)\nonumber \\
{\rightarrow}_{\rho \rightarrow 0}&& -{{\epsilon \, 4\sqrt{3} \pi
G }\over {c^3}} \phi^{-4}(\tau ,\vec \sigma )  \sum_u
\delta_{(a)u} e^{ {1\over {\sqrt{3}}}\sum_{\bar c}\gamma_{\bar
cu}r_{\bar c}(\tau ,\vec \sigma )}\nonumber \\
 &&\sum_s\sum_{\bar b}
\gamma_{\bar bs}\int d^3\sigma_1 {\cal K}^r_{(a)s}(\vec \sigma ,{\vec
\sigma}_1;\tau |\phi ,r_{\bar a}] (\phi^{-2}e^{-{1\over
{\sqrt{3}}}\sum_{\bar a}\gamma_{\bar as}r_{\bar a}}
\pi_{\bar b})(\tau ,{\vec \sigma}_1) ,\nonumber \\
 &&{}\nonumber \\
 &&{}\nonumber \\
 {}^3{\hat {\tilde \Pi}}^{rs}(\tau ,\vec \sigma )&=&{1\over
4}[{}^3{\hat e}^r
_{(a)}\, {}^3{\hat {\tilde \pi}}^s_{(a)} +{}^3{\hat e}^s_{(a)}\, {}^3{\hat
{\tilde \pi}}^r_{(a)}](\tau ,\vec \sigma )=\nonumber \\
 &=&{1\over 4}\phi^{-2}(\tau ,\vec \sigma )[e^{-{1\over {\sqrt{3}}}\sum_{\bar a}
\gamma_{\bar ar}r_{\bar a}} \delta^r_{(a)}\, {}^3{\hat {\tilde \pi}}^s_{(a)}+ \nonumber \\
 &+&e^{-{1\over {\sqrt{3}}}\sum_{\bar a}\gamma_{\bar as}r_{\bar a}}
\delta^s_{(a)}\, {}^3{\hat {\tilde \pi}}^r_{(a)}](\tau ,\vec \sigma
)=\nonumber \\
 &=&{1\over 4}\phi^{-2}(\tau ,\vec \sigma )\Big[ e^{-{1\over {\sqrt{3}}}\sum_{\bar a}
\gamma_{\bar ar}r_{\bar a}} \delta^r_{(a)}\nonumber \\
 && \sum_u \int d^3\sigma_1 {\cal K}^r_{(a)u}(\vec \sigma
 ,{\vec \sigma}_1,\tau |\phi ,r_{\bar a},\tilde \Pi ] +\nonumber \\
&+&e^{-{1\over {\sqrt{3}}}\sum_{\bar a}\gamma_{\bar as}r_{\bar a}}
\delta^s_{(a)}
 \sum_u \int d^3\sigma_1 {\cal K}^s_{(a)u}(\vec \sigma
 ,{\vec \sigma}_1,\tau |\phi ,r_{\bar a},\tilde \Pi ] \Big] \nonumber \\
 &&(\phi^{-2}e^{-{1\over {\sqrt{3}}}\sum_{\bar a}\gamma_{\bar au}r_{\bar
a}}) (\tau ,{\vec \sigma}_1) \Big[ {1\over 3}\rho +\sqrt{3} \sum_{\bar
b}\gamma_{\bar bu} \pi_{\bar b}\Big] (\tau ,{\vec
\sigma}_1),\nonumber \\
 &&{}\nonumber \\
 {}^3{\hat G}_{rsuv}(\tau ,\vec \sigma
) &=& [{}^3{\hat g}_{ru}\, {}^3{\hat g}
_{sv}+{}^3{\hat g}_{rv}\, {}^3{\hat g}_{su} -{}^3{\hat g}_{rs}\, {}^3{\hat g}
_{uv}](\tau ,\vec \sigma )=\nonumber \\
&=&\phi^8(\tau ,\vec \sigma )[e^{{2\over {\sqrt{3}}}\sum_{\bar
a}(\gamma_{\bar ar}+\gamma_{\bar as})r_{\bar a}}
(\delta_{ru}\delta_{sv}+\delta_{rv}\delta
_{su})-\nonumber \\
 &-&e^{{2\over {\sqrt{3}}}\sum_{\bar a}(\gamma_{\bar ar}+\gamma_{\bar au})
r_{\bar a}} \delta_{rs}\delta_{uv}](\tau ,\vec \sigma ).
\label{VI16}
\end{eqnarray}

\noindent The momenta ${}^3{\hat {\tilde \pi}}^r_{(a)}$ and
${}^3{\hat {\tilde \Pi}}^{rs}$ and the mean extrinsic curvature
${}^3\hat K$ are linear functions of the new momenta $\rho$ and
$\pi_{\bar c}$, but with a coordinate- and momentum-dependent
integral kernel. The determination of the {\it gravitomagnetic
potential} $W_{\pi}^r(\tau ,\vec \sigma )$, see Appendix C of II,
by solving the elliptic equations associated with the
supermomentum constraints in the conformal approach to metric
gravity, has been replaced here by the determination of the kernel
${\cal K}^r_{(a)s}(\vec \sigma ,{\vec \sigma}^{'};\tau )$
connecting the old momenta ${}^3{\hat {\tilde \pi}}^r_{(a)}(\tau
,\vec \sigma )$ to the new canonical ones ${\tilde \Pi}^r(\tau
,\vec \sigma )$, and therefore to the solution of the linear
partial differential equations (\ref{V23}) and (\ref{V24}).

Let us remark that at the level of the Dirac brackets for the
3-orthogonal gauges  we have the strong vanishing of the ADM
supermomentum constraints ${}^3{\tilde
\Pi}^{rs}(\tau ,\vec \sigma ){}_{|s}\equiv 0$ of Eqs.(4.16) of Ref.\cite{ru11},
so that (see Appendix C of II) the ADM momentum ${}^3{\tilde
\Pi}^{rs}(\tau ,\vec \sigma )$ of metric gravity becomes transverse
and is the sum of a TT-term and of a trace term\cite{york,yoyo,ciuf}.
The determination of ${}^3{\tilde \Pi}^{rs}_{TT}(\tau ,\vec
\sigma )$ can be done once one has found the solution of Eqs.(\ref{V23}),
(\ref{V24}).

 The variables $\rho$ and $\pi_{\bar a}$ replace ${}^3\hat K$ and
${}^3{\hat K}^{rs}_{TT}\,\,$ [or $\, {}^3{\hat {\tilde \Pi}}$ and
${}^3{\hat {\tilde \Pi}}^{rs}_{TT}\,\,$] of the conformal approach
respectively (see Appendix C of II) after the solution of the
supermomentum constraints (i.e. after the determination of the
gravitomagnetic potential) in the 3-orthogonal gauges. It would be
important to find the expression of $\rho$ and $\pi_{\bar a}$ in terms
of ${}^3{\hat g}_{rs}$ and ${}^3{\hat K}_{rs}\,\,$ [or ${}^3{\hat
{\tilde
\Pi}}^{rs}$]. The equation for ${}^3\hat K$ can be read as an integral
equation to get $\rho (\tau ,\vec
\sigma )$, the momentum conjugate to the conformal factor, in terms of
${}^3{\hat K}(\tau ,\vec \sigma )$, $q(\tau ,\vec \sigma )= {1\over 6}
ln\, {}^3{\hat g}(\tau ,\vec \sigma )$ and $r_{\bar a}(\tau ,\vec
\sigma )={{\sqrt{3}}\over 2} \sum_r \gamma_{\bar ar} ln\, [{}^3{\hat g}
_{rr}/{}^3\hat g](\tau ,\vec \sigma )$ [see Eqs.(\ref{VI13})],
in the 3-orthogonal gauges.

Let us remark that if we would have added only the gauge-fixing
$\alpha_{(a)}(\tau ,\vec \sigma ) \approx 0$ \footnote{So that the
associated Dirac brackets would coincide with the ADM Poisson
brackets for metric gravity.}, the four variables $\vec \xi (\tau
,\vec \sigma )$, $q(\tau ,\vec \sigma )$ \footnote{With conjugate
momenta ${\tilde \pi}^{\vec \xi}_r(\tau ,\vec \sigma )\approx 0$,
$\rho (\tau ,\vec \sigma )$.} of the canonical basis (\ref{VI11})
would correspond to the variables used in Ref.\cite{ish} to label
the points of the spacetime $M^4$ (assumed compact), following the
suggestion of Ref.\cite{dc11}, if $q(\tau ,\vec \sigma )$ is
interpreted as a time variable. However, we do not follow this
interpretation.

In Appendix  B there is the expression of the weak and strong
Poincar\'e charges of Eqs.(\ref{II25})  and (\ref{II23}) in the new
canonical basis in  3-orthogonal gauges.

The shift functions $n_r(\tau ,\vec \sigma  )$ of Eq.(\ref{VI8}) in
the rest-frame instant form on the WSW hypersurfaces and in
3-orthogonal gauges become

\begin{eqnarray}
 n_r(\tau ,\vec \sigma ) &\approx& -{{\epsilon \, 4\pi G}\over {c^3}}
\Big[ \phi^4 e^{{2\over {\sqrt{3}}}\sum_{\bar a}\gamma_{\bar ar}r_{\bar a}}
\Big] (\tau ,\vec \sigma ) \int d^3\sigma_1 [\epsilon - n(\tau ,{\vec \sigma}_1)]
\nonumber \\
 &&\phi^{-2}(\tau ,{\vec \sigma}_1)\, \sum_{wu} e^{{1\over {\sqrt{3}}}\sum_{\bar a}
 (\gamma_{\bar aw}+\gamma_{\bar au})r_{\bar a}}\, {}^3G_{o(a)(b)(c)(d)}\nonumber \\
 &&\delta_{(a)w} \sum_v \int d^3\sigma_2 {\cal K}^w_{(b)v}({\vec \sigma}_1, {\vec \sigma}_2;\tau )
 \Big[ \phi^{-2} e^{-{1\over {\sqrt{3}}}\sum_{\bar a}
 \gamma_{\bar av}r_{\bar a}}\Big] (\tau ,{\vec \sigma}_2 )\nonumber \\
 &&\Big[ {1\over 3} \rho +\sqrt{3}\sum_{\bar b}\gamma_{\bar bv}\pi_{\bar b}
 \Big] (\tau ,{\vec \sigma}_2)\,\, \delta_{(c)u} G^{ur}_{(d)}({\vec \sigma}_1,
 \vec \sigma ;\tau ),\nonumber \\
 &&{}\nonumber \\
 {\rightarrow}_{\rho \rightarrow 0}&& -{{\epsilon \, 4\sqrt{3} \pi G}\over {c^3}}
  \Big[ \phi^2 e^{{1\over {\sqrt{3}}}\sum_{\bar a}\gamma_{\bar ar}r_{\bar a}}
\Big] (\tau ,\vec \sigma ) \int d^3\sigma_1 [\epsilon - n(\tau ,{\vec \sigma}_1)]
\phi^{-2}(\tau ,{\vec \sigma}_1)\nonumber \\
 &&\sum_{wu}e^{{1\over {\sqrt{3}}}\sum_{\bar a}(\gamma_{\bar aw}+\gamma_{\bar au})
 r_{\bar a}(\tau ,{\vec \sigma}_1)} \, \Big( \delta_{wu}\delta_{(b)(d)}+\delta_{(b)u}
 \delta_{(d)w}-\delta_{(b)w}\delta_{(d)u}\Big)\nonumber \\
 &&\sum_v \int d^3\sigma_2 {\cal K}^w_{(b)v}({\vec \sigma}_1, {\vec \sigma}_2;\tau )
 \Big[ \phi^{-2} e^{-{1\over {\sqrt{3}}}\sum_{\bar a}
 \gamma_{\bar au}r_{\bar a}}\Big] (\tau ,{\vec \sigma}_2 )\nonumber \\
 &&\sum_{\bar b}\gamma_{\bar bv}\pi_{\bar b}(\tau ,{\vec \sigma}_2) \, G^{ur}_{(d)}({\vec
 \sigma}_1, \vec \sigma ;\tau ).
\label{VI17}
\end{eqnarray}

\subsection{The Superhamiltonian Constraint as the Reduced Lichnerowicz Equation.}

 By using the new canonical basis and Eq.(\ref{VI15})
the superhamiltonian constraint (\ref{II11}), (\ref{V27}) restricted
to 3-orthogonal gauges becomes [$\rho = \pi_{\phi}/2\phi$]

\begin{eqnarray}
{\hat {\cal H}}_R(\tau ,\vec \sigma )&=&-\epsilon {{2\pi G\,
\phi^{-2}(\tau ,\vec \sigma )}\over {c^3}}\Big[ \Big( \phi^{-4}[6
\sum_{\bar a}\pi^2 _{\bar a}-{1\over 3}\rho^2]\Big)(\tau ,\vec
\sigma
 )+\nonumber \\
  &+&2 \Big( \phi^{-2}\sum_ue^{{1\over {\sqrt{3}}}\sum_{\bar
a}\gamma_{\bar au}r_{\bar a}} [2\sqrt{3}\sum_{\bar b}\gamma_{\bar
bu}\pi_{\bar b}-{1\over 3}\rho ]\Big)(\tau ,\vec \sigma )\times
\nonumber \\
 &&\int d^3\sigma_1 \sum_r
\delta^u_{(a)} {\cal T}^u_{(a)r}(\vec \sigma ,{\vec
\sigma}_1,\tau ) \Big( \phi^{-2}e^{-{1\over {\sqrt{3}}}\sum_{\bar a}
\gamma_{\bar ar}r_{\bar a}}[{{\rho}\over 3}+\sqrt{3}\sum_{\bar b}\gamma_{\bar
br} \pi_{\bar b}]\Big) (\tau ,{\vec \sigma}_1)+\nonumber \\
&+&\int d^3\sigma_1d^3\sigma_2 \Big( \sum_u e^{{2\over {\sqrt{3}}}\sum_{\bar a}
\gamma_{\bar au}r_{\bar a}(\tau ,\vec \sigma )} \times \nonumber \\
&&\sum_r{\cal T}^u_{(a)r}(\vec \sigma ,{\vec
\sigma}_1,\tau ) \Big( \phi^{-2}e^{-{1\over {\sqrt{3}}}\sum_{\bar a}
\gamma_{\bar ar}r_{\bar a}}[{{\rho}\over 3}+\sqrt{3}\sum_{\bar b}\gamma_{\bar
br} \pi_{\bar b}]\Big) (\tau ,{\vec \sigma}_1)\times \nonumber \\
&&\sum_s {\cal T}^u_{(a)s}(\vec \sigma ,{\vec
\sigma}_2,\tau ) \Big( \phi^{-2}e^{-{1\over {\sqrt{3}}}\sum_{\bar a}
\gamma_{\bar as}r_{\bar a}}[{{\rho}\over 3}+\sqrt{3}\sum_{\bar c}\gamma_{\bar
cs} \pi_{\bar c}]\Big) (\tau ,{\vec \sigma}_2)+\nonumber \\
&+&\sum_{uv} e^{{1\over {\sqrt{3}}}\sum_{\bar a}(\gamma_{\bar au}+\gamma_{\bar
av})r_{\bar a}(\tau ,\vec \sigma )} (\delta^u_{(b)}\delta^v_{(a)}-\delta^u_{(a)}
\delta^v_{(b)})\times \nonumber \\
&&\sum_r {\cal T}^u_{(a)r}(\vec \sigma ,{\vec
\sigma}_1,\tau ) \Big(\phi^{-2}e^{-{1\over {\sqrt{3}}}\sum_{\bar a}
\gamma_{\bar ar}r_{\bar a}}[{{\rho}\over 3}+\sqrt{3}\sum_{\bar b}\gamma_{\bar
br} \pi_{\bar b}]\Big) (\tau ,{\vec \sigma}_1)\nonumber \\
&&\sum_s {\cal T}^v_{(b)s}(\vec \sigma ,{\vec
\sigma}_2,\tau ) \Big( \phi^{-2}e^{-{1\over {\sqrt{3}}}\sum_{\bar a}
\gamma_{\bar as}r_{\bar a}}[{{\rho}\over 3}+\sqrt{3}\sum_{\bar c}\gamma_{\bar
cs} \pi_{\bar c}]\Big) (\tau ,{\vec \sigma}_2)\, \Big)\, \Big]\,
+\nonumber \\
 &+&{{\epsilon \, c^3}\over {16\pi G}} \sum_{r,s}\Big[ \phi^2e^{-{1\over
{\sqrt{3}}}\sum_{\bar a}(\gamma
_{\bar ar}+\gamma_{\bar as})r_{\bar a}}\Big] (\tau, \vec
\sigma ) \epsilon_{(a)(b)(c)}\delta_{(a)r}\delta_{(b)s} \,
{}^3{\hat \Omega}_{rs(c)}[\phi ,r_{\bar c}](\tau ,\vec \sigma )\approx
0,\nonumber \\
 &&{}\nonumber \\
  &&\lbrace {\hat {\cal H}}_R(\tau ,\vec
\sigma ),{\hat {\cal H}}_R(\tau , {\vec \sigma }^{'})\rbrace {}^{*}
\equiv \lbrace {\hat {\cal H}}(\tau ,\vec
\sigma ),{\hat {\cal H}}(\tau ,{\vec \sigma}^{'})\rbrace {}^{*}\equiv
\nonumber \\
&&\equiv \Big( -{{\partial}\over {\partial \sigma^s}} \Big[\,
\lbrace \xi^r(\tau ,\vec \sigma ),{\hat {\cal H}}(\tau ,{\vec \sigma}^{'})
\rbrace \, {\hat {\cal H}}_R(\tau ,\vec \sigma ) \Big] +\nonumber \\
&&+{{\partial}\over {\partial \sigma^{{'}s}}}
\Big[\, \lbrace \xi^r(\tau ,{\vec \sigma}^{'}),{\hat {\cal H}}(\tau ,\vec \sigma )
\rbrace \, {\hat {\cal H}}_R(\tau ,{\vec \sigma}^{'}) \Big]
\Big)_{\xi^r=\sigma^r} \approx 0.
\label{VI18}
\end{eqnarray}

\noindent The last line of ${\hat {\cal H}}_R$ is equal to
${{\epsilon \, c^3}\over {16\pi G}} \phi^6\, {}^3\hat R[\phi
,r_{\bar a}]$.

The constraint is no more an algebraic relation among the final
variables, but rather an integro-differential equation, the {\it
reduced Lichnerowicz equation}, whose unknown is the conformal factor
$\phi (\tau ,\vec \sigma )= e^{{1\over 2}q(\tau ,\vec \sigma )}$ as
said in Subsection D of Section III for the case of asymptotically
flat spacetimes. Its solution in 3-orthogonal coordinates gives  $\phi
=e^{q/2}$ as a functional $e^{F[r_{\bar a}, \pi_{\bar a}, \pi_{\phi}]}$ of
the canonical variables $r_{\bar a}(\tau ,\vec \sigma )$, $\pi_{\bar
a}(\tau ,\vec \sigma )$, and of the last gauge variable: the momentum
$\pi_{\phi}(\tau ,\vec \sigma )=2\, e^{-q(\tau \vec \sigma )/2} \rho
(\tau ,\vec \sigma )$ conjugate to the conformal factor.

In the 3-orthogonal gauges the functions $r_{\bar a}(\tau ,\vec \sigma
)$, $\bar a=1,2$, give a parametrization of the {\it Hamiltonian
physical degrees of freedom of the gravitational field} and of the
{\it space of conformal 3-geometries}
\footnote{The quotient of {\it superspace} by the group $Weyl\,
\Sigma_{\tau}$, if by varying $\rho$ the solution $\phi =e^{q/2}
\approx e^{F[r_{\bar a},\pi_{\bar a}, \rho ]}$ of the reduced Lichnerowicz
equation spans all the {\it Weyl rescalings}.}: it turns out that a
point (a  {\it conformal 3-geometry}) in this space, i.e. a ${}^3{\hat
g}^{diag}_{rs}$ \footnote{It is simultaneously the York\cite{york}
reduced metric and the Misner's one\cite{misner} in 3-orthogonal
coordinates.}, is an equivalence class of conformally related
3-metrics ({\it conformal gauge orbit}).

The solution $\phi =e^{q/2}\approx e^{F[r_{\bar a},
\pi_{\bar a}, \pi_{\phi} ]}$ of the reduced Lichnerowicz equation
in these gauges just determines an equivalence class of 3-geometries
(i.e. a {\it conformal 3-geometry}) parametrized by the gauge variable
$\rho (\tau ,\vec \sigma )=2[\phi^{-1}\pi_{\phi}](\tau ,\vec \sigma )$
(it is the coordinate in the {\it conformal gauge orbit}); the natural
representative of an equivalence class is obtained with the
gauge-fixing $\rho (\tau ,\vec \sigma )\approx 0$: ${}^3{\hat
g}_{rs}=e^{4F[r_{\bar a}, \pi_{\bar a}, 0]}\, {}^3{\hat g}
^{diag}_{rs}[r_{\bar a}, \pi_{\bar a}]$.

When we add the {\it natural gauge-fixing} $\rho (\tau ,\vec
\sigma )\approx 0$ to the reduced superhamiltonian constraint in
3-orthogonal coordinates on $\Sigma^{(WSW)}_{\tau}$ and go to
Dirac brackets eliminating the conjugate variables $q(\tau ,\vec
\sigma )$, $\rho(\tau ,\vec \sigma )$ [or $\phi$, $\pi_{\phi}$],
the functions $r_{\bar a}(\tau ,\vec \sigma )$ and $\pi_{\bar
a}(\tau ,\vec \sigma )$ become the {\it physical canonical
variables for the gravitational field} in this {\it special
3-orthogonal gauge, because they have vanishing Poisson brackets
with $\rho$}. This does not happens with the gauge-fixing ({\it
maximal slicing condition}) ${}^3K(\tau ,\vec \sigma )\approx 0\,
(or\, const.)$ when we use 3-orthogonal coordinates.

Let us remark that usually (see for instance Ref.\cite{ako}) in
metric gravity  one introduces four gauge-fixing constraints on
${}^3g_{rs}$, ${}^3{\tilde \Pi}^{rs}$ (they can be used also in
tetrad gravity) instead of our gauge fixings $\vec \xi (\tau ,\vec
\sigma )-\vec \sigma \approx 0$, $\rho (\tau ,\vec \sigma )\approx
0$, whose functional form is oriented to the parametrizations
needed for the Shanmugadhasan canonical transformations.

The evolution in $\tau$ (the time parameter labelling the leaves
$\Sigma_{\tau}^{(WSW)}$ of the foliation associated with the 3+1
splitting of $M^4$) is instead generated by the weak ADM energy
${\hat P}^{\tau}_{ADM}$, absent in closed spacetimes. The {\it ADM
energy}, which, in this special gauge, depends only on $r_{\bar
a}$, $\pi_{\bar a}$ is {\it the Hamiltonian generating the
$\tau$-evolution of the physical (non covariant) gravitational
field degrees of freedom} \footnote{This corresponds to the two
dynamical equations contained in the 10 Einstein equations in this
gauge.}.

However, since a closed form of the conformal factor in terms of
$r_{\bar a}$, $\pi_{\bar a}$ as a solution of the superhamiltonian
constraint (after having put $\rho (\tau ,\vec \sigma )=0$ in it)
is not known, the ADM energy \footnote{Weakly coinciding with the
ADM invariant mass in the rest-frame instant form.} {\it cannot}
be explicitly expressed in terms of the physical degrees of
freedom of the gravitational field in 3-orthogonal coordinates.

It seems quite difficult to be able to implement the last step of
the program, namely to find the final Shanmugadhasan canonical
transformation $\phi$, $\pi_{\phi}$, $r_{\bar a}$, $\pi_{\bar a}$
$\mapsto$ ${\hat {\cal H}}^{'}_R$, $\pi_{\phi}^{'}$, $r^{'}_{\bar
a}$, $\pi^{'}_{\bar a}$  \footnote{${\hat {\cal H}}^{'}_R(\tau
,\vec \sigma )\approx 0$ equivalent to ${\hat {\cal H}}_R(\tau
,\vec \sigma )\approx 0$ but with $\{ {\hat {\cal H}}^{'}_R(\tau
,\vec \sigma), {\hat {\cal H}}^{'}_R(\tau ,{\vec \sigma}^{'}) \}
=0$.}, so that all the first class constraints of tetrad gravity
appear in the final canonical basis in Abelianized form (this
would implement Kuchar's program defined in
Refs.\cite{dc10,kuchar1}). Indeed, if $e^{F[r_{\bar a}, \pi_{\bar
a},\pi_{\phi}]}$ is the solution of the reduced Lichnerowicz
equation, a functional form of such a ${\hat {\cal H}}_R^{'}$ is
${\hat {\cal H}}_R^{'}=\phi -e^{F[r_{\bar a}, \pi_{\bar a},
\pi_{\phi}]} \approx 0$. The existence of this final canonical
transformation is connected to the integrability problem of
Einstein equations: in Ref.\cite{smolin} it is reported that they
could admit chaotic solutions and, according to Newman, this would
be an obstruction to the existence of Dirac observables in the
final canonical basis.

Equally difficult is to find the analogue of the York
map\cite{yorkmap} in the 3-orthogonal gauges: $\phi$,
$\pi_{\phi}$, $r_{\bar a}$, $\pi_{\bar a}$ $\mapsto$ ${\cal
T}=-{{c^3}\over {12\pi G}} \epsilon \, {}^3{\hat K}$, ${\cal
P}_{\cal T}=- \phi^{12}$, $r^{(K)}_{\bar a}$, $\pi^{(K)}_{\bar
a}$.

To transform the superhamiltonian constraint in the reduced
Lichnerowicz equation for the conformal factor, we shall use the
canonical variable $\phi = e^{q/2}$ but we shall go on to use the
notation $\rho$ for ${1\over 2} \phi \pi_{\phi}$ for notational
simplicity. By using Eq.(\ref{a2}) of Appendix A we get

\begin{eqnarray}
{}^3{\hat g}_{rs}&=&\phi^4 e^{{2\over {\sqrt{3}}}\sum_{\bar
a}\gamma_{\bar ar}r
_{\bar a}}\, \delta_{rs}\equiv \phi^4\, {}^3{\tilde
g}_{rs},\quad\quad {}^3{\tilde g}_{rs}=e^{{2\over {\sqrt{3}}}\sum_{\bar a}
\gamma_{\bar ar}r_{\bar a}} \delta_{rs},\nonumber \\
&&\Rightarrow \sum_r \, {}^3{\tilde \Gamma}^r_{rs}=0,\nonumber \\
{}^3{\hat R}&=&\phi^{-4}[-8\, {}^3{\tilde
g}^{rs}\partial_r\partial_sln\, \phi -8\, {}^3{\tilde
g}^{rs}\partial_rln\, \phi
\partial_sln\, \phi -8 \partial_r\, {}^3{\tilde g}^{rs}
\partial_sln\, \phi +{}^3{\tilde R}[r_{\bar a}]]=\nonumber \\
&=&\phi^{-4}[\phi^{-1}(-8{\tilde \triangle}[r_{\bar a}] \phi +8\,
{}^3{\tilde \Gamma}^r_{rs}\, {}^3{\tilde g}^{su}
\partial_u\phi )+{}^3{\tilde R}[r_{\bar a}]]=\nonumber \\
&=&\phi^{-5}[-8 {\tilde \triangle}[r_{\bar a}] \phi +{}^3{\tilde R}[r_{\bar a}]
 \phi ],
\label{VI19}
\end{eqnarray}

\noindent where ${}^3\tilde R={}^3\tilde R[r_{\bar a}]$ and
$\tilde \triangle =\tilde \triangle [r_{\bar
a}]=\partial_r({}^3{\tilde g}^{rs}\partial_s)$ are the scalar
curvature and the Laplace-Beltrami operator associated with the
3-metric ${}^3{\tilde g}_{rs}$ respectively \footnote{$\tilde
\triangle -{1\over 8}{}^3\tilde R$ is a conformally invariant
operator \cite{conf}.}. From Eqs.(\ref{a3}) of Appendix A, we have
[$\tilde \gamma = det\, |{}^3{\tilde g}_{rs}|=1$]

\begin{eqnarray}
{}^3\hat R [\phi ,r_{\bar a}]&=&
-\sum_{uv}\{ (2\partial_vln\, \phi +{1\over {\sqrt{3}}}\sum_{\bar a}\gamma_{\bar au}
\partial_vr_{\bar a})(4\partial_vln\, \phi -{1\over {\sqrt{3}}}\sum_{\bar b}\gamma
_{\bar bu}\partial_vr_{\bar b})+\nonumber \\
&&+\phi^{-4}e^{{2\over {\sqrt{3}}}\sum_{\bar c}\gamma_{\bar cv}r_{\bar
c}} [2\partial^2_vln\, \phi +{1\over {\sqrt{3}}}\sum_{\bar
a}\gamma_{\bar au}\partial_v^2r_{\bar a}+\nonumber \\
 &&+{2\over {\sqrt{3}}}(2\partial_vln\, \phi +{1\over {\sqrt{3}}}\sum_{\bar a}\gamma
_{\bar au}\partial_vr_{\bar a})\sum_{\bar b}(\gamma_{\bar bu}-\gamma_{\bar bv})
\partial_vr_{\bar b}-\nonumber \\
&&-(2\partial_vln\, \phi +{1\over {\sqrt{3}}}\sum_{\bar a}\gamma_{\bar
av}\partial_vr_{\bar a})(2\partial_vln\, \phi +{1\over
{\sqrt{3}}}\sum_{\bar b}\gamma_{\bar bu}
\partial_vr_{\bar b}] \} +\nonumber \\
&&+\phi^{-4}\sum_ue^{{2\over {\sqrt{3}}}\sum_{\bar c}\gamma_{\bar
cu}r_{\bar c}} [-2\partial^2_uln\, \phi +{2\over {\sqrt{3}}}\sum_{\bar
a}\gamma_{\bar au} \partial^2_ur_{\bar a}+\nonumber \\
&&+(2\partial_uln\, \phi +{1\over {\sqrt{3}}}\sum_{\bar a}\gamma_{\bar
au}\partial_ur_{\bar a})(2\partial_uln\, \phi -{2\over
{\sqrt{3}}}\sum_{\bar b}\gamma_{\bar bu}
\partial_ur_{\bar b})]\nonumber \\
&&\Big[ {\rightarrow}_{r_{\bar a}\, \rightarrow 0}\,
-24\sum_u(\partial_uln\, \phi )^2-8\phi^{-4}\sum_u[\partial^2_uln\, \phi
-2 (\partial_uln\, \phi )^2]\, {\rightarrow}_{q\,
\rightarrow 0}\, 0\Big],\nonumber \\
&&{}\nonumber \\
{\rightarrow}_{q\, \rightarrow 0}\,&& {}^3\tilde R[r_{\bar a}]=\nonumber \\
&&=-{1\over {\sqrt{3}}}\sum_{uv}\{
-{1\over {\sqrt{3}}}\sum_{\bar a\bar b}\gamma_{\bar au}\gamma_{\bar bu}
\partial_vr_{\bar a}\partial_vr_{\bar b}+e^{-{2\over {\sqrt{3}}}\sum_{\bar c}
\gamma_{\bar cv}r_{\bar c}}\sum_{\bar a}\gamma_{\bar au}\cdot \nonumber \\
&&[\partial^2_vr_{\bar a}+{2\over {\sqrt{3}}}\sum_{\bar b}(\gamma_{\bar bu}-
\gamma_{\bar bv})\partial_vr_{\bar a}\partial_vr_{\bar b}-{1\over {\sqrt{3}}}
\sum_{\bar b}\gamma_{\bar bv}\partial_vr_{\bar a}\partial_vr_{\bar b}]\}+
\nonumber \\
&&+{2\over {\sqrt{3}}}\sum_ue^{-{2\over {\sqrt{3}}}\sum_{\bar c}\gamma
_{\bar cu}r_{\bar c}}\sum_{\bar a}\gamma_{\bar au}[\partial^2_ur_{\bar a}+
{1\over {\sqrt{3}}}\sum_{\bar b}\gamma_{\bar bu}\partial_ur_{\bar a}
\partial_ur_{\bar b}]=\nonumber \\
&&={1\over 3}\sum_u (1-2e^{-{2\over {\sqrt{3}}}\sum_{\bar a}\gamma_{\bar au}
r_{\bar a}}) \sum_{\bar b}(\partial_ur_{\bar b})^2+\nonumber \\
&&+{2\over {\sqrt{3}}}\sum_ue^{-{2\over {\sqrt{3}}}\sum_{\bar c}\gamma
_{\bar cu}r_{\bar c}}\sum_{\bar a}\gamma_{\bar au}[\partial^2_ur_{\bar a}+
{1\over {\sqrt{3}}}\sum_{\bar b}\gamma_{\bar bu}\partial_ur_{\bar a}
\partial_ur_{\bar b}],\nonumber \\
&&{}\nonumber \\
\tilde \triangle [r_{\bar a}]&=& \partial_r [{}^3{\tilde g}^{rs}\, \partial_s]
={}^3{\tilde g}^{rs}\, {}^3{\tilde \nabla}_r\, {}^3{\tilde \nabla}_s=
\nonumber \\
&=&\sum_re^{-{2\over {\sqrt{3}}}\sum_{\bar a}\gamma_{\bar ar} r_{\bar a}}
[\partial_r^2-{2\over {\sqrt{3}}} \sum_{\bar b}\gamma_{\bar br}
\partial_rr_{\bar b} \partial_r].
\label{VI20}
\end{eqnarray}

Using Eq.(\ref{VI18}), the reduced superhamiltonian constraint becomes
the following reduced Lichnerowicz equation

\begin{eqnarray}
{\tilde {\cal H}}_R(\tau ,\vec \sigma )&=&\epsilon \Big[
{{c^3}\over {16\pi G}}\phi^6\, {}^3{\hat R}- {{2\pi G\,
\phi^{-6}}\over {c^3}} {}^3G_{o(a)(b)(c)(d)}\, {}^3{\hat
e}_{((a)r}\, {}^3{\hat {\tilde \pi}}^r_{(b))}\, {}^3{\hat
e}_{((c)s}\, {}^3{\hat {\tilde \pi}}^s_{(d))} \Big] (\tau ,\vec
\sigma )=\nonumber \\
 &=&\epsilon \phi (\tau ,\vec \sigma ) \,
\Big[ \, {{c^3}\over {16\pi G}}(-8{\tilde \triangle}[r_{\bar a}]
 +{}^3{\tilde R}[r_{\bar a}])\phi -\nonumber \\
 &-&{{2\pi G\, \phi^{-7}}\over {c^3}} {}^3G_{o(a)(b)(c)(d)}\, {}^3{\hat
e}_{((a)r}\, {}^3{\hat {\tilde \pi}}^r_{(b))}\, {}^3{\hat
e}_{((c)s}\, {}^3{\hat {\tilde \pi}}^s_{(d))} \Big] (\tau ,\vec
\sigma )=\nonumber \\ &=&\epsilon \phi (\tau ,\vec \sigma )\,
\Big[ \, {{c^3}\over {16\pi G}} (-8{\tilde \triangle}[r_{\bar a}]
+{}^3{\tilde R}[r_{\bar a}])\phi -\nonumber \\
 &-&{{2\pi G}\over {c^3}}
\Big[ \Big( \phi^{-7} (6 \sum_{\bar a} \pi^2_{\bar a}-{1\over 3}\rho^2)\Big)(\tau ,\vec
\sigma )+\nonumber \\
&+&2\Big( \phi^{-5}
\sum_ue^{{1\over {\sqrt{3}}}\sum_{\bar a}\gamma_{\bar au}r_{\bar a}}
[2\sqrt{3}\sum_{\bar b}\gamma_{\bar bu}\pi_{\bar b}-{1\over 3}\rho
]\Big)(\tau ,\vec \sigma )\times \nonumber \\
 &&\int d^3\sigma_1 \sum_r
\delta^u_{(a)} {\cal T}^u_{(a)r}(\vec \sigma ,{\vec
\sigma}_1,\tau ) \Big( \phi^{-2}
e^{-{1\over {\sqrt{3}}}\sum_{\bar a}
\gamma_{\bar ar}r_{\bar a}}[{{\rho}\over 3}+\sqrt{3}\sum_{\bar b}\gamma_{\bar
br} \pi_{\bar b}]\Big) (\tau ,{\vec \sigma}_1)+\nonumber \\
&+&\phi^{-3}(\tau ,\vec \sigma )
\int d^3\sigma_1d^3\sigma_2 \Big( \sum_u e^{{2\over {\sqrt{3}}}\sum_{\bar a}
\gamma_{\bar au}r_{\bar a}(\tau ,\vec \sigma )} \times \nonumber \\
&&\sum_r{\cal T}^u_{(a)r}(\vec \sigma ,{\vec
\sigma}_1,\tau ) \Big( \phi^{-2}
e^{-{1\over {\sqrt{3}}}\sum_{\bar a}
\gamma_{\bar ar}r_{\bar a}}[{{\rho}\over 3}+\sqrt{3}\sum_{\bar b}\gamma_{\bar
br} \pi_{\bar b}]\Big) (\tau ,{\vec \sigma}_1)\times \nonumber \\
&&\sum_s {\cal T}^u_{(a)s}(\vec \sigma ,{\vec
\sigma}_2,\tau ) \Big( \phi^{-2}
e^{-{1\over {\sqrt{3}}}\sum_{\bar a}
\gamma_{\bar as}r_{\bar a}}[{{\rho}\over 3}+\sqrt{3}\sum_{\bar c}\gamma_{\bar
cs} \pi_{\bar c}]\Big) (\tau ,{\vec \sigma}_2)+\nonumber \\
&+&\sum_{uv} e^{{1\over {\sqrt{3}}}\sum_{\bar a}(\gamma_{\bar au}+\gamma_{\bar
av})r_{\bar a}(\tau ,\vec \sigma )} (\delta^u_{(b)}\delta^v_{(a)}-\delta^u_{(a)}
\delta^v_{(b)})\times \nonumber \\
&&\sum_r {\cal T}^u_{(a)r}(\vec \sigma ,{\vec
\sigma}_1,\tau ) \Big( \phi^{-2}
e^{-{1\over {\sqrt{3}}}\sum_{\bar a}
\gamma_{\bar ar}r_{\bar a}}[{{\rho}\over 3}+\sqrt{3}\sum_{\bar b}\gamma_{\bar
br} \pi_{\bar b}]\Big) (\tau ,{\vec \sigma}_1)\nonumber \\
&&\sum_s {\cal T}^v_{(b)s}(\vec \sigma ,{\vec
\sigma}_2,\tau ) \Big( \phi^{-2}
e^{-{1\over {\sqrt{3}}}\sum_{\bar a}
\gamma_{\bar as}r_{\bar a}}[{{\rho}\over 3}+\sqrt{3}\sum_{\bar c}\gamma_{\bar
cs} \pi_{\bar c}]\Big) (\tau ,{\vec \sigma}_2)\, \Big)\, \Big]\,
\Big] \approx 0,\nonumber \\
&&{} \nonumber \\
 {\tilde {\cal H}}_R(\tau ,\vec \sigma )\, &{\rightarrow}_{\rho
\rightarrow 0}\,&
\epsilon \phi (\tau ,\vec \sigma )\, \Big[ \, {{c^3}\over {16\pi
G}} (-8{\tilde \triangle}[r_{\bar a}] +{}^3{\tilde R}[r_{\bar a}])\phi
-\nonumber \\
 &-&{{6\pi G}\over {c^3}}\Big[ \Big( 2\phi^{-7}  \sum_{\bar a}
 \pi^2_{\bar a}\Big)(\tau ,\vec \sigma )+\nonumber \\
 &+&4\Big( \phi^{-5}
\sum_ue^{{1\over {\sqrt{3}}}\sum_{\bar a}\gamma_{\bar au}r_{\bar a}}
\sum_{\bar b}\gamma_{\bar bu}\pi_{\bar b}\Big)(\tau
,\vec \sigma )\times \nonumber \\
 &&\int d^3\sigma_1 \sum_r
\delta^u_{(a)} {\cal T}^u_{(a)r}(\vec \sigma ,{\vec
\sigma}_1,\tau ) \Big( \phi^{-2}
e^{-{1\over {\sqrt{3}}}\sum_{\bar a}
\gamma_{\bar ar}r_{\bar a}}\sum_{\bar b}\gamma_{\bar
br} \pi_{\bar b}\Big) (\tau ,{\vec \sigma}_1)+\nonumber \\
 &+&\phi^{-3}(\tau ,\vec \sigma )
\int d^3\sigma_1d^3\sigma_2 \Big( \sum_u e^{{2\over {\sqrt{3}}}\sum_{\bar a}
\gamma_{\bar au}r_{\bar a}(\tau ,\vec \sigma )} \times \nonumber \\
&&\sum_r{\cal T}^u_{(a)r}(\vec \sigma ,{\vec
\sigma}_1,\tau ) \Big( \phi^{-2}
e^{-{1\over {\sqrt{3}}}\sum_{\bar a}
\gamma_{\bar ar}r_{\bar a}}\sum_{\bar b}\gamma_{\bar
br} \pi_{\bar b}\Big) (\tau ,{\vec \sigma}_1)\times \nonumber \\
&&\sum_s {\cal T}^u_{(a)s}(\vec \sigma ,{\vec
\sigma}_2,\tau ) \Big( \phi^{-2}
e^{-{1\over {\sqrt{3}}}\sum_{\bar a}
\gamma_{\bar as}r_{\bar a}}\sum_{\bar c}\gamma_{\bar
cs} \pi_{\bar c}\Big) (\tau ,{\vec \sigma}_2)+\nonumber \\
&+&\sum_{uv} e^{{1\over {\sqrt{3}}}\sum_{\bar a}(\gamma_{\bar
au}+\gamma_{\bar av})r_{\bar a}(\tau ,\vec \sigma )}
(\delta^u_{(b)}\delta^v_{(a)}-\delta^u_{(a)}
\delta^v_{(b)})\times \nonumber \\
&&\sum_r {\cal T}^u_{(a)r}(\vec \sigma ,{\vec
\sigma}_1,\tau ) \Big( \phi^{-2}
e^{-{1\over {\sqrt{3}}}\sum_{\bar a}
\gamma_{\bar ar}r_{\bar a}}\sum_{\bar b}\gamma_{\bar
br} \pi_{\bar b}\Big) (\tau ,{\vec \sigma}_1)\nonumber \\ &&\sum_s
{\cal T}^v_{(b)s}(\vec \sigma ,{\vec
\sigma}_2,\tau ) \Big( \phi^{-2}
e^{-{1\over {\sqrt{3}}}\sum_{\bar a}
\gamma_{\bar as}r_{\bar a}}\sum_{\bar c}\gamma_{\bar
cs} \pi_{\bar c}\Big) (\tau ,{\vec \sigma}_2)\, \Big)\, \Big]\,
\Big] \approx 0.
\label{VI21}
\end{eqnarray}

\subsection{The Natural Gauge Replacing the Maximal Slicing Condition
         in 3-Orthogonal Gauges.}

As already said the canonical basis (\ref{VI11}) suggests that in
the 3-orthogonal gauges the natural gauge fixing  to the
superhamiltonian constraint is $\rho (\tau ,\vec \sigma ) \approx
0$ and not the maximal slicing condition ${}^3K(\tau ,\vec \sigma
) \approx 0$. This gauge fixing selects a well defined 3+1
splitting whose leaves are a well defined family of WSW
hypersurfaces $\Sigma_{\tau}^{(WSW) \rho =0}$. For $\rho (\tau
,\vec \sigma ) \approx 0$ we get from Eqs.(\ref{VI16})

\bea {}^3{\hat K}(\tau ,\vec \sigma ) &=& -{{\epsilon \, 4
\sqrt{3} \pi G}\over {c^3}} \phi^{-4}(\tau ,\vec \sigma )  \sum_u
\delta_{(a)u}\nonumber \\
 &&\sum_s\sum_{\bar b}
\gamma_{\bar bs}\int d^3\sigma_1 {\cal K}^r_{(a)s}(\vec \sigma ,{\vec
\sigma}_1;\tau ) (\phi^{-2}e^{-{1\over
{\sqrt{3}}}\sum_{\bar a}\gamma_{\bar as}r_{\bar a}}
\pi_{\bar b})(\tau ,{\vec \sigma}_1).
\label{VI22}
\eea

From Eq.(\ref{VI20}) the final reduced form of the Lichnerowicz
equation in the special 3-orthogonal gauge identified by the natural
gauge fixing $\rho (\tau ,\vec \sigma )\approx 0$ is

\begin{eqnarray}
(-{\tilde \triangle}[r_{\bar a}] &+&{1\over 8}{}^3{\tilde R}[r_{\bar a}])(\tau
,\vec \sigma ) \phi (\tau ,\vec \sigma )={{12 \pi^2 G^2}\over {c^6}}
\Big[ 2(\phi^{-7}  \sum_{\bar a} \pi^2_{\bar a})(\tau ,\vec
\sigma )+\nonumber \\
&+&4\Big( \phi^{-5}
\sum_ue^{{1\over {\sqrt{3}}}\sum_{\bar a}\gamma_{\bar au}r_{\bar a}}
\sum_{\bar b}\gamma_{\bar bu}\pi_{\bar b}\Big)(\tau
,\vec \sigma )\times \nonumber \\
&&\int d^3\sigma_1 \sum_r \delta^u_{(a)} {\cal T}^u_{(a)r}(\vec \sigma ,{\vec
\sigma}_1,\tau ) \Big( \phi^{-2}
e^{-{1\over {\sqrt{3}}}\sum_{\bar a}
\gamma_{\bar ar}r_{\bar a}}\sum_{\bar b}\gamma_{\bar
br} \pi_{\bar b}\Big) (\tau ,{\vec \sigma}_1)+\nonumber \\
&+&\phi^{-3}(\tau ,\vec \sigma )
\int d^3\sigma_1d^3\sigma_2 \Big( \sum_u e^{{2\over {\sqrt{3}}}\sum_{\bar a}
\gamma_{\bar au}r_{\bar a}(\tau ,\vec \sigma )} \times \nonumber \\
&&\sum_r{\cal T}^u_{(a)r}(\vec \sigma ,{\vec
\sigma}_1,\tau ) \Big( \phi^{-2}
e^{-{1\over {\sqrt{3}}}\sum_{\bar a}
\gamma_{\bar ar}r_{\bar a}}\sum_{\bar b}\gamma_{\bar
br} \pi_{\bar b}\Big) (\tau ,{\vec \sigma}_1)\times \nonumber \\
&&\sum_s {\cal T}^u_{(a)s}(\vec \sigma ,{\vec
\sigma}_2,\tau ) \Big( \phi^{-2}
e^{-{1\over {\sqrt{3}}}\sum_{\bar a}
\gamma_{\bar as}r_{\bar a}}\sum_{\bar c}\gamma_{\bar
cs} \pi_{\bar c}\Big) (\tau ,{\vec \sigma}_2)+\nonumber \\
&+&\sum_{uv} e^{{1\over {\sqrt{3}}}\sum_{\bar a}(\gamma_{\bar au}+\gamma_{\bar
av})r_{\bar a}(\tau ,\vec \sigma )} (\delta^u_{(b)}\delta^v_{(a)}-\delta^u_{(a)}
\delta^v_{(b)})\times \nonumber \\
&&\sum_r {\cal T}^u_{(a)r}(\vec \sigma ,{\vec
\sigma}_1,\tau ) \Big( \phi^{-2}
e^{-{1\over {\sqrt{3}}}\sum_{\bar a}
\gamma_{\bar ar}r_{\bar a}}\sum_{\bar b}\gamma_{\bar
br} \pi_{\bar b}\Big) (\tau ,{\vec \sigma}_1)\nonumber \\
&&\sum_s {\cal T}^v_{(b)s}(\vec \sigma ,{\vec
\sigma}_2,\tau ) \Big( \phi^{-2}
e^{-{1\over {\sqrt{3}}}\sum_{\bar a}
\gamma_{\bar as}r_{\bar a}}\sum_{\bar c}\gamma_{\bar
cs} \pi_{\bar c}\Big) (\tau ,{\vec \sigma}_2)\, \Big)\, \Big] .
\label{VI23}
\end{eqnarray}

Let us remark that, if this integro-differential equation for $\phi
(\tau ,\vec \sigma )= e^{{1\over 2}q(\tau ,\vec \sigma )}\, > 0$ would
admit different solutions $\phi_1[r_{\bar a},\pi_{\bar a}]$,
$\phi_2[r_{\bar a},\pi_{\bar a}]$ ,..., they would correspond to
inequivalent gravitational fields in vacuum (there are no more gauge
transformations for correlating them) evolving according to the
associated ADM energies. But it is hoped that Lichnerowicz's results
in the case of maximal slicing imply the unicity of the solution
\footnote{With the boundary condition $\phi (\tau ,\vec \sigma )\,
{\rightarrow}_{r\,
\rightarrow \infty}\, 1+{M\over {4r}}+O(r^{-3/2})$ of Eqs.(\ref{VI11}).}
also in tetrad gravity with 3-orthogonal coordinates and with the
natural gauge fixing $\rho (\tau ,\vec \sigma ) \approx 0$.

If we add the natural gauge-fixing $\rho (\tau ,\vec \sigma )={1\over
2}\phi (\tau ,\vec \sigma ) \pi_{\phi}(\tau ,\vec \sigma ) \approx 0$
to ${\hat {\cal H}}_R(\tau ,\vec \sigma )\approx 0$, its time
constancy implies \footnote{With ${\hat H}_{(D)ADM,R}$ from
Eq.(\ref{VI9}), ${\hat P}^A_{ADM,R}$ from Eq.(\ref{b2}) of Appendix B
and ${\hat {\cal H}}_R(\tau ,\vec \sigma )$ from Eq.(\ref{VI22}).}

\begin{eqnarray}
\partial_{\tau} \rho (\tau ,\vec \sigma )\, &{\buildrel \circ \over =}\,&
\{ \rho (\tau ,\vec \sigma ), {\hat H}_{(D)ADM,R} \} = \int
d^3\sigma_1 n(\tau ,{\vec \sigma}_1) \{ \rho (\tau ,\vec \sigma ), {\hat {\cal
H}}_R(\tau ,{\vec \sigma}_1) \} +\nonumber \\
&+& {\tilde \lambda}_{\tau}(\tau ) \{ \rho (\tau ,\vec \sigma ),{\hat P}^{\tau}
_{ADM,R} \} + {\tilde \lambda}_r(\tau ) \{ \rho (\tau ,\vec \sigma ), {\hat P}^r
_{ADM,R} \} \approx \nonumber \\
&\approx& -{1\over 2}\phi (\tau ,\vec \sigma ) \Big[ \int d^3\sigma_1
n(\tau ,{\vec \sigma}_1) {{\delta {\hat {\cal H}}_R(\tau ,{\vec
\sigma}_1)}\over {\delta \phi (\tau ,\vec \sigma )}}+\nonumber \\ &+&
{\tilde \lambda}_{\tau} {{\delta {\hat P}^{\tau}_{ADM,R}}\over {\delta
\phi (\tau ,\vec \sigma )}}+{\tilde \lambda}_r(\tau ) {{\delta {\hat
P}^r_{ADM,R}}\over {\delta \phi (\tau ,\vec \sigma )}}  \Big]\approx
0,\nonumber \\
 &&{}\nonumber \\
\Rightarrow&& \quad n(\tau ,\vec \sigma ) - \hat n(\tau ,\vec \sigma |r_{\bar a}
,\pi_{\bar a},{\tilde \lambda}_A] \approx 0,\nonumber \\
&&{}\nonumber \\
\partial_{\tau}&& \Big[ n(\tau ,\vec \sigma ) - \hat n(\tau ,\vec \sigma
|r_{\bar a},\pi_{\bar a},{\tilde \lambda}_A] \Big] =\nonumber \\
&=&\lambda_n(\tau ,\vec \sigma )- \{ \hat n(\tau ,\vec \sigma
|r_{\bar a},\pi_{\bar a},{\tilde \lambda}_A] , {\hat
H}^{(WSW)}_{(D)ADM,R} \} \approx 0,\nonumber \\ &&{}\nonumber \\
\Rightarrow&& \quad \lambda_n(\tau ,\vec \sigma )\quad
\text{determined};\nonumber \\
 &&{}\nonumber \\
 && \text{the rest-frame instant form expression of this
 equation is}\nonumber \\
 &&{}\nonumber \\
 \int d^3\sigma_1 && n(\tau ,{\vec \sigma}_1){{\delta {\hat {\cal H}}_R(\tau ,{\vec
\sigma}_1)}\over {\delta \phi (\tau ,\vec \sigma )}} =-\epsilon
{{\delta {\hat P}^{\tau}_{ADM,R}}\over {\delta
\phi (\tau ,\vec \sigma )}}.
\label{VI24}
\end{eqnarray}

Therefore we find an integral equation for the lapse function
$n(\tau ,\vec \sigma )$ implying its being different from zero
(this avoids a finite time breakdown), even in the rest-frame
instant form where ${\tilde \lambda}_r(\tau )=0$, ${\tilde
\lambda}_{\tau }(\tau )= \epsilon$. It is hoped that the boundary
condition $n(\tau ,\vec \sigma ) {\rightarrow}_{r\, \rightarrow
\infty}\, 0 + O(r^{-(2+\epsilon )})$ in a direction-independent
way [see Eq.(\ref{II55})] implies a unique solution of this
integral equation. Then, Eq.(\ref{VI17}) would imply a unique
determination of the shift functions.

If we now go to the final Dirac brackets with respect to the
second class constraints $\rho (\tau ,\vec \sigma )\approx 0$,
${\hat {\cal H}}_R(\tau ,\vec \sigma )\approx 0$, $n(\tau ,\vec
\sigma ) - \hat n(\tau ,\vec \sigma |r_{\bar a},\pi_{\bar
a},{\tilde \lambda}_A] \approx 0$, ${\tilde \pi}^n (\tau ,\vec
\sigma )\approx 0$, on the WSW hypersurfaces $\Sigma^{(WSW) \rho
=0 }_{\tau}$, asymptotically orthogonal to ${\hat P}^{(\mu
)}_{ADM,R}$ at spatial infinity, we remain only with the canonical
variables $r_{\bar a}$, $\pi_{\bar a}$ and with the following form
of the Dirac-Hamiltonian and of the remaining four first class
constraints

\begin{eqnarray}
{\hat H}^{(WSW)}_{(D)ADM,R\rho =0}&=&
-{\tilde \lambda}_{\tau }(\tau ) \Big( \epsilon
_{(\infty )}-{\hat P}^{\tau}_{ADM,R}[r_{\bar a},\pi_{\bar a},\phi (r_{\bar a},
\pi_{\bar a})]\Big)-\nonumber \\
 &-&{\tilde \lambda}_r(\tau ) {\hat P}^r_{ADM,R}
[r_{\bar a},\pi_{\bar a},\phi (r_{\bar a},\pi_{\bar a})],\nonumber
\\
 &&{}\nonumber \\
 &&\epsilon_{(\infty )}-{\hat P}^{\tau}_{ADM,R}
[r_{\bar a},\pi_{\bar a},\phi (r_{\bar a},\pi_{\bar a})] \approx
0,\nonumber \\
 &&{\hat P}^r_{ADM,R} [r_{\bar a},\pi_{\bar a},\phi
(r_{\bar a},\pi_{\bar a})] \approx 0.
\label{VI25}
\end{eqnarray}

\noindent where $\phi (r_{\bar a},\pi_{\bar a})$ is the solution
of the reduced Lichnerowicz equation ${\hat {\cal H}}_R(\tau ,\vec
\sigma ){|}_{\rho (\tau ,\vec \sigma )=0} =0$ and the weak ADM
energy is given in Eq.(\ref{b3}) of Appendix B.

After the gauge-fixing $T_{(\infty )}-\tau \approx 0$, one gets
${\tilde \lambda}_{\tau}(\tau )=\epsilon$ and Eq.(\ref{II49}) imply

\begin{eqnarray}
{\hat H}^{(WSW)}_{(D)ADM}&=& -\epsilon {\hat P}^{\tau}_{ADM,R}
[r_{\bar a},\pi_{\bar a},\phi (r_{\bar a},\pi_{\bar a})]
+\nonumber \\
 &+&{\tilde \lambda}_r(\tau ){\hat P}^r_{ADM,R}[r_{\bar a},\pi_{\bar a},\phi (r_{\bar a},
\pi_{\bar a})],\nonumber \\ &&{}\nonumber \\
 {\hat P}^r_{ADM,R}[r_{\bar a},\pi_{\bar a},\phi (r_{\bar a}, \pi_{\bar
a})] &\approx& 0. \label{VI26}
\end{eqnarray}

In the gauge ${\tilde \lambda}_r(\tau )=0$, implied by the gauge
fixings ${\hat J}^{\tau r}_{ADM,R}[r_{\bar a},\pi_{\bar a}, \phi
(r_{\bar a},\pi_{\bar a})] \approx 0$  [see Eq.(\ref{b3})] on the
{\it internal} 3-center-of-mass, we get the final Dirac
Hamiltonian in the asymptotic rest-frame instant form of dynamics
for tetrad gravity.

\beq
{\hat H}^{(WSW){'}}_{(D)ADM}= -\epsilon {\hat P}^{\tau}_{ADM,R},
\label{VI27}
\eeq

\noindent and that the Hamilton equations imply the following
normal form (namely solved in the accelerations) of the two dynamical
Einstein equations for the gravitational field Dirac observables in
the 3-orthogonal gauge with $\rho (\tau ,\vec \sigma )
\approx 0$ and in the rest frame

\begin{eqnarray}
\partial_{\tau} r_{\bar a}(\tau ,\vec \sigma )\, &{\buildrel \circ \over =}\,&
\{ r_{\bar a}(\tau ,\vec \sigma ),-\epsilon {\hat P}_{ADM,R}^{\tau}[r_{\bar b},\pi_{\bar
b},\phi (r_{\bar b},\pi_{\bar b})]\} ,\nonumber \\
\partial_{\tau} \pi_{\bar a}(\tau ,\vec \sigma )\, &{\buildrel \circ \over =}\,&
\{ \pi_{\bar a}(\tau ,\vec \sigma ),-\epsilon {\hat P}_{ADM,R}^{\tau}[r_{\bar b},\pi
_{\bar b},\phi (r_{\bar b},\pi_{\bar b})] \} ,\nonumber \\
&&{}\nonumber \\
 &&{\hat P}^r_{ADM,R}[r_{\bar a},\pi_{\bar a},\phi
(r_{\bar a},\pi_{\bar a})] \approx 0,\nonumber \\
 &&{\hat J}^{\tau r}_{ADM,R}[r_{\bar a},\pi_{\bar a},
\phi (r_{\bar a},\pi_{\bar a})]\approx 0.
\label{VI28}
\end{eqnarray}

The ADM Hamilton equations of metric gravity, equivalent to the
spatial Einstein equations, are \cite{ru11} $\partial_{\tau}\,
{}^3g_{rs}(\tau ,\vec \sigma )\, {\buildrel \circ \over =}\, \Big[
n_{r|s} + n_{s|r}-2N\, {}^3K_{rs}\Big] (\tau ,\vec \sigma ) $ and
$\partial_{\tau}\, {}^3K_{rs}(\tau ,\vec \sigma )\, {\buildrel
\circ \over =}\, \Big( N \Big[ {}^3R_{rs}+ {}^3K\, {}^3K_{rs}-2\,
{}^3K_{ru}\, {}^3K^u{}_s\Big] - n_{|s|r} + n^u{}_{|s}\,
{}^3K_{ur}+ N^u\, {}^3K_{rs|u}\Big) (\tau ,\vec \sigma) $: their
restriction to our completely fixed gauge is satisfied due to the
Hamilton equations (\ref{VI28}).

The 4-metric and the line element in adapted coordinates $\sigma^A$
on the WSW hypersurfaces are

\begin{eqnarray}
{}^4g_{\tau\tau} &=& \epsilon \Big[
 \Big(-\epsilon + n[r_{\bar a},\pi_{\bar a},
{\vec {\lambda}}]\Big)^2-\phi^{-4}[r_{\bar a},\pi_{\bar a}]\sum_r
e^{-{2\over {\sqrt{3}}}\sum_{\bar a}\gamma_{\bar ar}r_{\bar a}}
\Big({\lambda}_r(\tau )-{n}_r[r_{\bar a},\pi_{\bar a}, {\vec
{\lambda}}]\Big)^2 \Big] \nonumber \\ &\rightarrow&_{{\vec
{\lambda}}(\tau ,\vec \sigma )=0}\,\, \epsilon \Big[
\Big(-\epsilon +n[r_{\bar a},\pi_{\bar a},0]\Big)^2-
\phi^{-4}[r_{\bar a},\pi_{\bar a}]\sum_r e^{-{2\over
{\sqrt{3}}}\sum_{\bar a}\gamma_{\bar ar}r_{\bar a}}
{n}^2_r[r_{\bar a},\pi_{\bar a},0] \Big],\nonumber \\
 &&{}\nonumber \\
 {}^4g_{\tau r} &=& \epsilon \Big[
 -{\lambda}_r(\tau )+{n}_r[r_{\bar a},
\pi_{\bar a},{\vec {\lambda}}] \Big] \nonumber \\
&\rightarrow&_{{\vec {\lambda}}(\tau ,\vec \sigma )=0}\,\,
 \epsilon  {n}_r[r_{\bar a},\pi_{\bar a},0],\nonumber \\
 &&{}\nonumber \\
 {}^4g_{rs} &=& -\epsilon
\phi^4[r_{\bar a},\pi_{\bar a}]e^{{2\over {\sqrt{3}}} \sum_{\bar
a}\gamma_{\bar ar}r_{\bar a}}\delta_{rs}\, , \nonumber \\
 &&{}\nonumber \\
 &&{}\nonumber \\
ds^2&=&{}^4{\hat g}_{\tau\tau}(d\tau )^2 +2\, {}^4{\hat g}_{\tau r}
d\tau d\sigma^r +\sum_r {}^4{\hat g}_{rr}(d\sigma^r)^2.
\label{VI29}
\end{eqnarray}

\noindent Even for ${\lambda}_r(\tau )=0$ we do not get vanishing
shift functions ({\it synchronous} coordinates), like instead it is
assumed by Christodoulou and Klainermann for their singularity-free
solutions.

Let us remark that Eqs.(\ref{VI17}) and (\ref{VI24}) imply that both
$n$ and ${n}_r$ depend on $G/c^3$ and $c^3/G$ simultaneously, so that
both their post-Newtonian (expansion in $1/c$) and post-Minkowskian
(formal expansion in powers of G) may be non trivial after having done
the gauge fixings. From Eq.(\ref{VI23}) it is clear that $\phi
[r_{\bar a},\pi_{\bar a}]$ depends on $G^2/c^6$, while the previous
equation implies that ${\hat H}^{(WSW) '}_{(D)ADM}
= {\hat P}^{\tau}_{ADM,R}$ depends a priori on both $G/c^3$ and
$c^3/G$.

\vfill\eject

\section{The Embedding into Spacetime of the Wigner-Sen-Witten
         Hypersurfaces.}

We will see in this Section that the special WSW spacelike
hypersurfaces $\Sigma_{\tau}^{(WSW)}$, needed for the rest-frame
instant form of tetrad gravity in our class of spacetimes,
asymptotically flat at spatial infinity and without supertranslations,
and corresponding to the Wigner hyperplanes orthogonal to the
4-momentum of an isolated system, can be defined by special embeddings
$z^{\mu}(\tau ,\vec
\sigma )$ \footnote{Generalizing the embeddings $z^{(\mu )}(\tau ,\vec
\sigma )=x^{(\mu )}(\tau )+\epsilon^{(\mu )}_r(u(p_s))
\sigma^r$ for Minkowski Wigner spacelike hyperplanes.}.

Moreover, we will see that they enjoy the same formal properties
of spacelike hyperplanes in Minkowski spacetime,  namely that,
given an origin on each one of them and an adapted tetrad at this
origin, there is a {\it natural parallel transport} so that one
can uniquely define the adapted tetrads in all points of the
hyperplane starting from the given adapted one at the origin.
Namely due to the property of tending asymptotically to Minkowski
Wigner spacelike hyperplanes in a direction-independent way at
spatial infinity, the WSW spacelike hypersurfaces allow the
definition of asymptotic (angle-independent) adapted tetrads with
the timelike component parallel to the weak ADM 4-momentum. Then
an adaptation to tensors of the Sen-Witten spinorial
equation\cite{p27,sen,spinor,spinor1,rindler} based on the Sen
connection\footnote{See Ref.\cite{soluz} for the existence of
solutions on noncompact spacetimes including the
Christodoulou-Klinermann ones.} allows to define preferred adapted
tetrads in each point of $\Sigma^{(WSW)}_{\tau}$ tending to the
given ones at spatial infinity: this can be reinterpreted as a
special form of {\it parallel transport} generalizing the trivial
Euclidean one on Minkowski spacelike hyperplanes.

In Ref. \cite{p28} Frauendiener , exploiting the fact that there is a
unique 2-1 (up to a global sign) correspondence between a SU(2) spinor
and a triad on a spacelike hypersurface, derives the necessary and
sufficient conditions that have to be satisfied by a triad in order to
correspond to a spinor that satisfies the Sen-Witten equation. In this
way it is possible to eliminate completely any reference to spinors
and to speak only of triads ${}^3e^{(WSW)}{}^r_{(a)}$ on
$\Sigma^{(WSW)}_{\tau}$ and $\Sigma^{(WSW)}_{\tau}$-adapted tetrads on
$M^4$.

These triads ${}^3e^{(WSW)}{}^r_{(a)}$ are built in terms of the SU(2)
spinors solutions of the Sen-Witten equation and, as a consequence of
this equation, they are shown \cite{p28} to satisfy the following
equations

\begin{eqnarray}
&&{}^3\nabla_r\, {}^3e^{(WSW) r}_{(1)}={}^3\nabla_r\, {}^3e^{(WSW)
r}_{(2)}=0,\nonumber \\
 &&{}^3\nabla_r\, {}^3e^{(WSW) r}_{(3)}=-\alpha
{}^3K,\nonumber \\
 &&{}^3e^{(WSW) r}_{(1)}\, {}^3e^{(WSW) s}_{(3)}\,
{}^3\nabla_r\, {}^3e^{(WSW)}_{(2)s} +{}^3e^{(WSW) r}_{(3)}\,
 {}^3e^{(WSW) s}_{(2)}\, {}^3\nabla_r\, {}^3e^{(WSW)}_{(1)s}+\nonumber \\
 &+&{}^3e^{(WSW) r}_{(2)}\, {}^3e^{(WSW) s}_{(1)}\, {}^3\nabla_r\,
{}^3e^{(WSW)}_{(3)s}=0.
\label{VII1}
\end{eqnarray}

\noindent Here for ${}^3K$ one uses Eq.(\ref{VI17}) in the 3-orthogonal gauge or
(\ref{VI23}) when $\rho (\tau ,\vec \sigma )\approx 0$.

Therefore, these triads are formed by 3 vector fields with the
properties: i) two vector fields are divergence free; ii) the
third one has a non-vanishing divergence proportional to the trace
of the extrinsic curvature of $\Sigma^{(WSW)}_{\tau}$ (on  a
maximal slicing hypersurface (${}^3K=0$) all three vectors would
be divergence free); iii) the vectors satisfy a cyclic condition.

In Ref.\cite{p28} it is shown: 1) these triads do not exist for
compact $\Sigma_{\tau}$; 2) with nontrivial topology for
$\Sigma_{\tau}$ there can be less than 4 real solutions and the triads
cannot be build; 3) the triads exist for asymptotically null surfaces,
but the corresponding tetrad will be degenerate in the limit of null
infinity.

Moreover, in Ref.\cite{p28}, using the results of Ref.\cite{p32},
it is noted that the Einstein energy-momentum
pseudo-tensor\cite{emp} is a canonical object only in the frame
bundle over $M^4$, where it coincides with the Sparling 3-form. In
order to bring this 3-form back to a 3-form (and then to an
energy-momentum tensor) over the spacetime $M^4$, one needs a
section (i.e. a tetrad) in the frame bundle. Only with the 3+1
decomposition of $M^4$ with WSW foliations one gets that (after
imposition of Einstein's equations together with the local energy
condition) one has a {\it preferred (geometrical and dynamical)
adapted tetrad} on the initial surface $\Sigma^{(WSW)}_{\tau}$.

A triad satisfying Eqs.(\ref{VII1}) is unique up to global frame
{\it rotations} and {\it homotheties}. But Eqs.(\ref{II55}) imply
that we must select the solutions of Eqs.(\ref{VII1}) with the
same asymptotic behaviour of ordinary triads ${}^3e^r_{(a)}$, i.e.
${}^3e^{(WSW)}_{(\infty )}{}^r_{(a)}={}^3e^r_{(\infty
)(a)}=\delta^r_{(a)}$ and ${}^3e^{(WSW)}{}^r_{(a)}(\tau ,\vec
\sigma )\, {\rightarrow}_{r\, \rightarrow \infty}\, (1-{M\over
{2r}})\delta^r_{(a)}+O(r^{-3/2})$. In this sense, the geometry of
an initial data set {\it uniquely} determines a {\it triad} on
$\Sigma^{(WSW)}_{\tau}$ (called {\it geometrical}  in
Ref.\cite{p28}) and hence together with the normal an {\it adapted
tetrad} ${}^4_{(\Sigma )}{\check E}^{(W) (\mu )}_A$ in spacetime
according to Eq.(\ref{II28}).

Therefore, we can define the $\Sigma^{(WSW)}_{\tau}$-adapted {\it
preferred tetrads} of the rest-frame instant form

\begin{eqnarray}
{}^4_{(\Sigma )}{\check {\tilde E}}^{(WSW)}{}^A_{(o)}&=& {1\over
{-\epsilon +n}} (1; -n^r),\nonumber \\
  {}^4_{(\Sigma )}{\check {\tilde E}}^{(WSW)}{}^A_{(a)} &=& (0; {}^3e^{(WSW)}{}^r_{(a)}),
  \nonumber \\
  &&{}\nonumber \\
  {}^4_{(\Sigma )}{\check E}^{(WSW)}{}^{\mu}_{(o)} &=& l^{\mu},\nonumber \\
  {}^4_{(\Sigma )}{\check E}^{\mu}_{(a)} &=& b^{\mu}_s\,\,\, {}^3e^{(WSW)}{}^s_{(a)}.
\label{VII2}
\end{eqnarray}

Since the WSW hypersurfaces and the 3-metric on them are dynamically
determined
\footnote{The solution of Einstein equations is needed to find the
physical 3-metric, the allowed WSW hypersurfaces and the Sen
connection.}, one has neither a static background on
system-independent hyperplanes like in parametrized Newton theories
nor a static one on the system-dependent Wigner hyperplanes like in
parametrized Minkowski theories. Now both the WSW hyperplanes and the
metric on it are system dependent.

These preferred tetrads  correspond to the {\it non-flat preferred
observers} of Bergmann\cite{be}: they are a set of {\it privileged
observers} (privileged tetrads adapted to $\Sigma^{(WSW)}_{\tau}$) of
{\it geometrical nature} \footnote{Since they depend on the intrinsic
and extrinsic geometry of $\Sigma_{\tau}^{(WSW)}$; on the solutions of
Einstein's equations they also acquire a {\it dynamical nature}
depending on the configuration of the gravitational field itself.} and
not of {\it static nature} like in the approaches of M\o
ller\cite{p23}, Pirani \cite{p24} and Goldberg\cite{gold}. These
privileged observers are associated with the existence of the
asymptotic Poincar\'e charges, since their asymptotic 4-velocity is
determined by the weak ADM 4-momentum. A posteriori, namely after
having solved Einstein's equations, one could try to use these {\it
geometrical and dynamical} privileged observers \footnote{Privileged
non-holonomic coordinate systems replacing the rectangular Minkowski
coordinates of the flat case.} in the same way as, in metric gravity,
are used the {\it bimetric theories}, like the one of Rosen\cite{p25},
with a set of privileged static non-flat background metrics. This
congruence of timelike preferred observers \footnote{With asymptotic
inertial observers in the rest-frame instant form with ${\tilde
\lambda}_A(\tau )= (\epsilon ;\vec 0)$ and ${\tilde
\lambda}_{AB}(\tau )=0$.} is a
non-Machian element of these noncompact spacetimes. The asymptotic
worldlines of the congruence may replace the static concept of {\it
fixed stars} in the study of the precessional effects of
gravitomagnetism on gyroscopes ({\it dragging of inertial frames}) and
seem to be naturally connected with the definition of post-Newtonian
coordinates \cite{mtw,soffel}.

With the asymptotic triads ${}^3e^r_{(\infty
)(a)}={}^3e^{(WSW)}_{(\infty )}{}^r_{(a)}=\delta^r_{(a)}$ as boundary
conditions at spatial infinity for the Frauendiener equations, their
solution defines a set of {\it preferred triads}
${}^3e^{(WSW)}{}^r_{(a)}(\tau ,\vec \sigma )$ in each point of
$\Sigma^{(WSW)}_{\tau}$ ({\it Sen-Witten parallel transport} of the
asymptotic triads), which will be connected by a rotation of angle
$\alpha^{(WSW)}_{(a)}(\tau ,\vec
\sigma )$ to the ordinary triads (analogous formulas are valid for cotriads)

\begin{equation}
{}^3e^r_{(a)}(\tau ,\vec \sigma ) =
R_{(a)(b)}(\alpha_{(c)}^{(WSW)}(\tau ,\vec \sigma ))\,
{}^3e^{(WSW)}{}^r_{(b)}(\tau ,\vec \sigma ).
\label{VII3}
\end{equation}

The asymptotic transition functions from arbitrary coordinates on
$M^4$ to WSW hypersurfaces $\Sigma^{(WSW)}_{\tau}$ are [see
Eqs.(\ref{II39})-(\ref{II45})]

\begin{eqnarray}
 {\hat b}^{(\mu )}_{(\infty )l}&\approx& \epsilon l^{(\mu )}_{(\infty
 )}\approx {{b^{(\mu )}_{(\infty )\tau}}\over {-\epsilon +n}},
 \nonumber \\
 {\hat b}^{(\mu )}_{(\infty ) r}&=& b^{(\mu )}_{(\infty )r}=
 \epsilon^{(\mu )}_r(u(p_{(\infty )})),\nonumber \\
 {\hat b}^l_{(\infty )(\mu )}&=&  l_{(\infty )(\mu )}=-\epsilon b^{\tau}_{(\infty )(\mu )},
 \nonumber \\
 {\hat b}^s_{(\infty )(\mu )}&=&b^s_{(\infty )(\mu )}-{\tilde \lambda}_s(\tau )b^{\tau}
 _{(\infty )(\mu )} \approx b^s_{(\infty )(\mu )},\nonumber \\
  &&with\nonumber \\
  b^{(\mu )}_{(\infty )A}(\tau ) &\equiv& L^{(\mu )}{}_{(\nu
)=A}(p_{(\infty )},{\buildrel \circ \over p}_{(\infty )}) =
\epsilon^{(\mu )}_A(u(p_{(\infty )})),\nonumber \\
 &&{}\nonumber \\
 \epsilon l^{(\mu )}_{(\infty )}&=&
 \epsilon^{(\mu )}_o(u(p_{(\infty )}))=u^{(\mu )}(p_{(\infty )})
 \approx {{{\hat P}^{(\mu )}_{ADM}}\over {\epsilon_{(\infty )}}},\nonumber \\
 \epsilon_{(\infty )}\approx M_{ADM}&=&\sqrt{\epsilon {\hat P}^2_{ADM}},\quad\quad
 S^{(\mu )(\nu )}_{(\infty )}\equiv {\hat S}^{(\mu )(\nu )}_{ADM}.
\label{VII4}
\end{eqnarray}

Given the previous boundary conditions on the triads [${}^3e^r_{(a)}
\rightarrow {}^3e^r_{(\infty )(a)}={}^3e^{(WSW)}_{(\infty )}{}^r_{(a)}
=\delta_{(a)}^r$] and cotriads [${}^3e_{(a)r} \rightarrow {}^3e_{(\infty )(a)r}=
{}^3e^{(WSW)}_{(\infty ) (a)r}=\delta_{(a)r}$], we have the following
associated asymptotic tetrads on $\Sigma^{(WSW)}_{\tau}$  (we assumed
$\varphi_{(a)}(\tau ,\vec \sigma )\, {\rightarrow}_{r\, \rightarrow
\infty}\, 0$ in Eq.(\ref{II55}) as a standard of asymptotic inertiality)

\begin{eqnarray}
{}^4E^{(WSW)}_{(\infty )}{}^{(\mu )}_{(o)} \delta^{\mu}_{(\mu )}&=&
{}^4_{(\Sigma )}{\check E}^{(WSW)}_{(\infty )}{}^{(\mu )}_{(o)}
\delta^{\mu}_{(\mu )}=
\delta^{\mu}_{(\mu )} l^{(\mu )}_{(\infty )}=\delta^{\mu}_{(\mu )} {\hat b}^{(\mu )}_{(\infty )l}
 \equiv \delta^{\mu}_{(\mu )} b^{(\mu )}_{(\infty )\tau}= \delta^{\mu}_{(\mu )}
u^{(\mu )}(p_{(\infty )}),\nonumber \\
 {}^4E^{(WSW)}_{(\infty)}{}^{(\mu )}_{(a)} \delta^{\mu}_{(\mu )} &=&
 {}^4_{(\Sigma )}{\check E}^{(WSW)}_{(\infty )}{}^{(\mu )}_{(a)} \delta^{\mu}_{(\mu )}=
 \delta^{\mu}_{(\mu )} {\hat b}^{(\mu )}_{(\infty )s}\, {}^3e^{(WSW)}_{(\infty )}{}^s_{(a)}
 \equiv \delta^{\mu}_{(\mu )} b^{(\mu )}_{(\infty )s}\, {}^3e^{(WSW)}_{(\infty )}{}^s_{(a)}=
 \nonumber \\
 &=&\delta^{\mu}_{(\mu )} b^{(\mu )}_{(\infty )s} \delta^s_{(a)}=
 \delta^{\mu}_{(\mu )}
\epsilon^{(\mu )}_s(u(p_{(\infty )})) \delta^s_{(a)},\nonumber \\
&&{}\nonumber \\
 {}^4E^{(WSW)}_{(\infty )}{}^A_{(o)}&=& {}^4_{(\Sigma )}{\check {\tilde E}}^{(WSW)}
 _{(\infty )}{}^A_{(o)}= (-\epsilon ; 0),\nonumber \\
 {}^4E^{(WSW)}_{(\infty )}{}^a_{(a)} &=& {}^4_{(\Sigma )}{\check {\tilde E}}^{(WSW)}
 _{(\infty )}{}^A_{(a)}= (0; \delta^r_{(a)}),\nonumber \\
 &&{}\nonumber \\
 {}^4E^{(WSW)}_{(\infty )}{}^{(o)}_{(\mu )} \delta^{(\mu )}_{\mu} &=&
 {}^4_{(\Sigma )}{\check E}^{(WSW)}_{(\infty )}{}^{(o)}_{(\mu )} \delta^{(\mu )}_{\mu}=
 \delta^{(\mu )}_{\mu}  l_{(\infty )(\mu )} =\delta^{(\mu )}_{\mu}
 {\hat b}^l_{(\infty )(\mu )}= -\epsilon \delta^{(\mu )}_{\mu} b^{\tau}_{(\infty )(\mu )},
 \nonumber \\
 {}^4E^{(WSW)}_{(\infty )}{}^{(a)}_{(\mu )} \delta^{(\mu )}_{\mu} &=&
 {}^4_{(\Sigma )}{\check E}^{(WSW)}_{(\infty )}{}^{(a)}_{(\mu )} \delta^{(\mu )}_{\mu}=
 \delta^{(\mu )}_{\mu} {\hat b}^s_{(\infty )(\mu )}\,
 {}^3e^{(WSW)}_{(\infty )}{}^{(a)}_s \equiv \delta^{(\mu )}_{\mu}
 b^s_{(\infty )(\mu )} \delta^{(a)}_s,\nonumber \\
 &&{}\nonumber \\
 {}^4E^{(WSW)}_{(\infty )}{}^{(o)}_A &=& {}^4_{(\Sigma )}{\check {\tilde E}}^{(WSW)}
 _{(\infty )}{}^{(o)}_A \equiv (-\epsilon ; 0),\nonumber \\
 {}^4E^{(WSW)}_{(\infty )}{}^{(a)}_A &=& {}^4_{(\Sigma )}{\check {\tilde E}}^{(WSW)}
 _{(\infty )}{}^{(a)}_A \equiv (0; {}^3e^{(WSW)}_{(\infty )}{}_{(a)r}=\delta_{(a)r}).
\label{VII5}
\end{eqnarray}

The final form of arbitrary tetrads and cotetrads obtained starting
from the adapted ones of Eqs.(\ref{VII2}) is [see Eqs.(\ref{II2}),
(\ref{II4})]

\begin{eqnarray}
{}^4E^{\mu}_{(o)} &=& \sqrt{1+\sum_{(c)} \varphi_{(c)}^2} l^{\mu}
 +\epsilon \sum_{(b)} \varphi_{(b)}\, {}^3e^s_{(b)} b^{\mu}_s=
 \nonumber \\
 &=& \sqrt{1+\sum_{(c)}\varphi_{(c)}^2} l^{\mu} +\epsilon \sum_{(b),(c)} \varphi_{(b)}
 R_{(b)(c)}(\alpha_{(a)}^{(WSW)})\, {}^3e^{(WSW)}{}^s_{(c)} b^{\mu}_s,\nonumber \\
 {}^4E^{\mu}_{(a)} &=& \epsilon \varphi_{(a)} l^{\mu} +\sum_{(b)}\Big[ \delta_{(a)(b)}
 +{{\varphi_{(a)}\varphi_{(b)}}\over {1+\sqrt{1+\sum_{(c)}\varphi_{(c)}^2}}}\Big] {}^3e^s_{(b)}
 b^{\mu}_s=\nonumber \\
 &=& \epsilon \varphi_{(a)} l^{\mu} +\sum_{(b),(c)}\Big[ \delta_{(a)(b)}
 +{{\varphi_{(a)}\varphi_{(b)}}\over {1+\sqrt{1+\sum_{(c)}\varphi_{(c)}^2}}}\Big]
 R_{(b)(c)}(\alpha_{(a)}^{(WSW)}) \, {}^3e^{(WSW)}{}^s_{(b)}
 b^{\mu}_s,\nonumber \\
 &&{}\nonumber \\
 {}^4E^A_{(o)} &\equiv& \Big( {{\sqrt{1+\sum_{(c)}\varphi_{(c)}^2}}\over {-\epsilon +n}};
 -\sqrt{1+\sum_{(c)}\varphi_{(c)}^2} {{n^r}\over {-\epsilon +n}} +\epsilon \sum_{(b)}
 \varphi_{(b)}\, {}^3e^r_{(b)}=\nonumber \\
 &=& -\sqrt{1+\sum_{(c)}\varphi_{(c)}^2} {{n^r}\over {-\epsilon +n}} +\epsilon \sum_{(b),(c)}
 \varphi_{(b)} R_{(b)(c)}(\alpha_{(a)}^{(WSW)})\, {}^3e^{(WSW)}{}^r_{(c)} \Big) ,\nonumber \\
 {}^4E^A_{(a)} &\equiv& \Big( {{\epsilon \varphi_{(a)}}\over {-\epsilon +n}};
 -\epsilon \varphi_{(a)} {{n^r}\over {-\epsilon +n}} +\sum_{(b)}\Big[ \delta_{(a)(b)}+
 {{\varphi_{(a)}\varphi_{(b)}}\over {1+\sqrt{1+\sum_{(c)}\varphi_{(c)}^2}}}\Big]
 \, {}^3e^r_{(b)}=\nonumber \\
 &=&  -\epsilon \varphi_{(a)} {{n^r}\over {-\epsilon +n}} +\sum_{(b),(c)}\Big[ \delta_{(a)(b)}+
 {{\varphi_{(a)}\varphi_{(b)}}\over {1+\sqrt{1+\sum_{(c)}\varphi_{(c)}^2}}}\Big]
 R_{(b)(c)}(\alpha_{(a)}^{(WSW)})\, {}^3e^{(WSW)}{}^r_{(c)} \Big) ,\nonumber \\
 &&{}\nonumber \\
 &&{}\nonumber \\
 {}^4E^{(o)}_{\mu} &=& \epsilon \sqrt{1+\sum_{(c)}\varphi_{(c)}^2} l_{\mu} -\epsilon
 \sum_{(b)} \varphi_{(b)}\, {}^3e_{(b)s} b^s_{\mu}=\nonumber \\
 &=& \epsilon \sqrt{1+\sum_{(c)}\varphi_{(c)}^2} l_{\mu} -\epsilon
 \sum_{(b),(c)} \varphi_{(b)} R_{(b)(c)}(\alpha_{(a)}^{(WSW)})\,
 {}^3e^{(WSW)}{}_{(b)s} b^s_{\mu},\nonumber \\
 {}^4E^{(a)}_{\mu} &=& -\varphi_{(a)} l_{\mu} =\sum_{(b)}\Big[ \delta_{(a)(b)}
 + {{\varphi_{(a)}\varphi_{(b)}}\over {1+\sqrt{1+\sum_{(c)}\varphi_{(c)}^2}}}\Big]
 {}^3e_{(b)s} b^s_{\mu}=\nonumber \\
 &=& -\varphi_{(a)} l_{\mu} =\sum_{(b),(c)}\Big[ \delta_{(a)(b)}
 + {{\varphi_{(a)}\varphi_{(b)}}\over {1+\sqrt{1+\sum_{(c)}\varphi_{(c)}^2}}}\Big]
 R_{(b)(c)}(\alpha_{(a)}^{(WSW)})\, {}^3e^{(WSW)}{}_{(b)s} b^s_{\mu},\nonumber \\
 &&{}\nonumber \\
 {}^4E^{(o)}_A&=& \Big( \sqrt{1+\sum_{(c)}\varphi_{(c)}^2}(-\epsilon +n)-
 \epsilon \sum_{(a)}\varphi_{(a)} n_{(a)}; \nonumber \\
 &&-\epsilon \sum_{(a)} \varphi_{(a)}\, {}^3e_{(a)r}=-\epsilon \sum_{(a),(b)}
 \varphi_{(a)} R_{(a)(b)}(\alpha_{(c)}^{(WSW)})\, {}^3e^{(WSW)}{}_{(b)r} \Big) ,
 \nonumber \\
 {}^4E^{(a)}_A &=& \Big( -\epsilon (-\epsilon +n) \varphi_{(a)}+\sum_{(b)}\Big[
 \delta_{(a)(b)}+{{\varphi_{(a)}\varphi_{(b)}}\over {1+\sqrt{1+\sum_{(c)}\varphi_{(c)}^2}}}
 \Big] n_{(b)};\nonumber \\
 &&\sum_{(b)}\Big[ \delta_{(a)(b)}+{{\varphi_{(a)}\varphi_{(b)}}\over {1+
 \sqrt{1+\sum_{(c)}\varphi_{(c)}^2}}} \Big] {}^3e_{(b)r}= \nonumber \\
 &=&\sum_{(b),(c)}\Big[ \delta_{(a)(b)}+{{\varphi_{(a)}\varphi_{(b)}}\over {1+
 \sqrt{1+\sum_{(c)}\varphi_{(c)}^2}}} \Big] R_{(b)(c)}(\alpha_{(a)}^{(WSW)})\,
 {}^3e^{(WSW)}{}_{(c)r} \Big) .
\label{VII6}
\end{eqnarray}

Since ${}^4{\check E}^{(\alpha )}_{\mu} dz^{\mu}= {}^4\theta^{(\alpha
)}$ are non-holonomic coframes, there are not coordinate hypersurfaces
and lines for the associated non-holonomic coordinates $z^{(\alpha )}$
\cite{pet} on $M^4$; as shown in Ref.\cite{rob} for them we have
${}^4\theta^{(\alpha )}=dz^{(\alpha )} +z^{(\beta )}
\Big[ {}^4{\check E}^{(\alpha )}_{\mu}\, {{\partial \,
{}^4{\check E}^{\mu}_{(\beta )}}\over {\partial z^{(\gamma )}}}
\Big] dz^{(\gamma )}$.

The embeddings $z^{\mu}(\tau ,\vec \sigma )$ of $R^3$ into $M^4$
associated with WSW spacelike hypersurfaces $\Sigma^{(WSW)}_{\tau}$ in
the rest-frame instant form of tetrad gravity are restricted to assume
the same form at spatial infinity of those in Minkowski spacetime
identifying the Wigner hyperplanes in the rest-frame instant form

\begin{eqnarray}
z^{\mu}(\tau ,\vec \sigma ) &{\rightarrow}_{r \rightarrow \infty}&\,
\delta^{\mu}_{(\mu )} z^{(\mu )}_{(\infty )}(\tau ,\vec \sigma ),\nonumber \\
 &&{}\nonumber \\
 z^{(\mu )}_{(\infty )}(\tau ,\vec \sigma ) &=& x^{(\mu )}_{(\infty )}(\tau )
 + \epsilon^{(\mu )}_r(u(p_{(\infty )})) \sigma^r =\nonumber \\
 &=& x^{(\mu )}_{(\infty )}(0) +u^{(\mu )}(p_{(\infty )}) \tau +
 \epsilon^{(\mu )}_r(u(p_{(\infty )})) \sigma^r.
\label{VII7}
\end{eqnarray}

By using the notation

\begin{eqnarray}
l^{\mu}&=& \epsilon {\hat b}^{\mu}_l ={{\epsilon}\over {-\epsilon +n}}
[ b^{\mu}_{\tau}- n^r b^{\mu}_r]={1\over {\sqrt{{}^3g}}}
\epsilon^{\mu}_{\alpha\beta\gamma}\, {}^4_{(\Sigma )}{\check E}^{(WSW)}{}^{\alpha}_{(1)}\,
{}^4_{(\Sigma )}{\check E}^{(WSW)}{}^{\beta}_{(2)}\, {}^4_{(\Sigma
)}{\check E}^{(WSW)}{}^{\gamma}_{(3)},\nonumber \\
\epsilon^{\mu}_r &=& b^{\mu}_s \, {}^3e^{(WSW)}{}^s_{(a)} \delta_{(a)r}\, \rightarrow \,
\delta^{\mu}_{(\mu )} b^{(\mu )}_{(\infty )s} \delta^s_{(a)}\delta_{(a)r}=
\delta^{\mu}_{(\mu )} b^{(\mu )}_{(\infty ) r},\nonumber \\
 &&{}\nonumber \\
 {\hat b}^{\mu}_r&=& b^{\mu}_r,\nonumber \\
 {\hat b}^l_{\mu}&=& l_{\mu} =(-\epsilon +n) b^{\tau}_{\mu}=(-\epsilon +n)
 \partial_{\mu} \tau (z),\nonumber \\
 {\hat b}^r_{\mu}&=& b^r_{\mu} +n^r b^{\tau}_{\mu},
\label{VII8}
\end{eqnarray}

\noindent we get the following expression for the embedding

\begin{eqnarray}
z^{\mu}_{(WSW)}(\tau ,\vec \sigma)&=& \delta^{\mu}_{(\mu )} x^{(\mu
)}_{(\infty )}(0)+ l^{\mu}(\tau ,\vec \sigma ) \tau
+\epsilon^{\mu}_r(\tau ,\vec \sigma ) \sigma^r =\nonumber \\
 &=&x^{\mu}_{(\infty )}(0)+ l^{\mu}(\tau ,\vec \sigma ) \tau + b^{\mu}_s(\tau ,\vec \sigma )
 \, {}^3e^{(WSW)}{}^s_{(a)}(\tau ,\vec \sigma ) \delta_{(a)r} \sigma^r=\nonumber \\
 &=& x^{\mu}_{(\infty )}(0) +b^{\mu}_A(\tau ,\vec \sigma ) F^A(\tau ,\vec \sigma ),\nonumber \\
  &&{}\nonumber \\
  &&F^{\tau}(\tau ,\vec \sigma )= {{\tau}\over {-\epsilon +n(\tau ,\vec \sigma )}},
  \nonumber \\
  && F^s(\tau ,\vec \sigma )={}^3e^{(WSW)}{}^s_{(a)}(\tau ,\vec \sigma ) \delta
  _{(a)r} \sigma^r -{{n^s(\tau ,\vec \sigma )}\over {-\epsilon +n(\tau ,\vec \sigma )}} \tau ,
\label{VII9}
\end{eqnarray}

\noindent
with $x^{(\mu )}_{(\infty )}(0)$ arbitrary \footnote{It reflects the
arbitrariness of the absolute location of the origin of asymptotic
coordinates (and, therefore, also of the {\it external} center of mass
${\tilde x}^{(\mu )}_{(\infty )}(0)$) near spatial infinity.}. See
Ref. \cite{bai} and its interpretation of the center of mass in
general relativity \footnote{This paper contains the main references
on the problem starting from Dixon's definition\cite{dixon},}:
$x^{(\mu )}_{(\infty )}(\tau )$ may be interpreted as the arbitrary
{\it reference} (or {\it central}) timelike worldline of this paper.

{}From Eqs.(\ref{VII9}) we can find the equations for determining the
transition coefficients $b^{\mu}_A(\tau ,\vec \sigma )={{\partial
z^{\mu}_{(WSW)}(\tau ,\vec \sigma )}\over {\partial \sigma^A}}$ and
therefore the coordinate transformation $x^{\mu} \mapsto \sigma^A$
from general 4-coordinates to adapted 4-coordinates

\begin{eqnarray}
b^{\mu}_A &=& {{\partial z^{\mu}_{(WSW)}}\over {\partial \sigma^A}}=
b^{\mu}_B {{\partial F^B}\over {\partial \sigma^A}} + {{\partial
b^{\mu}_B}\over {\partial \sigma^A}} F^B,\nonumber \\
 &&{}\nonumber \\
 &&A_A{}^B = \delta_A^B - {{\partial F^B}\over {\partial \sigma^A}},\nonumber \\
 &&{}\nonumber \\
 F^B {{\partial b^{\mu}_B}\over {\partial \sigma^A}} &=& A_A{}^B b^{\mu}_B,\nonumber \\
 &&or\quad b^{\mu}_b = (A^{-1})_B{}^a F^C {{\partial b^{\mu}_C}\over {\partial \sigma^A}}.
\label{VII10}
\end{eqnarray}

The coordinates $\sigma^A$ (our special 3-orthogonal coordinates) for
the 3+1 splitting of $M^4$ with leaves $\Sigma^{(WSW)}_{\tau}$ replace
the {\it standard PN coordinates} ($x^{\mu}(o)$ is the arbitrary
origin) and should tend to them in the Post-Newtonian approximation!

Moreover, from the equation $\partial_{\mu} \tau
(z)=l_{\mu}(z)/[-\epsilon +n(z)]$ we could determine the function
$\tau (z)$ associated with this class of globally hyperbolic
spacetimes. The WSW hypersurface $\Sigma^{(WSW)}_{\tau}$
associated with the given solution is the set of points
$z^{\mu}(\tau ,\vec \sigma )$ such that $\tau (z) = \tau$.

In conclusion, there are {\it preferred ADM geometrical and
dynamical  Eulerian observers}

\bea
&&{}^4_{(\Sigma )}{\check E}^{(WSW)}{}^{\mu}_{(\alpha )}=\Big(
l^{\mu}; b^{\mu}_s\, {}^3e^{(WSW)}{}^s_{(a)} \Big) ,\nonumber \\
  &&{}\nonumber \\
 && {}^4_{(\Sigma )}{\check{\tilde E}}^A_{(\alpha )}
= \Big( {1\over {-\epsilon +n}} (1; -n^r) ;
(0; {}^3e^{(WSW)}{}^r_{(a)}) \Big)  .
 \label{VII11}
\eea

They should be used as {\it conventional celestial reference
system (CCRS)} $S_I$ based on an extragalactic radio-source
catalogue system \footnote{Or , somewhat less accurate, by a star
catalogue system such as the FK5; or else by using the cosmic
microwave background.}\cite{frame} : this is a conventional
definition of inertiality with respect to rotations \footnote{The
tabulated {\it right ascensions} and {\it declinations} and, in
the case of a star catalogue, the {\it proper motions
(ephemerides)} define the reference axes of CCRS; the axes are
chosen in such a way that at a basic epoch  they coincide in
optimal approximation with the {\it mean equatorial frame} defined
by the {\it mean celestial pole} and the {\it mean dynamical
equinox}; these are non-relativistic definitions which can be
applied to the asymptotic triads; in the relativistic case one
considers the {\it proper reference frame of a single observer},
represented as a tetrad propagated along the worldline of the
observer by Fermi-Walker transport: the time axis of the tetrad is
the timelike worldline of the observer, while the three space axes
are spacelike geodesics (Fermi normal coordinates).}. Here the
Fermi-Walker transport is replaced by the $\tau$-evolution of the
WSW preferred tetrads. In this way one construct a {\it
kinematical} reference frame in our special 3-orthogonal
coordinates \footnote{It is {\it dynamical} if referred to the
ephemerides of some body in the solar system.}. Let us remember
that, given a {\it reference (coordinate) system} (one gives the
form of the 4-metric), to construct a {\it reference frame} is to
prescribe ({\it materialization}) some definite values of
coordinates for reference astronomical objects.

Let us remark that in presence of matter Eq.(\ref{VII9}) can be
used to reformulate Dixon's theory of multipoles for extended
objects \cite{dix,dixon} on WSW hypersurfaces by using the results
of Refs.\cite{dixx,bai}.

\vfill\eject

\section{Void Spacetimes in the 3-Orthogonal Gauges.}

Let us remark that  Minkowski spacetime in Cartesian coordinates is a
solution of Einstein equations, which in the 3-orthogonal gauges
corresponds to $q=\rho =r_{\bar a}=\pi_{\bar a}=0$ [$\phi =e^{q/2}=1$]
and $n=n_r=N_{(as)r}=0$, $N_{(as)}
=-\epsilon$. For $q=\rho =r_{\bar a}=0$ Eq.(\ref{VI15}) implies ${}^3{\hat
{\tilde \pi}}^r_{(a)}$ proportional to $\pi_{\bar a}$; the condition $\Sigma
_{\tau}=R^3$ implies ${}^3K_{rs}=0$ and then $\pi_{\bar a}=0$ as it will
be shown in Eq.(\ref{VIII6}).

Therefore, it is consistent with Einstein equations to
add by hand the two pairs of second class constraints
$r_{\bar a}(\tau ,\vec \sigma )\approx 0$, $\pi_{\bar a}(\tau ,\vec \sigma )
\approx 0$, to the Dirac Hamiltonian (\ref{VI5}) in the 3-orthogonal gauges
with arbitrary multipliers,

\beq
H^{'}_{(D)ADM,R}=H_{(D)ADM,R}+\int d^3\sigma [
\sum_{\bar a}(\mu_{\bar a}r_{\bar a}+\nu_{\bar a}\pi_{\bar a})](\tau ,\vec
\sigma ).
\label{VIII1}
\eeq

 The time constancy of these second class constraints
determines the multipliers

\bea
\partial
_{\tau}r_{\bar a}(\tau ,\vec \sigma )\, &{\buildrel \circ \over =}\,& \nu_{\bar a}
(\tau ,\vec \sigma )+\int d^3\sigma_1 n(\tau ,{\vec \sigma}_1)
\lbrace r_{\bar a}(\tau ,\vec \sigma ),{\hat {\cal H}}_R(\tau ,{\vec \sigma}_1)
\rbrace {}^{*}\approx 0,\nonumber \\
 \partial_{\tau} \pi_{\bar a}(\tau ,\vec \sigma )\,
&{\buildrel \circ \over =}\,& -\mu_{\bar a}(\tau ,\vec \sigma )+\int
d^3\sigma_1 n(\tau ,{\vec \sigma}_1) \lbrace \pi_{\bar a}(\tau ,\vec
\sigma ), {\hat {\cal H}}_R(\tau ,{\vec \sigma}_1)\rbrace
{}^{*}\approx 0.
\label{VIII2}
\eea

By going to new Dirac brackets, we remain with the only conjugate
pair $q(\tau ,\vec \sigma )$, $\rho(\tau ,\vec \sigma )$,
constrained by the first class constraint ${\hat {\cal H}}_R(\tau
,\vec \sigma ){|}_{r_{\bar a}=\pi_{\bar a} =0} \approx 0$. In this
way we get the description of a {\it family of gauge equivalent
spacetimes $M^4$ without gravitational field} (see III), which
could be called {\it void spacetimes}, with 3-orthogonal
coordinates for $\Sigma_{\tau}$. They turn out to be {\it
3-conformally flat} because ${}^3{\hat g}_{rs}=\phi^4\,
\delta_{rs}$, but with the conformal factor determined by the
Lichnerowicz equation as a function of $\rho$ (therefore it is
gauge dependent). Now, the line of Eqs.(\ref{VI16}) giving ${}^3K$
(with $r_{\bar a} =\pi_{\bar a}=0$) is an integral equation to get
$\rho$ in terms of ${}^3{\hat {\tilde \Pi}}\,$ (or ${}^3\hat K$)
and $q=2ln\, \phi ={1\over 6}\, ln\, {}^3\hat g$.

As we shall see, if we add the extra gauge-fixing $\rho (\tau ,\vec
\sigma )\approx 0$, we get the 3-Euclidean metric $\delta_{rs}$ on $\Sigma
_{\tau}$, since the superhamiltonian constraint (\ref{VI21}) has $q(\tau ,\vec \sigma )
\approx 0$ [$\phi (\tau ,\vec \sigma )\approx 1$]
as a solution in absence of matter. The time constancy of $\rho (\tau
,\vec \sigma )\approx 0$ implies $n(\tau ,\vec \sigma ) \approx 0$.
Indeed, for the reduction to Minkowski spacetime, besides the solution
$q(\tau ,\vec \sigma )\approx 0$ of the superhamiltonian constraint
(vanishing of the so called internal intrinsic (many-fingered) time
\cite{kku}), we also need the gauge-fixings $N_{(as)}(\tau ,\vec
\sigma )\approx \epsilon$, $N_{(as)r}(\tau ,\vec \sigma )
\approx 0$, $n_r=0$.

The members of the equivalence class of void spacetimes represent
flat Minkowski spacetimes in the most arbitrary coordinates
compatible with Einstein theory with the associated inertial
effects: they should correspond to the relativistic generalization
(in absence of matter) of the class of Galilean non inertial
frames (with their inertial forces) obtainable from an inertial
frame of the nonrelativistic Galileo spacetime \footnote{For
example the (maybe time-dependent) pseudo-diffeomorphisms in
$Diff\, \Sigma_{\tau}$ replace the Galilean coordinate
transformations generating the inertial forces.}. Therefore, they
seem to represent the most general {\it pure acceleration effects
without gravitational field (i.e. without tidal effects) but with
a control on the boundary conditions} compatible with Einstein's
general relativity for globally hyperbolic, asymptotically flat at
spatial infinity spacetimes \footnote{See also the discussion on
general covariance and on the various formulations of the
equivalence principle (homogeneous gravitational fields = absence
of tidal effects) in Norton's papers \cite{norton}. See also
Refs.\cite{abram} for the definition of {\it inertial forces} in
general relativity: the criticism to this approach saying that the
separation of inertial forces from the gravitational field is
arbitrary due to the equivalence principle, is solved in our
approach by the Shanmugadhasan canonical transformation which
separates the conformal factor from the Dirac observables for the
gravitational field. In this way the inertial forces become {\it
gauge} quantities as expected, to be determined by the gauge
fixing to the superhamiltonian constraint, which determines a
unique 3+1 splitting of spacetime. These forces depend on the
embedding $z^{\mu}(\tau ,\vec \sigma )$ of the leaves
$\Sigma_{\tau}$.}.

The concept of void spacetime implements the viewpoint of Synge\cite{synge}
that, due to tidal (i.e. curvature) effects, there is a difference between
true gravitational fields and accelerated motions, even if, as shown in
Ref.\cite{norton}, Einstein arrived at general relativity through the
intermediate step of showing the equivalence of uniform acceleration with
special homogeneous gravitational fields. It is only in the not generally
covariant Hamiltonian approach that one is able to identify the genuine
physical degrees of freedom of the gravitational field.

Since in void spacetimes without matter there are no physical
degrees of freedom of the gravitational field but only gauge
degrees of freedom, we expect that this equivalence class of
spacetimes is not described by scenario b) of Chapter IV, but that
it corresponds to scenario a) with vanishing Poincar\'e charges
(the exceptional Poincar\'e orbit). Indeed, in this way Minkowski
spacetime (and its gauge copies) would be selected as the static
background for special relativity with zero energy (after the
regularization of Ref.\cite{hh} which however influences only the
ADM boosts), starting point for parametrized Minkowski theories
where the special relativistic energy would be generated {\it
only} by the added matter (and/or fields). Tetrad gravity with
matter is  described by scenario b) (with the WSW hypersurfaces,
defined by the gravitational field and by matter, corresponding to
Wigner hyperplanes, defined by matter) and in the limit
$G\rightarrow 0$ the weak ADM energy [present in scenario b) but
not in scenario a)] tends to the special relativistic energy of
that matter system with no trace left of the {\it gravitational
field energy}. However, starting from parametrized Minkowski
theories on the background given by Minkowski spacetime in
Cartesian coordinates, we can introduce much more allowed 3+1
splittings (much more types of inertial forces) than in general
relativity, because in special relativity there is no restriction
on the spacelike hypersurfaces like being conformally flat.

To define {\it void spacetimes} independently from the
3-orthogonal gauge, let us remark that, since the conditions
$r_{\bar a}(\tau ,\vec \sigma )=0$ imply the vanishing of the
3-conformal Cotton-York tensor (see  Appendix A after
Eq.(\ref{a3}) for the definition of this tensor and Eq.(\ref{a4})
for its vanishing). Therefore, the general theory of void
spacetimes could be reformulated in arbitrary gauges by adding
with Lagrange multipliers the two independent components of the
Cotton-York tensor ${}^3{\cal Y} _{rs}(\tau ,\vec \sigma )$ (which
is a function only of cotriads) to the tetrad ADM Lagrangian of
Eq.(\ref{II8}) for tetrad gravity. In this way one should get two
extra holonomic constraints equivalent to $r_{\bar a}(\tau ,\vec
\sigma )\approx 0$. Their time constancy should produce two
secondary (momentum dependent) constraints equivalent to
$\pi_{\bar a}(\tau ,\vec \sigma )\approx 0$.

Deferring to a future paper the study of the general case, let us explore
the properties of void spacetimes in the 3-orthogonal gauges.

Void spacetimes depend only on the following variables:\hfill\break i)
$\phi(\tau ,\vec
\sigma ) =e^{q(\tau ,\vec \sigma )/2}$, to be determined by the
reduced Lichnerowicz equation;\hfill\break ii) $\pi_{\phi}(\tau ,\vec
\sigma )=2\phi^{-1}(\tau ,\vec \sigma ) \rho(\tau ,\vec \sigma )$, the
conjugate gauge variable.

We have

\begin{eqnarray}
&&r_{\bar a}=\pi_{\bar a}=0,\quad\quad \phi =e^{q/2}\quad
[or\, q=2 ln\, \phi],\nonumber \\
&&{}^3{\hat e}_{(a)r}=\phi^2 \delta_{(a)r},\quad\quad {}^3{\hat e}^r_{(a)}=
\phi^{-2} \delta^r_{(a)},\nonumber \\
&&{}^3{\hat g}_{rs}=\phi^4\delta_{rs},\quad\quad {}^3{\tilde g}_{rs}[r_{\bar
a}=0]=\delta_{rs},\nonumber \\
&&\tilde \triangle [r_{\bar a}=0]={\triangle}_{FLAT},\quad\quad
{}^3\tilde R[r_{\bar a}=0]=0.
\label{VIII3}
\end{eqnarray}

From Eqs.(\ref{VI14}), (\ref{VI15}), (\ref{VI16}) and (\ref{VI21}) one
has in void spacetimes (before putting $\rho
={1\over 2}\phi \pi_{\phi}=0$)

\begin{eqnarray}
{}^3{\hat g}_{rs}(\tau ,\vec \sigma )&=&=\phi^4(\tau ,\vec \sigma
)\delta_{rs} ,\nonumber \\ {}^3{\hat {\tilde \pi}}^r_{(a)}(\tau ,\vec
\sigma )&=& {1\over 3}\int d^3\sigma_1 {\tilde {\cal K}}
^r_{(a)s}(\vec \sigma ,{\vec \sigma}_1;\tau |\phi ,0]
\rho (\tau ,{\vec \sigma}_1),
\label{VIII4}
\end{eqnarray}

\noindent and

\begin{eqnarray}
{}^3{\hat K}_{rs}(\tau ,\vec \sigma )&=&{{\epsilon \, 4\pi G}\over
{c^3}} [ \sum_u(\delta_{ru}\delta_{(a)s}
+\delta_{su}\delta_{(a)r}-\delta_{rs}\delta_{(a)u})\, {}^3{\hat
{\tilde \pi}}^u_{(a)}](\tau ,\vec \sigma ),\nonumber \\
 {}^3{\hat K}(\tau ,\vec \sigma ) &=&-{{\epsilon \, 4\pi G}\over {c^3}} [\phi^{-4}
\sum_u \delta_{(a)u} \, {}^3{\hat {\tilde \pi}} ^u_{(a)}](\tau
,\vec \sigma )=\nonumber \\
 &=&-{{\epsilon \, 4\pi G}\over
{c^3}}\phi^{-6}(\tau ,\vec \sigma ) \{ \rho(\tau ,\vec \sigma
)+{1\over 3}\phi^2(\tau ,\vec \sigma ) \sum_u \int d^3\sigma_1
\delta_{(a)u} \nonumber \\ &&\sum_s{\cal T}^u_{(a)s}(\vec \sigma
,{\vec \sigma}_1;\tau |\phi ,0] \phi^{-2}(\tau, {\vec
\sigma}_1)\rho (\tau ,{\vec \sigma}_1) \} ,\nonumber \\
&&{}\nonumber \\
 {\hat {\cal H}}^{'}_R(\tau ,\vec \sigma )&=&
{\hat {\cal H}}_R(\tau ,\vec \sigma ){|}_{r_{\bar a}=\pi_{\bar
a}=0}= \nonumber \\ &=&\epsilon \phi (\tau ,\vec \sigma ) \Big(
-{{c^3}\over {2\pi G}} \triangle_{FLAT} \phi (\tau ,\vec \sigma
)+{{2\pi G}\over {3c^3}}\Big[ (\phi^{-7} \rho )(\tau ,\vec \sigma
)+\nonumber \\
 &+&{2\over 3} (\phi^{-5} \rho )(\tau ,\vec \sigma )
\int d^3\sigma_1 \sum_r \delta^u_{(a)}{\cal T}^u_{(a)r}(\vec
\sigma ,{\vec \sigma}_1;\tau |\phi ,0] (\phi^{-2} \rho )(\tau
,{\vec \sigma}_1)-\nonumber \\ &-&{1\over 3} \phi^{-3}(\tau ,\vec
\sigma ) \int d^3\sigma_1 d^3\sigma_2 \Big( \sum_u \nonumber \\
 &&\sum_r{\cal T}^u_{(a)r}(\vec \sigma ,{\vec \sigma}_1;\tau |\phi
,0] (\phi^{-2} \rho )(\tau ,{\vec \sigma}_1)\nonumber \\
 &&\sum_s{\cal T}^u_{(a)s}(\vec \sigma ,{\vec \sigma}_2;\tau |\phi ,0]
(\phi^{-2} \rho )(\tau ,{\vec \sigma}_2)+\nonumber \\
&+&\sum_{uv}(\delta^u_{(b)}\delta^v_{(a)}-\delta^u_{(a)}\delta^v_{(b)})
\nonumber \\
 &&\sum_r{\cal T}^u_{(a)r}(\vec \sigma ,{\vec
\sigma}_1;\tau |\phi ,0] (\phi^{-2} \rho )(\tau ,{\vec
\sigma}_1)\nonumber \\ &&\sum_s {\cal T}^v_{(a)s}(\vec \sigma
,{\vec \sigma}_2;\tau |\phi ,0] (\phi^{-2} \rho )(\tau ,{\vec
\sigma}_2) \Big) \Big] \Big) \approx 0. \label{VIII5}
\end{eqnarray}

With the natural gauge $\rho (\tau ,\vec \sigma )\approx 0$, one has

\begin{eqnarray}
{}^3{\hat {\tilde \pi}}^r_{(a)}(\tau ,\vec \sigma )&\approx 0,\nonumber \\
{}^3{\hat K}_{rs}(\tau ,\vec \sigma )&\approx& 0,\quad\quad
\Rightarrow  {}^3\hat K(\tau ,\vec \sigma )\approx 0,
\label{VIII6}
\end{eqnarray}

\noindent and
the reduced superhamiltonian constraint becomes the reduced Lichnerowicz
equation [$\triangle_{FLAT}={\vec \partial}^2$]

\begin{equation}
\triangle_{FLAT} \phi (\tau ,\vec \sigma )\approx 0,\quad \Rightarrow
\phi (\tau ,\vec \sigma ) = 1, \Rightarrow {}^3{\hat g}_{rs}=\delta_{rs},
\label{VIII7}
\end{equation}

\noindent where we have shown the solution corresponding to the boundary
condition of Eq.(\ref{VI11}).

In scenario a) with the Dirac Hamiltonian (\ref{II22}) we add the
gauge fixing constraints ${\tilde \lambda}_{\tau}(\tau )\approx
\epsilon$, ${\tilde \lambda}_r(\tau )\approx 0$, ${\tilde
\lambda}_{AB}(\tau )\approx 0$ to the primary constraints ${\tilde
\pi}^A(\tau )\approx 0$, ${\tilde \pi}^{AB}(\tau )\approx 0$. This
implies $\zeta_A(\tau )\approx 0$, $\zeta_{AB}(\tau )=0$ and
$N_{(as)}(\tau ,\vec \sigma )\approx -\epsilon$, $N_{(as)r}(\tau
,\vec \sigma )\approx 0$. Then we add the gauge fixings
$\alpha_{(a)}(\tau ,\vec \sigma )\approx 0$, $\varphi_{(a)}(\tau
,\vec \sigma )\approx 0$ to the rotation and boost primary
constraints (they imply $\lambda^{\vec \varphi}_{(a)}(\tau ,\vec
\sigma )={\hat \mu}_{(a)}(\tau ,\vec \sigma )=0$). Then we add the
second class constraints $r_{\bar a}(\tau ,\vec \sigma )\approx
0$, $\pi_{\bar a}(\tau ,\vec \sigma )\approx 0$ by hand. We choose
the 3-orthogonal gauges for parametrizing the cotriads and we add
the gauge fixings $\xi^r(\tau ,\vec \sigma )-\vec \sigma \approx
0$: this determines the shift functions $N_r(\tau ,\vec \sigma
)=n_r(\tau ,\vec \sigma )$ and the Dirac multipliers
$\lambda^{\vec n}_r(\tau ,\vec \sigma )$. We have $N(\tau ,\vec
\sigma )=-\epsilon +n(\tau ,\vec \sigma )$.

 The Dirac Hamiltonian becomes

\beq
H^{(1)}_{(D)ADM}=\int d^3\sigma [n{\hat {\cal
H}}^{'}_R+\lambda_n{\tilde
\pi}^n](\tau ,\vec \sigma )-\epsilon {\hat P}^{\tau \, {'}}_{ADM},
\label{VIII8}
\eeq

\noindent
with ${\hat {\cal H}}^{'}_R={\hat {\cal H}}_R{|}_{r_{\bar a}=\pi_{\bar
a}=0}$, ${\hat P}^{\tau \, {'}}_{ADM}={\hat
P}^{\tau}_{ADM}{|}_{r_{\bar a}=\pi_{\bar a}=0}$.

The natural gauge fixing $\rho (\tau ,\vec \sigma )\approx 0$ implies

\beq
\partial_{\tau}\rho (\tau ,\vec \sigma )\, {\buildrel \circ \over =}\, \int
d^3\sigma_1 n(\tau ,{\vec \sigma}_1) \{ \rho (\tau ,\vec \sigma
),{\hat {\cal H}}^{'}_R(\tau ,{\vec \sigma}_1) \} -\epsilon \{ \rho
(\tau ,\vec \sigma ), {\hat P}^{\tau \, {'}}_{ADN} \} ;
\label{VIII9}
\eeq

\noindent
but from Eq.(\ref{II25}) and from ${}^3{\tilde \Pi}^{rs}\approx 0$ we
see that only the term bilinear in the Christoffel symbols contributes
to $\{ \rho (\tau ,\vec \sigma ), {\hat P}^{\tau \, {'}}_{ADN} \}$ for
$\rho (\tau ,\vec \sigma )
\approx 0$. Now from Eq.(\ref{VI16}) we get ${}^3{\hat \Gamma}^r_{uv}=2\phi^{-1}
[\delta_{uv}\partial_r\phi +\delta_{ru}\partial_v\phi +\delta_{rv}\partial_u
\phi ] \rightarrow_{\phi \rightarrow const.}\, 0$. Since $\phi (\tau ,\vec
\sigma )=1$ is the solution of the reduced Lichnerowicz equation for $\rho (\tau
,\vec \sigma )\approx 0$, we get $\{ \rho (\tau ,\vec \sigma ), {\hat P}^{\tau
\, {'}}_{ADN} \} \approx 0$ and then $n(\tau ,\vec \sigma )\approx 0$ and
$\lambda_n(\tau ,\vec \sigma )\approx 0$. Therefore, at the end the
lapse function is $N(\tau ,\vec \sigma ) \approx
-\epsilon$.

Since, as we shall see, ${\hat P}^{\tau
\, {'}}_{ADN}$ vanishes for $\rho (\tau ,\vec \sigma )\approx 0$,
$\phi (\tau ,\vec \sigma )=1$, the final Dirac Hamiltonian vanishes:
$H^{(1)}_{(D)ADM}\approx 0$, and the final 4-metric becomes

\beq
{}^4{\hat g}_{AB}(\tau ,\vec \sigma )=\epsilon \,
\left( \begin{array}{cc} 1 & 0\\ 0 & -\delta_{rs} \end{array} \right).
\label{VIII10}
\eeq

In void spacetimes the two gauge-fixings $\rho (\tau ,\vec \sigma
)\approx 0$ and ${}^3\hat K(\tau ,\vec \sigma )\approx 0$ are
equivalent and one has $\phi (\tau ,\vec \sigma )=1$ (i.e. $q(\tau
,\vec \sigma )=0$); in this gauge one has ${}^3\hat R=0$ for the
3-hypersurfaces $\Sigma_{\tau}$ (they have both the scalar
curvature and the trace of the extrinsic one vanishing), but in
other gauges the 3-curvature and the trace of the extrinsic one
may be not vanishing because the solution $\phi (\tau ,\vec \sigma
)$ of the reduced Lichnerowicz equation may become nontrivial.

In Appendix  B there are the weak and strong Poincar\'e charges
for void spacetimes. It is shown in Eqs.(\ref{b4}) that at the
level of Dirac brackets with respect to the natural gauge fixing
$\rho (\tau ,\vec \sigma )\approx 0$ (i.e. with respect to the
pair of second class constraints $\rho \approx 0$, $\phi - 1
\approx 0$) the ten weak and strong Poincar\'e charges {\it
vanish} for the solution $\phi (\tau ,\vec \sigma )=1$ selected by
the boundary conditions (\ref{VI11}) (so that they must vanish in
all the others gauges connected with this solution, being
conserved gauge invariant quantities). In connection with this
result let us remark that only the boosts depend on the
regularization.  The result is consistent with parametrized
Minkowski theories whose Poincar\'e charges are not defined in
absence of matter.

As said void spacetimes describe pure acceleration effects without
dynamical gravitational field (no tidal effects), allowed in flat
spacetimes as the relativistic generalization of Galilean non
inertial observers, they cannot be used to describe {\it test
matter in flat spacetimes} in some post-Minkowskian approximation.

As we shall see in a future paper, where we shall study the
action-at-a- distance instantaneous effects on scalar particles
implied by Einstein theory in the ideal limit of a negligible
gravitational field \footnote{A more realistic situation with
tidal effects will be possible only after having studied the
linearization of tetrad gravity in 3-orthogonal gauges.} , in
presence of matter, in the static ideal limit $|r_{\bar a}(\tau
,\vec \sigma )| << 1$, $|\pi_{\bar a}(\tau ,\vec \sigma )| << 1$,
Eq.(\ref{VIII7}) becomes the Poisson equation $-\triangle_{FLAT}
\phi =\rho_{matter}+O(r_{\bar a},\pi_{\bar a})$, showing that
$\phi (\tau ,\vec \sigma )$ is the general relativistic
generalization of the Newton potential [see also the term M in the
boundary conditions (\ref{VI11})]. But now the Poincar\'e charges
are not vanishing so that we cannot use void spacetimes as
approximations of spacetimes $M^4$ for extremely weak
gravitational fields: already this kind of post-Minkowskian
approximation lives in non trivial spacetimes (with non trivial
WSW hypersurfaces) not gauge equivalent to Minkowski spacetime in
Cartesian coordinates.
 This is contrary to the expectation that for weak
gravitational fields a spacetime can be approximated by a void
spacetime with {\it test} matter, but it is consistent with
parametrized Minkowski theory for that {\it test} matter: its
arbitrary spacelike hypersurfaces embedded in Minkowski spacetime
describe a family of accelerated observers much bigger of the one
allowed in tetrad gravity (namely the WSW hypersurfaces). However,
as noted in footnote 39 of Section II, to avoid 3+1 splittings
geometrically ill-defined at spatial infinity one has to make the
restriction ${\tilde \lambda}_{AB}(\tau )=0$ also in Minkowski
spacetime, and this is a strong restriction on the allowed
accelerated observers. The implications of this fact, which have
still to be investigated, are that most of the possible
accelerated reference systems of Minkowski spacetime (which are
the starting point for the classical basis of the Unruh effect
\cite{vallis}) are unrelated with general relativity, at least
with its canonical ADM formulation presented in this paper. Also,
more study will be needed to clarify the conceptual difference
between a {\it test particle} following a geodesic of an external
gravitational field and a {\it dynamical particle plus the
gravitational field} (such a particle will not follow a priori a
geodesic of the resulting dynamical gravitational field).

\vfill\eject

\section{Conclusions.}

In this paper we have defined the rest-frame instant form and have
studied the Hamiltonian group of gauge transformations of tetrad
gravity, following what had been done in Ref.\cite{restmg} for
metric gravity, for a class of spacetimes of the
Christodoulou-Klainermann type. We have clarified the meaning of
the gauge transformations generated by the 14 first class
constraints of tetrad gravity. In particular, like in metric
gravity, the action of the superhamiltonian constraint is to
transform the leaves of the foliation associated with an allowed
3+1 splitting of spacetime into the leaves of another allowed 3+1
splitting. Since in Ref.\cite{restmg} there is also the
distinction between {\it Hamiltonian kinematical and dynamical
gravitational fields}(the second ones are solutions of the
Einstein equations), which is needed because at the Hamiltonian
level the Dirac observables are determined before solving the
Hamilton equations, we have not touched this point here.

We have then found a quasi-Shanmugadhasan canonical transformation
adapted to 3-orthogonal gauges and identified the Hamiltonian
(kinematical) Dirac observables for the gravitational field in a
special 3-orthogonal gauge defined by the vanishing of the
momentum conjugate to the conformal factor of the 3-metric, which
in turn is determined by the reduced Lichnerowicz equation, even
if we are not able to solve this equation. The evolution in the
mathematical time labelling the WSW spacelike hypersurfaces of the
rest-frame instant form, namely the leaves of the foliation in
this special 3-orthogonal gauge, is governed by the weak ADM
energy restricted to this gauge. We have also shown that the WSW
hypersurfaces imply the existence of asymptotic preferred
geometrical and dynamical inertial observers to be identified with
the fixed stars.

Therefore for the first time we can visualize the (non covariant)
physical degrees of freedom of the gravitational field in a
completely fixed gauge. This allows to define special spacetimes,
void spacetimes, in which there is no gravitational field but only
inertial effects like in the non-inertial frames in Newtonian
gravity. Minkowski spacetime without matter, with arbitrary
coordinates and with the allowed 3+1 splittings restricted to
those  whose leaves are 3-conformally flat (with the conformal
factor determined by the Lichnerowicz equation) seems to be the
only void spacetime: it is the background with vanishing
Poincar\'e charges where parametrized Minkowski theories live
\footnote{Parametrized Minkowski theories considered as autonomous
special relativistic theories and not as a limiting case of
general relativity have no restriction on the allowed 3+1
splittings. Therefore the class of their allowed inertial forces
is much bigger, since the leaves are not restricted to be
3-conformally flat.}.

The results of this paper seems to be the best which can be
obtained with the existing technology.

Referring to the Conclusions of Ref.\cite{restmg} for some comments on
the quantization of gravity in a completely fixed gauge, we list here
the main open problem to be studied:

1) To define the linearized theory: to find  a solution of a
linearization of the Lichnerowicz equation, which put inside a
linearization of the weak ADM energy to a quadratic expression in
$r_{\bar a}$ and $\pi_{\bar a}$ will imply a linear equation for
the physical canonical variables $r_{\bar a}(\tau ,\vec \sigma )$
describing the gravitational field, so to make contact with the
theory of gravitational waves. Let us remark that such a
linearized theory is not a {\it spin-2 theory on Minkowski
background} but an {\it approximate linearized Einstein spacetime}
with non-vanishing shift functions and gravitomagnetic effects.

2) To study other 3-coordinate gauges like the 3-normal
coordinates with respect to a point or systems of 3-coordinates
implying the reduction of the 3+1 splitting to an effective 2+2
splitting. It could be that the Lichnerowicz equation has some
simplification in these gauges.

3) To add matter (either scalar particles or perfect
fluids\cite{nowak}) and extract the action-at-a-distance potential
hidden in gravity in a post-Minkowskian approximation.

4) To add the electromagnetic field and to study the canonical
reduction in its presence as a first step in the direction of
unifying tetrad gravity with the standard SU(3)xSU(2)xU(1)
elementary particle model.

5) To understand better the observables the non-tensorial Dirac
observables versus the diffeomorphism-invariant ones), the time
problem and the physical identification of the points of spacetime
in general relativity (see Ref.\cite{restmg}). In particular to
make contact with measurements in a completely fixed gauge of
tidal effects: which kind of observables are selected by them?

6) To study tetrad gravity as the dynamical  theory of a
congruence of timelike observers: reformulate gravitomagnetism,
accelerated observers and their simultaneity surface and similar
problems in the tetrad gravity language.

\vfill\eject

\appendix

\section{3-Tensors in the special 3-orthogonal gauge.}

The 3-dimensional spin connection 1-form
${}^3\omega^{(a)}_{r(b)}d\sigma^r$, Christoffel coefficients
${}^3\Gamma^r_{uv}$, field strength ${}^3\Omega_{rs(a)}$ and the
curvature tensor ${}^3R^r{}_{suv}$ of the Riemannian 3-manifold
$(\Sigma_{\tau}, {}^3g_{rs}={}^3e_{(a)r}\, {}^3e_{(a)s})$ have the
following expression in terms of the   cotriads ${}^3e_{(a)r}$ (see
Ref.\cite{ru11} for the definitions)

\begin{eqnarray}
{}^3\omega^{(a)}_{r(b)}&=&{}^3\omega^{(a)}_{(c)(b)}\, {}^3e^{(c)}_r={}^3e_s
^{(a)}\, {}^3\nabla_r\, {}^3e^s_{(b)}=\nonumber \\
&=&{}^3e^{(a)}_s\, {}^3e^s_{(b) | r}={}^3e^{(a)}_s [\partial_r\, {}^3e^s_{(b)}+
{}^3\Gamma^s_{ru}\, {}^3e^u_{(b)}],\nonumber \\
&&{}\nonumber \\
{}^3\omega_{(a)(b)}&=&\delta_{(a)(c)}\, {}^3\omega^{(c)}_{r(b)} d\sigma^r=-
{}^3\omega_{(b)(a)},\quad\quad
{}^3\omega_{r(a)}={1\over 2}\epsilon_{(a)(b)(c)}\, {}^3\omega_{r(b)(c)},
\nonumber \\
{}^3\omega_{r(a)(b)}&=&\epsilon_{(a)(b)(c)}\, {}^3\omega_{r(c)}=[{\hat R}^{(c)}
{}^3\omega_{r(c)}]_{(a)(b)}=[{}^3\omega_r]_{(a)(b)},\nonumber \\
 &&{}\nonumber \\
 {}^3\Gamma^u_{rs}&=&{}^3\Gamma^u_{sr}=
{1\over 2}\, {}^3e^u_{(a)} \Big[ \partial_r\, {}^3e_{(a)s}+
\partial_s\, {}^3e_{(a)r}+\nonumber \\
&+&{}^3e^v_{(a)} \Big( {}^3e_{(b)r}(\partial_s\, {}^3e_{(b)v}-\partial_v\,
{}^3e_{(b)s})+{}^3e_{(b)s}(\partial_r\, {}^3e_{(b)v}-\partial_v\, {}^3e_{(b)r})
\Big) \Big] =\nonumber \\
&=&{1\over 2}\epsilon_{(a)(b)(c)}\, {}^3e^u_{(a)}({}^3e_{(b)r}\, {}^3\omega
_{s(c)}+{}^3e_{(b)s}\, {}^3\omega_{r(c)})-{1\over 2}({}^3e_{(a)r}\partial_s\,
{}^3e^u_{(a)}+{}^3e_{(a)s}\partial_r\, {}^3e^u_{(a)}),\nonumber \\
{}^3\omega_{r(a)(b)}&=&-{}^3\omega_{r(b)(a)}=
{1\over 2}\Big[ {}^3e^s_{(a)}(\partial_r\, {}^3e_{(b)s}-
\partial_s\, {}^3e_{(b)r})+\nonumber \\
&+&{}^3e^s_{(b)}(\partial_s\, {}^3e_{(a)r}-\partial_r\, {}^3e_{(a)s})+{}^3e^u
_{(a)}\, {}^3e^v_{(b)}\, {}^3e_{(c)r}(\partial_v\, {}^3e_{(c)u}-\partial_u\,
{}^3e_{(c)v}) \Big] =\nonumber \\
&=&{1\over 2} \Big[ {}^3e_{(a)u} \partial_r\, {}^3e^u_{(b)}-{}^3e_{(b)u}
\partial_r\, {}^3e^u_{(a)}+{}^3\Gamma^u_{rs} ({}^3e_{(a)u}\, {}^3e^s_{(b)}-
{}^3e_{(b)u}\, {}^3e^s_{(a)})\Big] ,\nonumber \\
{}^3\omega_{r(a)}&=&{1\over 2} \epsilon_{(a)(b)(c)} \Big[ {}^3e^u_{(b)}
(\partial_r\, {}^3e_{(c)u}-\partial_u\, {}^3e_{(c)r})+\nonumber \\
&+&{1\over 2}\, {}^3e^u_{(b)}\, {}^3e^v_{(c)}\, {}^3e_{(d)r}(\partial_v\,
{}^3e_{(d)u}-\partial_u\, {}^3e_{(d)v})\Big] ,\nonumber \\
 &&{}\nonumber \\
 &&[{}^3e_{(a)},
{}^3e_{(b)}]=({}^3\omega^{(c)}_{(a)(b)}-{}^3\omega^{(c)}_{(b)(a)}){}^3e_{(c)},
\nonumber \\
 &&{}\nonumber \\
{}^3\Omega^{(a)}{}_{(b)(c)(d)}&=&{}^3e_{(c)}({}^3\omega^{(a)}_{(d)(b)})-{}^3e
_{(d)}({}^3\omega^{(a)}_{(c)(b)})+\nonumber \\
&+&{}^3\omega^{(n)}_{(d)(b)}\, {}^3\omega^{(a)}_{(c)(n)}-{}^3\omega^{(n)}
_{(c)(b)}\, {}^3\omega^{(a)}_{(d)(n)}-({}^3\omega^{(n)}_{(c)(d)}-{}^3\omega
^{(n)}_{(d)(c)}){}^3\omega^{(a)}_{(a)(b)}=\nonumber \\
&=&{}^3e^{(a)}_r\, {}^3R^r{}_{stw}\, {}^3e^s_{(b)}\, {}^3e^t_{(c)}\, {}^3e^w
_{(d)},\nonumber \\
&&{}\nonumber \\
{}^3\Omega_{rs}{}^{(a)}{}_{(b)}&=&{}^3e^{(c)}_r\, {}^3e^{(d)}_s\, {}^3\Omega
^{(a)}{}_{(b)(c)(d)}={}^3R^t{}_{wrs}\, {}^3e_t^{(a)}\, {}^3e^w_{(b)}=
\nonumber \\
&=&\partial_r\, {}^3\omega^{(a)}_{s(b)} -\partial_s\, {}^3\omega^{(a)}_{r(b)}
+{}^3\omega^{(a)}_{r(c)}\, {}^3\omega^{(c)}_{s(b)} -{}^3\omega^{(a)}_{s(c)}\,
{}^3\omega^{(c)}_{r(b)}=\nonumber \\
&=&\delta^{(a)(c)}\, {}^3\Omega_{rs(c)(b)}=\delta^{(a)(c)}\, \epsilon_{(c)(b)
(d)}\, {}^3\Omega_{rs(d)},\nonumber \\
&&{}\nonumber \\
{}^3\Omega_{rs(a)}&=&{1\over 2}\epsilon_{(a)(b)(c)}\, {}^3\Omega_{rs(b)(c)}=
\partial_r\, {}^3\omega_{s(a)}-\partial_s\, {}^3\omega_{r(a)} -\epsilon
_{(a)(b)(c)}\, {}^3\omega_{r(b)}\, {}^3\omega_{s(c)}=\nonumber \\
&=&{1\over 2} \epsilon_{(a)(b)(c)} \Big[
\partial_r\, {}^3e^u_{(b)}\partial_s\, {}^3e_{(c)u}-\partial_s\,
{}^3e^u_{(b)}\partial_r\, {}^3e_{(c)u}+\nonumber \\
&+&{}^3e^u_{(b)}(\partial_u\partial_s\,
{}^3e_{(c)r}-\partial_u\partial_r\, {}^3e_{(c)s})+\nonumber \\
&+&{1\over 2}\Big( {}^3e^u_{(b)}\, {}^3e^v_{(c)}(\partial_r\,
{}^3e_{(d)s}-
\partial_s\, {}^3e_{(d)r})(\partial_v\, {}^3e_{(d)u}-\partial_u\, {}^3e_{(d)v})
+\nonumber \\
&+&({}^3e_{(d)s}\partial_r-{}^3e_{(d)r}\partial_s)[{}^3e^u_{(b)}\, {}^3e^v
_{(c)}(\partial_v\, {}^3e_{(d)u}-\partial_u\, {}^3e_{(d)v})]\Big) \Big] -
\nonumber \\
&-&{1\over 8}[\delta_{(a)(b_1)}\epsilon_{(c_1)(c_2)(b_2)}+\delta_{(a)(b_2)}
\epsilon_{(c_1)(c_2)(b_1)}+\delta_{(a)(c_1)}\epsilon_{(b_1)(b_2)(c_2)}+\delta
_{(a)(c_2)}\epsilon_{(b_1)(b_2)(c_1)}]\times \nonumber \\
&&{}^3e^{u_1}_{(b_1)}\, {}^3e^{u_2}_{(b_2)}\Big[ (\partial_r\, {}^3e_{(c_1)u_1}
-\partial_{u_1}\, {}^3e_{(c_1)r})(\partial_s\, {}^3e_{(c_2)u_2}-\partial_{u_2}\,
{}^3e_{(c_2)s})+\nonumber \\
&+&{1\over 2}\Big( {}^3e^{v_2}_{(c_2)}\, {}^3e_{(d)s}(\partial_r\, {}^3e
_{(c_1)u_1}-\partial_{u_1}\, {}^3e_{(c_1)r})(\partial_{v_2}\, {}^3e_{(d)u_2}-
\partial_{u_2}\, {}^3e_{(d)v_2})+\nonumber \\
&+&{}^3e^{v_1}_{(c_1)}\, {}^3e_{(d)r}(\partial_s\, {}^3e_{(c_2)u_2}-\partial
_{u_2}\, {}^3e_{(c_2)s})(\partial_{v_1}\, {}^3e_{(d)u_1}-\partial_{u_1}\,
{}^3e_{(d)v_1})\Big) +\nonumber \\
&+&{1\over 4}\, {}^3e^{v_1}_{(c_1)}\, {}^3e^{v_2}_{(c_2)}\, {}^3e_{(d_1)r}\,
{}^3e_{(d_2)s}(\partial_{v_1}\, {}^3e_{(d_1)u_1}-\partial_{u_1}\, {}^3e
_{(d_1)v_1})(\partial_{v_2}\, {}^3e_{(d_2)u_2}-\partial_{u_2}\, {}^3e
_{(d_2)v_2}) \Big] ,\nonumber \\
{}^3\Omega_{rs(a)(b)}&=&\epsilon_{(a)(b)(c)}\, {}^3\Omega_{rs(c)},\nonumber \\
{}^3R_{rsuv}&=&\epsilon_{(a)(b)(c)}\, {}^3e_{(a)r}\, {}^3e_{(b)s}\, {}^3\Omega
_{uv(c)},\nonumber \\
{}^3R_{rs}&=&{1\over 2} \epsilon_{(a)(b)(c)}\, {}^3e^u_{(a)} \Big[ {}^3e_{(b)r}
\, {}^3\Omega_{us(c)}+{}^3e_{(b)s}\, {}^3\Omega_{ur(c)}\Big] ,\nonumber \\
{}^3R&=&\epsilon_{(a)(b)(c)}\, {}^3e^r_{(a)}\, {}^3e^s_{(b)}\, {}^3\Omega
_{rs(c)}.
\label{a1}
\end{eqnarray}

\noindent where $\epsilon_{(a)(b)(c)}$ is the standard Euclidean antisymmetric
tensor and $({\hat R}^{(c)})_{(a)(b)}=\epsilon_{(a)(b)(c)}$ is the
adjoint representation of SO(3) generators. The first Bianchi identity
${}^3R^t{}_{rsu}+{}^3R^t{}_{sur}+{}^3R^t{}
_{urs}\equiv 0$ implies the cyclic identity ${}^3\Omega_{rs(a)}\, {}^3e^s_{(a)}
\equiv 0$. Under local SO(3) rotations R [$R^{-1}=R^t$] we have
${}^3\omega^{(a)}_{r(b)} \mapsto [R\, {}^3\omega_r\, R^T-R
\partial_r\, R^T]^{(a)}{}_{(b)}$,
${}^3\Omega_{rs}{}^{(a)}{}_{(b)} \mapsto [R\, {}^3\Omega_{rs}\,
R^T]^{(a)} {}_{(b)}$.

 For the conformal transformations in 3-manifolds one has (${}^3{\nabla}_r$ is the covariant
derivative associated with the 3-metric ${}^3{g}_{rs}$):

\bea
 {}^3g_{rs} &\mapsto& {}^3{\hat g}_{rs}=\phi^4\, {}^3g_{rs},\nonumber \\
 {}^3\Gamma^u_{rs} &\mapsto& {}^3{\hat \Gamma}^u_{rs}={}^3\Gamma^u_{rs}+2 \phi^{-1}
(\delta^u_r\, {}^3{\nabla}_s\phi+\delta^u_s\, {}^3{\nabla}_r\phi-
{}^3{g}_{rs}\, {}^3{g}^{uv}\, {}^3{\nabla}_v\phi ),\nonumber \\
 {}^3R &\mapsto& {}^3{\hat R}=\phi^{-4}\, {}^3R -8 \phi^{-5} ({}^3{g}^{uv}\,
{}^3{\nabla}_u\, {}^3{\nabla}_v)\phi.
\label{a2}
\eea

Eqs.(\ref{VI14}) imply the following expressions for the field
strengths and curvature tensors of $(\Sigma_{\tau}, {}^3g)$ given in
Eqs.(\ref{a1}) (we also give their limits for $r_{\bar a}\,
\rightarrow \, 0$ and $q=2 ln\, \phi \, \rightarrow \, 0$)

\begin{eqnarray}
{}^3{\hat \Omega}_{rs(a)}&=&
\epsilon_{(a)(b)(c)}\sum_u\delta_{(c)u}\nonumber \\
&&\Big( \delta_{(b)s}e^{{1\over {\sqrt{3}}}\sum_{\bar a}(\gamma_{\bar as}-\gamma
_{\bar au})r_{\bar a}}
\Big[ {1\over {\sqrt{3}}}(2\partial_uln\, \phi +{1\over {\sqrt{3}}}\sum
_{\bar b}\gamma_{\bar bs}\partial_ur_{\bar b})\sum_{\bar c}(\gamma_{\bar cs}-
\gamma_{\bar cu})\partial_rr_{\bar c}+\nonumber \\
&&+2\partial_u\partial_rln\, \phi +{1\over {\sqrt{3}}}\sum_{\bar
b}\gamma_{\bar bs}
\partial_u\partial_rr_{\bar b}\Big] -\nonumber \\
&&-\delta_{(b)r}e^{{1\over {\sqrt{3}}}\sum_{\bar a}(\gamma_{\bar ar}-\gamma
_{\bar au})r_{\bar a}}
\Big[ {1\over {\sqrt{3}}}(2\partial_uln\, \phi +{1\over {\sqrt{3}}}\sum
_{\bar b}\gamma_{\bar br}\partial_ur_{\bar b})\sum_{\bar c}(\gamma_{\bar cr}-
\gamma_{\bar cu})\partial_sr_{\bar c}+\nonumber \\
&&+2\partial_u\partial_sln\, \phi +{1\over {\sqrt{3}}}\sum_{\bar
b}\gamma_{\bar br}
\partial_u\partial_sr_{\bar b}\Big] \Big) +\nonumber \\
&&+{1\over 2}\sum_{uv}
\Big[ \delta_{(a)(b)}\epsilon_{(c)(d)(e)}-\delta_{(a)(c)}\epsilon
_{(b)(d)(e)}+\delta_{(a)(d)}\epsilon_{(e)(c)(b)}-\delta_{(a)(e)}\epsilon
_{(d)(c)(b)}\Big]\nonumber \\
&&e^{{1\over {\sqrt{3}}}\sum_{\bar a}(\gamma_{\bar ar}+\gamma_{\bar as}-\gamma
_{\bar au}-\gamma_{\bar av})r_{\bar a}}\Big( 2\partial_uln\, \phi +{1\over {\sqrt{3}}}\sum
_{\bar b}\gamma_{\bar br}\partial_ur_{\bar b}\Big) \Big( 2\partial_vln\, \phi +{1\over
{\sqrt{3}}}\sum_{\bar c}\gamma_{\bar cs}\partial_vr_{\bar
c}\Big)\nonumber \\ &&{}\nonumber \\
 {\rightarrow}_{r_{\bar a}\rightarrow 0}&& 2\epsilon_{(a)(b)(c)}\sum_u\delta
_{(c)u}\Big[ \delta_{(b)s} \partial_u\partial_rln\, \phi -\delta_{(b)r} \partial_u
\partial_sln\, \phi \Big] +\nonumber \\
&&+2\Big[
\delta_{(a)(b)}\epsilon_{(c)(d)(e)}-\delta_{(a)(c)}\epsilon
_{(b)(d)(e)}+\delta_{(a)(d)}\epsilon_{(e)(c)(b)}-\delta_{(a)(e)}\epsilon
_{(d)(c)(b)}\Big] \nonumber \\
 &&\partial_uln\, \phi \partial_vln\, \phi \, {\rightarrow}_{q\, \rightarrow 0}\,
0,\nonumber \\
 {\rightarrow}_{q\, \rightarrow 0}&& {1\over {\sqrt{3}}}\epsilon_{(a)(b)(c)}
\sum_u\delta_{(c)u}\Big( \delta_{(b)s}e^{{1\over {\sqrt{3}}}\sum_{\bar a}(\gamma
_{\bar as}-\gamma_{\bar au})r_{\bar a}}\cdot \nonumber \\
&&\sum_{\bar b}\gamma_{\bar bs}\Big[ \partial_r
\partial_ur_{\bar b}+{1\over {\sqrt{3}}}\partial_ur_{\bar b}\sum_{\bar c}(\gamma
_{\bar cs}-\gamma_{\bar cu})\partial_rr_{\bar c}\Big] -\nonumber \\
&&-\delta_{(b)r}e^{{1\over {\sqrt{3}}}\sum_{\bar a}(\gamma_{\bar ar}-\gamma
_{\bar au})r_{\bar a}}\sum_{\bar b}\gamma_{\bar br}\Big[ \partial_s\partial_ur
_{\bar b}+{1\over {\sqrt{3}}}\partial_ur_{\bar b}\sum_{\bar c}(\gamma
_{\bar cr}-\gamma_{\bar cu})\partial_sr_{\bar c}\Big] \Big) +\nonumber \\
&&+{1\over 6}\sum_{uv}\Big[ \delta_{(a)(b)}\epsilon_{(c)(d)(e)}-\delta_{(a)(c)}
\epsilon_{(b)(d)(e)}+\delta_{(a)(d)}\epsilon_{(e)(c)(b)}-\delta_{(a)(e)}\epsilon
_{(d)(c)(b)}\Big]\nonumber \\
&&e^{{1\over {\sqrt{3}}}\sum_{\bar a}(\gamma_{\bar ar}+\gamma_{\bar as}-\gamma
_{\bar au}+\gamma_{\bar av})r_{\bar a}}\sum_{\bar b}\gamma_{\bar br}\partial_ur
_{\bar b}\sum_{\bar c}\gamma_{\bar cs}\partial_vr_{\bar c},\nonumber \\
&&{}\nonumber \\
 {}^3{\hat R}_{rusv}&=&(\delta_{rv}\delta_{su}-\delta_{rs}\delta_{uv})\phi^8e^{{2\over
{\sqrt{3}}}\sum_{\bar c}(\gamma_{\bar cr}+\gamma_{\bar cu})r_{\bar
c}}\nonumber \\
 &&\sum_n\Big( 2\partial_nln\, \phi +{1\over
{\sqrt{3}}}\sum_{\bar a}\gamma_{\bar ar}\partial_nr_{\bar a}\Big)
\Big( 2\partial_nln\, \phi +{1\over {\sqrt{3}}}\sum_{\bar b}\gamma_{\bar bu}\partial_nr
_{\bar b}\Big) +\nonumber \\
 &&+\phi^4e^{{2\over {\sqrt{3}}}\sum_{\bar c}\gamma_{\bar cr}r_{\bar c}}
\Big(
\delta_{rv}\Big[ 2\partial_s\partial_uln\, \phi +{1\over {\sqrt{3}}}\sum_{\bar a}\gamma
_{\bar ar}\partial_s\partial_ur_{\bar a}+\nonumber \\
&&+{1\over {\sqrt{3}}}\Big( 2\partial_uln\, \phi +{1\over
{\sqrt{3}}}\sum_{\bar a}\gamma_{\bar ar}\partial_ur_{\bar a}\Big)
\sum_{\bar b} (\gamma_{\bar br}-\gamma_{\bar bu})\partial_sr_{\bar
b}-\nonumber \\
 &&-\Big( 2\partial_uln\, \phi +{1\over {\sqrt{3}}}\sum_{\bar
a}\gamma_{\bar as}\partial_ur
_{\bar a}\Big) \Big( 2\partial_sln\, \phi +{1\over {\sqrt{3}}}\sum_{\bar b}\gamma_{\bar
br}\partial_sr_{\bar b}\Big) \Big]-\nonumber \\
 &&-\delta_{rs}\Big[
2\partial_v\partial_uln\, \phi +{1
\over {\sqrt{3}}}\sum_{\bar ar} \partial_v\partial_ur_{\bar a}+\nonumber \\
&&+{1\over {\sqrt{3}}}\Big( 2\partial_uln\, \phi +{1\over
{\sqrt{3}}}\sum_{\bar a}\gamma
_{\bar ar}\partial_ur_{\bar a}\Big) \sum_{\bar b}(\gamma_{\bar br}-\gamma
_{\bar bu})\partial_vr_{\bar b}-\nonumber \\
&&-\Big( 2\partial_uln\, \phi +{1\over {\sqrt{3}}}\sum_{\bar a}
\gamma_{\bar av}\partial_ur_{\bar a}\Big) \Big( 2\partial_vln\, \phi +{1\over {\sqrt{3}}}
\sum_{\bar b}\gamma_{\bar br}\partial_vr_{\bar b}\Big) \Big] \Big) +\nonumber \\
&&+\phi^4e^{{2\over {\sqrt{3}}}\sum_{\bar c}\gamma_{\bar cu}r_{\bar
c}}\Big( \delta_{su}\Big[ 2\partial_v\partial_rln\, \phi +{1\over
{\sqrt{3}}}\sum_{\bar a}\gamma
_{\bar au}\partial_v\partial_rr_{\bar a}+{1\over {\sqrt{3}}}\Big( 2\partial_rln\, \phi +
\nonumber \\
&&+{1\over {\sqrt{3}}}\sum_{\bar a}\gamma_{\bar au}\partial_rr_{\bar a}\Big)
\sum_{\bar b}(\gamma_{\bar bu}-\gamma_{\bar br})\partial_vr_{\bar b}-
\nonumber \\
&&-\Big( 2\partial_rln\, \phi +{1\over {\sqrt{3}}}\sum_{\bar
a}\gamma_{\bar ar}\partial
_rr_{\bar a}\Big) \Big( 2\partial_vln\, \phi +{1\over {\sqrt{3}}}\sum_{\bar b}\gamma
_{\bar bu}\partial_vr_{\bar b}\Big) \Big] -\nonumber \\
 &-&\delta_{uv}\Big[ 2\partial_s\partial
_rln\, \phi +{1\over {\sqrt{3}}}
\sum_{\bar a}\gamma_{\bar au}\partial_s\partial_rr_{\bar a}+\nonumber \\
&&+{1\over {\sqrt{3}}}\Big( 2\partial_rln\, \phi +{1\over
{\sqrt{3}}}\sum_{\bar a}\gamma
_{\bar au}\partial_rr_{\bar a}\Big) \sum_{\bar b}(\gamma_{\bar bu}-\gamma
_{\bar br})\partial_sr_{\bar b}-\nonumber \\
&&-\Big( 2\partial_rln\, \phi +{1\over {\sqrt{3}}}\sum_{\bar a}\gamma
_{\bar as}\partial_rr_{\bar a}\Big) \Big( 2\partial_sln\, \phi +{1\over {\sqrt{3}}}\sum
_{\bar b}\gamma_{\bar bu}\partial_sr_{\bar b}\Big) \Big] \Big) \nonumber \\
&&{}\nonumber \\
 {\rightarrow}_{r_{\bar a}\, \rightarrow 0}&& 4(\delta_{rv}\delta_{su}-\delta
_{rs}\delta_{uv}) \phi^8 \sum_n (\partial_nln\, \phi )^2+\nonumber \\
&&+2\phi^4\Big( \delta_{rv}[\partial_s\partial_uln\, \phi
-2\partial_sln\, \phi \partial_uln\, \phi ]-
\delta_{rs}[\partial_v\partial_uln\, \phi -2\partial_vln\, \phi
 \partial_uln\, \phi ]+\nonumber \\
&&+\delta_{su}[\partial_v\partial_rln\, \phi -2\partial_vln\, \phi
\partial_rln\, \phi ]-\delta_{uv} [\partial_s\partial_rln\, \phi -2\partial_sln\, \phi
\partial_rln\, \phi ] \Big) {\rightarrow}_{q\,
\rightarrow 0}\, 0,\nonumber \\
 {\rightarrow}_{q\, \rightarrow 0}&& {1\over 3}(\delta_{rv}\delta_{su}-\delta
_{rs}\delta_{uv}) e^{ {2\over {\sqrt{3}}}\sum_{\bar c}(\gamma_{\bar cr}+\gamma
_{\bar cu})r_{\bar c}} \sum_n\sum_{\bar a\bar b}\gamma_{\bar ar}\gamma_{\bar bu}
\partial_nr_{\bar a}\partial_nr_{\bar b}+\nonumber \\
&&+{1\over {\sqrt{3}}}e^{{2\over {\sqrt{3}}}\sum_{\bar c}\gamma_{\bar cr}
r_{\bar c}} \sum_{\bar a}\gamma_{\bar ar} \Big( \delta_{rv}\Big[ \partial_s
\partial_ur_{\bar a}+\nonumber \\
&&+{1\over {\sqrt{3}}}\sum_{\bar b}(\gamma_{\bar br}-\gamma_{\bar bu})\partial
_ur_{\bar a}\partial_sr_{\bar b}-{1\over {\sqrt{3}}}\sum_{\bar b}\gamma
_{\bar bs}\partial_sr_{\bar a}\partial_ur_{\bar b}\Big] -\nonumber \\
&&-\delta_{rs}\Big[ \partial_v\partial_ur_{\bar a}+{1\over {\sqrt{3}}}\sum
_{\bar b}(\gamma_{\bar br}-\gamma_{bar bu})\partial_ur_{\bar a}\partial_vr
_{\bar b}-{1\over {\sqrt{3}}}\sum_{\bar b}\gamma_{\bar bv}\partial_vr_{\bar a}
\partial_ur_{\bar b}\Big] \Big) +\nonumber \\
&&+{1\over {\sqrt{3}}} e^{{2\over {\sqrt{3}}}\sum_{\bar c}\gamma_{\bar cu}
r_{\bar c}} \sum_{\bar a}\gamma_{\bar au} \Big( \delta_{su} \Big[ \partial_v
\partial_rr_{\bar a}+\nonumber \\
&&+{1\over {\sqrt{3}}}\sum_{\bar b}(\gamma_{\bar bu}-\gamma_{\bar br})
\partial_rr_{\bar a}\partial_vr_{\bar b}-{1\over {\sqrt{3}}}\sum_{\bar b}
\gamma_{\bar bv}\partial_vr_{\bar a}\partial_rr_{\bar b}\Big] -\nonumber \\
&&-\delta_{uv}\Big[
\partial_s\partial_rr_{\bar a}+{1\over {\sqrt{3}}}\sum_{\bar b}
(\gamma_{\bar bu}-\gamma_{\bar br})\partial_rr_{\bar a}\partial_sr_{\bar b}-
{1\over {\sqrt{3}}}\sum_{\bar b}\gamma_{\bar bs}\partial_sr_{\bar a}\partial
_rr_{\bar b}\Big] \Big) ,\nonumber \\
&&{}\nonumber \\
{}^3{\hat R}_{uv}&=&
-2\partial_u\partial_vln\, \phi +{1\over {\sqrt{3}}}\sum_{\bar a}(\gamma_{\bar au}+
\gamma_{\bar av})\partial_u\partial_vr_{\bar a}+\nonumber \\
&&+{1\over 2}\Big[ \Big( 2\partial_uln\, \phi +{1\over
{\sqrt{3}}}\sum_{\bar a}\gamma
_{\bar av}\partial_ur_{\bar a}\Big) \Big( 2\partial
_vln\, \phi -{2\over {\sqrt{3}}}\sum_{\bar b}\gamma_{\bar bu}\partial_vr_{\bar b}\Big) +
\nonumber \\
&&+\Big( 2\partial_vln\, \phi +{1\over {\sqrt{3}}}\sum_{\bar
a}\gamma_{\bar u}\partial_vr
_{\bar a}\Big) \Big( 2\partial_uln\, \phi -{2\over {\sqrt{3}}}\sum_{\bar b}\gamma_{\bar
bv}r_{\bar b}\Big) \Big] -\nonumber \\ &&-{1\over
{2\sqrt{3}}}\sum_n\Big[ \Big( 2\partial_uln\, \phi +{1\over
{\sqrt{3}}}\sum
_{\bar a}\gamma_{\bar ar}\partial_yr_{\bar a}\Big) \sum_{\bar b}(\gamma
_{\bar bn}-\gamma_{bar bu})\partial_vr_{\bar b}+\nonumber \\
&&+\Big( 2\partial_vln\, \phi +{1\over {\sqrt{3}}}\sum_{\bar
a}\gamma_{\bar ar}\partial_vr
_{\bar a}\big) \sum_{\bar b}(\gamma_{\bar bn}-\gamma_{\bar bv})\partial_ur
_{\bar b}\Big] -\delta_{uv} \phi^4e^{{2\over {\sqrt{3}}}\sum_{\bar c}\gamma
_{\bar cu}r_{\bar c}}\nonumber \\
&&\sum_n \Big( (2\partial_nln\, \phi +{1\over {\sqrt{3}}}\sum_{\bar
a}\gamma_{\bar au}
\partial_nr_{\bar a})(4\partial_nln\, \phi -{1\over {\sqrt{3}}}\sum_{\bar b}\gamma
_{\bar bu}\partial_nr_{\bar b})+\nonumber \\
&&+\phi^{-4}e^{-{2\over {\sqrt{3}}}\sum_{\bar c}\gamma
_{\bar cn}r_{\bar c}}\Big[ 2\partial^2_nln\, \phi +\nonumber \\
&&+{1\over {\sqrt{3}}}\sum_{\bar a}\gamma
_{\bar au}\partial^2_nr_{\bar a}+{1\over {\sqrt{3}}}(2\partial_nln\, \phi +{1\over
{\sqrt{3}}}\sum_{\bar au}\partial_nr_{\bar a})\sum_{\bar b}(\gamma_{\bar bu}-2
\gamma_{\bar bn})\partial_nr_{\bar b}\Big] \Big) \nonumber \\
&&{}\nonumber \\
 {\rightarrow}_{r_{\bar a}\, \rightarrow 0} && -2\partial_u\partial_vln\, \phi+
4\partial_uln\, \phi \partial_vln\, \phi -\nonumber \\
 &&-2\delta_{uv}\phi^4\sum_n\Big[ 4\phi^4(\partial_nln\, \phi )^2+
\partial^2_nln\, \phi -2(\partial_nln\, \phi )^2\Big]\, {\rightarrow}_{q\, \rightarrow 0}\, 0,
\nonumber \\
 {\rightarrow}_{q\, \rightarrow 0}&& {1\over {\sqrt{3}}}\sum_{\bar a}
(\gamma_{\bar au}+\gamma_{\bar av})\partial_u\partial_vr_{\bar a}-{2\over 3}
\sum_{\bar a\bar b}\gamma_{\bar au}\gamma_{\bar bv}\partial_vr_{\bar a}
\partial_ur_{\bar b}-\nonumber \\
&&-{1\over 6}\sum_n\sum_{\bar a\bar b}\gamma_{\bar an}\Big[ (\gamma_{\bar br}-
\gamma_{\bar bu})\partial_ur_{\bar a}\partial_vr_{\bar b}+(\gamma_{\bar bn}-
\gamma_{\bar bv})\partial_vr_{\bar a}\partial_ur_{\bar b}\Big] +\nonumber \\
&&+{1\over {\sqrt{3}}}\delta_{uv}e^{{2\over {\sqrt{3}}}\sum_{\bar c}\gamma
_{\bar cu}r_{\bar c}}\sum_n\Big( {1\over {\sqrt{3}}}\sum_{\bar a\bar b}\gamma
_{\bar au}\gamma_{\bar bu}\partial_nr_{\bar a}\partial_nr_{\bar b}-\nonumber \\
&&-e^{-{2\over {\sqrt{3}}}\sum_{\bar c}\gamma_{\bar cn}r_{\bar c}}\sum_{\bar a}
\gamma_{\bar au}\Big[ \partial^2_nr_{\bar a}+{1\over {\sqrt{3}}}\sum_{\bar b}
(\gamma_{\bar bu}-2\gamma_{\bar bn})\partial_nr_{\bar a}\partial_nr_{\bar b}
\Big] \Big) , \nonumber \\
&&{}\nonumber \\
{}^3\hat R&=&
-\sum_{uv}\Big( (2\partial_vln\, \phi +{1\over {\sqrt{3}}}\sum_{\bar a}\gamma_{\bar au}
\partial_vr_{\bar a})(4\partial_vln\, \phi -{1\over {\sqrt{3}}}\sum_{\bar b}\gamma
_{\bar bu}\partial_vr_{\bar b})+\nonumber \\
&&+\phi^{-4}e^{-{2\over {\sqrt{3}}}\sum_{\bar c}\gamma_{\bar
cv}r_{\bar c}}
\Big[ 2\partial^2_vln\, \phi +{1\over {\sqrt{3}}}\sum_{\bar a}\gamma_{\bar au}\partial_v
^2r_{\bar a}+\nonumber \\
&&+{2\over {\sqrt{3}}}(2\partial_vln\, \phi +{1\over
{\sqrt{3}}}\sum_{\bar a}\gamma
_{\bar au}\partial_vr_{\bar a})\sum_{\bar b}(\gamma_{\bar bu}-\gamma_{\bar bv})
\partial_vr_{\bar b}-\nonumber \\
&&-(2\partial_vln\, \phi +{1\over {\sqrt{3}}}\sum_{\bar a}\gamma_{\bar
av}\partial_vr
_{\bar a})(2\partial_vln\, \phi +{1\over {\sqrt{3}}}\sum_{\bar b}\gamma_{\bar bu}
\partial_vr_{\bar b}\Big] \Big) +\nonumber \\
&&+\phi^{-4}\sum_ue^{-{2\over {\sqrt{3}}}\sum_{\bar c}\gamma_{\bar
cu}r_{\bar c}}
\Big[ -2\partial^2_uln\, \phi +{2\over {\sqrt{3}}}\sum_{\bar a}\gamma_{\bar au} \partial
^2_ur_{\bar a}+\nonumber \\
&&+(2\partial_uln\, \phi +{1\over {\sqrt{3}}}\sum_{\bar a}\gamma_{\bar
au}\partial_ur
_{\bar a})(2\partial_uln\, \phi -{2\over {\sqrt{3}}}\sum_{\bar b}\gamma_{\bar bu}
\partial_ur_{\bar b})\Big]\nonumber \\
&&{}\nonumber \\
 {\rightarrow}_{r_{\bar a}\, \rightarrow 0}&& -24\sum_u(\partial_uln\, \phi)^2-
8\phi^{-4}\sum_u\Big[ \partial_u^2ln\, \phi -2(\partial_uln\,
\phi )^2\Big]\, {\rightarrow}_{q\,
\rightarrow 0}\, 0,\nonumber \\
 {\rightarrow}_{q\, \rightarrow 0}&& -{1\over {\sqrt{3}}}\sum_{uv}\Big(
-{1\over {\sqrt{3}}}\sum_{\bar a\bar b}\gamma_{\bar au}\gamma_{\bar bu}
\partial_vr_{\bar a}\partial_vr_{\bar b}+e^{-{2\over {\sqrt{3}}}\sum_{\bar c}
\gamma_{\bar cv}r_{\bar c}}\sum_{\bar a}\gamma_{\bar au}\cdot \nonumber \\
&&\Big[ \partial^2_vr_{\bar a}+{2\over {\sqrt{3}}}\sum_{\bar b}(\gamma_{\bar
bu}-\gamma_{\bar bv})\partial_vr_{\bar a}\partial_vr_{\bar b}-{1\over
{\sqrt{3}}}\sum_{\bar b}\gamma_{\bar bv}\partial_vr_{\bar a}\partial_vr_{\bar
b}\Big] \Big) +\nonumber \\
&&+{2\over {\sqrt{3}}}\sum_ue^{-{2\over {\sqrt{3}}}\sum_{\bar c}\gamma
_{\bar cu}t_{\bar c}}\sum_{\bar a}\gamma_{\bar au}\Big[ \partial^2_ur_{\bar a}+
{1\over {\sqrt{3}}}\sum_{\bar b}\gamma_{\bar bu}\partial_ur_{\bar a}
\partial_ur_{\bar b}\Big] .
\label{a3}
\end{eqnarray}

The Weyl-Schouten tensor  ${}^3C_{rsu}={}^3\nabla_u\,
{}^3R_{rs}-{}^3\nabla_s\, {}^3R_{ru}-{1\over
4}({}^3g_{rs}\partial_u\, {}^3R-{}^3g_{ru}\partial_s\, {}^3R)$  of
the 3-manifold $(\Sigma_{\tau},{}^3g)$ (see Ref.\cite{naka})
satisfies ${}^3C^r{}_{ru}=0$, ${}^3C_{rsu}=-{}^3C _{rus}$,
${}^3C_{rsu}+{}^3C_{urs}+{}^3C_{sur}=0$ and has 5 independent
components. The related York's conformal tensor\cite{york,mtw}
${}^3Y^{rs}=\gamma^{1/3}\, \epsilon^{ruv}({}^3R_v{}^s- {1\over 4}
\delta^s_v\, {}^3R){}_{|u}=-{1\over 2} \gamma^{1/3}
\epsilon^{ruv}\, {}^3g^{sm}\, {}^3C_{muv}$  \footnote{It is a
tensor density of weight 5/3 and involves the third derivatives of
the metric.} is symmetric [${}^3Y^{rs}={}^3Y^{sr}$], traceless
[${}^3Y^r{}_r=0$] and transverse [${}^3Y^{rs}{}_{|s}=0$] besides
being invariant under 3-conformal transformations; therefore, it
has only 2 independent components \footnote{$Y^{rs}=Y^{rs}_{TT}$
according to York's decomposition of Appendix C of II.} and
provides what York calls the pure spin-two representation of the
3-geometry intrinsic to $\Sigma_{\tau}$. Its explicitly symmetric
form is the Cotton-York tensor  given by ${}^3{\cal
Y}^{rs}={1\over 2} ({}^3Y^{rs}+{}^3Y^{sr})={1\over 2} \gamma^{1/3}
(\epsilon^{ruv}\, {}^3g^{sc}+\epsilon^{suv}\, {}^3g^{rc})
{}^3R_{vc|u}= -{1\over 4}\gamma^{1/3}(\epsilon^{ruv}\,
{}^3g^{sm}+\epsilon^{suv}\, {}^3g^{rm}) {}^3C_{muv}$.

A 3-manifold is {\it conformally flat} if and only if either the
Weyl-Schouten or the Cotton-York tensor vanishes \cite{mtw,york,naka}.
We have

\begin{eqnarray}
{}^3C_{rsu}&=&{}^3\nabla_u\, {}^3R_{rs}-{}^3\nabla_s\, {}^3R_{ru}-{1\over 4}
({}^3g_{rs}\partial_u\, {}^3R-{}^3g_{ru}\partial_s\, {}^3R)
\mapsto \nonumber \\
&&\mapsto {}^3{\hat C}_{rsu}={}^3{\hat R}_{rs|u}-{}^3{\hat R}_{ru|s}-{1\over 4}
e^{2q_r}(\delta_{rs}\partial_u\, {}^3\hat R-\delta_{ru}\partial_s\, {}^3\hat R)
,\nonumber \\
&&{}\nonumber \\
{}^3{\cal Y}_{mn} &=&{1\over 2}\gamma^{1/3}\sum_{rsu}
(\epsilon_{mur}\, {}^3g_{ns}+
\epsilon_{nur}\, {}^3g_{ms}) {}^3R_{rs|u} \mapsto \nonumber \\
&&\mapsto {}^3{\hat {\cal Y}}_{mn}={1\over 2}e^{{2\over 3}\sum_vq_v}
\sum_{rsu}e^{-2q_s}(\epsilon_{mur}\delta_{ns}+\epsilon_{nur}\delta_{ms})
{}^3{\hat R}_{rs|u}=\nonumber \\
&&={1\over 2}e^{2q}\sum_{rsu}e^{-2q_s}
(\epsilon_{mur}\delta_{ns}+\epsilon_{nur}\delta_{ms})\cdot \nonumber \\
&&\Big( \, \partial_u\partial_r\partial_s(q_r+q_s-\sum_tq_t)-
\partial_u(q_r+q_s)\partial_r\partial_s(q_r+q_s-\sum_tq_t)-\nonumber \\
&&-\partial_rq_u \partial_u\partial_s(q_u+q_s-\sum_tq_t)-\partial_sq_u
\partial_u\partial_r(q_u+q_r-\sum_tq_t)-\nonumber \\
&&-{1\over 2}\partial_u[\partial_rq_s\partial_s(2q_r-\sum_tq_t)+\partial_sq_r
\partial_r(2q_s-\sum_tq_t)]-\nonumber \\
&&-{1\over 2}\sum_n \partial_u[\partial_rq_n\partial_s(q_n-q_r)+\partial_sq_n
\partial_r(q_n-q_s)]+\nonumber \\
&&+{1\over 2}\partial_u(q_r+q_s) [\partial_rq_s\partial_s(2q_r-\sum_tq_t)+
\partial_sq_r \partial_r(2q_s-\sum_tq_t)]+\nonumber \\
&&+{1\over 2}\partial_rq_u [\partial_uq_s \partial_s(2q_u-\sum_tq_t)+
\partial_sq_u \partial_u(2q_s-\sum_tq_t)]+\nonumber \\
&&+{1\over 2}\partial_sq_u [\partial_uq_r \partial_r(2q_u-\sum_tq_t)+
\partial_rq_u \partial_u(2q_r-\sum_tq_t)]+\nonumber \\
&&+{1\over 2}\partial_u(q_r+q_s) \sum_n[\partial_rq_n \partial_s(q_n-q_r)+
\partial_sq_n \partial_r(q_n-q_s)]+\nonumber \\
&&+{1\over 2}\partial_rq_u \sum_n[\partial_uq_n \partial_s(q_n-q_u)+
\partial_sq_n \partial_u(q_n-q_s)]+\nonumber \\
&&+{1\over 2}\partial_sq_u \sum_n[\partial_uq_n \partial_r(q_n-q_u)+
\partial_rq_n \partial_u(q_n-q_r)]+\nonumber \\
&&+\delta_{rs}\, e^{2q_r}\sum_n
\Big[ 2\partial_uq_r[\partial_nq_r\partial_n(q_r-
\sum_tq_t)-e^{-2q_n}(\partial^2_nq_r+\partial_nq_r\partial_n(q_r-2q_n))]+
\nonumber \\
&&+\partial_u[\partial_nq_r\partial_n(q_r-\sum_tq_t)]+e^{-2q_n}[2\partial_uq_n
(\partial_n^2q_r+\partial_nq_r\partial_n(q_r-2q_n))-\nonumber \\
&&-\partial_u(\partial_n^2q_r+\partial_nq_r\partial_n(q_r-2q_n))]-\nonumber \\
&&-2\partial_uq_r[\partial_nq_r\partial_n(q_r-\sum_tq_t)-e^{-2q_n}(\partial^2_n
q_r+\partial_nq_r\partial_n(q_r-2q_n))] \Big] +\nonumber \\
&&+\delta_{ru}\, e^{2q_u}\Big[ \sum_ve^{-2q_v} \partial_vq_u \Big( \partial_v
\partial_s(q_v+q_s-\sum_tq_t)-\nonumber \\
&&-{1\over 2}[\partial_vq_s\partial_s(2q_v-\sum_tq_t)+\partial_sq_v\partial_v
(2q_s-\sum_tq_t)]-\nonumber \\
&&-{1\over 2}\sum_n[\partial_vq_n\partial_s(q_n-q_v)+
\partial_sq_n\partial_v(q_n-q_s)] \Big) +\nonumber \\
&&+\sum_n (\partial_sq_u [\partial_nq_s\partial_n(q_s-\sum_tq_t)-
e^{-2q_n}(\partial_n^2q_s+\partial_nq_s\partial_n(q_s-2q_n))]-\nonumber \\
&&-\partial_sq_u[\partial_nq_r\partial_n(q_r-\sum_tq_t)-e^{-2q_n}(\partial_n
^2q_r+\partial_nq_r\partial_n(q_r-2q_n))] ) \Big] +\nonumber \\
&&+\delta_{su}\, e^{2q_u} \Big[ \sum_ve^{-2q_v} \partial_vq_u
\Big( \partial_v\partial_r(q_v+q_r-\sum_tq_t)-\nonumber \\
&&-{1\over 2}[\partial_vq_r\partial_r(2q_v-\sum_tq_t)+\partial_rq_v\partial_v
(2q_r-\sum_tq_t)]-\nonumber \\
&&-{1\over 2}\sum_n[\partial_vq_n\partial_r(q_n-q_v)+
\partial_rq_n\partial_v(q_n-q_r)] \Big) +\nonumber \\
&&+\sum_n (\partial_rq_u [\partial_nq_r\partial_n(q_r-\sum_tq_t)-
e^{-2q_n}(\partial_n^2q_r+\partial_nq_r\partial_n(q_r-2q_n)]-\nonumber \\
&&-\partial_rq_u[\partial_nq_s\partial_n(q_s-\sum_tq_t)-e^{-2q_n}(\partial_n
^2q_s+\partial_nq_s\partial_n(q_s-2q_n))] ) \Big] \, \Big) \nonumber \\
&&{}\nonumber \\
 {\rightarrow}_{r_{\bar a}\, \rightarrow 0}&& {1\over 2}\sum_{rsu}(\epsilon
_{mur}\delta_{ns}+\epsilon_{nur}\delta_{ms})\cdot \nonumber \\
&&-\partial_u\partial_r\partial_sq
+2(\partial_uq \partial_r\partial_sq+\partial_rq \partial_s\partial_uq+
\partial_sq \partial_u\partial_rq)-4\partial_uq \partial_rq \partial_sq+
\nonumber \\
&&+\delta_{rs}\sum_n \Big(
\partial_u[\partial_n^2q-(\partial_nq)^2]+2\partial_uq
[\partial^2_nq-(\partial_nq)^2]-2e^{2q}\partial_u(\partial_nq)^2 \Big) -
\nonumber \\
&&-\delta_{ru}\sum_v\partial_vq (\partial_v\partial_sq-\partial_vq \partial_sq)-
\delta_{su}\sum_v\partial_vq (\partial_v\partial_rq-\partial_vq \partial_rq)=0.
\label{a4}
\end{eqnarray}

\noindent Since in the 3-orthogonal gauges
the condition $r_{\bar a}=0$ corresponds to conformally flat
3-manifolds $\Sigma_{\tau}$, the Cotton-York conformal tensor vanishes
in the limit $r_a \rightarrow 0$ in these gauges.

\vfill\eject

\section{The strong and weak ADM Poincar\'e charges in the special 3-orthogonal gauge.}

In the canonical basis of   Section VI for the 3-orthogonal gauge,
by  using Eqs.(\ref{VI14}) and (\ref{VI16}) the {\it weak}
Poincar\'e charges of Eqs.(\ref{II25})  take the form

\begin{eqnarray}
 {\hat P}^{\tau}_{ADM,R}&=&\epsilon \int d^3\sigma
\Big( {{c^3}\over {16\pi G}} \Big[ \phi^2 \sum_r e^{-{2\over
{\sqrt{3}}}\sum_{\bar a}\gamma_{\bar ar}r_{\bar a}}
\times\nonumber \\
 &&\Big( 8(\partial_rln\, \phi )^2-{1\over 3}\sum_{\bar b}(\partial_rr_{\bar b})^2-
 \nonumber \\
 &-&{4\over {\sqrt{3}}} \partial_rln\, \phi \sum_{\bar b}\gamma_{\bar br}\partial_rr_{\bar b}+
 {2\over 3}(\sum_{\bar b}\gamma_{\bar br}\partial_rr_{\bar b})^2\Big)
  \Big] (\tau ,\vec \sigma )-\nonumber \\
  &&{}\nonumber \\
&-&{{2\pi G}\over {c^3}} \phi^{-2}(\tau ,\vec \sigma )\Big[
(\phi^{-4}[6 \sum_{\bar a}\pi^2 _{\bar a}-{1\over 3}\rho^2])(\tau
,\vec \sigma )+\nonumber \\
 &+&2 (\phi^{-2}\sum_ue^{{1\over {\sqrt{3}}}\sum_{\bar
a}\gamma_{\bar au}r_{\bar a}} [2\sqrt{3}\sum_{\bar b}\gamma_{\bar
bu}\pi_{\bar b}-{1\over 3}\rho ])(\tau ,\vec \sigma )\times
\nonumber \\
 &&{}\nonumber \\
 &&\int d^3\sigma_1 \sum_r
\delta^u_{(a)} {\tilde {\cal T}}^u_{(a)r}(\vec \sigma ,{\vec
\sigma}_1,\tau |\phi ,r_{\bar a}] \Big( \phi^{-2}e^{-{1\over {\sqrt{3}}}\sum_{\bar a}
\gamma_{\bar ar}r_{\bar a}}[{{\rho}\over 3}+\sqrt{3}\sum_{\bar b}\gamma_{\bar
br} \pi_{\bar b}]\Big) (\tau ,{\vec \sigma}_1)+\nonumber \\
&+&\int d^3\sigma_1d^3\sigma_2 \Big( \sum_u e^{{2\over {\sqrt{3}}}\sum_{\bar a}
\gamma_{\bar au}r_{\bar a}(\tau ,\vec \sigma )} \times \nonumber \\
&&\sum_r{\tilde {\cal T}}^u_{(a)r}(\vec \sigma ,{\vec
\sigma}_1,\tau |\phi ,r_{\bar a}] \Big( \phi^{-2}e^{-{1\over {\sqrt{3}}}\sum_{\bar a}
\gamma_{\bar ar}r_{\bar a}}[{{\rho}\over 3}+\sqrt{3}\sum_{\bar b}\gamma_{\bar
br} \pi_{\bar b}]\Big) (\tau ,{\vec \sigma}_1)\times \nonumber \\
&&\sum_s {\tilde {\cal T}}^u_{(a)s}(\vec \sigma ,{\vec
\sigma}_2,\tau |\phi ,r_{\bar a}] \Big( \phi^{-2}e^{-{1\over {\sqrt{3}}}\sum_{\bar a}
\gamma_{\bar as}r_{\bar a}}[{{\rho}\over 3}+\sqrt{3}\sum_{\bar c}\gamma_{\bar
cs} \pi_{\bar c}]\Big) (\tau ,{\vec \sigma}_2)+\nonumber \\
&+&\sum_{uv} e^{{1\over {\sqrt{3}}}\sum_{\bar a}(\gamma_{\bar au}+\gamma_{\bar
av})r_{\bar a}(\tau ,\vec \sigma )} (\delta^u_{(b)}\delta^v_{(a)}-\delta^u_{(a)}
\delta^v_{(b)})\times \nonumber \\
&&\sum_r {\tilde {\cal T}}^u_{(a)r}(\vec \sigma ,{\vec
\sigma}_1,\tau |\phi ,r_{\bar a}] \Big( \phi^{-2}e^{-{1\over {\sqrt{3}}}\sum_{\bar a}
\gamma_{\bar ar}r_{\bar a}}[{{\rho}\over 3}+\sqrt{3}\sum_{\bar b}\gamma_{\bar
br} \pi_{\bar b}]\Big) (\tau ,{\vec \sigma}_1)\nonumber \\ &&\sum_s
{\tilde {\cal T}}^v_{(b)s}(\vec \sigma ,{\vec
\sigma}_2,\tau |\phi ,r_{\bar a}] \Big( \phi^{-2}e^{-{1\over {\sqrt{3}}}\sum_{\bar a}
\gamma_{\bar as}r_{\bar a}}[{{\rho}\over 3}+\sqrt{3}\sum_{\bar c}\gamma_{\bar
cs} \pi_{\bar c}]\Big) (\tau ,{\vec \sigma}_2)\, \Big)\, \Big]\,
\Big) ,\nonumber \\
 &&{}\nonumber \\
 {\hat P}^r_{ADM,R}&=&-\int d^3\sigma \phi^{-2}(\tau ,\vec \sigma )
\Big( \phi^{-2}(\tau ,\vec \sigma ) \nonumber \\
 &&\Big[ e^{-{2\over {\sqrt{3}}}\sum_{\bar a}\gamma_{\bar ar}r_{\bar a}}
(2\partial_rln\, \phi \rho +\sum_{\bar b}\partial_rr_{\bar b}\pi_{\bar
b})+\nonumber \\
 &+&2\sum_u e^{-{2\over {\sqrt{3}}}\sum_{\bar a}\gamma_{\bar au}r_{\bar a}}
  \delta^r_u (2\partial_uln\, \phi +{1\over {\sqrt{3}}}\sum_{\bar b}\gamma_{\bar
bu}\partial_ur_{\bar b})\nonumber \\
 &&({{\rho}\over 3}+\sqrt{3}\sum_{\bar c}\gamma_{\bar cu}\pi_{\bar c})
\Big] (\tau ,\vec \sigma )+\nonumber \\
 &&{}\nonumber \\
 &+& \int d^3\sigma_1 \sum_s \Big[ -\sum_u\Big( e^{{1\over {\sqrt{3}}}\sum_{\bar a}(\gamma
 _{\bar au}-2\gamma_{\bar ar})r_{\bar a}}(2\partial_rln\, \phi +{1\over {\sqrt{3}}}
 \sum_{\bar b}\gamma_{\bar bu}\partial_rr_{\bar b})\Big)(\tau ,\vec \sigma )\nonumber \\
 &&\delta^u_{(a)}{\tilde {\cal T}}^u_{(a)s}(\vec \sigma ,{\vec \sigma}_1,\tau |\phi ,r_{\bar a}]
 \nonumber \\
 &+&\sum_{uv}\Big(  e^{-{1\over {\sqrt{3}}}\sum_{\bar a}\gamma_{\bar au}r_{\bar
a}} \Big( \delta^r_u(2\partial_vln\, \phi +{1\over
{\sqrt{3}}}\sum_{\bar b}\gamma_{\bar bu}\partial_vr_{\bar
b})+\nonumber \\
 &+&\delta^r_v(2\partial_uln\, \phi +{1\over {\sqrt{3}}}
 \sum_{\bar b}\gamma_{\bar bv}\partial_ur_{\bar b})\Big)\Big) (\tau ,\vec \sigma )
 \nonumber \\
 &&\delta^u_{(a)}{\tilde {\cal T}}^v_{(a)s}
(\vec \sigma ,{\vec \sigma}_1,\tau |\phi ,r_{\bar a}] \Big]
\Big( \phi^{-2}e^{-{1\over {\sqrt{3}}}\sum_{\bar a}\gamma_{\bar as}r_{\bar a}}
[{{\rho}\over 3}+\sqrt{3}\sum_{\bar c}\gamma_{\bar cs}\pi_{\bar
c}]\Big) (\tau ,{\vec \sigma}_1)\Big),\nonumber \\
 &&{}\nonumber \\
 {\hat J}^{rs}_{ADM,R}&=&\int d^3\sigma \phi^{-2}(\tau ,\vec \sigma )
\Big( \phi^{-2}(\tau ,\vec \sigma ) \nonumber \\
 &&\Big[ e^{-{2\over {\sqrt{3}}}\sum_{\bar a}\gamma_{\bar ar}r_{\bar a}}
\Big( \sigma^r (2\partial_sln\, \phi \rho +\sum_{\bar b}\partial_sr_{\bar b}\pi_{\bar
b})-\nonumber \\
 &-&\sigma^s(2\partial_rln\, \phi \rho +\sum_{\bar b}\partial_rr_{\bar b}\pi_{\bar b})\Big)
+\nonumber \\
 &+&2\sum_u e^{-{2\over {\sqrt{3}}}\sum_{\bar a}\gamma_{\bar au}r_{\bar a}}
  (\sigma^r \delta^s_u-\sigma^s\delta^r_u)
  (2\partial_uln\, \phi +{1\over {\sqrt{3}}}\sum_{\bar b}\gamma_{\bar
bu}\partial_ur_{\bar b})\nonumber \\
 &&({{\rho}\over 3}+\sqrt{3}\sum_{\bar c}\gamma_{\bar cu}\pi_{\bar c})
\Big] (\tau ,\vec \sigma )-\nonumber \\
 &&{}\nonumber \\
 &-& \int d^3\sigma_1 \sum_w \Big[ \Big(-\sigma^r
 \sum_u\Big( e^{{1\over {\sqrt{3}}}\sum_{\bar a}(\gamma
 _{\bar au}-2\gamma_{\bar as})r_{\bar a}}(2\partial_sln\, \phi +{1\over {\sqrt{3}}}
 \sum_{\bar b}\gamma_{\bar bu}\partial_sr_{\bar b})\Big)(\tau ,\vec \sigma )+\nonumber \\
 &+&\sigma^s
 \sum_u\Big( e^{{1\over {\sqrt{3}}}\sum_{\bar a}(\gamma
 _{\bar au}-2\gamma_{\bar ar})r_{\bar a}}(2\partial_rln\, \phi +{1\over {\sqrt{3}}}
 \sum_{\bar b}\gamma_{\bar bu}\partial_rr_{\bar b})\Big)(\tau ,\vec \sigma )\Big)+\nonumber \\
 &&\delta^u_{(a)}{\tilde {\cal T}}^u_{(a)w}(\vec \sigma ,{\vec \sigma}_1,\tau |\phi ,r_{\bar a}]
 \nonumber \\
 &+&\sum_{uv}\Big(  e^{-{1\over {\sqrt{3}}}\sum_{\bar a}\gamma_{\bar au}r_{\bar
a}} \Big( (\sigma^r\delta^s_u-\sigma^s\delta^r_u)(2\partial_vln\, \phi
+{1\over {\sqrt{3}}}\sum_{\bar b}\gamma_{\bar bu}\partial_vr_{\bar
b})+\nonumber \\
 &+&(\sigma^r\delta^s_v-\sigma^s\delta^r_v)(2\partial_uln\, \phi +{1\over {\sqrt{3}}}
 \sum_{\bar b}\gamma_{\bar bv}\partial_ur_{\bar b})\Big)\Big) (\tau ,\vec \sigma )
 \nonumber \\
 &&\delta^u_{(a)}{\tilde {\cal T}}^v_{(a)w}
(\vec \sigma ,{\vec \sigma}_1,\tau |\phi ,r_{\bar a}] \Big]
\Big( \phi^{-2}e^{-{1\over {\sqrt{3}}}\sum_{\bar a}\gamma_{\bar as}r_{\bar a}}
[{{\rho}\over 3}+\sqrt{3}\sum_{\bar c}\gamma_{\bar cs}\pi_{\bar
c}]\Big) (\tau ,{\vec \sigma}_1)\Big),\nonumber \\
 &&{}\nonumber \\
 {\hat J}^{\tau r}_{ADM,R}&=& \epsilon \int d^3\sigma \sigma^r
 \Big( {{c^3}\over {16\pi G}} \Big[ \phi^2 \sum_r e^{-{2\over {\sqrt{3}}}\sum_{\bar
a}\gamma_{\bar ar}r_{\bar a}} \times\nonumber \\
 &&\Big( 8(\partial_rln\, \phi )^2-{1\over 3}\sum_{\bar b}(\partial_rr_{\bar b})^2-
 \nonumber \\
 &-&{4\over {\sqrt{3}}} \partial_rln\, \phi \sum_{\bar b}\gamma_{\bar br}\partial_rr_{\bar b}+
 {2\over 3}(\sum_{\bar b}\gamma_{\bar br}\partial_rr_{\bar b})^2\Big)
  \Big] (\tau ,\vec \sigma )-\nonumber \\
&-&{{\phi^{-2}(\tau ,\vec \sigma )}\over {8k}}\Big[ (\phi^{-4}[6
\sum_{\bar a}\pi^2
_{\bar a}-{1\over 3}\rho^2])(\tau ,\vec \sigma )+\nonumber \\
 &&{}\nonumber \\
 &+&2 (\phi^{-2}\sum_ue^{{1\over {\sqrt{3}}}\sum_{\bar a}\gamma_{\bar
au}r_{\bar a}} [2\sqrt{3}\sum_{\bar b}\gamma_{\bar bu}\pi_{\bar
b}-{1\over 3}\rho ])(\tau ,\vec \sigma )\times \nonumber \\
 &&\int d^3\sigma_1 \sum_r
\delta^u_{(a)} {\tilde {\cal T}}^u_{(a)r}(\vec \sigma ,{\vec
\sigma}_1,\tau |\phi ,r_{\bar a}] \Big( \phi^{-2}e^{-{1\over {\sqrt{3}}}\sum_{\bar a}
\gamma_{\bar ar}r_{\bar a}}[{{\rho}\over 3}+\sqrt{3}\sum_{\bar b}\gamma_{\bar
br} \pi_{\bar b}]\Big) (\tau ,{\vec \sigma}_1)+\nonumber \\
 &&{}\nonumber \\
 &+&\int d^3\sigma_1d^3\sigma_2 \Big( \sum_u e^{{2\over {\sqrt{3}}}\sum_{\bar a}
\gamma_{\bar au}r_{\bar a}(\tau ,\vec \sigma )} \times \nonumber \\
&&\sum_r{\tilde {\cal T}}^u_{(a)r}(\vec \sigma ,{\vec
\sigma}_1,\tau |\phi ,r_{\bar a}] \Big( \phi^{-2}e^{-{1\over {\sqrt{3}}}\sum_{\bar a}
\gamma_{\bar ar}r_{\bar a}}[{{\rho}\over 3}+\sqrt{3}\sum_{\bar b}\gamma_{\bar
br} \pi_{\bar b}]\Big) (\tau ,{\vec \sigma}_1)\times \nonumber \\
&&\sum_s {\tilde {\cal T}}^u_{(a)s}(\vec \sigma ,{\vec
\sigma}_2,\tau |\phi ,r_{\bar a}] \Big( \phi^{-2}e^{-{1\over {\sqrt{3}}}\sum_{\bar a}
\gamma_{\bar as}r_{\bar a}}[{{\rho}\over 3}+\sqrt{3}\sum_{\bar c}\gamma_{\bar
cs} \pi_{\bar c}]\Big) (\tau ,{\vec \sigma}_2)+\nonumber \\
&+&\sum_{uv} e^{{1\over {\sqrt{3}}}\sum_{\bar a}(\gamma_{\bar au}+\gamma_{\bar
av})r_{\bar a}(\tau ,\vec \sigma )} (\delta^u_{(b)}\delta^v_{(a)}-\delta^u_{(a)}
\delta^v_{(b)})\times \nonumber \\
&&\sum_r {\tilde {\cal T}}^u_{(a)r}(\vec \sigma ,{\vec
\sigma}_1,\tau |\phi ,r_{\bar a}] \Big( \phi^{-2}e^{-{1\over {\sqrt{3}}}\sum_{\bar a}
\gamma_{\bar ar}r_{\bar a}}[{{\rho}\over 3}+\sqrt{3}\sum_{\bar b}\gamma_{\bar
br} \pi_{\bar b}]\Big) (\tau ,{\vec \sigma}_1)\nonumber \\ &&\sum_s
{\tilde {\cal T}}^v_{(b)s}(\vec \sigma ,{\vec
\sigma}_2,\tau |\phi ,r_{\bar a}] \Big( \phi^{-2}e^{-{1\over {\sqrt{3}}}\sum_{\bar a}
\gamma_{\bar as}r_{\bar a}}[{{\rho}\over 3}+\sqrt{3}\sum_{\bar c}\gamma_{\bar
cs} \pi_{\bar c}]\Big) (\tau ,{\vec \sigma}_2)\, \Big)\, \Big]\,
\Big) -\nonumber \\
 &&{}\nonumber \\
 &-&{{\epsilon c^3}\over {8\pi G}}  \int d^3\sigma \Big[ \phi^{-2}
\sum_{uv}\delta^r_u(\delta_{uv}-1) (\phi^4e^{{2\over {\sqrt{3}}}\sum_{\bar a}
\gamma_{\bar au}r_{\bar a}}-1)\nonumber \\
 &&e^{-{2\over {\sqrt{3}}}\sum_{\bar a}(\gamma_{\bar av}-\gamma_{\bar au})r_{\bar a}}
 (\partial_uln\, \phi +{1\over {\sqrt{3}}}\sum_{\bar b}(\gamma_{\bar bv}-
 \gamma_{\bar bu})\partial_ur_{\bar b})\Big] (\tau ,\vec \sigma ),
\label{b1}
\eea

\noindent while the {\it strong} Poincar\'e charges of
Eq.(\ref{II23}) are

\bea P^{\tau}_{ADM,R}&=&{\hat P}^{\tau}_{ADM,R}+\int d^3\sigma
{\hat {\cal H}}_R(\tau ,\vec \sigma )=\nonumber\\
 &=& {{\epsilon c^3}\over {8\pi G}}
\sum_u \int_{S^2_{\tau ,\infty}} d^2\Sigma_u \Big(
\phi^{-2}e^{-{2\over {\sqrt{3}}}\sum_{\bar a}\gamma_{\bar
au}r_{\bar a}}[-4\partial_uln\, \phi +{1\over {\sqrt{3}}}
\sum_{\bar b}\gamma_{\bar bu}
\partial_ur_{\bar b}] \Big) (\tau ,\vec \sigma ),\nonumber \\
 &&{}\nonumber \\
 P^r_{ADM,R}&=&{\hat P}^r_{ADM,R}+\int d^3\sigma {\hat {\cal H}}^r(\tau
,\vec
\sigma )\equiv {\hat P}^r_{ADM,R}=\nonumber \\
&=&-\int_{S^2_{\tau ,\infty}} d^2\Sigma_r \Big[ \phi^{-4}e^{-{2\over
{\sqrt{3}}}\sum_{\bar a}\gamma
_{\bar ar}r_{\bar a}}[{{\rho}\over 3}+\sqrt{3}\sum_{\bar c}\gamma_{\bar cr}
\pi_{\bar c}]\Big] (\tau ,\vec \sigma )-\nonumber \\
&-&{1\over 2}\sum_{u} \int_{S^2_{\tau ,\infty}} d^2\Sigma_u \phi (\tau
,\vec \sigma ) \sum_v \int d^3\sigma_1 \Big( e^{-{1\over
{\sqrt{3}}}\sum_{\bar a}
\gamma_{\bar ar}r_{\bar a}(\tau ,\vec \sigma )} \delta^r_{(a)}{\tilde {\cal T}}^u_{(a)v}+
\nonumber \\
&+&e^{-{1\over {\sqrt{3}}}\sum_{\bar a}\gamma_{\bar au}r_{\bar a}(\tau ,\vec
\sigma )} \delta^u_{(a)}{\tilde {\cal T}}^r_{(a)v}\Big) (\vec \sigma ,{\vec \sigma}_1,\tau
|\phi ,r_{\bar a}]\nonumber \\
 &&(\phi^{-2}e^{-{1\over {\sqrt{3}}}\sum_{\bar
a}\gamma_{\bar av}r_{\bar a}}[{{\rho}
\over 3}+\sqrt{3}\sum_{\bar c}\gamma_{\bar cv}\pi_{\bar c}])(\tau ,{\vec
\sigma}_1)  ,\nonumber \\
 &&{}\nonumber \\
 J^{rs}_{ADM,R}&=&{\hat J}^{rs}_{ADM,R}+{1\over 4}\int d^3\sigma
[\sigma^s {\hat {\cal H}}^r-\sigma^r {\hat {\cal H}}^s](\tau ,\vec
\sigma )=\nonumber \\
 &=&-{1\over 2} \sum_{u} \int_{S^2_{\tau
,\infty}} d^2\Sigma_u \phi^{-4}(\tau ,\vec \sigma ) \nonumber \\
&&\Big( \delta^r_u \sigma^s [e^{-{1\over {\sqrt{3}}}\sum_{\bar
a}\gamma_{\bar ar}r_{\bar a}}({{\rho}\over 3}+\sqrt{3}\sum_{\bar
b}\gamma_{\bar br}\pi_{\bar b} )]-\nonumber \\
 &-&\delta^s_u \sigma^r
[e^{-{1\over {\sqrt{3}}}\sum_{\bar a}\gamma_{\bar as}r_{\bar
a}}({{\rho}\over 3}+\sqrt{3}\sum_{\bar b}\gamma_{\bar bs}\pi_{\bar b}
)] \Big) (\tau ,\vec \sigma ) -\nonumber \\
 &&{}\nonumber \\
 &-&{1\over 4}\sum_u
\int_{S^2_{\tau ,\infty}} d^2\Sigma_u \phi^{-2}(\tau ,\vec
\sigma ) \sum_v \int d^3\sigma_1 \nonumber \\
&&\Big[ \sigma^s \Big( e^{-{1\over {\sqrt{3}}}\sum_{\bar
a}\gamma_{\bar ar} r_{\bar a}(\tau ,\vec \sigma )}\delta_{r(a)}{\tilde
{\cal T}}^u_{(a)v}+ e^{-{1\over {\sqrt{3}}}\sum_{\bar a}\gamma_{\bar
au} r_{\bar a}(\tau ,\vec \sigma )}\delta_{u(a)}{\tilde {\cal
T}}^r_{(a)v}\Big)-\nonumber \\
 &-& \sigma^r \Big( e^{-{1\over
{\sqrt{3}}}\sum_{\bar a}\gamma_{\bar as} r_{\bar a}(\tau ,\vec \sigma
)}\delta_{s(a)}{\tilde {\cal T}}^u_{(a)v}+ e^{-{1\over
{\sqrt{3}}}\sum_{\bar a}\gamma_{\bar au} r_{\bar a}(\tau ,\vec \sigma
)}\delta_{u(a)}{\tilde {\cal T}}^s_{(a)v}\Big) \Big]
\nonumber \\
&&(\vec \sigma ,{\vec \sigma}_1,\tau |\phi ,r_{\bar a}]
(\phi^{-2}e^{-{1\over {\sqrt{3}}}\sum_{\bar a}\gamma_{\bar av}r_{\bar
a}}[{{\rho}
\over 3}+\sqrt{3}\sum_{\bar c}\gamma_{\bar cv}\pi_{\bar c}](\tau ,{\vec
\sigma}_1) \} ,\nonumber \\
 &&{}\nonumber \\
 J^{\tau r}_{ADM,R}&=&{\hat J}^{\tau r}_{ADM,R}+{1\over 2} \int d^3\sigma
\sigma^r {\hat {\cal H}}_R(\tau ,\vec \sigma )=\nonumber \\
&=&{{\epsilon c^3}\over {8\pi G}} \sum_u \int_{S^2_{\tau ,\infty}}
d^2\Sigma_u \sigma^r \Big( \phi^2e^{-{2\over {\sqrt{3}}}\sum_{\bar
a}\gamma_{\bar au}r_{\bar a}}[-4\partial_uln\, \phi +{1\over
{\sqrt{3}}} \sum_{\bar b}\gamma_{\bar bu}
\partial_ur_{\bar b}] \Big) (\tau ,\vec \sigma )+\nonumber \\
&+&{{\epsilon C^3}\over {16\pi G}}  \int_{S^2_{\tau ,\infty}}
d^2\Sigma_r \Big( \phi^{-2}\sum_se^{-{2\over {\sqrt{3}}}\sum_{\bar
a}(\gamma_{\bar ar} +\gamma_{\bar as})r_{\bar a}}(\phi^4e^{{2\over
{\sqrt{3}}} \sum_{\bar a}\gamma_{\bar as}r_{\bar
a}}-1)(1-\delta^r_s) \Big) (\tau,\vec \sigma ).\nonumber \\ &&{}
\label{b2}
\end{eqnarray}

From Eqs.(\ref{b1}) evaluated with the gauge fixing $\rho (\tau
,\vec \sigma )\approx 0$, we get the {\it weak} ADM Poincar\'e
charges in this special 3-orthogonal gauge

\begin{eqnarray}
{\hat P}^{\tau}_{ADM,R}&=&\epsilon \int d^3\sigma \Big( {{c^3}\over {16\pi G}}
\Big[ \phi^2 \sum_r e^{-{2\over {\sqrt{3}}}\sum_{\bar a}\gamma_{\bar ar}r_{\bar
a}} \times \nonumber \\
 &&\Big( 8(\partial_rln\, \phi )^2 -{1\over 3}\sum_{\bar b}(\partial_rr_{\bar b})^2-\nonumber \\
 &-&{4\over {\sqrt{3}}} \partial_rln\, \phi \sum_{\bar b}\gamma_{\bar br}\partial_rr_{\bar b}+
 {2\over 3}(\sum_{\bar b}\gamma_{\bar br}\partial_rr_{\bar b})^2
\Big) \Big] (\tau ,\vec \sigma )-\nonumber \\
 &&{}\nonumber \\
 &-&{{6\pi G}\over {c^3}} \phi^{-2}(\tau ,\vec \sigma )\Big[
 2(\phi^{-4}\sum_{\bar a} \pi^2_{\bar a})(\tau ,\vec \sigma )+\nonumber \\
 &+&4(\phi^{-2}\sum_ue^{{1\over {\sqrt{3}}}\sum_{\bar a}\gamma_{\bar
au}r_{\bar a}} \sum_{\bar b}\gamma_{\bar bu}\pi_{\bar b})(\tau ,\vec
\sigma )\times \nonumber \\
&&\int d^3\sigma_1 \sum_r \delta^u_{(a)} {\tilde {\cal
T}}^u_{(a)r}(\vec
\sigma ,{\vec
\sigma}_1,\tau |\phi ,r_{\bar a}] \Big( \phi^{-2}
e^{-{1\over {\sqrt{3}}}\sum_{\bar a}
\gamma_{\bar ar}r_{\bar a}}\sum_{\bar b}\gamma_{\bar
br} \pi_{\bar b}\Big) (\tau ,{\vec \sigma}_1)+\nonumber \\
 &&{}\nonumber \\
 &+&\int d^3\sigma_1d^3\sigma_2 \Big( \sum_u e^{{2\over {\sqrt{3}}}\sum_{\bar a}
\gamma_{\bar au}r_{\bar a}(\tau ,\vec \sigma )} \times \nonumber \\
&&\sum_r{\tilde {\cal T}}^u_{(a)r}(\vec \sigma ,{\vec
\sigma}_1,\tau |\phi ,r_{\bar a}] \Big( \phi^{-2}
e^{-{1\over {\sqrt{3}}}\sum_{\bar a}
\gamma_{\bar ar}r_{\bar a}}\sum_{\bar b}\gamma_{\bar
br} \pi_{\bar b}\Big) (\tau ,{\vec \sigma}_1)\times \nonumber \\
&&\sum_s {\tilde {\cal T}}^u_{(a)s}(\vec \sigma ,{\vec
\sigma}_2,\tau |\phi ,r_{\bar a}] \Big( \phi^{-2}
e^{-{1\over {\sqrt{3}}}\sum_{\bar a}
\gamma_{\bar as}r_{\bar a}}\sum_{\bar c}\gamma_{\bar
cs} \pi_{\bar c}\Big) (\tau ,{\vec \sigma}_2)+\nonumber \\
&+&\sum_{uv} e^{{1\over {\sqrt{3}}}\sum_{\bar a}(\gamma_{\bar au}+\gamma_{\bar
av})r_{\bar a}(\tau ,\vec \sigma )} (\delta^u_{(b)}\delta^v_{(a)}-\delta^u_{(a)}
\delta^v_{(b)})\times \nonumber \\
&&\sum_r {\tilde {\cal T}}^u_{(a)r}(\vec \sigma ,{\vec
\sigma}_1,\tau |\phi ,r_{\bar a}] \Big( \phi^{-2}
e^{-{1\over {\sqrt{3}}}\sum_{\bar a}
\gamma_{\bar ar}r_{\bar a}}\sum_{\bar b}\gamma_{\bar
br} \pi_{\bar b}\Big) (\tau ,{\vec \sigma}_1)\nonumber \\ &&\sum_s
{\tilde {\cal T}}^v_{(b)s}(\vec \sigma ,{\vec
\sigma}_2,\tau |\phi ,r_{\bar a}] \Big( \phi^{-2}
e^{-{1\over {\sqrt{3}}}\sum_{\bar a}
\gamma_{\bar as}r_{\bar a}}\sum_{\bar c}\gamma_{\bar
cs} \pi_{\bar c}\Big) (\tau ,{\vec \sigma}_2)\, \Big)\, \Big]\, \Big)
,\nonumber \\
 &&{}\nonumber \\
 {\hat P}^r_{ADM,R}&=&
-\int d^3\sigma \phi^{-2}(\tau ,\vec \sigma )
\Big( \phi^{-2}(\tau ,\vec \sigma ) \nonumber \\
 &&\Big[ e^{-{2\over {\sqrt{3}}}\sum_{\bar a}\gamma_{\bar ar}r_{\bar a}}
\sum_{\bar b}\partial_rr_{\bar b}\pi_{\bar b}+\nonumber \\
 &+&2\sqrt{3}  e^{-{2\over {\sqrt{3}}}\sum_{\bar a}\gamma_{\bar ar}r_{\bar a}}
   (2\partial_r ln\, \phi +{1\over {\sqrt{3}}}\sum_{\bar b}\gamma_{\bar
br}\partial_r r_{\bar b})\nonumber \\
 &&\sum_{\bar c}\gamma_{\bar cr}\pi_{\bar c}
\Big] (\tau ,\vec \sigma )+\nonumber \\
 &&{}\nonumber \\
 &+& \sqrt{3} \int d^3\sigma_1 \sum_s
 \Big[ -\sum_u\Big( e^{{1\over {\sqrt{3}}}\sum_{\bar a}(\gamma
 _{\bar au}-2\gamma_{\bar ar})r_{\bar a}}(2\partial_rln\, \phi +{1\over {\sqrt{3}}}
 \sum_{\bar b}\gamma_{\bar bu}\partial_rr_{\bar b})\Big)(\tau ,\vec \sigma )\nonumber \\
 &&\delta^u_{(a)}{\tilde {\cal T}}^u_{(a)s}(\vec \sigma ,{\vec \sigma}_1,\tau |\phi ,r_{\bar a}]
 \nonumber \\
 &+&\sum_{uv}\Big(  e^{-{1\over {\sqrt{3}}}\sum_{\bar a}\gamma_{\bar au}r_{\bar
a}} \Big( \delta^r_u(2\partial_vln\, \phi +{1\over
{\sqrt{3}}}\sum_{\bar b}\gamma_{\bar bu}\partial_vr_{\bar
b})+\nonumber \\
 &+&\delta^r_v(2\partial_uln\, \phi +{1\over {\sqrt{3}}}
 \sum_{\bar b}\gamma_{\bar bv}\partial_ur_{\bar b})\Big)\Big) (\tau ,\vec \sigma )
 \nonumber \\
 &&\delta^u_{(a)}{\tilde {\cal T}}^v_{(a)s}
(\vec \sigma ,{\vec \sigma}_1,\tau |\phi ,r_{\bar a}]
\Big]\nonumber \\
 &&\Big( \phi^{-2}e^{-{1\over {\sqrt{3}}}\sum_{\bar a}\gamma_{\bar as}r_{\bar a}}
\sum_{\bar c}\gamma_{\bar cs}\pi_{\bar c}\Big) (\tau ,{\vec
\sigma}_1)\Big),\nonumber \\
 &&{}\nonumber \\
 {\hat J}^{rs}_{ADM,R}&=&\int d^3\sigma \phi^{-2}(\tau ,\vec \sigma )
\Big( \phi^{-2}(\tau ,\vec \sigma ) \nonumber \\
 &&\Big[ e^{-{2\over {\sqrt{3}}}\sum_{\bar a}\gamma_{\bar ar}r_{\bar a}}
\sum_{\bar b}(\sigma^r \partial_s - \sigma^s \partial_r) r_{\bar
 b}\pi_{\bar b})+\nonumber \\
 &+&2\sqrt{3}\sum_u e^{-{2\over {\sqrt{3}}}\sum_{\bar a}\gamma_{\bar au}r_{\bar a}}
  (\sigma^r \delta^s_u-\sigma^s\delta^r_u)
  (2\partial_uln\, \phi +{1\over {\sqrt{3}}}\sum_{\bar b}\gamma_{\bar
bu}\partial_ur_{\bar b})
 \sum_{\bar c}\gamma_{\bar cu}\pi_{\bar c}
\Big] (\tau ,\vec \sigma )+\nonumber \\
 &&{}\nonumber \\
 &+& \sqrt{3}\int d^3\sigma_1 \sum_w \Big[ \Big(-\sigma^r
 \sum_u\Big( e^{{1\over {\sqrt{3}}}\sum_{\bar a}(\gamma
 _{\bar au}-2\gamma_{\bar as})r_{\bar a}}(2\partial_sln\, \phi +{1\over {\sqrt{3}}}
 \sum_{\bar b}\gamma_{\bar bu}\partial_sr_{\bar b})\Big)(\tau ,\vec \sigma )-\nonumber \\
 &-&\sigma^s
 \sum_u\Big( e^{{1\over {\sqrt{3}}}\sum_{\bar a}(\gamma
 _{\bar au}-2\gamma_{\bar ar})r_{\bar a}}(2\partial_rln\, \phi +{1\over {\sqrt{3}}}
 \sum_{\bar b}\gamma_{\bar bu}\partial_rr_{\bar b})\Big)(\tau ,\vec \sigma )\Big)\nonumber \\
 &&\delta^u_{(a)}{\tilde {\cal T}}^u_{(a)w}(\vec \sigma ,{\vec \sigma}_1,\tau |\phi ,r_{\bar a}]
 -\nonumber \\
 &-&\sum_{uv}\Big(  e^{-{1\over {\sqrt{3}}}\sum_{\bar a}\gamma_{\bar au}r_{\bar
a}} \Big( (\sigma^r\delta^s_u-\sigma^s\delta^r_u)(2\partial_vln\, \phi
+{1\over {\sqrt{3}}}\sum_{\bar b}\gamma_{\bar bu}\partial_vr_{\bar
b})+\nonumber \\
 &+&(\sigma^r\delta^s_v-\sigma^s\delta^r_v)(2\partial_uln\, \phi +{1\over {\sqrt{3}}}
 \sum_{\bar b}\gamma_{\bar bv}\partial_ur_{\bar b})\Big)\Big) (\tau ,\vec \sigma )
 \nonumber \\
 &&\delta^u_{(a)}{\tilde {\cal T}}^v_{(a)w}
(\vec \sigma ,{\vec \sigma}_1,\tau |\phi ,r_{\bar a}]
\Big]\nonumber \\
 &&\Big( \phi^{-2}e^{-{1\over {\sqrt{3}}}\sum_{\bar a}\gamma_{\bar as}r_{\bar a}}
\sum_{\bar c}\gamma_{\bar cs}\pi_{\bar c}\Big) (\tau ,{\vec
\sigma}_1)\Big),\nonumber \\
 &&{}\nonumber \\
 {\hat J}^{\tau r}_{ADM,R}&=&\epsilon \int d^3\sigma   \sigma^r
\Big(  {{c^3}\over {16\pi G}}
\Big[ \phi^2 \sum_r e^{-{2\over {\sqrt{3}}}\sum_{\bar a}\gamma_{\bar ar}r_{\bar
a}} \times \nonumber \\
 &&\Big( 8(\partial_rln\, \phi )^2 -{1\over 3}\sum_{\bar b}(\partial_rr_{\bar b})^2-\nonumber \\
 &-&{4\over {\sqrt{3}}} \partial_r\phi \sum_{\bar b}\gamma_{\bar br}\partial_rr_{\bar b}+
 {2\over 3}(\sum_{\bar b}\gamma_{\bar br}\partial_rr_{\bar b})^2
\Big) \Big] (\tau ,\vec \sigma )-\nonumber \\
 &&{}\nonumber \\
 &-&{{6\pi G}\over {c^3}} \phi^{-2}(\tau ,\vec \sigma )\Big[
2(\phi^{-4}\sum_{\bar a} \pi^2_{\bar a})(\tau ,\vec \sigma )+\nonumber
\\
 &+&4(\phi^{-2}\sum_ue^{{1\over {\sqrt{3}}}\sum_{\bar a}\gamma_{\bar
au}r_{\bar a}} \sum_{\bar b}\gamma_{\bar bu}\pi_{\bar b})(\tau ,\vec
\sigma )\times \nonumber \\
&&\int d^3\sigma_1 \sum_r \delta^u_{(a)} {\tilde {\cal
T}}^u_{(a)r}(\vec
\sigma ,{\vec
\sigma}_1,\tau |\phi ,r_{\bar a}] \Big( \phi^{-2}
e^{-{1\over {\sqrt{3}}}\sum_{\bar a}
\gamma_{\bar ar}r_{\bar a}}\sum_{\bar b}\gamma_{\bar
br} \pi_{\bar b}\Big) (\tau ,{\vec \sigma}_1)+\nonumber \\
 &&{}\nonumber \\
 &+&\int d^3\sigma_1d^3\sigma_2 \Big( \sum_u e^{{2\over {\sqrt{3}}}\sum_{\bar a}
\gamma_{\bar au}r_{\bar a}(\tau ,\vec \sigma )} \times \nonumber \\
&&\sum_r{\tilde {\cal T}}^u_{(a)r}(\vec \sigma ,{\vec
\sigma}_1,\tau |\phi ,r_{\bar a}] \Big( \phi^{-2}
e^{-{1\over {\sqrt{3}}}\sum_{\bar a}
\gamma_{\bar ar}r_{\bar a}}\sum_{\bar b}\gamma_{\bar
br} \pi_{\bar b}\Big) (\tau ,{\vec \sigma}_1)\times \nonumber \\
&&\sum_s {\tilde {\cal T}}^u_{(a)s}(\vec \sigma ,{\vec
\sigma}_2,\tau |\phi ,r_{\bar a}] \Big( \phi^{-2}
e^{-{1\over {\sqrt{3}}}\sum_{\bar a}
\gamma_{\bar as}r_{\bar a}}\sum_{\bar c}\gamma_{\bar
cs} \pi_{\bar c}\Big) (\tau ,{\vec \sigma}_2)+\nonumber \\
&+&\sum_{uv} e^{{1\over {\sqrt{3}}}\sum_{\bar a}(\gamma_{\bar au}+\gamma_{\bar
av})r_{\bar a}(\tau ,\vec \sigma )} (\delta^u_{(b)}\delta^v_{(a)}-\delta^u_{(a)}
\delta^v_{(b)})\times \nonumber \\
&&\sum_r {\tilde {\cal T}}^u_{(a)r}(\vec \sigma ,{\vec
\sigma}_1,\tau |\phi ,r_{\bar a}] \Big( \phi^{-2}
e^{-{1\over {\sqrt{3}}}\sum_{\bar a}
\gamma_{\bar ar}r_{\bar a}}\sum_{\bar b}\gamma_{\bar
br} \pi_{\bar b}\Big) (\tau ,{\vec \sigma}_1)\nonumber \\ &&\sum_s
{\tilde {\cal T}}^v_{(b)s}(\vec \sigma ,{\vec
\sigma}_2,\tau |\phi ,r_{\bar a}] \Big( \phi^{-2}
e^{-{1\over {\sqrt{3}}}\sum_{\bar a}
\gamma_{\bar as}r_{\bar a}}\sum_{\bar c}\gamma_{\bar
cs} \pi_{\bar c}\Big) (\tau ,{\vec \sigma}_2)\, \Big)\, \Big]\, \Big)
-\nonumber \\
 &&{}\nonumber \\
 &-&{{\epsilon c^3}\over {8\pi G}} \int d^3\sigma \Big[ \phi^{-2}
\sum_{uv}\delta^r_u(\delta_{uv}-1) (\phi^4e^{{2\over {\sqrt{3}}}\sum_{\bar a}
\gamma_{\bar au}r_{\bar a}}-1)\nonumber \\
 &&e^{-{2\over {\sqrt{3}}}\sum_{\bar a}(\gamma_{\bar av}-\gamma_{\bar au})r_{\bar a}}
 (\partial_uln\, \phi +{1\over {\sqrt{3}}}\sum_{\bar b}(\gamma_{\bar bv}-
 \gamma_{\bar bu})\partial_ur_{\bar b})\Big] (\tau ,\vec \sigma ).
\label{b3}
\end{eqnarray}

The {\it weak} Poincar'e charges of {\it void spacetimes} in the
3-orthogonal gauges can be obtained from Eqs.(\ref{b1}) by putting
$r_{\bar a}=\pi_{\bar a}=0$  (the explicit form of the kernal
${\tilde {\cal T}}^r_{(a)u}$ is not needed)

\begin{eqnarray}
{\hat P}^{\tau}_{ADM,R}&=&\epsilon \int d^3\sigma \Big( {{2\pi
G}\over {3 c^3}} \phi^{-2}(\tau ,\vec \sigma ) [(\phi^{-4}
\rho^2)(\tau ,\vec \sigma )- \nonumber \\
 &-&{2\over 3}(\phi^{-2} \rho )(\tau ,\vec
\sigma )\int d^3\sigma_1 \sum_{nr}\delta_{(a)}^r {\tilde {\cal
T}}^r_{(a)n}(\vec \sigma ,{\vec \sigma}_1,\tau |\phi ,0]
(\phi^{-2} \rho ) (\tau ,{\vec \sigma}_1)-\nonumber \\
 &&{}\nonumber \\
 &-&{1\over
3}\sum_{rs}(\delta_{rs}\delta_{(a)(b)}+\delta_{r(b)}\delta_{s(a)}-
\delta_{r(a)}\delta_{s(b)})\nonumber \\
&&\int d^3\sigma_1 \sum_m {\tilde {\cal T}}^r_{(a)m}(\vec \sigma
,{\vec
\sigma}_1,\tau |\phi ,0] (\phi^{-2} \rho )(\tau ,{\vec
\sigma}_1)\nonumber \\ &&\int d^3\sigma_2 \sum_n {\cal
T}^s_{(b)n}(\vec \sigma ,{\vec \sigma}_2,\tau |\phi ,0] (\phi^{-2}
\rho )(\tau ,{\vec \sigma}_2) ]+\nonumber \\
 &&{}\nonumber \\
 &+&{{c^3}\over {2\pi G}} [\phi^2\sum_r(\partial_r ln\, \phi )^2](\tau ,\vec \sigma ) \Big) \,
   \nonumber \\
 &&{\rightarrow}_{\rho \, \rightarrow 0}\, {{\epsilon c^3}\over {2\pi G}} \int d^3\sigma
[\phi^2\sum_r(\partial_rln\, \phi )^2](\tau ,\vec \sigma )\nonumber \\
&&{\rightarrow}_{\phi \rightarrow 1}\, 0,\nonumber \\
 &&{}\nonumber \\
 {\hat P}^r_{ADM,R}&=&-{2\over 3} \int d^3\sigma \phi^{-2}(\tau ,\vec \sigma )
\Big( 5\Big( \rho \phi^{-2}\partial_rln\, \phi \Big) (\tau ,\vec \sigma )+
\nonumber \\
 &&{}\nonumber \\
 &+&\sum_s\int d^3\sigma_1 \Big[ -\partial_rln\, \phi (\tau ,\vec
\sigma )\sum_u\delta^u_{(a)}
{\tilde {\cal T}}^u_{(a)s}(\vec \sigma ,{\vec \sigma}_1,\tau |\phi
,0]+\nonumber \\
 &+&(\delta^r_u\partial_vln\, \phi +\delta^r_v\partial_uln\, \phi )(\tau ,\vec \sigma )
 \delta^u_{(a)}{\tilde {\cal T}}^v_{(a)s}(\vec \sigma ,{\vec \sigma}_1,\tau |\phi ,0]
\Big] (\phi^{-2}\rho )(\tau ,{\vec \sigma}_1) \Big) \nonumber \\
 &&{\rightarrow}_{\rho \rightarrow 0, \phi \rightarrow 1}\, 0,\nonumber \\
 &&{}\nonumber \\
 {\hat J}^{rs}_{ADM,R}&=&{{10}\over 3}\int d^3\sigma [\phi^{-4} \rho
](\tau ,\vec \sigma ) \Big[ \sigma^r\partial_sln\, \phi
-\sigma^s\partial_rln\, \phi \Big] (\tau ,\vec \sigma )+\nonumber \\
 &&{}\nonumber \\
 &+&{2\over 3}\sum_w \int d^3\sigma d^3\sigma_1 \phi^{-2}(\tau ,\vec
\sigma ) \nonumber \\
&&\Big[ (\sigma^r\partial_sln\, \phi -\sigma^s\partial_rln\, \phi )
(\tau ,\vec \sigma )\sum_u\delta^u_{(a)}{\tilde {\cal
T}}^u_{(a)w}(\vec
\sigma ,{\vec \sigma}_1,\tau |\phi ,0]+\nonumber \\
 &+&\sum_{uv}\Big( (\sigma^r\delta^s_u-\sigma^s\delta^r_u)\partial_vln\, \phi +
 (\sigma^r\delta^s_v-\sigma^s\delta^r_v)\partial_uln\, \phi
(\tau ,\vec \sigma )\nonumber \\
 &&\delta^u_{(a)}{\tilde {\cal T}}^v_{(a)w} (\vec
\sigma ,{\vec \sigma}_1,\tau |\phi ,0] \Big] (\phi^{-2}\rho )(\tau
,{\vec \sigma}_1)\nonumber \\
 &&{\rightarrow}_{\rho \rightarrow 0, \phi \rightarrow 1}\, 0,\nonumber \\
 &&{}\nonumber \\
 {\hat J}^{\tau r}_{ADM,R}&=&\epsilon \int d^3\sigma
\sigma^r\Big( {{2\pi G}\over {3 c^3}} \phi^{-2}(\tau ,\vec \sigma
) [(\phi^{-4} \rho^2)(\tau ,\vec \sigma )- \nonumber \\
 &&{}\nonumber \\
 &-&{2\over 3}(\phi^{-2} \rho )(\tau ,\vec \sigma )\int d^3\sigma_1
\sum_{nr}\delta_{(a)}^r {\tilde {\cal T}}^r_{(a)n}(\vec \sigma
,{\vec \sigma}_1,\tau |\phi ,0] (\phi^{-2} \rho ) (\tau ,{\vec
\sigma}_1)-\nonumber \\
 &-&{1\over
3}\sum_{rs}(\delta_{rs}\delta_{(a)(b)}+\delta_{r(b)}\delta_{s(a)}-
\delta_{r(a)}\delta_{s(b)})\nonumber \\
&&\int d^3\sigma_1 \sum_m {\tilde {\cal T}}^r_{(a)m}(\vec \sigma
,{\vec
\sigma}_1,\tau |\phi ,0] (\phi^{-2} \rho )(\tau ,{\vec
\sigma}_1)\nonumber \\ &&\int d^3\sigma_2 \sum_n {\cal
T}^s_{(b)n}(\vec \sigma ,{\vec \sigma}_2,\tau |\phi ,0] (\phi^{-2}
\rho )(\tau ,{\vec \sigma}_2) ]+\nonumber \\
 &&{}\nonumber \\
 &+&{{c^3}\over {2\pi G}} [\phi^2\sum_r(\partial_r ln\, \phi )^2](\tau ,\vec \sigma )
 \Big)+\nonumber \\
 &+&{{\epsilon c^3}\over {4\pi G}}  \int d^3\sigma \Big[ \phi^{-2} (\phi^4-1)
\partial_rln\, \phi \Big] (\tau ,\vec \sigma )\nonumber \\
&&{\rightarrow}_{\rho \rightarrow 0}\, {{\epsilon c^3}\over {4\pi
G}} \int d^3\sigma \Big[ \phi^{-2} (\phi^4-1)
\partial_rln\, \phi \Big] (\tau ,\vec \sigma )\nonumber \\
 &&{\rightarrow}_{\phi \rightarrow 1}\, 0.\nonumber \\
\label{b4}
\end{eqnarray}

Therefore the Poincar\'e charges vanish in the special 3-orthogonal
gauge $\rho (\tau ,\vec \sigma )=0$ by using the solution $\phi (\tau
,\vec \sigma )=1$ of the reduced Lichnerowicz equation in this gauge.
Since they are gauge invariant, as shown in Section II before
Eq.(\ref{II26}), they vanish in every gauge.

\vfill\eject

\end{document}